\documentclass[aps,prx,reprint,twocolumn,floatfix,superscriptaddress,amsmath,amssymb,amsfonts,nofootinbib,balancelastpage]{revtex4-2}

\usepackage{url}
\usepackage{bm}
\usepackage{graphicx}
\usepackage{amsmath}
\usepackage{amstext}
\usepackage{amssymb}
\usepackage{amsfonts}
\usepackage{amsbsy}
\usepackage{verbatim}
\usepackage{color}
\usepackage{floatrow}
\usepackage{float}
\usepackage{tikz}
\usepackage{gensymb}
\usepackage{textcomp}
\usepackage{enumitem}
\usepackage[version=3]{mhchem}
\usepackage{afterpage}
\usepackage{pbox}
\usepackage{dsfont}
\usepackage{siunitx}
\usepackage{stackengine,wasysym,scalerel}
\usepackage{latexsym}
\usepackage{cancel}
\usepackage{comment}
\usepackage{array, multirow, bigdelim, makecell, booktabs}
\usepackage[misc]{ifsym}
\usepackage{sidecap}
\usepackage{mathrsfs}

\usepackage{color}
\usepackage{pifont}%
\newcommand{\cmark}{\ding{51}}%
\newcommand{\xmark}{\ding{55}}%

\usepackage[normalem]{ulem}
\newcommand{\pdagger}{\phantom{\dagger}}
\newcommand{\rev}[1]{#1}

\definecolor{darkblue}{HTML}{004D6B}
\definecolor{darkred}{HTML}{8c1515}
\definecolor{darkgreen}{HTML}{006400}
\usepackage[colorlinks=true,
            urlcolor=blue,
            citecolor=blue,
            linkcolor=blue]{hyperref}

\allowdisplaybreaks

\begin{document}
\title{Evidence for a $\mathbb{Z}_{2}$ Dirac spin liquid in the generalized Shastry-Sutherland model}
\author{Atanu Maity}
\affiliation{Institut f\"ur Theoretische Physik und Astrophysik and W\"urzburg-Dresden Cluster of Excellence ctd.qmat, Julius-Maximilians-Universit\"at W\"urzburg, Am Hubland, Campus S\"ud, W\"urzburg 97074, Germany}
\affiliation{Department of Physics and Quantum Centre of Excellence for Diamond and Emergent Materials (QuCenDiEM), Indian Institute of Technology Madras, Chennai 600036, India}
\author{Francesco Ferrari}
\affiliation{Department of Physics and Quantum Centre of Excellence for Diamond and Emergent Materials (QuCenDiEM), Indian Institute of Technology Madras, Chennai 600036, India}
\affiliation{Institut für Theoretische Physik, Goethe Universität Frankfurt, Max-von-Laue-Straße 1, 60438 Frankfurt am Main, Germany}
\author{Sylvain Capponi}
\affiliation{Univ Toulouse, CNRS, Laboratoire de Physique Th\'eorique, Toulouse,  France.}
\author{Karlo Penc}
\affiliation{Institute for Solid State Physics and Optics, Wigner Research Centre for Physics, H-1525 Budapest, P.O. Box 49, Hungary}
\affiliation{Department of Physics and Quantum Centre of Excellence for Diamond and Emergent Materials (QuCenDiEM), Indian Institute of Technology Madras, Chennai 600036, India}
\author{Janik Potten}
\affiliation{Institut f\"ur Theoretische Physik und Astrophysik and W\"urzburg-Dresden Cluster of Excellence ctd.qmat, Julius-Maximilians-Universit\"at W\"urzburg, Am Hubland, Campus S\"ud, W\"urzburg 97074, Germany}
\author{Tobias M\"uller}
\affiliation{Institut f\"ur Theoretische Physik und Astrophysik and W\"urzburg-Dresden Cluster of Excellence ctd.qmat, Julius-Maximilians-Universit\"at W\"urzburg, Am Hubland, Campus S\"ud, W\"urzburg 97074, Germany}
\author{Andreas Feuerpfeil}
\affiliation{Institut f\"ur Theoretische Physik und Astrophysik and W\"urzburg-Dresden Cluster of Excellence ctd.qmat, Julius-Maximilians-Universit\"at W\"urzburg, Am Hubland, Campus S\"ud, W\"urzburg 97074, Germany}
\author{Ronny Thomale}
\affiliation{Institut f\"ur Theoretische Physik und Astrophysik and W\"urzburg-Dresden Cluster of Excellence ctd.qmat, Julius-Maximilians-Universit\"at W\"urzburg, Am Hubland, Campus S\"ud, W\"urzburg 97074, Germany}
\affiliation{Department of Physics and Quantum Centre of Excellence for Diamond and Emergent Materials (QuCenDiEM), Indian Institute of Technology Madras, Chennai 600036, India}
\author{Jong Yeon Lee}
\email{jongyeon@illinois.edu}
\affiliation{Department of Physics and Anthony J. Leggett Institute for Condensed Matter Theory, University of Illinois at Urbana-Champaign, Urbana, IL 61801, USA}
\author{Rhine Samajdar}
\email{rhine\_samajdar@princeton.edu}
\affiliation{Department of Physics, Princeton University, Princeton, NJ 08544, USA}
\affiliation{Princeton Center for Theoretical Science, Princeton University, Princeton, NJ 08544, USA}
\author{Yasir Iqbal}
\email{yiqbal@physics.iitm.ac.in}
\affiliation{Department of Physics and Quantum Centre of Excellence for Diamond and Emergent Materials (QuCenDiEM), Indian Institute of Technology Madras, Chennai 600036, India}

\begin{abstract}
We present a multimethod investigation into the nature of the recently reported quantum spin liquid (QSL) phase in the spin-$1/2$ Heisenberg antiferromagnet on the Shastry-Sutherland lattice.
A comprehensive projective symmetry group classification of fermionic mean-field \textit{Ans\"atze} on this lattice yields 46 U(1) and 80 $\mathbb{Z}_2$ states. Using density-matrix renormalization group (DMRG) and exact diagonalization calculations, we find  that the Shastry-Sutherland model and the square-lattice $J_1$--$J_2$ Heisenberg antiferromagnet share the same QSL phase.
Motivated by this observation, we establish an explicit mapping of our \textit{Ans\"atze} to those on the square lattice, and identify the counterpart of the square-lattice $\mathbb{Z}_2$ Dirac QSL (Z2A$zz$13) in the Shastry-Sutherland system. Employing state-of-the-art variational Monte Carlo calculations with Gutzwiller-projected wavefunctions, further improved by L\'anczos steps, we demonstrate excellent agreement in both energies and correlation functions between a gapless (Dirac) $\mathbb{Z}_2$ spin liquid---characterized by only a few variational parameters---and results obtained from neural quantum states and DMRG.
Working within a unified variational family that continuously connects the spin liquid to the adjacent N\'eel and plaquette valence-bond-crystal states, we locate the two phase boundaries in an unbiased way from the finite-size scaling of the respective order parameters and from correlation-ratio crossings, and find that both order parameters vanish continuously.
Finally, we apply the recently developed Keldysh formulation of the pseudo-fermion functional renormalization group to compute the dynamical spin structure factor. The resulting spectra exhibit features consistent with Dirac cones in the excitation spectrum, providing strong independent evidence for a Dirac QSL ground state. Our identification of a $d$-wave pairing $\mathbb{Z}_2$ Dirac QSL is consistent with recently observed signatures of QSL behavior in Pr$_2$Ga$_2$BeO$_7$ and \rev{makes} predictions for future experiments.

\end{abstract}

\maketitle

\section{Introduction}
The celebrated spin-$1/2$ Heisenberg antiferromagnet on the Shastry-Sutherland lattice presents a rare example of a two-dimensional spin model \rev{that} admits an \textit{exact} solution for its ground state~\cite{Shastry-1981}. This Archimedean lattice, which is also called the snub square or $(3^2, 4,3,4)$ lattice, features two symmetry-inequivalent bonds, and the Shastry-Sutherland model (SSM) is defined on this lattice as
\begin{equation}
\hat{H}=J^{}_{s}\sum_{\langle i,j\rangle_{\rm square}}\mathbf{\hat S}_{i} \cdot \mathbf{\hat S}_{j}+J^{}_{d}\sum_{\langle i,j\rangle_{\rm dimer}}\mathbf{\hat S}_{i} \cdot \mathbf{\hat S}_{j}\,.
\label{eq:mod-ham}
\end{equation}
Here, $\mathbf{\rev{\hat S}}_{i}$ denotes the SU(2) spin operator acting on the spin $S\,{=}\,1/2$ representation at site $i$. The nearest-neighbor antiferromagnetic exchange is given by $J_{s}$, with the first sum in Eq.~\eqref{eq:mod-ham} running over all the bonds of the square lattice. The next-nearest-neighbor interaction is represented by $J_{d}$ and the second sum runs over the additional diagonal bonds introduced on every other plaquette of the square lattice, creating a pattern of alternating dimers [see Fig.~\ref{fig:schematic}(b)]. In the limit $J_{s}/J_{d}=0$, the Hamiltonian reduces to one consisting of decoupled dimers and the ground state is given by a tensor product of singlets with an energy of $-3/8\, J_{d}$ per site. This state is always an exact eigenstate of Eq.~\eqref{eq:mod-ham} and remains the ground state \rev{up to} a finite critical $J_{s}/J_{d}$. The opposite limit, $J_{s}/J_{d}\to\infty$, yields the Heisenberg model on the square lattice, which has a two-sublattice N\'eel-ordered ground state.

Over the past few decades, the SSM has been the focus of extensive theoretical efforts to map out the phase diagram of Eq.~\eqref{eq:mod-ham}, with interest intensifying in particular after the discovery of SrCu$_2$(BO$_3$)$_2$~\cite{Kageyama-1999,Miyahara-2003}, which is well approximated by this model\footnote{The role of interlayer couplings in accurately describing its behavior has lately been appreciated~\cite{Vlaar-2023}.}.
Other intriguing compounds in the family of Shastry-Sutherland lattice materials have also attracted much attention lately \rev{ owing to their promise of hosting} exotic nonmagnetic ground states, possibly QSLs, as well as exciting possibilities \rev{of realizing} altermagnetism~\cite{Ferrari-2024}. These include rare-earth magnets such as BaCe$_2$ZnS$_5$~\cite{Ma-2024}, Pr$_2$Ga$_2$BeO$_7$~\cite{Li-2024}, Pr$_2$Be$_2$GeO$_7$~\cite{Liu-2024_exp}, Yb$_2$Be$_2$GeO$_7$~\cite{Pula-2024}, Er$_2$Be$_2$SiO$_7$~\cite{Brassington-2024}, and Er$_2$Be$_2$GeO$_7$~\cite{Yadav-2024,Pula-2024_er}.

With regard to the ground-state phase diagram of Eq.~\eqref{eq:mod-ham}, while most early works advocated for a direct transition between the dimer and N\'eel phases~\cite{Miyahara-1999,Weihong-1999,Hartmann-2000}, one of the first studies based on Schwinger-boson mean-field theory~\cite{Albrecht-1996}, and subsequent variational calculations~\cite{Munehisa-2003}, found an intermediate state with helimagnetic order, which is known to be the classical $(S\to\infty)$ ground state~\cite{Shastry-1981}. Later investigations based on series expansion, stochastic state selection~\cite{Munehisa-2004}, coupled-cluster~\cite{Darradi-2005}, and perturbative~\cite{Knetter-2000,Hajj-2005} methods introduced the possibility of a gapped nonmagnetic intermediate phase, with either plaquette~\cite{Koga-2000,Takushima-2001} or columnar~\cite{Zheng-2001} valence\rev{-}bond\rev{-}crystal (VBC) order. The presence of plaquette order was subsequently confirmed by a large-$N$ expansion based on a generalization to $Sp(2N)$ symmetry~\cite{Chung-2001}, exact diagonalization and a combination of dimer- and quadrumer-boson methods~\cite{Lauchli-2002}, triplon mean-field analysis~\cite{Kumar_2017}, two-step density-matrix renormalization group (DMRG)~\cite{Moukouri-2008}, and $D$\,$=$\,$2$ projected entangled pair states~\cite{Isacsson-2006}. More recently, numerical quantum many-body approaches using tensor networks~\cite{Corboz-2013}, DMRG~\cite{Lee-2019}, and the pseudo-fermion functional renormalization group (pf-FRG)~\cite{Keles-2022} have established the presence of a plaquette-ordered ground state.

Today, the existence of an intermediate plaquette\rev{-}VBC phase between the dimer and N\'eel orders has found broad consensus. This VBC breaks reflection symmetry across the lines containing the $J_d$ bonds, implying a twofold degenerate ground state, and features resonating singlets on ``empty'' plaquettes (with no $J_{d}$ bonds). In SrCu$_2$(BO$_3$)$_2$, a plaquette VBC, likely ascribed to broken fourfold rotation symmetry about the centers of the empty squares, has been realized under the application of hydrostatic pressure~\cite{Waki-2007,Zayed-2017}. Thermodynamic measurements under high pressure have also provided evidence for the exciting possibility of a deconfined quantum critical point (DQCP) between N\'eel order and the plaquette VBC~\cite{Guo-2020}.
Theoretical evidence for a continuous transition between the plaquette VBC and N\'eel-ordered states, and a resultant DQCP exhibiting emergent O(4) symmetry, has been provided by a combination of infinite density-matrix renormalization group (iDMRG) calculations and field-theoretic analysis~\cite{Lee-2019}, as well as recent tensor-network studies~\cite{Liu-2023}. Other works have argued for a weak first-order transition between these phases~\cite{Qian-2024,Chen-2024,Yang-2024b,lv2026} and suggested that the DQCP might instead be proximate in parameter space~\cite{Qian-2024, Xi-2021}. A recent field-theoretic study~\cite{Feuerpfeil-2026} by some of us formulated a fermionic gauge theory for the transition, characterizing it as a pseudocritical point with emergent $\mathrm{SO}(5)$ symmetry that naturally accommodates a gapless $\mathbb{Z}_2$ Dirac spin liquid between the adjacent VBC and N\'eel phases.

In contrast to both \rev{of} these scenarios, recent numerical investigations using DMRG~\cite{Yang-2022}, exact diagonalization~\cite{Wang-2022,Nakano-2018,Ronquillo-2014}, pf-FRG~\cite{Keles-2022}, and state-of-the-art neural\rev{-}network quantum states~\cite{Luciano-2024,Mezera-2023} have all reported the presence of a QSL phase sandwiched between the plaquette VBC and N\'eel-ordered states. This QSL phase occupies a narrow parameter range, $0.77 \lesssim J_s/J_d \lesssim 0.82$~\cite{Yang-2022,Luciano-2024}. This raises the question of whether one can (i) explore an extended parameter space to widen the window of QSL behavior, and (ii) identify its precise microscopic character, i.e., \rev{whether} it is gapped or gapless, as well as the nature of its low-energy gauge group.

In this work, using a combination of DMRG and exact diagonalization calculations, we demonstrate that a broader ``river'' of spin liquidity can be accessed by activating antiferromagnetic exchange couplings on all the ``other'' diagonal bonds (within the square plaquettes) that were originally excluded in the Shastry-Sutherland geometry. As the strength of these additional interactions is increased, we observe no signs of a phase transition all the way to the square-lattice $J_1$--$J_2$ limit. This suggests the intriguing possibility that the quantum spin liquids realized in the two models are adiabatically connected and therefore belong to the same quantum phase. Our analysis rigorously establishes this continuity using both numerical diagnostics and the framework of projective symmetry groups (PSGs). Unlike conventional ordered phases, which are characterized by spontaneous symmetry breaking and local order parameters, quantum spin liquids exhibit a more subtle form of organization known as ``quantum order\rev{.}'' In this case, global symmetries are preserved, but they are implemented projectively on emergent fractionalized excitations such as spinons.
This projective symmetry structure, captured systematically by the PSG classification, determines the allowed operators in the low-energy gauge theory and thereby fixes the universality class of the spin liquid.

Quantifying the width of the spin-liquid regime, we find that not only does it extend continuously from the square lattice to the Shastry-Sutherland \rev{lattice, }but it also retains a finite extent in the parameter space of the latter, thus yielding new evidence for a robust QSL phase in the SSM. We find that the shared phase is a gapless $\mathbb{Z}_2$ Dirac quantum spin liquid, characterized by linearly dispersing spinon excitations at symmetry-protected Dirac nodes and an emergent gapped $\mathbb{Z}_2$ gauge field. Physically, such a state exhibits algebraically decaying spin correlations, broad continuum features in the dynamical structure factor rather than sharp magnon modes, and fractionalized spin-$1/2$ quasiparticles as its elementary excitations.

Importantly, the PSG construction does not merely enumerate symmetric parton \textit{Ans\"atze}; as mentioned above, it determines how lattice symmetries act projectively on fractionalized spinons and thereby fixes the symmetry-allowed structure of the emergent gauge theory.
Within this framework, the Dirac spin liquid naturally emerges as a parent state from which N\'eel and VBC phases can be obtained via Higgs instabilities, placing the Shastry-Sutherland and square-lattice models within a unified description of fermionic deconfined criticality.

\section{Connection to the square-lattice $J_{1}$--$J_{2}$ Heisenberg model}

\begin{figure}[t]
\includegraphics[width=1.0\linewidth]{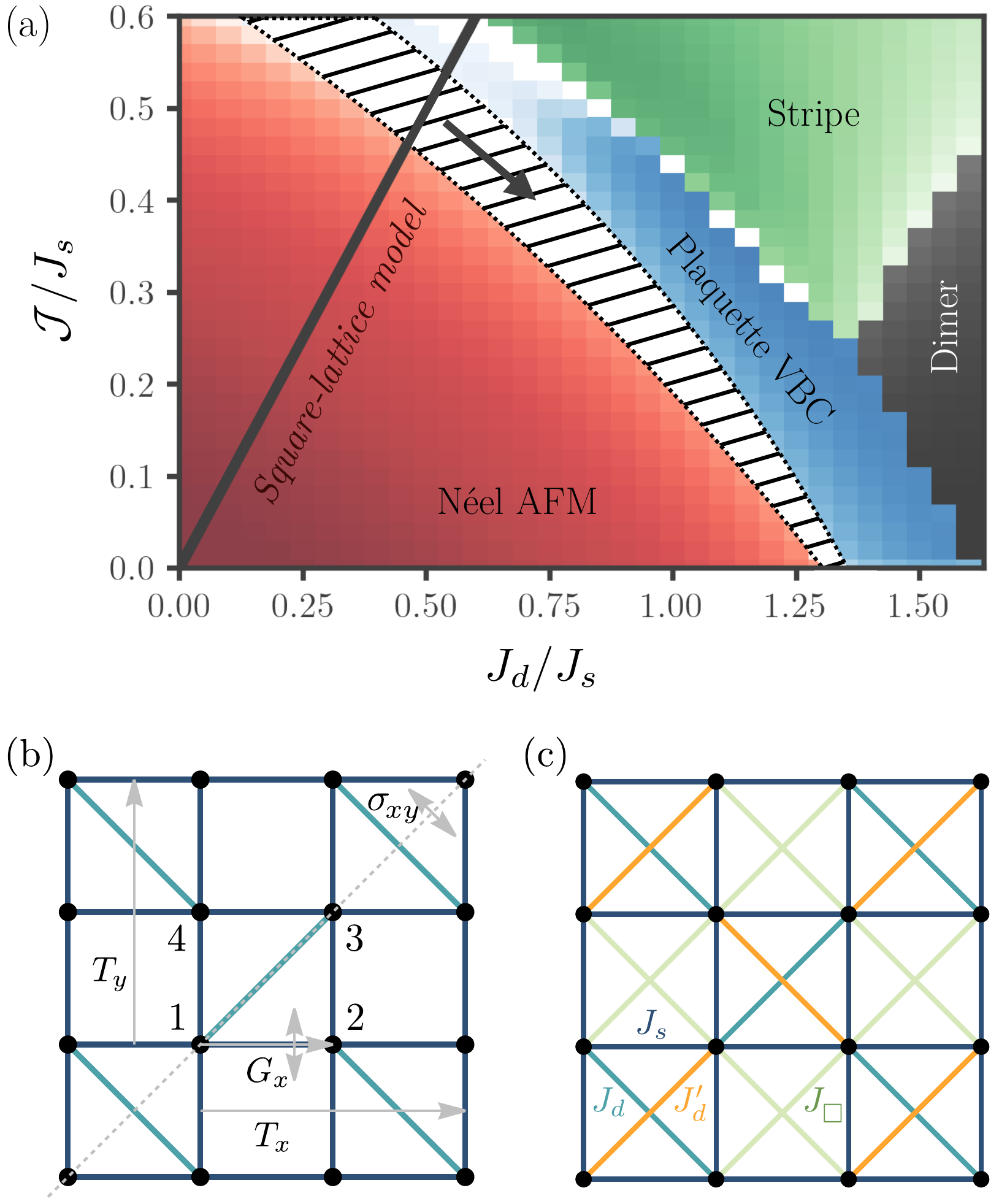}
	\caption{(a) Schematic quantum ground-state  phase diagram of the Hamiltonian~\eqref{eq:mod2} interpolating between the square-lattice $J_1$--$J_2$ Heisenberg and Shastry-Sutherland models. The solid line in the figure indicates the parameters of the square-lattice model, while the horizontal axis corresponds to the pure SSM. The \rev{color shading} conveys the strength of the appropriate order parameter as computed using iDMRG (Sec.~\ref{sec:iDMRG}). The hatched region indicates the extent of the gapless spin\rev{-}liquid phase identified in our exact diagonalization calculations  (Sec.~\ref{sec:ED}).  (b) Structure of the Shastry-Sutherland lattice, highlighting its space-group symmetries. (c) Mapping of the Shastry-Sutherland lattice to the square lattice. The two additional types of symmetry-inequivalent bonds are denoted by orange and light\rev{-}green lines, respectively. In the square-lattice limit, i.e., $J^{}_d=J'_d=J^{}_{\square}$, all the diagonal bonds are related by symmetry.}
	\label{fig:schematic}
\end{figure}

Previous studies using variational Monte Carlo (VMC)~\cite{Hu-2013,Ferrari-2020}, many-variable VMC~\cite{Morita-2015}, neural\rev{-}network wavefunctions~\cite{Nomura-2021}, DMRG~\cite{Wang-2018}, and tensor networks~\cite{Liu-2018_peps,Liu-2022_bull} have offered evidence for the presence of a gapless spin liquid sandwiched between the N\'eel and VBC phases in the $S=1/2$ square-lattice Heisenberg model with nearest-neighbor ($J_1$) and next-nearest-neighbor ($J_2$) couplings. On the other hand, as noted above, numerical simulations on the $J_{s}$--$J_{d}$ SSM point to a variety of possibilities, including weakly first-order or continuous phase transitions or a gapless QSL phase. Given the geometric similarity between these two models, we introduce the following Hamiltonian \rev{that} straightforwardly interpolates between them~\cite{Qian-2024}:
\begin{alignat}{1}
\nonumber
\hat{H}&=J^{}_{s}  \sum_{\langle i,j\rangle}  \mathbf{\hat S}^{}_{i} \cdot \mathbf{\hat S}^{}_{j}+J^{}_{d} \hspace{-6pt} \sum_{ \langle \hspace{-1pt}\langle i,j\rangle\hspace{-1pt}\rangle_\textrm{dimer}} \hspace{-6pt} \mathbf{\hat S}^{}_{i} \cdot \mathbf{\hat S}^{}_{j}\\
&+J'_{d} \hspace{-6pt} \sum_{ \langle \hspace{-1pt}\langle i,j\rangle\hspace{-1pt}\rangle_\textrm{orthogonal}} \hspace{-6pt} \mathbf{\hat S}^{}_{i} \cdot \mathbf{\hat S}^{}_{j} +J^{}_{\square} \hspace{-6pt} \sum_{ \langle \hspace{-1pt}\langle i,j\rangle\hspace{-1pt}\rangle_\textrm{other}} \hspace{-6pt} \mathbf{\hat S}^{}_{i} \cdot \mathbf{\hat S}^{}_{j}\,.
\label{eq:mod2}
\end{alignat}
These couplings are \rev{depicted} in Fig.~\ref{fig:schematic}(c). In addition to the standard nearest-neighbor and dimer interactions of the SSM, denoted by $J_s$ and $J_d$, respectively, we introduce two additional exchange couplings. Specifically, $J'_d$ labels the interactions on dimers that are orthogonal to those present in the original Shastry-Sutherland lattice, while $J_\square$ denotes the interactions on the square plaquettes that are vacant in the SSM.

With these definitions, the limit $J'_d = J^{}_\square = 0$ reduces to the SSM, whereas the isotropic limit $J'_d = J^{}_d = J^{}_\square$ corresponds to the square-lattice $J_1$--$J_2$ model. In the following, to interpolate continuously between the Shastry-Sutherland and square-lattice geometries, we vary a single tuning parameter defined as $\mathcal{J} \equiv J'_d = J^{}_\square$.

\subsection{Density-matrix renormalization group}
\label{sec:iDMRG}

In Fig.~\ref{fig:schematic}, we illustrate the schematic phase diagram for this Hamiltonian, where the approximate phase boundaries are obtained from iDMRG simulations on an infinite cylinder with a circumference \rev{of} $L\,{=}\,10$ and a bond dimension $\chi\,{=}\,2000$. In the hatched white region---which is studied more thoroughly in the next subsection---the simulation converges very slowly (or not at all, to within the conventionally accepted truncation error of $10^{-5}$), with N\'eel and VBC order parameters that are both small and of similar magnitude.
Furthermore, along the transverse direction indicated by the arrow in Fig.~\ref{fig:schematic}, we find no signatures of any phase transitions. This suggests that if a gapless spin\rev{-}liquid phase exists in the SSM as proposed in Refs.~\cite{Yang-2022, Luciano-2024}, it should be adiabatically connected to the gapless spin\rev{-}liquid phase of the square-lattice $J_1$--$J_2$ model.
For the chosen bond dimension, the phase transition between the VBC and N\'eel phases in the SSM appears to be weakly first-order.

In the following sections, we provide extensive evidence \rev{for} this spin\rev{-}liquid phase in the SSM based on exact diagonalization numerics.

\subsection{Exact diagonalization}
\label{sec:ED}

\subsubsection{Low-energy levels and level spectroscopy}

Long-range order in quantum many-body systems is characterized by the spontaneous breaking of a symmetry, manifested in nonvanishing correlation functions of the corresponding order parameters at large distances. Detecting such correlations reliably requires large system sizes, which are beyond the reach of exact diagonalization. However, a symmetry-broken phase is characterized \rev{not only} by long-range correlations \rev{ but also} by a particular structure of low-energy states: in the thermodynamic limit, symmetry breaking implies a degenerate ground-state manifold, while on finite clusters, this degeneracy is lifted into a set of low-lying states distinguished by symmetry quantum numbers. Appropriate linear combinations of these states form the symmetry-broken ground states. Exact diagonalization excels at computing symmetry-resolved low-energy spectra for small clusters and is ideal for detecting signatures of symmetry breaking.

\begin{figure}[t]
\centering
\includegraphics[width=0.95\linewidth]{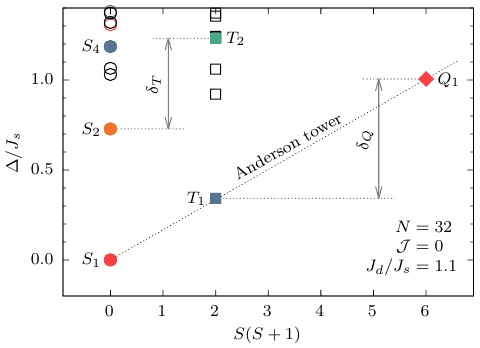}
\caption{Low-energy excitation spectrum of the SSM in the N\'eel phase, obtained from L\'anczos exact diagonalization on a cluster of $N=32$ sites at $\rev{J^{}_{d}/J^{}_{s}=1.1}$. The characteristic Anderson tower of states is clearly visible in the lowest singlet ($S_1$), triplet ($T_1$), and quintuplet ($Q_1$) excitations. As the level of frustration is increased toward the plaquette-singlet regime, the second singlet excitation ($S_2$) is observed to decrease in energy. Also indicated are the composite excitation gaps $\delta_T$ and $\delta_Q$, as defined by \citet{Wang-2022}. The \rev{labeled} states (filled symbols) are identified by their symmetry quantum numbers, as listed in Table~\ref{tab:symmetry}, and not by their position in the energy ordering; the open symbols are further eigenstates belonging to other symmetry sectors, among them the triplets lying between $T_1$ and $T_2$.
}
\label{fig:gaps_32_AT}
\end{figure}

A paradigmatic example is the N\'eel antiferromagnet, which breaks SU(2) spin-rotation symmetry. Its finite-size spectrum exhibits the Anderson tower of states: the lowest states in each total-spin sector $S=0,1,2,\dots$ form a rotor spectrum with excitation energies scaling approximately as $S(S+1)/N$, as illustrated in Fig.~\ref{fig:gaps_32_AT}. These states are well separated from other excitations and have known quantum numbers, making the Anderson tower a robust diagnostic of magnetic long-range order in exact diagonalization studies.

Another example is the plaquette-singlet phase of the SSM, which breaks lattice symmetries. The ground-state manifold consists of two quasidegenerate singlets, denoted $S_1$ and $S_2$, distinguished by lattice quantum numbers. The lowest magnetic excitation is a triplet $T_1$, followed by a second triplet $T_2$ related to $S_2$, and a quintuplet $Q_1$; here and below, $T_2$ is specified by its symmetry quantum numbers (Table~\ref{tab:symmetry}) and is in general not the second-lowest triplet in energy.

Phase transitions between competing orders can therefore be detected via level crossings in the low-energy spectrum. In particular, one of the singlet states ($S_1$) is common to both the N\'eel and plaquette phases, while the relative ordering of the second singlet associated with plaquette order and the lowest triplet of the Anderson tower determines whether a N\'eel or a plaquette state is favored. Tracking these crossings and their finite-size scaling provides a controlled method for locating phase boundaries.

\begin{table}[b]
\caption{Symmetry quantum numbers of the low-energy states used in the level-spectroscopy analysis. All the states are considered at zero wavevector. The labels $S_1$, $S_2$, $S_3$, and $S_4$ denote the lowest singlets, $T_1$ and $T_2$ the lowest triplets, and $Q_1$ the lowest quintuplet, in each case \emph{within the symmetry sector listed in the corresponding row}: their relative ordering may vary, and eigenstates belonging to other sectors may lie in between---see, e.g., the triplets located between $T_1$ and $T_2$ in Fig.~\ref{fig:gaps_32_AT}. These symmetry labels remain valid throughout the interpolation from the SSM to the $J_1$--$J_2$ limit, with symmetry enlargement occurring only at the latter point. The N\'eel state is even under the
$\sigma_{xy}$ and $\sigma_{x\bar{y}}$ diagonal reflections, so the states $S_1$, $T_1$, and $Q_1$ in the Anderson tower have character $+1$. $G_x$ and $G_y$ denote glide reflections (see Sec.~\ref{sec:symmetry} below).}
\label{tab:symmetry}
\begin{ruledtabular}
\begin{tabular}{ccccc}
State & $S$ & $G^{}_x,G^{}_y$ & $\sigma^{}_{xy}, \sigma^{}_{x\bar{y}} $ & Label\\
\hline
$S_1$ & 0 & $+1$ & $+1$ & $A_e$\\
$S_2$ & 0 & $-1$ & $-1$ & $A_o$ \\
$S_3$ & 0 & $+1$ & $-1$ & $B_o$ \\
$S_4$ & 0 & $-1$ & $+1$ & $B_e$ \\
$T_1$ & 1 & $-1$ & $+1$ & $B_e$ \\
$T_2$ & 1 & $+1$ & $-1$ & $B_o$ \\
$Q_1$ & 2 & $+1$ & $+1$ & $A_e$ \\
\end{tabular}
\end{ruledtabular}
\end{table}

While such level crossings are sufficient to identify transitions into magnetically or lattice-ordered phases, the identification of an intermediate quantum-disordered phase requires additional care. In particular, one must distinguish the absence of order from finite-size artifacts in small clusters. A recent work by \citet{Wang-2022} introduced an improved level-spectroscopic framework for the SSM, demonstrating that composite gaps between excited states (e.g., $\delta_T = E(T_2)-E(S_2)$ or $\delta_Q = E(Q_1)-E(T_1)$) allow unusually precise identification of phase boundaries even on small periodic clusters.

In this work, we extend that framework to a family of Hamiltonians interpolating continuously between the Shastry-Sutherland and $J_1$--$J_2$ limits. Crucially, all lattice and spin symmetries of the SSM are preserved throughout the interpolation, with an enlargement of symmetry occurring only at the $J_1$--$J_2$ point. This allows the same symmetry assignments and identification of low-energy states used \rev{in} Ref.~\cite{Wang-2022} to be applied uniformly across the full parameter space.

We focus on a small set of symmetry-resolved low-energy states that evolve continuously across all phases. In the N\'eel phase, the states $S_1$, $T_1$, and $Q_1$ become the lowest members of the Anderson tower, while $S_2$ and $T_2$ move to higher energies. Table~\ref{tab:symmetry} summarizes the symmetry quantum numbers of the relevant states. As all wallpaper-group and spin symmetries of the Shastry-Sutherland lattice are preserved along the interpolation, these labels remain valid across the entire parameter range, allowing unambiguous tracking of the evolution and crossings of low-energy states.

\subsubsection{Symmetry-resolved spectra}

\begin{figure}[!t]
\centering
\includegraphics[width=0.78\linewidth]{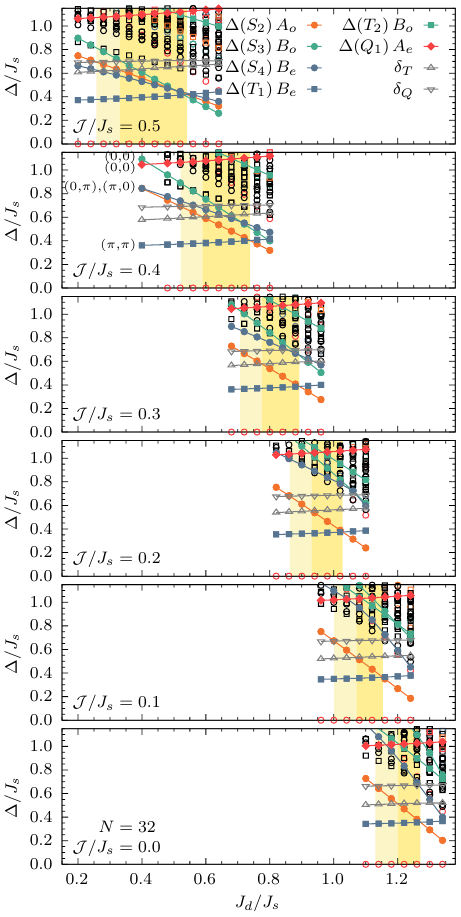}
\caption{Symmetry-resolved low-energy excitation spectrum obtained from L\'anczos exact diagonalization for the $N=32$ cluster as the Hamiltonian is interpolated from the Shastry-Sutherland limit toward the $J_1$--$J_2$ model. Shown are the gaps $\Delta(S_2)$, $\Delta(T_1)$, $\Delta(T_2)$, and $\Delta(Q_1)$, together with the composite gaps $\delta_T$ and $\delta_Q$, using the same symmetry labeling as in Fig.~4(b) of Ref.~\cite{Wang-2022}. The evolution and crossings of these levels illustrate the competition between the Anderson-tower spectrum of the N\'eel phase and the singlet structure [$\Delta(S_1)=0$ here and $\Delta(S_2)$] associated with plaquette ordering, and form the basis for identifying phase boundaries along the interpolation. In the $\mathcal{J}/J_s=0.4$ panel, the $J_d/J_s=0.4$ point corresponds to the $J_1$--$J_2$ model with enhanced symmetry; we indicate the
$(0,0)$, $(\pi,\pi)$, $(0,\pi)$, and $(\pi,0)$ momenta (the latter two forming a two-dimensional irreducible representation in which $S_2$ and $S_4$ become degenerate). The fact that the singlets $\Delta(S_3)$ and $\Delta(S_4)$ come down upon approaching the $J_{1}$--$J_{2}$ limit indicates the onset of columnar VBC order, which is known to be one of the competing phases~\cite{Ralko-2009}.}
\label{fig:spectra32}
\end{figure}

\begin{figure}[!t]
\centering
\includegraphics[width=0.78\linewidth]{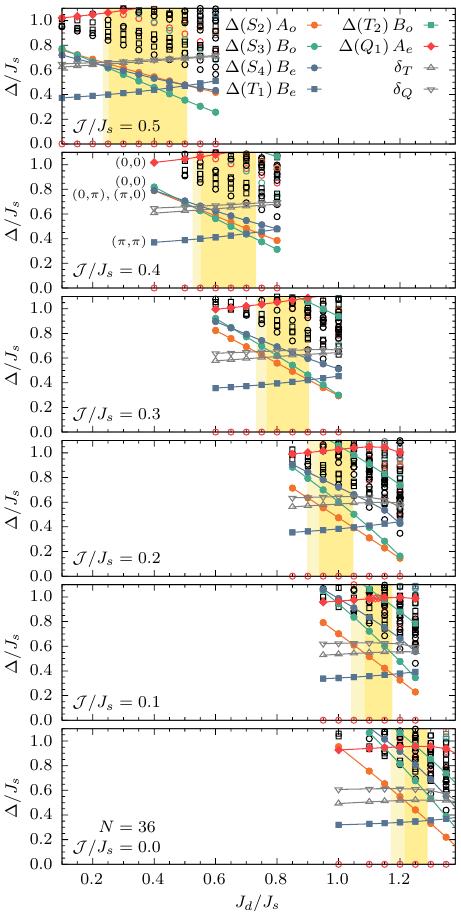}
\caption{Symmetry-resolved low-energy excitation spectrum obtained from L\'anczos exact diagonalization for the $N=36$ cluster as the Hamiltonian is interpolated from the Shastry-Sutherland limit to the $J_1$--$J_2$ model. The notation and symmetry labeling are the same as in Fig.~\ref{fig:spectra32}. The spectrum exhibits the same qualitative structure as for $N=32$, including the evolution and crossings of the states $S_1$, $S_2$, $T_1$, $T_2$, and $Q_1$. Upon approaching the $J_1$--$J_2$ limit, an additional singlet state belonging to the $B_o$ irreducible representation moves down in energy and eventually becomes the lowest singlet, signaling the proximity to a fourfold degenerate singlet\rev{-}ordered phase on the larger cluster. For the values $\mathcal{J} = J_d =0.4 J_s $, our Hamiltonian reduces to the $J_1$--$J_2$ model with enhanced symmetry, and we indicate the momenta
$(0,0)$ of the $S_3$ and $Q_1$ states (as well as the $S_1$ ground state at $\Delta =0$, unmarked), $(\pi,\pi)$ of the $T_1$ state, and the two-dimensional irreducible representation with $(0,\pi)$ and $(\pi,0)$ momenta \rev{in which} $S_2$ and $S_4$ become degenerate.}
\label{fig:spectra36}
\end{figure}

Figures~\ref{fig:spectra32} and \ref{fig:spectra36} show the symmetry-resolved low-energy spectra obtained from L\'anczos exact diagonalization for clusters of $N$\,$=$\,$32$ and $N$\,$=$\,$36$ sites (see also App.~\ref{sec:ed_others} for corresponding data on $N=20,24,28$, and $40$\rev{-}site clusters), as the model is interpolated from the Shastry-Sutherland limit to the $J_1$--$J_2$ model. The evolution of the states $S_1$, $S_2$, $T_1$, $T_2$, and $Q_1$ is shown explicitly, together with the level crossings separating the N\'eel, spin-liquid, and singlet-ordered regimes.
In Fig.~\ref{fig:spectra36}, the downward motion of the $B_o$ ($S_3$) and $B_e$ ($S_4$) singlets near the $J_1$--$J_2$ limit, together with $S_1$ and $S_2$, signals the emergence on finite clusters of the quasidegenerate four-state manifold associated with columnar VBC order, marking the termination of the spin-liquid regime on the singlet-ordered side.
In the high-symmetry $J_1$--$J_2$ limit, these four singlets occur at momenta $(0,0)$ ($\times 2$), $(\pi,0)$, and $(0,\pi)$. This momentum structure is the finite-size fingerprint of columnar VBC order, distinguishing it from a plaquette VBC, which decomposes into a different set of lattice irreducible representations; see Table~II of Ref.~\cite{Ralko-2009}.

Our central result is the identification of an intermediate spin-liquid regime that separates the N\'eel phase from the space-group-symmetry-breaking singlet-ordered phases. Spectroscopically, within this regime we do not observe clear signatures of either SU(2) spin-rotation symmetry breaking---as inferred via the Anderson tower of states---or quasidegenerate singlet manifolds signaling plaquette or columnar order. Instead, the lowest singlet and triplet excitations evolve smoothly with system size, and the relevant conventional and composite gap crossings delimit a finite-width region in parameter space.

\begin{figure}[!t]
\centering
\includegraphics[width=0.95\linewidth]{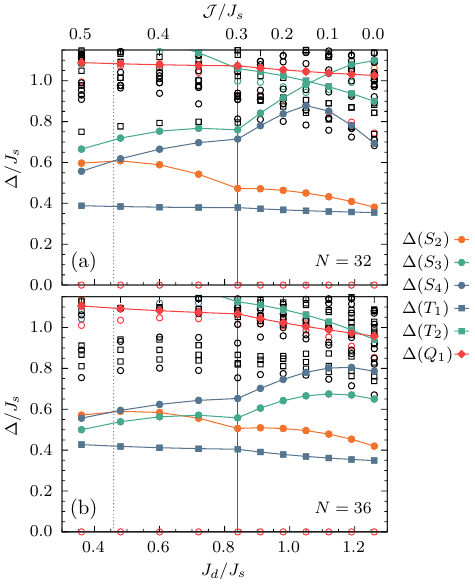}
\caption{Symmetry-resolved low-energy excitation spectrum obtained from L\'anczos exact diagonalization along representative paths inside the river of liquidity for clusters of size (a) $N=32$ and (b) $N=36$. The path is defined in the $(J^{}_{d}/J^{}_{s},\,\mathcal{J}/J^{}_{s})$ plane and consists of two straight segments, running from $(1.26,0)$---i.e., the pure Shastry-Sutherland model---to $(0.84,0.3)$, and from there to $(0.36,0.5)$; it is the same path that is drawn on the phase diagram in Fig.~\ref{fig:overlap}. The solid vertical line marks the intermediate point $(0.84,0.3)$ at which the path changes direction, while the dotted vertical line marks the crossing of the $J_{1}$--$J_{2}$ limit, $\mathcal{J}=J^{}_{d}$, at $\approx(0.459,0.459)$. We plot the excitation gaps associated with the low-energy singlet, triplet, and quintuplet states using the same symmetry labeling as in Fig.~\ref{fig:spectra32}. Along these trajectories, the lowest singlet and triplet levels evolve smoothly as a function of the coupling. The finite-size precursors of the columnar dimer phase, $\Delta(S_3)$ and $\Delta(S_4)$, gradually move down in energy as the $J_1$--$J_2$ limit is approached, but do not disrupt the continuous adiabatic evolution of the spectrum within the spin-liquid regime.}
\label{fig:river_32_36}
\end{figure}

\begin{figure}[!t]
\centering
\includegraphics[width=0.95\linewidth]{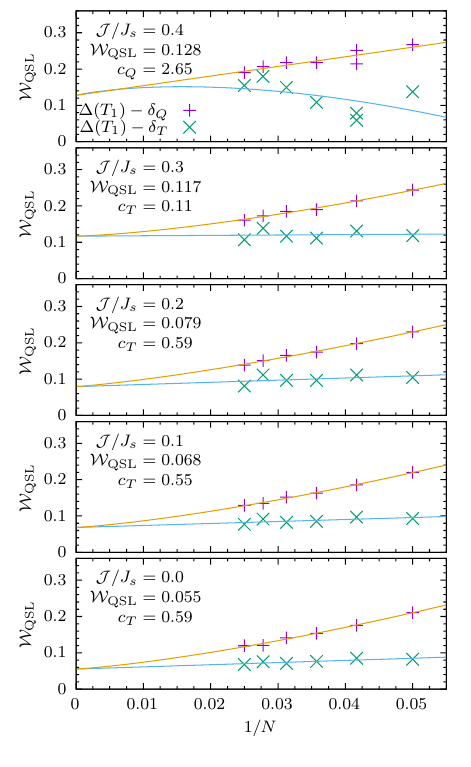}
\caption{Finite-size scaling of the width of the spin-liquid window, $\mathcal{W}^{}_{\mathrm{QSL}}=(J^{}_{d}/J^{}_{s})^{}_{c,2}-(J^{}_{d}/J^{}_{s})^{}_{c,1}$, extracted from level-crossing points as a function of $1/N$ for several values of the interpolating coupling $\mathcal{J}/J^{}_{s}$. The quantities shown are directly analogous to those used in Fig.~5 of Ref.~\cite{Wang-2022}, where crossings between excitation gaps were found to display only weak system-size dependence in the SSM. Here, we employ the same composite-gap constructions while continuously deforming the Hamiltonian toward the square-lattice $J_1$--$J_2$ limit. For each set of parameters, $\mathcal{W}^{}_{\mathrm{QSL}}$ remains finite upon extrapolation to the thermodynamic limit, indicating that the spin-liquid regime identified in the SSM persists throughout the interpolation. \rev{See} App.~\ref{sec:ed_others} for ED data on $N=20,24,28$, and $40$ sites. Following Ref.~\cite{Wang-2022}, the crossing points are extrapolated to the thermodynamic limit with the linear form $g^{\ast}(N)=g^{}_{c}+c/N$; $c^{}_{T}$ and $c^{}_{Q}$ are the coefficients of the $1/N$ term in the linear fits to the $\delta^{}_{T}$- and $\delta^{}_{Q}$-based crossing data, respectively.
}
\label{fig:crossings}
\end{figure}

\subsubsection{Composite gaps and the river of liquidity}

To test the internal consistency of the spin-liquid regime, we next follow the evolution of the low-energy spectrum along paths contained entirely  within this extended  region.

In Fig.~\ref{fig:crossings}, the plotted quantity is the width of the spin-liquid window, $\mathcal{W}^{}_{\mathrm{QSL}}\equiv(J^{}_{d}/J^{}_{s})^{}_{c,2}-(J^{}_{d}/J^{}_{s})^{}_{c,1}$, i.e.\rev{,} the separation in $J^{}_{d}/J^{}_{s}$ between the two crossing points delimiting the spin-liquid regime at fixed $\mathcal{J}/J^{}_{s}$ (we use $\mathcal{W}$ rather than $\Delta$ here so as not to confuse this width in coupling space with the excitation gaps $\Delta(S^{}_{i})$, $\Delta(T^{}_{i})$\rev{,} and $\Delta(Q^{}_{1})$), evaluated with either of the two composite-gap constructions indicated in the legend, $\Delta(T^{}_{1})-\delta^{}_{Q}$ and $\Delta(T^{}_{1})-\delta^{}_{T}$. Each panel quotes the coefficient of the $1/N$ term for whichever of the two data sets follows the linear form over the accessible cluster sizes ($c^{}_{T}$ for $\mathcal{J}/J^{}_{s}\le0.3$ and $c^{}_{Q}$ at $\mathcal{J}/J^{}_{s}=0.4$), together with the extrapolated width, to which both constructions converge within our resolution.

As the Hamiltonian is deformed from the Shastry-Sutherland limit toward the $J_1$--$J_2$ model, these crossings flow continuously without qualitative reorganization, demonstrating adiabatic connectivity of the intermediate phase. The resulting spin-liquid region therefore forms a continuous \emph{river of liquidity} along the boundary of the N\'eel phase (see Figs.~\ref{fig:river_32_36} and \ref{fig:crossings}), extending across the full interpolation and terminating naturally at the $J_1$--$J_2$ point.

On the opposite boundary, adjoining the plaquette-singlet phase of the SSM and the columnar dimer phase of the $J_1$--$J_2$ model, the dominant low-energy levels also evolve adiabatically. The only qualitative change is the gradual appearance of additional singlet states associated with the enlarged degeneracy of the columnar dimer phase. These states ``creep in'' from higher energies on finite clusters and do not signal a breakdown of adiabatic continuity of the liquid regime, but rather the onset of lattice-symmetry breaking.

\begin{figure}[t]
\centering
\includegraphics[width=0.95\linewidth]{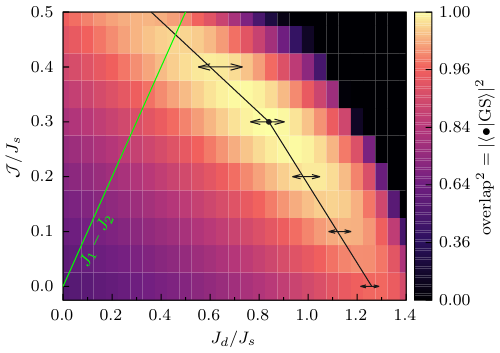}
\caption{Density plot of the overlap between a reference spin-liquid state (the ground state $|\bullet\rangle$ at $(J_d/J_s,\mathcal{J}/J_s)=(0.84,0.30)$ is indicated by the small black circle) and the exact ground states across the space of exchange couplings for the 36-site cluster. The continuous region of high overlap coincides with the river of liquidity. The arrows indicate the extent of the spin-liquid regime as determined independently from the level crossings shown in Fig.~\ref{fig:spectra36}. The green straight line marks the parameters corresponding to the $J_1$--$J_2$ square-lattice Heisenberg model. The dark straight lines indicate cuts within the river of liquidity along which the excitation spectra are computed in Fig.~\ref{fig:river_32_36}.}
\label{fig:overlap}
\end{figure}

 The accuracy of composite level spectroscopy in the SSM was demonstrated by \citet{Wang-2022}\rev{,} and our work extends this framework to reveal an adiabatically connected spin-liquid regime extending across distinct frustrated lattices.

\subsubsection{Wavefunction overlap}

Figure~\ref{fig:overlap} provides complementary evidence for the adiabatic connectivity identified above by showing the overlap between a representative spin-liquid ground state and the exact ground states across the space of exchange couplings as a density plot. A continuous region of large overlap traces the river of liquidity along the N\'eel boundary, confirming that the spin-liquid states are adiabatically connected across the interpolation pathway.

By combining symmetry-resolved level spectroscopy with wave\rev{-}function overlap diagnostics, we demonstrate that the spin-liquid regime identified in the SSM extends adiabatically to the $J_1$--$J_2$ limit on the square lattice. This river of liquidity forms a continuous boundary between N\'eel order and lattice-symmetry-breaking singlet phases, providing a unified spectral picture of quantum disorder in frustrated square-lattice antiferromagnets.

\subsection{Summary of theoretical approach and results}

\rev{Before turning to the parton description, we summarize what the numerics of this section have established and outline the route taken in the rest of the paper. Infinite DMRG locates a window in which the simulations converge poorly and the N\'eel and plaquette-VBC order parameters are both small (Sec.~\ref{sec:iDMRG}), and symmetry-resolved exact diagonalization resolves that window spectroscopically. Along the entire interpolation to the $J_{1}$--$J_{2}$ model the low-energy levels evolve without qualitative reorganization, and within the intermediate regime neither an Anderson tower nor a quasidegenerate singlet manifold develops (Sec.~\ref{sec:ED}). The composite-gap crossings delimiting this regime extrapolate to a \emph{finite} width in the thermodynamic limit at every value of $\mathcal{J}/J^{}_{s}$ studied (Fig.~\ref{fig:crossings}), and the wavefunction-overlap map of Fig.~\ref{fig:overlap} independently traces the same connected region---the river of liquidity. What these diagnostics cannot supply is the microscopic identity of the state: its gauge structure, the nature of its excitations, and whether the two ends of the river realize one and the same phase. The remainder of the paper addresses this question.}

To further shed light on the nature of the QSL in the SSM, we employ the framework of projective symmetry groups (PSGs)~\cite{Wen-2002,Wen-2007}, described in Sec.~\ref{sec:psg} (and extensively utilized on many two- and three-dimensional lattices~\cite{Wang-2006,Lawler-2008,Choy-2009,Yang-2010,Lu-2011a,Lu-2011b,Yang-2012,Messio-2013,Bieri-2015,Yang-2016,Lu-2016a,Bieri-2016,Huang-2017,Huang-2018,Lu-2018,Liu-2019,Jin-2020,Sonnenschein-2020,Sahoo-2020,Liu-2021,Chern-2021,Chern-2022,Benedikt-2022,Maity-2023,Chauhan-2023,Liu-2024,Sonnenschein-2024,Maity-2024a,Maity-2024b}), to perform a systematic classification of fermionic mean-field states with SU(2), U(1), and $\mathbb{Z}_{2}$ low-energy gauge groups. \rev{Since} previous studies~\cite{Yang-2022,Luciano-2024} have provided evidence for a gapless spin liquid, we choose to employ the fermionic representation of spin operators, as opposed to \rev{the bosonic one, because} the former enables a description of gapless QSLs while the latter is restricted to describing only gapped QSLs~\cite{Liu-2024}. Considering QSLs which respect the full space-group symmetry as well as QSLs compatible with the lower symmetry of the adjacent plaquette VBC, we present the microscopic \textit{Ans\"atze} describing these states in Sec.~\ref{sec:ansatz}. \rev{The classification yields 46 U(1) and 80 $\mathbb{Z}_{2}$ PSG classes for the $p4g$ wallpaper group of the Shastry-Sutherland lattice; restricting the mean-field amplitudes to NN and NNN bonds leaves twelve U(1) and eighteen $\mathbb{Z}_{2}$ \textit{Ans\"atze} (Fig.~\ref{fig:fig2} and Table~\ref{table:z2_ansatze}), while the reduced symmetry of the plaquette VBC---$C_{4}$ preserved, $G_{x}$ and $\sigma_{xy}$ broken---admits a further 48 short-range $\mathbb{Z}_{2}$ \textit{Ans\"atze} (App.~\ref{app:nematic}).} Then, we develop a mapping between the $\mathbb{Z}_{2}$ \textit{Ans\"atze} on the Shastry-Sutherland and square lattices by establishing a connection between the PSGs of the respective symmetry groups. \rev{Adjoining the reflection $\sigma_{x}$ enlarges $p4g$ to the square-lattice group $p4m$ and reproduces the 272 PSGs of Ref.~\cite{Wen-2002}; the correspondence is established for six square-lattice $\mathbb{Z}_{2}$ \textit{Ans\"atze} in all (Table~\ref{table:z2_ansatze_square_shastry}).} This approach allows us to systematically narrow down the candidate wavefunctions for the potential QSL in the generalized SSM by identifying the parton \textit{Ans\"atze} that can be adiabatically connected to those of the QSL phase in the $J_1$--$J_2$ model.

In Sec.~\ref{sec:mapping_square_lattice}, we analyze the \textit{Ansatz} on the Shastry-Sutherland lattice that maps to the Z2A$zz$13 state on the square lattice: a gapless (Dirac) $\mathbb{Z}_{2}$ QSL, which is believed to provide an excellent variational description of the QSL in the $J_{1}$--$J_{2}$ square-lattice Heisenberg model~\cite{Hu-2013,Nomura-2021}. To characterize this \textit{Ansatz}, we analyze its spinon band structure and compute the mean-field dynamical and equal-time spin structure factors at $J_{s}/J_{d}=0.80$, which lies within the putative QSL phase~\cite{Yang-2022,Wang-2022,Keles-2022,Luciano-2024}. We identify the state to be a gapless $\mathbb{Z}_2$ Dirac quantum spin liquid with $d$-wave pairing. \rev{Its} Dirac cones imply linearly dispersing spinon excitations, leading to algebraically decaying spin correlations in real space. In the dynamical spin structure factor, the fractionalized spinon pairs produce broad continua rather than sharp magnon modes, with enhanced spectral weight near twice the Dirac energy and characteristic fountain-like features emerging from the nodal points. The $\mathbb{Z}_2$ nature of the state reflects the gapping out of gauge fluctuations by pairing, leaving gapless fermionic quasiparticles coupled to a gapped gauge sector.

\rev{As the spinon band structure is what most directly distinguishes the Shastry-Sutherland realization from its square-lattice counterpart, it is worth setting out the relation between the two here. The parent U(1) state with uniform real hopping has a Fermi surface at half filling; adding the nearest-neighbor $d_{x^{2}-y^{2}}$ pairing $\Delta^{}_{s}$ gaps it down to four Dirac points pinned at $M(\pm\pi/2,\pm\pi/2)$, the IGG still being U(1). The empty-square pairing $\Delta^{}_{\square}$ then Higgses the gauge structure down to $\mathbb{Z}_{2}$, but leaves the nodes exactly where they are; their displacement along $\overline{\Gamma M}$ by $\varepsilon$ is produced instead by the amplitudes that the lower $p4g$ symmetry permits, namely the diagonal hoppings $\chi^{}_{d}$, $\chi^{}_{d'}$, $\chi^{}_{\square}$ and the chemical potential $\mu$~\footnote{At $J_s/J_d$\,$=$\,$0.8$ on a $14\times14$ cluster, the optimized \textit{Ansatz} gives $\varepsilon=0.091$~rad, i.e., $\varepsilon\approx0.029\,\pi$; the displacement is obtained in closed form in App.~\ref{app:epsilon}, and its evolution with system size is listed in Table~\ref{tab:epsilon_L}. The corresponding incommensurability of the gapless two-spinon momenta is $\delta=2\varepsilon\approx0.058\,\pi$.
In the Z2A$zz$13 state, the Dirac cone splits into four cones at $(\pi/2\pm\varepsilon,\pi/2\pm\varepsilon)$ due to the $d_{xy}$ pairing on the diagonal bonds connecting site $(0,0)$ to $(\pm1,\pm1)$\rev{, whereas} if we introduce the $d_{xy}$ pairing on the $(\pm2,\pm2)$ bonds, the Dirac cones remain intact at $(\pi/2,\pi/2)$.}. In App.~\ref{app:epsilon}, we reduce the nodal condition to a quartic polynomial whose coefficients are explicit functions of the mean-field amplitudes and solve it in closed form: the nodes stay exactly commensurate if and only if $\mu=\pm(\chi^{}_{d}+\chi^{}_{d'}-2\chi^{}_{\square})$, independently of both pairings. Evaluated on the optimized amplitudes, $\varepsilon$ decreases monotonically with cluster size---from $0.078\,\pi$ at $L=6$ to $0.008\,\pi$ at $L=60$---following $\varepsilon(L)\simeq1.2/L+\varepsilon^{}_{\infty}$ with $\varepsilon^{}_{\infty}\simeq0.002\,\pi$ (Table~\ref{tab:epsilon_L}). The optimization thus drives the state toward this pinning manifold, so that the incommensurability $\delta=2\varepsilon$ of the gapless two-spinon momenta is a finite-size feature and the Shastry-Sutherland dispersion approaches that of the square-lattice Z2A$zz$13 state in the thermodynamic limit.}

In Sec.~\ref{sec:numerics}, we carry out a state-of-the-art variational Monte Carlo study of the corresponding Gutzwiller-projected wavefunctions supplemented with L\'anczos steps, and demonstrate the excellent agreement of the energies and equal-time correlators of our $\mathbb{Z}_{2}$ Dirac QSL with those obtained from vision transformer (ViT) neural networks, DMRG, and exact diagonalization on small clusters. \rev{The projected Z3000 state attains the lowest energy of all the $\mathbb{Z}_{2}$ \textit{Ans\"atze} on every cluster from $L=6$ to $40$, its pairing amplitudes $\Delta^{}_{s}$ and $\Delta^{}_{\square}$ converge to finite values by $L\simeq50$, and the empty-square pairing responsible for the Higgsing yields an energy gain of order $1\%$ over the parent U(1) state---evidence of a robust $\mathbb{Z}_{2}$ gauge structure in the thermodynamic limit. After two L\'anczos steps and a zero-variance extrapolation, the energy per site at $J^{}_{s}/J^{}_{d}=0.80$ is $E^{\rm 2D}_{\infty}/J^{}_{d}=-0.4481(3)$, equal within error bars to the ViT and DMRG values but obtained with only eight variational parameters.}

\rev{The correlation functions furnish the sharper test, and they also show how the state evolves along the river. On the $14\times14$ cluster, the projected state reproduces the sign of \emph{every} ViT correlator with quantitatively matching magnitudes, and its equal-time structure factor at the antiferromagnetic wavevector, $\mathcal{S}(\pi,\pi)\approx10.19$, agrees with the ViT value of $9.89$. The real-space spin-spin correlations decay algebraically with $z+\eta=1.35(3)$, consistent with the $z+\eta=1.30(3)$ obtained independently from the finite-size scaling of $m^{2}(L)$, while the connected plaquette-plaquette correlations decay faster, with an exponent $1.62(5)$. Such algebraic decay is a hallmark of gapless quantum spin liquids and is consistent with Dirac spinons coupled to a $\mathbb{Z}_{2}$ gauge field, whereas a gapped $\mathbb{Z}_{2}$ state would give exponentially decaying correlations. Since the gauge structure is $\mathbb{Z}_{2}$, the exponent is not pinned to a universal value, and we verify explicitly that it varies smoothly along the river: the same analysis at the points marked in Figs.~\ref{fig:crossings} and~\ref{fig:overlap} gives $z+\eta=1.29(2)$, $1.40(2)$, $1.48(1)$, $1.60(2)$, $1.74(2)$, and $1.85(2)$ for $\mathcal{J}/J^{}_{s}=0$, $0.1$, $0.2$, $0.3$, $0.4$, and $0.5$, respectively (Fig.~\ref{fig:exponent_river}). The evolution is monotonic and free of discontinuities, and it terminates inside the spread of values reported for the gapless QSL window of the square-lattice $J_{1}$--$J_{2}$ model, a range that is itself coupling dependent and on which the various methods do not agree. Band structure and correlation exponents therefore both interpolate continuously between the two limits, as expected if a single $\mathbb{Z}_{2}$ Dirac phase spans the river.}

\rev{Section~\ref{sec:phasediagram} then fixes the boundaries of this phase. The spin-liquid \textit{Ansatz} is built from singlet hopping and pairing alone, so it cannot develop magnetic order by construction; we therefore embed it in a single variational family that treats the neighboring phases on the same footing, by adding an in-plane magnetic field to the spinon Hamiltonian (Sec.~\ref{sec:pfaffian}). Once both a transverse field and singlet pairing are present, no single-particle basis renders the mean-field Hamiltonian number conserving, so the projected amplitudes become Pfaffians rather than determinants. The plaquette VBC is treated analogously through the P5A \textit{Ansatz}, the symmetry-lowered descendant of Z3000 identified in App.~\ref{app:nematic}, which returns to the symmetric state when its symmetry-inequivalent amplitudes are set equal. Within this family, both order parameters vanish \emph{continuously} and each correlation ratio shows a single crossing rather than the two-branch structure of a first-order transition, placing the boundaries at $(J^{}_{s}/J^{}_{d})^{}_{\rm PVBC}\simeq0.78$ and $(J^{}_{s}/J^{}_{d})^{}_{\rm N\acute{e}el}\simeq0.82$ (Sec.~\ref{sec:orderparam}). The energetic competition of separately optimized trial states, used here only as a consistency check, gives $\approx0.78$ and $\approx0.83$ (Sec.~\ref{sec:energetics}); both determinations agree with the DMRG, ED, neural-network, and tensor-network estimates collected in Table~\ref{tab:transitions}.}

\rev{Independently of any variational input, the recently developed Keldysh formulation of the pseudo-fermion functional renormalization group yields the hitherto unreported dynamical spin structure factor of this QSL directly on the real-frequency axis, avoiding analytic continuation, at $T/J^{}_{d}=0.1$ on a $9\times9$ lattice. It exhibits two lobes of maximal low-energy intensity flanking $(\pi,\pi)$, appreciable weight at $(\pi,\pi)$ itself, and a broad continuum---precisely the pattern expected from scattering between Dirac cones displaced from $M$---from a correlator-based method that makes no assumption about the wavefunction or its symmetry.}

Taken together, these results provide compelling evidence for a $\mathbb{Z}_2$ Dirac quantum spin\rev{-}liquid phase in the generalized Shastry-Sutherland model. Moreover, the adiabatic continuity established numerically between the Shastry-Sutherland and square-lattice limits indicates that both systems realize the same universal Dirac spin-liquid fixed point, rather than distinct lattice-specific phases. The river of liquidity therefore reflects a shared underlying infrared gauge theory. We emphasize that the projective realizations of the Shastry-Sutherland lattice symmetries are already built into the \textit{Ans\"atze} and dictate \rev{their} sign and amplitude structure. Hence, such an approach  requires a much smaller number of  variational parameters as compared to methods based on neural\rev{-}network wavefunctions, DMRG, and many-variable VMC that rely on unconstrained optimizations.

In light of recent observations of gapless QSL behavior in Pr$_2$Ga$_2$BeO$_7$~\cite{Li-2024}, which can approximately be modeled by $S=1/2$ XXZ interactions on the Shastry-Sutherland lattice, our results find additional significance. In particular, our finding of a $d$-wave pairing $\mathbb{Z}_{2}$ Dirac QSL is compatible with the thermodynamic signatures of a $T^2$ dependence of the specific heat and a finite $\kappa_{0}/T$ arising \rev{from} a disorder-induced finite density of states~\cite{Li-2024}. Taken together with indications from Keldysh pf-FRG of the persistent signatures of Dirac cones upon introduction of XXZ anisotropy to the Heisenberg model, our work supports the scenario of a $\mathbb{Z}_{2}$ Dirac QSL\rev{,} thereby clarifying the microscopic identity of the QSL in Pr$_2$Ga$_2$BeO$_7$.

\section{Methodology}
\label{sec:method}

\subsection{Parton construction}
\label{sec:parton}
A powerful framework to construct mean-field theories of QSLs is provided by the ``parton'' representation of spin operators. The first step  in this scheme is to express the spin operators $\hat{S}^\gamma$ ($\gamma=x,y,z$) in terms of spin-$1/2$, charge-neutral quasiparticles called \emph{spinons}, which can be chosen to have either bosonic or fermionic statistics. As we are interested in describing gapped as well as gapless QSLs and the competition between them, we henceforth consider the spinons to be fermionic. In this language, the spin operator can be rewritten as a bilinear with two flavors of complex fermions~\cite{Abrikosov-1965,Baskaran-1988}:
\begin{align}\label{eq:spin_op_fermion}
\hat{S}^\gamma_{i} = \frac{1}{2} \sum_{\alpha\beta} \hat{f}^\dagger_{i\alpha} \tau^\gamma_{\alpha\beta} \hat{f}_{i\beta}^{\vphantom\dagger}.
\end{align}
Here, the $\hat{f}^{\pdagger}_{i\alpha}$ ($\hat{f}_{i\alpha}^{\dagger}$) are spinon operators which annihilate (create) a spin-$1/2$ fermion with flavor $\alpha \in \left\lbrace \uparrow, \downarrow \right\rbrace$ at site $i$ and $\tau^\gamma$ ($\gamma=x,y,z$) are the three Pauli matrices. In order to make the symmetries of the fermionic representation manifest, it proves instructive to re-express Eq.~\eqref{eq:spin_op_fermion} as
\begin{equation}
\label{eq:spin_op_doublet}
\hat{S}^{\gamma}_{i}=\frac{1}{\rev{4}}\text{Tr}\left[\hat{\psi}^\dagger_i\tau^\gamma\hat{\psi}^{\pdagger}_i\right]
\end{equation}
in terms of  the spinor doublet $\hat{\psi}^{\pdagger}_i$\,$\equiv$\,$(\hat{\textbf{f}}^{\pdagger}_i,\hat{\bar{\textbf{f}}}^{\pdagger}_i)$, where $\hat{\textbf{f}}^{\pdagger}_i=(\hat{f}^{\pdagger}_{i,\uparrow},\hat{f}^{\pdagger}_{i,\downarrow})^\mathrm{T}$ and $\hat{\bar{\textbf{f}}}^{\pdagger}_i=(\hat{f}^\dagger_{i,\downarrow},-\hat{f}^\dagger_{i,\uparrow})^\mathrm{T}$. Under a right multiplication $\hat{\psi}_i\rightarrow\hat{\psi}_i W_i$ with $W_i\in$ SU(2), the spin operators in Eq.~\eqref{eq:spin_op_doublet} and the fermionic anticommutation relations of the spinons remain unaffected, i.e., such a transformation does not affect the physical observables. Thus, the fermionization of the spin operators is accompanied by a local gauge freedom associated with an internal symmetry group (ISG)~\cite{Affleck-1988,Liu-2010}, which, in this case, is SU(2). \rev{In contrast to} right multiplications, a left SU(2) multiplication, i.e., $\hat{\psi}_i\rightarrow G\hat{\psi}_{i}$ effects a rotation in spin space. Moreover, it is important to note that the \rev{on-site} fermionic Hilbert space is \rev{$\mathbb{C}^{4}$} while the spin space is $\mathbb{C}^{2}$. This necessitates the introduction of an additional constraint requiring that there be exactly one fermion per site:
\begin{equation}\label{eq:constraint_doublet}
\hat{\psi}^{\pdagger}_{i}\tau^{\gamma}\hat{\psi}^{\dagger}_{i}=0, \;  \gamma\in\{x,y,z\},
\end{equation}
which ensures that the Hilbert space dimensions of the spin and spinon representations are in agreement.

Inserting Eq.~\eqref{eq:spin_op_fermion} into a generic Heisenberg model such as, say, Eq.~\eqref{eq:mod-ham}, yields a four-fermion interaction term in the Hamiltonian, following which we can perform a mean-field decoupling to obtain an effective quadratic Hamiltonian. Given that we intend to formulate a low-energy Hamiltonian for a spin-rotation-invariant quantum paramagnet, we restrict this decoupling to only the singlet channel. Since a spin rotation is implemented in the fermionic Hilbert space by a left SU(2) multiplication, the quadratic field $\hat{\psi}^\dagger_i\hat{\psi}^{\pdagger}_j$ is manifestly spin-rotation invariant. Accordingly, the mean field in the singlet channel can be identified as $u^{\pdagger}_{ij}=\langle\hat{\psi}^\dagger_i\hat{\psi}^{\pdagger}_j\rangle$, which, in turn, consists of two subchannels $\chi_{ij}$ and $\Delta_{ij}$:
\begin{eqnarray}
u^{\pdagger}_{ij} =
\begin{bmatrix}
{\chi}^{\pdagger}_{ij} & {\Delta}^\dagger_{ij}  \\
{\Delta}^{\pdagger}_{ij} & -{\chi}^\dagger_{ij}
\end{bmatrix}.
\label{eq:ansatz}
\end{eqnarray}
Here, $\chi_{ij}$ and $\Delta_{ij}$ are the hopping and pairing singlet mean fields, respectively, which reside on the bond connecting sites $i$ and $j$. The resulting quadratic spinon Hamiltonian for a quantum paramagnet generically takes the form
\begin{align}\label{eq:mf_ham}
    \hat{H}^{\pdagger}_\mathrm{MF} = & \sum_{\langle ij \rangle} \frac{3}{8} J^{\pdagger}_{ij} \left[\frac{1}{2}\text{Tr}(u_{ij}^{\dagger} u^{\pdagger}_{ij}) + \text{Tr}(\hat{\psi}^{\pdagger}_{i} u^{\pdagger}_{ij} \hat{\psi}^{\dagger}_{j} + \text{H.c.}) \right] \; \notag \\
    & + \sum_{i} \sum_\gamma a^{\pdagger}_{\gamma,i} \text{Tr}\left[\hat{\psi}^{\pdagger}_{i} \tau^{\gamma} \hat{\psi}^{\dagger}_{i} \right].\;
\end{align}
The term in the second line above incorporates the constraint forbidding empty and doubly occupied sites at the mean-field level, i.e., $\langle\hat{\psi}_{i}\tau^{\gamma}\hat{\psi}^{\dagger}_{i}\rangle=0$, using three Lagrange multipliers $a_\gamma$, with $\gamma=x,y,z$, at each site. As the spinon Hamiltonian is completely specified by ($u_{ij},a_{\gamma,i}$), these fields together constitute an \textit{Ansatz} for a given QSL phase. The mean-field Hamiltonian\rev{~}\eqref{eq:mf_ham} is now invariant under the following gauge transformations:
\begin{equation}
    \hat{\psi}^{\pdagger}_i\rightarrow \hat{\psi}^{\pdagger}_iW^{\pdagger}_i, \; u^{\pdagger}_{ij}\rightarrow W^\dagger_iu^{\pdagger}_{ij}W^{\pdagger}_j,\;a^{\pdagger}_{\gamma,i}\tau^\gamma \rightarrow a^{\pdagger}_{\gamma,i} W^\dagger_i\tau^\gamma W^{\pdagger}_i,
    \label{eq:gauge_symmetry}
\end{equation}
for $W_i\in$ SU(2).
Therefore, the ISG turns out to also be the gauge-symmetry group in the mean-field construction. Next, we describe how this property leads to nontrivial symmetry operations in the fermionic Hilbert space.

\subsection{Projective symmetry group framework}\label{sec:psg}
Fully symmetric QSLs are quantum paramagnets that do not spontaneously break lattice space-group and time-reversal symmetries and\rev{,} consequently\rev{,} cannot be characterized by local physical order parameters. In principle, QSLs may break one or more lattice symmetries and time reversal, but they still cannot be uniquely characterized by local order parameters since there can exist a plethora of QSLs breaking the very same symmetries---as occurs for chiral QSLs~\cite{Bieri-2016} and spin-/lattice-nematic QSLs~\cite{Shindou-2011,Reuther-2014b,Lu-2016a}. To establish the invariance of an \textit{Ansatz} under different symmetry operations, one therefore has to specify how a lattice or time-reversal symmetry operation acts in the fermionic Hilbert space.

In general, an \textit{Ansatz} given by $u^{}_{ij}$ is modified to $u'_{ij}$\,$=$\,$u^{}_{\mathcal{O}(i),\mathcal{O}(j)}$ under a symmetry operation denoted by $\mathcal{O}$, so, at first sight, it appears that the symmetry of the \textit{Ansatz} under $\mathcal{O}$ is broken. However, the SU(2) gauge symmetry in the fermionic representation implies that two \textit{Ans\"atze} related by \rev{an} SU(2) gauge transformation label the same physical spin state. Thus, if we choose an appropriate local SU(2) matrix $W^{\pdagger}_i$ that connects $u^{}_{ij}$ and $u'_{ij}$, the symmetry of the \textit{Ansatz} can be restored. As a result, lattice and time-reversal symmetries are realized projectively (i.e., up to SU(2) gauge transformations) in the fermionic Hilbert space. This lays the foundation for the idea of the projective symmetry group (PSG) framework~\cite{Wen-2002}, whereby an \textit{Ansatz} is invariant under a symmetry operation $\mathcal{O}$ \textit{iff} the following condition is satisfied:
\begin{align}\label{eq:sym_con_gauge}
W^{\dagger}_{\mathcal{O}}(\mathcal{O}(i)) u^{\pdagger}_{\mathcal{O}(i)\mathcal{O}(j)} W^{\pdagger}_{\mathcal{O}}(\mathcal{O}(j))= u^{\pdagger}_{ij}.
\end{align}
This combined operation of the physical symmetry element and the element of the ISG defines the PSG of an \textit{Ansatz}. Physically, if two \textit{Ans\"atze} have gauge-inequivalent PSGs, they correspond to two distinct QSL states. Accordingly, PSGs serve to characterize the internal quantum order of an \textit{Ansatz}~\cite{Wen-2007}, i.e., different \textit{Ans\"atze} can be distinguished and characterized by PSGs in a fashion akin to how classical solid-state phases are characterized by crystallographic space-group symmetries. Hence, a systematic classification of projective representations of given lattice space groups can serve as a powerful mathematical tool, enabling the construction of distinct QSL \textit{Ans\"atze} with a desired low-energy gauge group.

To characterize the \textit{Ans\"atze} in more detail, note that $u_{ij}$ in Eq.~\eqref{eq:ansatz} can be re-expressed in terms of four real parameters as
\begin{equation}\label{eq:link_singlet}
    u^{\pdagger}_{ij}=\dot{\iota}\chi^0_{ij} \tau^0 + \chi^3_{ij} \tau^z+\Delta^1_{ij} \tau^x+\Delta^2_{ij} \tau^y,
\end{equation}
where $\tau^0$ is the $2\times2$ identity matrix\rev{, so that the two complex mean fields of Eq.~\eqref{eq:ansatz} read $\chi^{}_{ij}=\dot\iota\chi^0_{ij}+\chi^3_{ij}$ and $\Delta^{}_{ij}=\Delta^1_{ij}+\dot\iota\Delta^2_{ij}$}. Due to this structure, there always exists a group $\mathbb{G}$ (where $\mathbb{G}$ is either SU(2)  or its subgroups U(1) and $\mathbb{Z}_2$) such that under gauge transformations by $\mathcal{G}_i \in \mathbb{G}$, the \textit{Ansatz} remains invariant. The group $\mathbb{G}$ is called the invariant gauge group (IGG) of the \textit{Ansatz}. Another interpretation of the IGG can be seen by comparing Eqs.~\eqref{eq:sym_con_gauge} and \eqref{eq:IGG} below, which implies that the IGG is a special subgroup of the PSG corresponding to an identity symmetry operation ($\mathcal{O}=\mathds{1}$).

When $u^{}_{ij}$\,$=$\,$\dot{\iota}\chi^0_{ij} \tau^0$, and  $\chi^3_{ij}$\,$=$\,$\Delta^1_{ij}=\Delta^2_{ij}$\,$=$\,$0$, i.e., the \textit{Ansatz} consists of only imaginary hoppings, the IGG is the global SU(2) gauge group $\{\exp({\dot\iota\phi\,\hat{\mathbf{n}}\cdot\rev{\bm{\tau}}})\}$. If we include real hopping terms as well, i.e., $\rev{\chi^3_{ij}\neq0}$ \rev{but} $\Delta^1_{ij}=\Delta^2_{ij}=0$, the IGG is the global U(1) gauge group $\{\exp({\dot\iota\phi\,\tau^z})\}$ while the inclusion of pairing terms breaks the IGG down to global $\mathbb{Z}_2$, $\{\pm1\}$. The three different IGGs discussed in this example only involve \textit{global} gauge elements;  these correspond to the so-called canonical structures of the \textit{Ans\"atze}. In canonical form, SU(2) \textit{Ans\"atze} contain only imaginary hopping terms, U(1) \textit{Ans\"atze} include only real and imaginary hopping amplitudes, and $\mathbb{Z}_2$  \textit{Ans\"atze} can take any generic form. However,  owing to the SU(2) gauge redundancy, the elements of the IGG need not always constitute a global gauge group, and the invariance of an \textit{Ansatz} may require the following generic local operation:
\begin{equation}    \mathcal{G}_{i}^{\dagger}u^{\pdagger}_{ij}\mathcal{G}^{\pdagger}_{j}=u^{\pdagger}_{ij},\;\mathcal{G}\in \mathrm{IGG}.
    \label{eq:IGG}
\end{equation}
Thus, in general, the IGG is a \textit{local} gauge group.

\begin{table*}
	\caption{The 46 U(1) PSG classes, characterized by different gauge-inequivalent choices of  $w^{}_\mathcal{O}$, $\bar{\phi}^{}_{\mathcal{O},u}$, and the U(1) phases $\xi_{\ldots}$. Each vector $\{\cdots\}$ lists the four values of $\bar{\phi}^{}_{\mathcal{O},u}$ for $u = 1,\ldots,4$. The parameters $p_{\ldots}$ are binary variables, which can be either 0 or 1.}
	\begin{ruledtabular}
		\begin{tabular}{cccccccc}
$w^{\pdagger}_{\mathcal{T}}$&$w^{\pdagger}_{G_x}$&$w^{\pdagger}_{\sigma_{xy}}$&$\xi^{\pdagger}_y$&$\xi^{\pdagger}_{G_{xy}}$&$\bar{\phi}^{\pdagger}_{G_{x},u}$&$\bar{\phi}^{\pdagger}_{\sigma_{xy},u}$ & \# of PSG classes\\
			\hline
$1$ & $0$ & $0$ & $0$ & $p^{}_{G_{xy}}\pi$ & $\{0,p^{}_{G_xT_x}\pi,0,p^{}_{G_xT_x}\pi\rev{-}p^{}_{G_{xy}}\pi\}$ & $\{0,0,p^{}_{\sigma_{xy}}\pi,0\}$ & $8$ \\
$1$ & $0$ & $1$ & $0$ & $p^{}_{G_{xy}}\pi$ & $\{0,p^{}_{G_xT_x}\pi,p^{}_{G_xT_x}\pi-p^{}_{G_{xy}}\pi,0\}$ & $\{0,0,p^{}_{\sigma_{xy}}\pi,0\}$ & $8$ \\
$1$ & $1$ & $0$ & $0$ & $p^{}_{G_{xy}}\pi$ & $\{0,p^{}_{G_xT_x}\pi,0,p^{}_{G_xT_x}\pi-p^{}_{G_{xy}}\pi\}$ & $\{0,0,p^{}_{\sigma_{xy}}\pi,0\}$ & $8$ \\
$1$ & $1$ & $1$ & $0$ & $p^{}_{G_{xy}}\pi$ & $\{0,p^{}_{G_xT_x}\pi,0,p^{}_{G_xT_x}\pi-p^{}_{G_{xy}}\pi\}$ & $\{0,0,p^{}_{\sigma_{xy}}\pi,0\}$ & $8$ \\
\hline
$0$ & $0$ & $0$ & $0$ & $0$ & $\{0, 0, 0, 0\}$ & $\{0,0,p^{}_{\sigma_{xy}}\pi,0\}$ & $2$\\
$0$ & $0$ & $1$ & $0$ & $0$ & $\{0, 0, 0, 0\}$ & $\{0,0,p^{}_{\sigma_{xy}}\pi,0\}$ & $2$ \\
$0$ & $1$ & $0$ & $0$ & $p^{}_{G_{xy}}\pi$ & $\{0,p^{}_{G_xT_x}\pi,0,p^{}_{G_xT_x}\pi-p^{}_{G_{xy}}\pi\}$ & $\{0,0,p^{}_{\sigma_{xy}}\pi,0\}$ & $8$ \\
$0$ & $1$ & $1$ & $-2\xi^{}_{G_{xy}}$ & $\xi^{}_{G_{xy}}$ & $\big\{0,p^{}_{G_xT_x}\pi,0,p^{}_{G_xT_x}\pi-\xi^{}_{G_{xy}}\big\}$ & $\rev{\{0,0,0,0\}}$ & $2$ \\
		\end{tabular}
	\end{ruledtabular}
	\label{table:u1_psg}
\end{table*}

\subsection{Symmetries of the Shastry-Sutherland model}
\label{sec:symmetry}

The symmetries of the Shastry-Sutherland lattice [see Fig.~\ref{fig:schematic}(b)] correspond to the $p4g $ wallpaper group. Let the positions of the lattice sites in the Cartesian coordinate system be denoted by
\begin{equation}
\mathbf{r} \equiv (x, y) \equiv (x\mathbf{\hat{x}} + y\mathbf{\hat{y}})a,
\end{equation}
where $a $ is the lattice parameter. The symmetry generators and their actions on a lattice site are then as follows~\cite{Lee-2019}:
\begin{equation}
\begin{aligned}\label{eq:ssl_generators}
    T^{}_x(x,y) & \rightarrow (x+2, y), \\
    T^{}_y(x,y) & \rightarrow (x, y+2), \\
    G^{}_x(x,y) & \rightarrow (x+1, -y), \\
    G^{}_y(x,y) & \rightarrow (-x, y+1), \\
    \sigma^{}_{xy}(x,y) & \rightarrow (y, x), \\
    \sigma^{}_{x\bar{y}}(x,y) & \rightarrow (-y+1, -x+1), \\
    C^{}_4(x,y) & \rightarrow (-y+2, x-1).
\end{aligned}
\end{equation}
Here, $T_x $ and $T_y $ stand for translations by $2a\mathbf{\hat{x}} $ and $2a\mathbf{\hat{y}} $, respectively, on the underlying Bravais (square) lattice. \rev{The operation} $G_x $ ($G_y $) \rev{is a glide reflection that combines} a translation $a\mathbf{\hat{x}} $ ($a\mathbf{\hat{y}} $) with a reflection across the $x$- ($y$-) axis. The symmetries $\sigma_{xy} $ and $\sigma_{x\bar{y}} $ denote reflections across the two diagonal axes, while $C_4 $ represents fourfold rotations about the centers of the empty squares.
Note that $T_x $, $T_y $, $G_y $, $\sigma_{x\bar{y}} $, and $C_4 $ can be expressed in terms of $G_x $ and $\sigma_{xy} $ as
\begin{equation}
\begin{aligned}
    G^{}_y & = \sigma^{}_{xy} G^{}_x \sigma_{xy}, \\
    T^{}_x & = G_x^2, \quad T^{}_y = G_y^2, \\
    \sigma^{}_{x\bar{y}} & = \sigma^{}_{xy} G^{}_x G_y^{-1}, \\
    C^{}_4 & = G^{}_x \sigma^{}_{xy} G^{}_x G_y^{-1} = G^{}_x \sigma^{}_{x\bar{y}}.
\end{aligned}
\end{equation}
Thus, $G_x $ and $\sigma_{xy} $ form a minimal set of symmetry generators for the $p4g $ wallpaper group. However, for simplicity, we include the translations $T_x $ and $T_y $ explicitly in the discussion hereafter.

Using the notation $(m, n, u) $, the position of a lattice site is expressed as
\begin{equation}\label{eq:coordinate}
\mathbf{r} \equiv (r, u) \equiv (m, n, u) \equiv 2(m\mathbf{\hat{x}} + n\mathbf{\hat{y}})a + \bm{\epsilon}^{}_u,
\end{equation}
where $u = 1, 2, 3, 4 $ labels the four distinct sites within the geometric unit cell, and
\begin{equation}
\bm{\epsilon}^{}_1 = 0, \quad \bm{\epsilon}^{}_2 = a\mathbf{\hat{x}}, \quad \bm{\epsilon}^{}_3 = a(\mathbf{\hat{x}} + \mathbf{\hat{y}}), \quad \bm{\epsilon}^{}_4 = a\mathbf{\hat{y}}.
\end{equation}
The symmetry operations, in this notation, are defined as
\begin{equation}
\begin{aligned}
    T^{}_x: (m, n, u) & \rightarrow (m+1, n, u), \\
    T^{}_y: (m, n, u) & \rightarrow (m, n+1, u), \\
    G^{}_x: (m, n, u) & \rightarrow \big(m + \delta^{}_{u,2/3}, -n - \delta^{}_{u,3/4}, G^{}_x(u)\big), \\
    \sigma^{}_{xy}: (m, n, u) & \rightarrow \big(n, m, \sigma^{}_{xy}(u)\big),
\end{aligned}
\end{equation}
where $\delta_{u,a/b}\equiv \delta_{u,a} + \delta_{u,b}$, $G_x(\{1,2,3,4\}) = \{2,1,4,3\}$, and $\sigma_{xy}(\{1,2,3,4\}) = \{1,4,3,2\}$.

In addition to the space-group symmetries, we need to consider time reversal $\mathcal{T}$ given that we are dealing with fully symmetric QSL \textit{Ans\"atze}. Being an internal symmetry, $\mathcal{T}$ leaves $(m, n, u) $ unchanged and commutes with all space-group operations. Taking $\mathcal{T}$ into account, the symmetry group of the Shastry-Sutherland lattice is then fully specified by the following identity relations:
\begin{equation}
\label{eq:id_relation}
\begin{aligned}
    T_x^{-1} T_y^{-1} T^{}_x T^{}_y & = \mathds{1}, \\
    G_x^{-1} T_x^{-1} G^{}_x T^{}_x & = \mathds{1}, \\
    G_x^{-1} \rev{T^{}_y} G^{}_x T^{}_y & = \mathds{1}, \\
    T_x^{-1} G_x^2 & = \mathds{1}, \\
    T_y^{-1} \sigma_{xy}^{-1} T_x \sigma^{}_{xy} & = \mathds{1}, \\
    T_x^{-1} \sigma_{xy}^{-1} T_y \sigma^{}_{xy} & = \mathds{1}, \\
    (\sigma^{}_{xy} G^{}_x)^4 & = \mathds{1}, \\
    \sigma_{xy}^2 & = \mathds{1}, \\
    \mathcal{T}^2 & = \mathds{1}, \\
    \mathcal{T} \mathcal{O} \mathcal{T}^{-1} \mathcal{O}^{-1} & = \mathds{1}, \quad \mathcal{O} \in \{T^{}_x, T^{}_y, G^{}_x, \sigma^{}_{xy}\}.
\end{aligned}
\end{equation}
Finally, time-reversal symmetry acts nontrivially on the mean fields, as $\mathcal{T}(u_{ij}, a_\gamma) = -(u_{ij}, a_\gamma) $~\cite{Wen-2002,Bieri-2016}. This imposes the condition
\begin{equation}
W^\dagger_{\mathcal{T}}(i) u^{\pdagger}_{ij} W^{\pdagger}_{\mathcal{T}}(j) = -u^{\pdagger}_{ij},
\end{equation}
where $W^{\pdagger}_{\mathcal{T}} \in$ SU(2)  is a site-dependent gauge transformation.

\label{sec:psg_solutions}
Since the physical symmetries act projectively in the fermionic Hilbert space, the group relations of Eq.~\eqref{eq:id_relation}, which are generically of the type
\begin{equation}
    \mathcal{O}_{1}\circ\mathcal{O}_{2}\circ\cdots=1,
\end{equation}
naturally translate to gauge-enriched group compositions of the type
\begin{equation}    \tilde{\mathcal{O}}_{1}\circ\tilde{\mathcal{O}}_{2}\circ\cdots=(W^{\pdagger}_{\mathcal{O}_{1}}\circ\mathcal{O}_{1})\circ(W^{\pdagger}_{\mathcal{O}_{2}}\circ\mathcal{O}_{2})\circ\cdots=\mathcal{G},
\end{equation}
where $\mathcal{G}\in {\rm IGG}$ is a pure gauge transformation which reduces to the identity operation for spins. The complete set of PSG solutions can be obtained using the relations in Eq.~\eqref{eq:id_relation}; the explicit gauge-enriched relations for the space group of the Shastry-Sutherland lattice are collected in App.~\ref{sec:genric_gauge_con}.

Let us first consider the case where the IGG is U(1). The generic form of a U(1) PSG solution  for any symmetry operator $\mathcal{O}$, in the canonical gauge, is given by $W_\mathcal{O}(m,n,u)=F_3(\phi^{\pdagger}_\mathcal{O}(m,n,u))(\dot{\iota}\tau^x)^{w^{\pdagger}_\mathcal{O}}$, where $F_3(\lambda) \equiv \exp({\dot{\iota}\lambda\tau^z})$, and $w^{}_\mathcal{O}$ takes the values $0$ and $1$. However, due to the consistency conditions on $G_x$ arising from the relations~\eqref{eq:id_relation}, there exist no solutions with $w^{\pdagger}_\mathcal{O}=1$ for $\mathcal{O}\in\{T_x,T_y\}$. The solutions for $\phi_\mathcal{O}(m,n,u)$ can be written as
\begin{align}
 \phi^{\pdagger}_{T_x}(m,n,u)&=\rev{n}\,\xi^{\pdagger}_y,\;\phi^{\pdagger}_{T_y}(m,n,u)=0,\label{eq:g_translation_u}\\
\phi^{\pdagger}_{G_x}(m,n,u)&=(-1)^{w^{\pdagger}_{G_x}}(n\,\xi^{\pdagger}_{G_{xy}}+m\,\xi^{\pdagger}_y\delta^{\pdagger}_{u,3/4})+\bar{\phi}^{\pdagger}_{G_x,u},\label{eq:g_G_u}\\
 \phi^{\pdagger}_{\sigma_{xy}}(m,n,u)&=mn\,\xi^{\pdagger}_{y}+\bar{\phi}^{\pdagger}_{\sigma_{xy},u},\label{eq:g_sigma_u}\\
 \phi^{\pdagger}_{\mathcal{T}}(m,n,u)&=u\pi\,\delta^{\pdagger}_{w^{}_\mathcal{T},0}
\label{eq:g_time_u},
\end{align}
with $\bar{\phi}^{\pdagger}_{\mathcal{O},u}$\,$\equiv$\,$\phi^{\pdagger}_{\mathcal{O}}(0,0,u)$; for details, refer to App.~\ref{app:u1_psg_derivation}. The different gauge-inequivalent choices of the phases $0 \le \xi_y$,\,$\xi_{G_{xy}}$,\,$\bar{\phi}_{G_x,u}$,\,$\bar{\phi}_{\rev{\sigma_{xy}},u} < 2\pi$ are listed in Table~\ref{table:u1_psg}. We find a total of $46$ U(1) solutions that can be realized on the Shastry-Sutherland lattice.

When the IGG is $\mathbb{Z}_2$, an identity can be defined up to a global sign in the canonical gauge. The solutions for the $\mathbb{Z}_2$ PSGs (for details, see App.~\ref{app:z2_psg_derivation}) can be expressed as
\begin{align}
  W^{\pdagger}_{T_x}(m,n,u)&=W^{\pdagger}_{T_y}(m,n,u)=\tau^0\ , \; \label{eq:solution_translationj_z2}\\
  W^{\pdagger}_{G_x}(m,n,u)&=\eta^n_{G_{xy}}\mathcal{W}^{\pdagger}_{G_x,u}\ , \; \label{eq:solution_glide_z2}\\
 W^{\pdagger}_{\sigma_{xy}}(m,n,u)&=\mathcal{W}^{\pdagger}_{\sigma_{xy},u}\
, \; \label{eq:solution_sigma_z2}\\
 W^{\pdagger}_{\mathcal{T}}(m,n,u)&=\mathcal{W}^{\pdagger}_{\mathcal{T},u}\, ,\label{eq:solution_time_z2}
\end{align}
where $\eta^{\pdagger}_{G_{xy}}=\pm\rev{1}$, and the matrices $\mathcal{W}_{\mathcal{O},u}=W^{\pdagger}_{\mathcal{O}}(0,0,u)$ can be read off from Table~\ref{table:z2_psg}. While there are $2^4\times5=80$ $\mathbb{Z}_2$ PSGs in total, for the $\eta^{\pdagger}_{G_x\mathcal{T}}=1$ class---in the first and fifth rows of Table~\ref{table:z2_psg}---\rev{every} mean-field amplitude vanishes \rev{identically, since $\mathcal{W}^{}_{\mathcal{T},u}\propto\tau^0$ on all four sublattices}, so we need to consider only $64$ $\mathbb{Z}_{2}$ PSGs.

\begin{table}[h]
\caption{ Different gauge-inequivalent choices of the matrices $\mathcal{W}_{\sigma_{xy}}$ and $\mathcal{W}_{\mathcal{T}}$ appearing in Eqs.~\eqref{eq:solution_sigma_z2} and \eqref{eq:solution_time_z2}. In addition, Eq.~\eqref{eq:solution_glide_z2} features a third such matrix, which is given by $\mathcal{W}_{G_x,u}=\{\tau^0,\eta_{G_xT_x}\tau^0,\tau^0,\eta_{G_xT_x}\eta_{G_{xy}}\tau^0\}$ for each row.}
\begin{ruledtabular}
\begin{tabular}{ccc}
No. &$\mathcal{W}^{\pdagger}_{\sigma_{xy},u}$&$\mathcal{W}^{\pdagger}_{\mathcal{T},u}$ \\
			\hline
1&$\{\tau^0,\tau^0,\eta^{\pdagger}_{\sigma}\tau^0,\tau^0\}$&$\{\tau^0,\eta^{\pdagger}_{G_x\mathcal{T}}\tau^0,\tau^0,\eta^{\pdagger}_{G_x\mathcal{T}}\tau^0\}$ \\
2&$\{\tau^0,\tau^0,\eta^{\pdagger}_{\sigma}\tau^0,\tau^0\}$&$\{\dot\iota\tau^y,\eta^{\pdagger}_{G_x\mathcal{T}}\dot\iota\tau^y,\dot\iota\tau^y,\eta^{\pdagger}_{G_x\mathcal{T}}\dot\iota\tau^y\}$ \\
3&$\{\dot\iota\tau^z,\dot\iota\tau^z,\eta^{\pdagger}_{\sigma}\dot\iota\tau^z,\dot\iota\tau^z\}$&$\{\dot\iota\tau^y,\eta^{\pdagger}_{G_x\mathcal{T}}\dot\iota\tau^y,\dot\iota\tau^y,\eta^{\pdagger}_{G_x\mathcal{T}}\dot\iota\tau^y\}$ \\
4&$\{\dot\iota\tau^z,\dot\iota\tau^z,\eta^{\pdagger}_{\sigma}\dot\iota\tau^z,\dot\iota\tau^z\}$&$\{\dot\iota\tau^z,\eta^{\pdagger}_{G_x\mathcal{T}}\dot\iota\tau^z,\dot\iota\tau^z,\eta^{\pdagger}_{G_x\mathcal{T}}\dot\iota\tau^z\}$ \\
5&$\{\dot\iota\tau^z,\dot\iota\tau^z,\eta^{\pdagger}_{\sigma}\dot\iota\tau^z,\dot\iota\tau^z\}$&$\{\tau^0,\eta^{\pdagger}_{G_x\mathcal{T}}\tau^0,\tau^0,\eta^{\pdagger}_{G_x\mathcal{T}}\tau^0\}$
		\end{tabular}
	\end{ruledtabular}
	\label{table:z2_psg}
\end{table}

\section{Mean-field Ans\"atze}
\label{sec:ansatz}

Having worked out the complete set of PSGs for the Shastry-Sutherland lattice, we are now in a position to derive the different realizable symmetric mean-field \textit{Ans\"atze}  from these PSG solutions. Restricting the mean-field amplitudes to square-lattice bonds (nearest neighbor, NN) and diagonal bonds (next-nearest neighbor, NNN) \emph{only}, we find a total of twelve \rev{distinct} U(1) and eighteen \(\mathbb{Z}_2\) \textit{Ans\"atze}. In addition to their PSGs, all the \textit{Ans\"atze} can be further characterized by defining SU(2) flux operators~\cite{Bieri-2016,Wen-2002}, which are associated with different loops originating from a fixed base point $i$:
\begin{align}
    P^{}_{\mathcal{C}_{i}} &= u^{}_{ij} u^{}_{jk} \cdots u^{}_{l i},
\end{align}
where $j, k, l, \ldots$ denote the sites included in the contour $\mathcal{C}_{i}$. Under a local SU(2) gauge transformation \(W(i)\), the flux operator transforms as \(P_{\mathcal{C}_{i}} \rightarrow W^\dagger(i) P_{\mathcal{C}_{i}} W(i)\), which affects all loop operators with the same base point only up to a global rotation in Pauli space. This property provides a powerful tool for diagnosing different QSL \textit{Ans\"atze}.
In general, for a \(q\)-sided loop, \(P_{\mathcal{C}_{i}}\) takes the form
\begin{equation}
    P^{\pdagger}_{\mathcal{C}_{i}}(\varphi^{\pdagger}_{\mathcal{C}_{i}}) \propto g^{\pdagger}_i e^{\rev{\dot\iota}\varphi^{}_{\mathcal{C}_{i}} \tau^z} (\tau^z)^q g_i^\dagger, \quad g^{\pdagger}_i \in \mathrm{SU(2)},
\end{equation}
where the phase \(\varphi^{}_{\mathcal{C}_{i}}\) can be interpreted as a magnetic flux threading the loop \(\mathcal{C}_{i}\).

\begin{figure}[tb]
\includegraphics[width=1.0\linewidth]{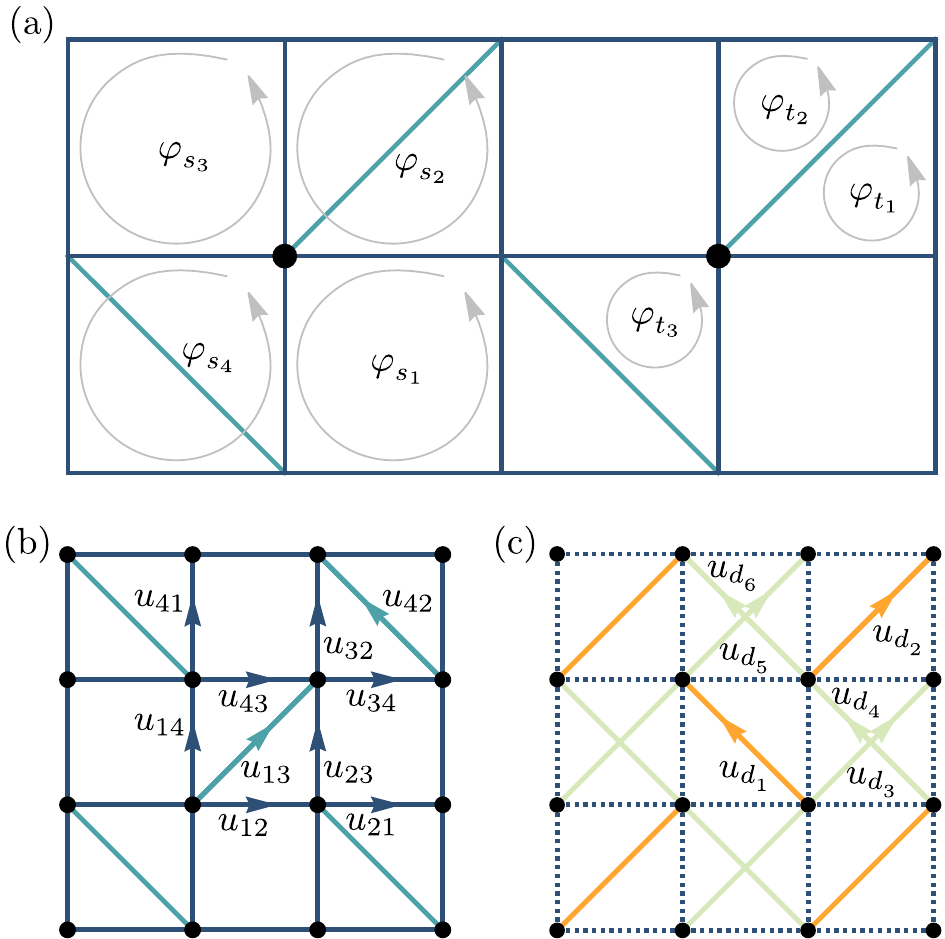}
	\caption{(a) Illustration of the different SU(2) fluxes which characterize the U(1) \textit{Ans\"atze}, defined with respect to a ``base site'' denoted by a black circle. The two such circles shown here are, in fact, equivalent. (b) Definition of mean-field amplitudes on the links of the Shastry-Sutherland lattice within a unit cell, and (c) the labeling convention for the additional diagonal bonds.}
	\label{fig:flux_bond_definition}
\end{figure}

Furthermore, note that the IGG of an \textit{Ansatz} may not be obvious if the \textit{Ansatz} is not in the canonical gauge. In that case, the disguised  IGG for an \textit{Ansatz} written in a generic gauge can be identified from the condition~\cite{Bieri-2016}:
\[
[\mathcal{G}, P_{\mathcal{C}_i}] = 0,
\]
for \(\mathcal{G} \in \text{IGG}\) and all nontrivial flux loops $\{P_{\mathcal{C}_i}\}$.
In the context of the Shastry-Sutherland lattice, starting from a base site at the origin, we identify four square loops and three triangular loops as illustrated in Fig.~\ref{fig:flux_bond_definition}(a). The associated gauge fluxes are denoted by \(\varphi_{s_1}\), \(\varphi_{s_2}\), \(\varphi_{s_3}\), \(\varphi_{s_4}\), \(\varphi_{t_1}\), \(\varphi_{t_2}\), and \(\varphi_{t_3}\), respectively.

In the following, we discuss the QSL mean-field \textit{Ans\"atze} for different IGGs. All these \textit{Ans\"atze} are defined by simply considering the reference bonds shown in Fig.~\ref{fig:flux_bond_definition}(b).

\subsection{U(1) \textit{Ans\"atze} on the Shastry-Sutherland lattice}
\begin{figure*}[tb]	\includegraphics[width=1.0\linewidth]{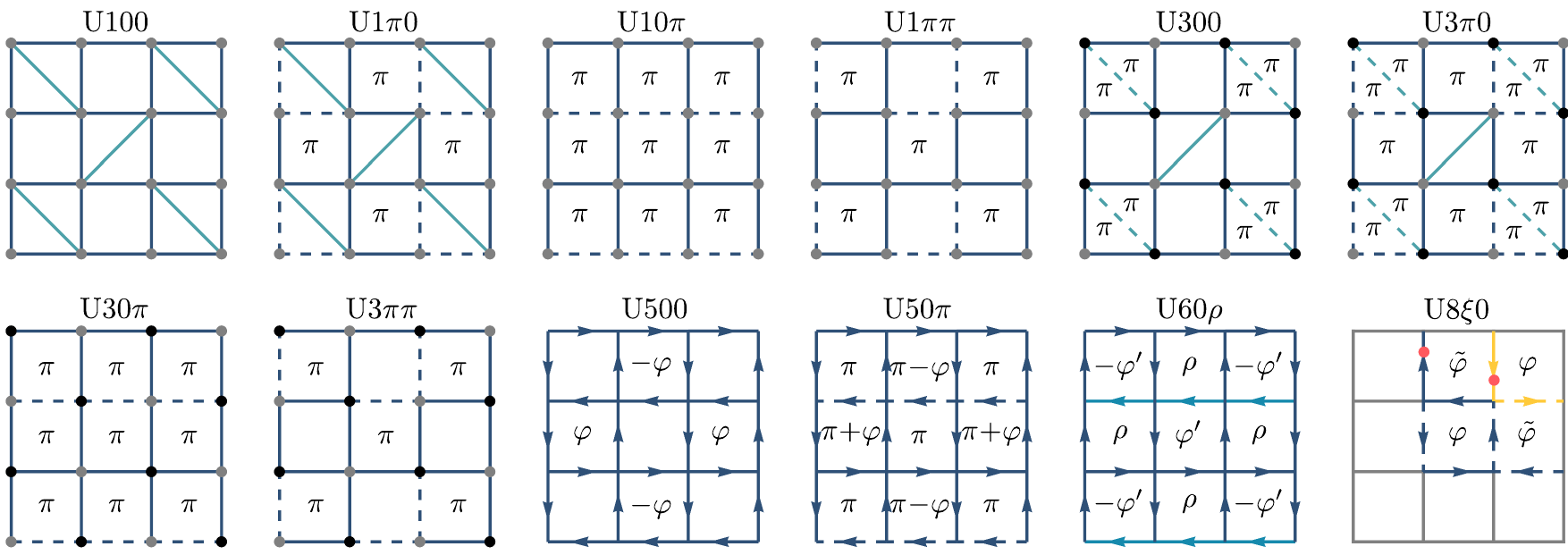}
	\caption{Sign structure and flux patterns for all U(1) \textit{Ans\"atze}. Gray (black) dots on the lattice sites denote positive (negative) signs of the onsite chemical potential. The solid (dashed) lines represent an overall positive (negative) \rev{sign} of the hoppings. Among these \textit{Ans\"atze}, the ones in the U1 and U3 classes consist of only real hoppings; the U5, U6, and U8 classes feature complex hoppings, so we label the directions of the link fields $u_{ij}$ by arrows. The link fields on dark-blue bonds are defined as $u^{}_s=\dot\iota\chi^0_s\tau^0+\chi^3_s\tau^z$. On the light-blue (in U60$\rho$) and orange (in U8$\xi$0) bonds,  the link fields are given by $F^{}_3(\rho=p_{\sigma_{xy}}\pi)u^{}_s$ and $F^{}_3(\xi=\xi_{G_{xy}})u^{}_s$, respectively, where $F_3(\lambda) \equiv \exp({\dot{\iota}\lambda\tau^z})$. For plaquettes with nonzero fluxes, we note the corresponding flux values $\varphi=\text{Arg}[u^{4}_s]$, $\varphi'=\rev{\varphi+\rho=\varphi+p^{}_{\sigma_{xy}}\pi}$, and $\tilde{\varphi}=-\varphi+\xi=-\varphi+\xi_{G_{xy}}$, where $\xi_{G_{xy}}=\mathcal{P}\pi/\mathcal{Q}$, for $\rev{\mathcal{P}\in\mathbb{Z}^{\geq0}}$ \rev{and} $\rev{\mathcal{Q}\in\mathbb{Z}^{+}}$. In the U$8\xi0$ \textit{Ansatz}, the two vertical bonds marked by a red dot are modified by a $2\times2$ U(1) matrix---specified in Eq.~\eqref{eq:uni_cell_enlarg}---under translations along $\hat{\mathbf{x}}$, while all other bonds are translation invariant.}
	\label{fig:fig2}
\end{figure*}

To begin, we consider the different \textit{Ans\"atze} realizable with  IGG U(1). In this family, we \rev{divide} the U(1) PSG solutions into eight different classes, denoted as U1, U2, $\dots$, U8, corresponding to the eight rows in Table~\ref{table:u1_psg}. The states are labeled by their respective classes and the parameters of the PSG as  ``$\text{Class }\xi_{G_{xy}}\bar{\phi}_{\sigma_{xy}}$'', where $\bar{\phi}_{\sigma_{xy}}$\,$=$\,$p_{\sigma_{xy}}\pi$. The flux patterns of all the U(1) \textit{Ans\"atze} are depicted in Fig.~\ref{fig:fig2}.

The class U1 yields four distinct \textit{Ans\"atze} labeled as U100, U1$\pi$0, U10$\pi$, and U1$\pi\pi$. Notably, for U10$\pi$, the mean-field amplitudes on the diagonal bonds vanish due to symmetry, and a uniform $\pi$-flux threads all the plaquettes, corresponding to a QSL with an SU(2) IGG realized on the square lattice~\cite{Wen-2002}.

Next, the four \textit{Ans\"atze}, labeled as U300, U3$\pi$0, U30$\pi$, and U3$\pi\pi$, belong to the class U3. The symmetry-allowed mean fields for U30$\pi$ and U3$\pi\pi$ are identical to those of U10$\pi$ and U1$\pi\pi$, respectively. However, the U3 class has a staggered chemical potential, whereas the chemical potential is uniform for the U1 class. Both the U1 and U3 classes of \textit{Ans\"atze} consist of real hopping terms only.

Additionally, we identify a few more classes of U(1) \textit{Ans\"atze}, labeled as U500, U50$\pi$, U600, U60$\pi$, and U8$\xi$\rev{0}. In these cases, the mean-field amplitudes vanish on the diagonal bonds, as does the onsite potential, while the square bonds include both real and imaginary hopping terms. Note that for the U5 class \textit{Ans\"atze}, a nontrivial flux, i.e., $\ne 0, \pi$\rev{,} threads the empty squares of the Shastry-Sutherland lattice, while for the U6 class a nontrivial flux threads those squares of the Shastry-Sutherland lattice which would have otherwise featured a diagonal bond. Up to NN bonds, these two classes of \textit{Ans\"atze} are equivalent.

The U8$\xi0$ class (with $\xi$\,$\equiv$\,$\xi_{G_{xy}}$), in particular, incorporates a series of mean-field \textit{Ans\"atze}. The explicit structure of the mean-field amplitudes in this case is given by
\begin{align}
    u^{\pdagger}_s &= \rev{\dot\iota}\chi^0_s\tau^0 + \chi^3_s\tau^z = u^{\pdagger}_{12} = -u^\dagger_{14} = -u^{\pdagger}_{23} \notag \\
    &= u^\dagger_{43} = -u^\dagger_{21} = u^{\pdagger}_{41} = F^{\pdagger}_3(\xi)u^\dagger_{32} = -F^{\pdagger}_3(\xi)u^{\pdagger}_{34}.
\end{align}
The bonds \(u_{41}\) and \(u_{32}\) are modified under a translation \(T_x\), while all other bonds remain invariant:
\begin{equation}\label{eq:uni_cell_enlarg}
\left.\begin{aligned}
&u^{\pdagger}_{41}(\rev{m,n}) = u^{\pdagger}_{(m,n,4),(m,n+1,1)} = F^{\pdagger}_3(2\rev{m}\xi)u^{\pdagger}_{41}, \\
&u^{\pdagger}_{32}(\rev{m,n}) = u^{\pdagger}_{(m,n,3),(m,n+1,2)} = F^{\pdagger}_3(2\rev{m}\xi)u^{\pdagger}_{32}.
\end{aligned}\right.
\end{equation}
To realize such an \textit{Ansatz} on a finite lattice, one must impose the constraint
\begin{equation}\label{eq:u8ansatze}
\xi = \xi^{\pdagger}_{G_{xy}} = \frac{\mathcal{P}}{\mathcal{Q}}\pi, \quad \text{with } \rev{\mathcal{P} \in \mathbb{Z}^{\geq 0},\; \mathcal{Q} \in \mathbb{Z}^{+}}.
\end{equation}
Thus, for each choice $\mathcal{P}/\mathcal{Q}\in\rev{[0,1]}$, Eq.~\eqref{eq:u8ansatze} provides a realization of at least one fully symmetric mean-field QSL within this PSG class, and in total, there \rev{exists} an infinite number of such QSLs~\cite{Wen-2002}.

Lastly, we note that U(1) \textit{Ans\"atze} in classes U2, U4, and U7 cannot be realized \rev{with the amplitudes restricted to} NNN bonds.

\begin{table}
\caption{Mean-field parameters on the reference bonds, $u_s$ and $u_d$, for all possible symmetric $\mathbb{Z}_2$ \textit{Ans\"atze} on the Shastry-Sutherland lattice. \rev{Each entry lists the Pauli matrices carrying a nonvanishing coefficient; for instance, $\tau^{z,x}$ indicates that only $\chi^{3}$ and $\Delta^{1}$ are nonzero, while $0$ denotes a bond field that vanishes identically.}}
\begin{ruledtabular}
\begin{tabular}{ccccc}
Label & $u^{}_s$ & $u^{}_d$ & Onsite & Parent U(1)\\
			\hline
Z2000&$\tau^{z,x}$& $\tau^{z,x}$ & $\tau^{z}$ & U1$00$, U3$00$\\
Z2100&$\tau^{z,x}$& $\tau^{z,x}$ & $\tau^{z}$ & U1$\pi0$, U3$\pi0$\\
Z2001&$\tau^{0,y}$& $\tau^{z,x}$ & $\tau^{z}$ & U1$00$, U3$00$, U5$00$\\
Z2101&$\tau^{0,y}$& $\tau^{z,x}$ & $\tau^{z}$ & U1$\pi0$, U3$\pi0$\\
Z2011&$\tau^{0,y}$& $0$ & $\tau^{z}$ & U1$0\pi$, U3$0\pi$, U5$0\pi$\\
Z2111&$\tau^{0,y}$& $0$ & $\tau^{z}$ & U1$\pi\pi$, U3$\pi\pi$\\
\hline
Z3000&$\tau^{\rev{z,x}}$& $\tau^{z}$ & $\tau^{z}$ & U1$00$, U3$00$, U8$00$\\
Z3100&$\tau^{\rev{z,x}}$& $\tau^{z}$ & $\tau^{z}$ & U1$\pi0$, U3$\pi0$, U8$\pi0$\\
Z3010&$\tau^{\rev{z,x}}$& $0$ & $\tau^{z}$ & U1$0\pi$, U3$0\pi$, U8$00$\\
Z3110&$\tau^{\rev{z,x}}$& $0$ & $\tau^{z}$ & U1$\pi\pi$, U3$\pi\pi$, U8$\pi0$\\
Z3001&$\tau^{0,y}$& $\tau^{z}$ & $\tau^{z}$ & U1$00$, U3$00$, U6$00$\\
Z3101&$\tau^{0,y}$& $\tau^{z}$ & $\tau^{z}$ & U1$\pi0$, U3$\pi0$\\
Z3011&$\tau^{0,y}$& $0$ & $\tau^{z}$ & U1$0\pi$, U3$0\pi$, U6$0\pi$\\
Z3111&$\tau^{0,y}$& $0$ & $\tau^{z}$ & U1$\pi\pi$, U3$\pi\pi$\\
\hline
Z5001&$\tau^{0,x,z}$& $0$ & $0$ & U5$00$, U6$00$, U8$00$\\
Z5101&$\tau^{0,x,z}$& $0$ & $0$ & U8$\pi0$\\
Z5011&$\tau^{0,x,z}$& $0$ & $0$ & U5$0\pi$, U6$0\pi$, U8$00$\\
Z5111&$\tau^{0,x,z}$& $0$ & $0$ & U8$\pi0$\\
		\end{tabular}
	\end{ruledtabular}
	\label{table:z2_ansatze}
\end{table}

\subsection{$\mathbb{Z}_2$ \textit{Ans\"atze} on the Shastry-Sutherland lattice}

Although there exist a total of 64 projective extensions of the symmetry group with IGG $\mathbb{Z}_2$, only 18 of these can be realized when the \textit{Ans\"atze} \rev{are restricted} to NNN mean-field amplitudes. The symmetric $\mathbb{Z}_2$ \textit{Ans\"atze} have the following structure:
\begin{align}
&u^{}_s = \dot\iota\chi^0_{s}\tau^0 + \Delta^1_{s}\tau^x + \Delta^2_{s}\tau^y + \chi^3_{s}\tau^z, \\
&u^{\pdagger}_{12} = \eta^{\pdagger}_{\sigma} u^\dagger_{43} = \eta^{\pdagger}_{G_xT_x}u^{\pdagger}_{21} = \eta^{\pdagger}_{G_xT_x}\eta^{\pdagger}_{\sigma}\eta^{\pdagger}_{G_{xy}}u^\dagger_{34} = u^{\pdagger}_s, \\
&u^{\pdagger}_{14} = u^\dagger_{23} = \eta^{\pdagger}_{G_xT_x}u^{\pdagger}_{41} = \eta^{\pdagger}_{G_xT_x}\eta^{\pdagger}_{G_{xy}}u^\dagger_{32} = (\tau^z)^p u^{\pdagger}_s (\tau^z)^p, \\
&u^{\pdagger}_{13} = u^{\pdagger}_{42} = u^{\pdagger}_d = \dot\iota\chi^0_{d}\tau^0 + \Delta^1_{d}\tau^x + \Delta^2_{d}\tau^y + \chi^3_{d}\tau^z, \\
&\rev{\sum_\gamma} a^{\pdagger}_\gamma(i)\tau^\gamma = a^{\pdagger}_{\rev{x}}\tau^x + a^{\pdagger}_{\rev{y}}\tau^y + a^{\pdagger}_{\rev{z}}\tau^z;
\end{align}
here, $p \equiv 0(1)$ for $\mathcal{W}_{\sigma_{xy},u}$\,$\propto$\,$\tau^{0(\rev{z})}$. In the equations above, the sign parameter $\eta_{G_xT_x}$ is irrelevant because\rev{,} up to NNNs\rev{,} \textit{Ans\"atze} with positive and negative $\eta_{G_xT_x}$ are related by a gauge transformation.

We label the \textit{Ans\"atze} using the scheme
\begin{equation}
\text{Z}\;\text{PSG}_\text{row}
\left(\frac{1-\eta^{}_{G_{xy}}}{2}\right)
\left(\frac{1-\eta^{}_{\sigma}}{2}\right)
\left(\frac{1-\eta^{}_{G_{x}\mathcal{T}}}{2}\right),
\end{equation}
where $\text{PSG}_\text{row}$ corresponds to the row number in Table~\ref{table:z2_psg}.

\begin{figure}[tb]	\includegraphics[width=1.0\linewidth]{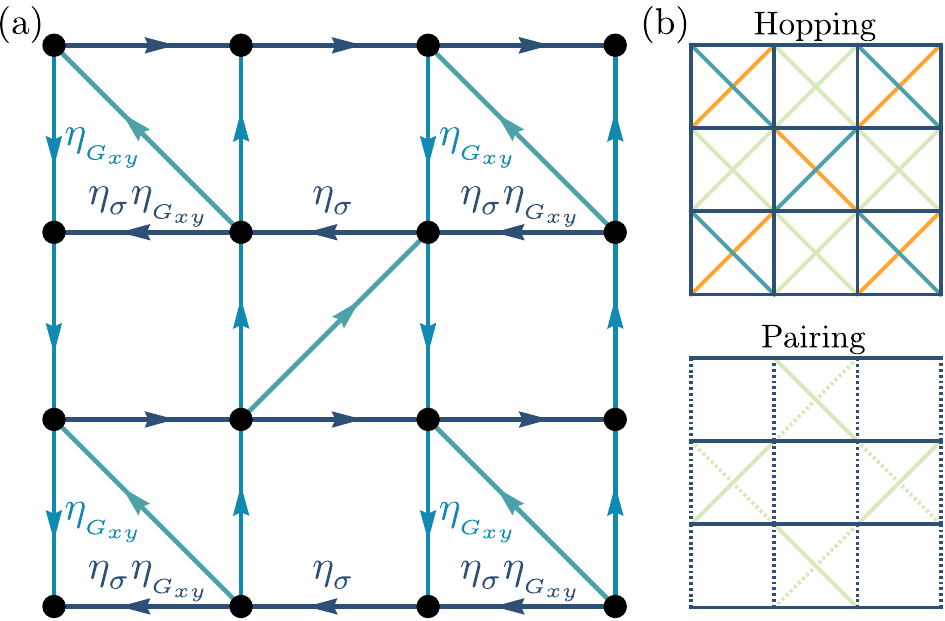}
	\caption{(a) Graphical representation of $\mathbb{Z}_2$ \textit{Ans\"atze}. Here, the link fields on the dark-blue (horizontal) bonds are $u^{}_s$, and the link fields $\Tilde{u}^{}_s$ on the light-blue (vertical) bonds are related to $u^{}_s$ as $\Tilde{u}^{}_s$\,$=$\,$(\tau^z)^pu^{}_s(\tau^z)^p$, where $p$\,$=$\,$0(1)$ for $\mathcal{W}^{}_{\sigma_{xy},u}\propto\tau^{0(z)}$. The teal (diagonal) bonds carry a link field $u^{}_d$. All the \textit{Ans\"atze} are translationally invariant. (b) Structure of the QSL \textit{Ansatz} Z3000. The top figure conveys the signs of the hopping fields with the hopping amplitudes on the blue, teal, orange, and green bonds given by $\chi^{}_{s}$, $\chi^{}_{d}$, $\chi^{}_{d'}$, and $\chi^{}_{\square}$, respectively. The bottom figure shows the sign structure of the pairing field, which has $d$-wave symmetry, and the pairing amplitudes on the blue and green bonds are given by $\Delta^{}_{s}$ and $\Delta^{}_{\square}$, respectively. Dashed bonds denote a negative sign for the pairing.}
	\label{fig:fig4}
\end{figure}

Upon restricting ourselves to only $J_{s}$ and $J_{d}$ bonds, a total of 18 different symmetric \textit{Ans\"atze} with IGG $\mathbb{Z}_2$ can be realized---these are listed in Table~\ref{table:z2_ansatze}. They are also graphically represented in Fig.~\ref{fig:fig4}(a), \rev{where the} parameter $\eta_{G_xT_x}$ \rev{is omitted} for simplicity. However, this parameter does become important when establishing a connection of our PSGs with those on the square lattice, as discussed in the following section.

\subsection{Mapping to square-lattice $\mathbb{Z}_2$ \textit{Ans\"atze}}
\label{sec:mapping_square_lattice}

In this section, we show how the $\mathbb{Z}_2$ QSL \textit{Ans\"atze} on the Shastry-Sutherland lattice can be continuously connected to $\mathbb{Z}_2$ \textit{Ans\"atze} realizable on the square lattice~\cite{Wen-2002}.

The square lattice has a higher symmetry than the Shastry-Sutherland \rev{lattice}.
The wallpaper group of the square lattice is $p4m$, which is generated by two translations \rev{$T_1$ and $T_2$ by $a\mathbf{\hat{x}}$ and $a\mathbf{\hat{y}}$}, respectively, and three reflections $\sigma_{xy}$, $\sigma_x$, and $\sigma_y$ about the $x=y$, $y=0$, and $x=0$ lines, respectively. As we show here, including $\sigma_x$ transforms the wallpaper group $p4g$ of the Shastry-Sutherland lattice into $p4m$. The reflection $\sigma_x$ transforms the coordinate $(x, y)$ to $(x, -y)$ and acts on $(m, n, u)$ as
\begin{equation}
    \label{eq:sig_x}
    \sigma^{}_x: (m, n, u) \rightarrow (m, -n - \delta^{}_{u,3/4}, u).
\end{equation}
One can verify that the other generators of the square lattice can be obtained as follows:
\begin{equation}\label{eq:square_gen}
	\left.\begin{aligned}
        &T^{}_1 = G^{}_x \sigma^{}_x,\\
        &T^{}_2 = \sigma^{}_{xy} G^{}_x \sigma^{}_x \sigma^{}_{xy} = \sigma^{}_{xy} T^{}_1 \sigma^{}_{xy},\\
        &\sigma^{}_y = \sigma^{}_{xy} \sigma^{}_x \sigma^{}_{xy}.
	\end{aligned}\right.
\end{equation}
Hence, the wallpaper group $p4m$ is generated by $\{G_x, \sigma_{xy}, \sigma_x\}$. Using this symmetry group, we establish a mapping from the Shastry-Sutherland lattice PSGs to the square-lattice PSGs in App.~\ref{app:mapping_square_lattice}. We find that upon including $\sigma_x$, the \textit{Ans\"atze} belonging to the $\eta_{G_{xy}} = -1$ class have to be excluded as we are incorporating a larger set of symmetries. Note that this class is also the one which yields $(0, \pi, 0, \pi)$- and $(\pi, 0, \pi, 0)$-flux phases in the Shastry-Sutherland lattice, while such staggered-flux phases are forbidden on the square lattice~\cite{Wen-2002}. \rev{In all}, we arrive at the 272 PSGs found in Ref.~\cite{Wen-2002}.

Next, we discuss how the $\mathbb{Z}_2$ \textit{Ans\"atze} in the class $\eta_{G_{xy}} = +1$ can be mapped to those on the square lattice. To do so, we first implement the geometric mapping between the two lattices as illustrated in Fig.~\ref{fig:schematic}(c). Beginning with the Shastry-Sutherland lattice, we have to include two additional symmetry-inequivalent diagonal interactions, denoted by orange and light\rev{-}green bonds\rev{, with coupling strengths $J'_d$ and $J_{\square}$,} respectively.  This introduces six new mean-field amplitudes $u_{d_1}, \ldots, u_{d_6}$, as shown in Fig.~\ref{fig:flux_bond_definition}(c).
The full \textit{Ansatz} can, in fact, be defined in terms of four fields: $u_s$, on the NN square bonds; $u_d$, on the diagonal bonds already present on the Shastry-Sutherland lattice; $u_{d'}$, on the other diagonal bonds in the same set of plaquettes as $u_d$; and $u_{\square}$, on \rev{both diagonals of} the empty Shastry-Sutherland plaquettes. The structure of the $\mathbb{Z}_2$ \textit{Ans\"atze} with $p4g$ symmetry on these reference bonds \rev{is} then given by
\begin{align}
&u^{}_{d'} = u^{}_{d_1} = u^{}_{d_2} =  \dot\iota \chi^0_{d'} \tau^0 + \Delta^1_{d'} \tau^x + \Delta^2_{d'} \tau^y + \chi^3_{d'} \tau^z,\label{eq:extended_ansatze_1}\\
&u^{}_{\square} = u^{}_{d_3} = (\tau^z)^p u^{}_{d_4} (\tau^z)^p = (\tau^z)^p u^{}_{d_5} (\tau^z)^p = u^{}_{d_6},\label{eq:extended_ansatze_2}
\end{align}
where
\begin{equation}\label{eq:extended_ansatze_3}
    u^{}_{\square} \equiv \dot\iota \chi^0_{\square} \tau^0 + \Delta^1_{\square} \tau^x + \Delta^2_{\square} \tau^y + \chi^3_{\square} \tau^z.
\end{equation}
The mean-field parameters $u^{}_s$, $u^{}_d$, $u^{}_{d'}$, and $u^{}_{\square}$ are listed in Table~\ref{table:z2_ansatze_extended}. Note that\rev{,} due to the inclusion of \emph{all} diagonal bonds, the different signs of $\eta_{G_xT_x}$ are no longer related by gauge transformations.

\begin{table}
\caption{ Extension of all the symmetric $\mathbb{Z}_2$ \textit{Ans\"atze} listed in Table~\ref{table:z2_ansatze} to the other diagonals---that were not originally included in the Shastry-Sutherland lattice---on which the reference bonds are denoted as $u^{}_{d'}$ and $u^{}_{\square}$ [see Eqs.~\eqref{eq:extended_ansatze_1}--\eqref{eq:extended_ansatze_3}]. Here, $u_{d'}$ corresponds to the diagonal bonds in the same set of plaquettes as $u_d$, and $u_{\square}$ represents the diagonal bonds in the empty Shastry-Sutherland plaquettes.}
\begin{ruledtabular}
\begin{tabular}{cccccc}
Label & $u^{}_s$ & $u^{}_d$ & $u^{}_{d'}$ &$u^{}_{\square}$ & Onsite\\
			\hline
Z2000&$\tau^{z,x}$& $\tau^{z,x}$ & $\tau^{z,x}$ & $\tau^{z,x}$ & $\tau^{z}$ \\
Z2100&$\tau^{z,x}$& $\tau^{z,x}$ &$\tau^{z,x}$ & $0$ & $\tau^{z}$ \\
Z2001&$\tau^{0,y}$& $\tau^{z,x}$ & $\tau^{z,x}$ & $\tau^{z,x}$ & $\tau^{z}$ \\
Z2101&$\tau^{0,y}$& $\tau^{z,x}$ & $\tau^{z,x}$ & $0$ & $\tau^{z}$ \\
Z2011&$\tau^{0,y}$& $0$ & $0$ & $0$ & $\tau^{z}$ \\
Z2111&$\tau^{0,y}$& $0$ & $0$ & $\tau^{z,x}$ & $\tau^{z}$ \\
\hline
Z3000&$\tau^{\rev{z,x}}$& $\tau^{z}$ & $\tau^{z}$ & $\tau^{z,x}$ & $\tau^{z}$ \\
Z3100&$\tau^{\rev{z,x}}$& $\tau^{z}$ & $\tau^{z}$ & $0$ & $\tau^{z}$ \\
Z3010&$\tau^{\rev{z,x}}$& $0$ & $0$ & $0$ & $\tau^{z}$ \\
Z3110&$\tau^{\rev{z,x}}$& $0$ & $0$ & $\tau^{z,x}$ & $\tau^{z}$ \\
Z3001&$\tau^{0,y}$& $\tau^{z}$ & $\tau^{z}$ & $\tau^{z,x}$ & $\tau^{z}$ \\
Z3101&$\tau^{0,y}$& $\tau^{z}$ & $\tau^{z}$ & $0$ & $\tau^{z}$ \\
Z3011&$\tau^{0,y}$& $0$ & $0$ & $0$ & $\tau^{z}$ \\
Z3111&$\tau^{0,y}$& $0$ & $0$ & $\tau^{z,x}$ & $\tau^{z}$ \\
\hline
Z5001&$\tau^{0,x,z}$& $0$ & $0$ & $0$ & $0$\\
Z5101&$\tau^{0,x,z}$& $0$ & $0$ & $0$ & $0$\\
Z5011&$\tau^{0,x,z}$& $0$ & $0$ & $0$ & $0$\\
Z5111&$\tau^{0,x,z}$& $0$ & $0$ & $0$ & $0$\\
		\end{tabular}
	\end{ruledtabular}
	\label{table:z2_ansatze_extended}
\end{table}

In general, $u^{}_d$, $u^{}_{d'}$, and $u^{}_{\square}$ are not related by symmetry. However, in the square-lattice limit (after including $\sigma_x$), i.e., $J_d = J'_d = J_{\square}$, all diagonal bonds do become symmetry equivalent. This allows us to construct $\mathbb{Z}_2$ \textit{Ans\"atze} on the square lattice starting from the ones on the Shastry-Sutherland lattice. For example, consider the \textit{Ansatz} labeled Z3000 and PSG 3(c) with $\eta_{G_xT_x} = -\eta_{\sigma_xG_x} = +1$ in Table~\ref{table:z2_psg_square_1}. After imposing the constraints associated with $\sigma_x$, we obtain the following \textit{Ansatz}:
\begin{subequations}
\begin{align}
&u^{}_{12} = u^{}_{21} = u^{}_{43} = u^{}_{34} = \chi^{}_s \tau^z + \Delta^{}_s \tau^x,\\
&u^{}_{14} = u^{}_{41} = u^{}_{23} = u^{}_{32} = \chi^{}_s \tau^z - \Delta^{}_s \tau^x,\\
&u^{}_{13} = u^{}_{42} = u^{}_{d_1} = u^{}_{d_2} = \chi^{}_d \tau^z,\; a^{}_\gamma(u) = 0,\\
&u^{}_{d_3} = u^{}_{d_4} = u^{}_{d_5} = u^{}_{d_6} = -\chi^{}_d \tau^z.
\end{align}
\end{subequations}
It is straightforward to see that after a gauge transformation $W(m, n, u) = \{\dot\iota \tau^y, -1, -\dot\iota \tau^y, 1\}$, this can be re-expressed as \rev{(note that this transformation rotates the nearest-neighbor link field by $\pi/2$ in the $(\tau^z,\tau^x)$ plane, so that the amplitudes multiplying $\tau^z$ and $\tau^x$ are interchanged on those bonds, while the diagonal bonds are unaffected)}:
\begin{subequations}
\begin{align}
&u^{}_{12} = u^{}_{21} = u^{}_{43} = u^{}_{34} = \rev{\Delta}^{}_{\rev{s}} \tau^z - \rev{\chi}^{}_{\rev{s}} \tau^x,\\
&u^{}_{14} = u^{}_{41} = u^{}_{23} = u^{}_{32} = \rev{\Delta}^{}_{\rev{s}} \tau^z + \rev{\chi}^{}_{\rev{s}} \tau^x,\\
&u^{}_{13} = u^{}_{d_2} = u^{}_{d_3} = u^{}_{d_5} = \chi^{}_d \tau^z,\; a^{}_\gamma(u) = 0,\\
&u^{}_{42} = u^{}_{d_1} = u^{}_{d_4} = u^{}_{d_6} = -\chi^{}_d \tau^z.
\end{align}
\end{subequations}
In the canonical square-lattice Cartesian notation, it can be written as
\begin{subequations}
\begin{align}
u^{}_{\mathbf{\hat{x}}} &= \rev{\Delta}^{}_{\rev{s}} \tau^z - \rev{\chi}^{}_{\rev{s}} \tau^x,\; u^{}_{\mathbf{\hat{y}}} = \rev{\Delta}^{}_{\rev{s}} \tau^z + \rev{\chi}^{}_{\rev{s}} \tau^x,\\
u^{}_{\mathbf{\hat{x}}+\mathbf{\hat{y}}} &= \chi^{}_d \tau^z,\; u^{}_{\mathbf{-\hat{x}}+\mathbf{\hat{y}}} = -\chi^{}_d \tau^z,\; a^{}_\gamma(u) = 0.
\end{align}
\end{subequations}
This corresponds to the Z2A$zz$13 \textit{Ansatz} in the PSG classification by \citet{Wen-2002}. Thus, the $\mathbb{Z}_2$ \textit{Ansatz} Z3000 on the Shastry-Sutherland lattice can be continuously connected to the $\mathbb{Z}_2$ \textit{Ansatz} Z2A$zz$13 on the square lattice. This connection is protected by the projective realization of $G_x$ and $\sigma_{xy}$. When considering up to NNN amplitudes, among the $\mathbb{Z}_2$ \textit{Ans\"atze} listed in this work, five more can be realized with IGG $\mathbb{Z}_2$, which are labeled \rev{Z2A0013, Z2B0013, Z2A$xx$0$z$, Z2A$x$2(12)$n$, and Z2B$x$2(12)$n$}. Their connections to the square lattice can be established following a procedure similar to that carried out above for the Z2A$zz$13 \textit{Ansatz}, as summarized in Table~\ref{table:z2_ansatze_square_shastry}.

\begin{table}
\caption{Mapping between the $p4g$-symmetric $\mathbb{Z}_2$ \textit{Ans\"atze} (left column) and $p4m$-symmetric $\mathbb{Z}_2$ \textit{Ans\"atze} (right column). The second and third columns tabulate the PSG parameters (see App.~\ref{app:mapping_square_lattice}) that make the connection between the two sets of \textit{Ans\"atze}.}
\begin{ruledtabular}
\begin{tabular}{cccc}
$\mathbb{Z}^{}_2$ \textit{Ansatz} ($p4g$) & PSG & $(\eta^{}_{G_xT_x},\eta^{}_{\sigma_xG_x})$ & $\mathbb{Z}^{}_2$ \textit{Ansatz} ($p4m$)\\
			\hline
Z3000& \rev{3(a)} & $(+,+)$ & \rev{Z2A0013} \\
Z3000& \rev{3(c)} & $(+,-)$ & \rev{Z2A$zz$13} \\
Z3010& \rev{3(a)} & $(+,+)$ & \rev{Z2B0013} \\
Z2000& \rev{2(b)} & $(+,-)$ & \rev{Z2A$xx$0$z$} \\
Z5011& \rev{5(d)} & $(-,+)$ & \rev{Z2A$x$2(12)$n$} \\
Z5001& \rev{5(d)} & $(-,+)$ & \rev{Z2B$x$2(12)$n$} \\
		\end{tabular}
	\end{ruledtabular}
	\label{table:z2_ansatze_square_shastry}
\end{table}

\section{Numerical Results}
\label{sec:numerics}

Unless stated otherwise, all the results of this section and of Sec.~\ref{sec:phasediagram} refer to the pure Shastry-Sutherland model, i.e., to $\mathcal{J}\equiv J'_{d}=J^{}_{\square}=0$ in Eq.~\eqref{eq:mod2}, so that $J^{}_{s}/J^{}_{d}$ is the only tuning parameter. The single exception is Fig.~\ref{fig:exponent_river}, where $\mathcal{J}$ is varied in order to follow the river of liquidity toward the $J_{1}$--$J_{2}$ limit; there, the value of $\mathcal{J}/J^{}_{s}$ is quoted explicitly for every data point.

\subsection{$\mathbb{Z}_{2}$ Dirac state: Mean-field properties}
\label{sec:mfn}

Given that the gapless Z3000 state serves as the most promising candidate to describe the QSL in the generalized SSM (as we will show below), we now \rev{present} its corresponding variational wavefunction in terms of Abrikosov pseudo-fermions~\cite{Abrikosov-1965}. First, we define a mean-field state in the fermionic Hilbert space by taking the ground state $|\phi^{\pdagger}_{\rm MF}\rangle$ of the following noninteracting spinon Hamiltonian, which
is of the generalized Bardeen-Cooper-Schrieffer form:
\begin{align}
{\cal \hat{H}}&^{\text{Z3000}}_{{\rm MF}} =
\chi^{\pdagger}_{d}\hspace*{-0.2cm}\sum_{\langle ij\rangle_\text{teal},\alpha}{\rm s}^{d}_{ij} \hat{f}_{i,\alpha}^{\dagger}\hat{f}^{\pdagger}_{j,\alpha}+
\chi^{\pdagger}_{d'}\hspace*{-0.2cm}\sum_{\langle ij\rangle_\text{orange},\alpha}{\rm s}^{d'}_{ij} \hat{f}_{i,\alpha}^{\dagger}\hat{f}^{\pdagger}_{j,\alpha}\nonumber\\
&+
\sum_{\langle ij\rangle_\text{blue}}\,\,\left[\chi^{\pdagger}_{s}\sum_\alpha{\rm s}^{s}_{ij} \hat{f}_{i,\alpha}^{\dagger}\hat{f}^{\pdagger}_{j,\alpha}+\Delta^{\pdagger}_{s}{\nu}^{s}_{ij} \big(\hat{f}_{i,\uparrow}^{\dagger}\hat{f}^{\dagger}_{j,\downarrow}+{\rm h.c.}\big)\right] \nonumber \\
&+
\sum_{\langle ij\rangle_\text{green}}\left[\chi^{\pdagger}_{\square}\sum_\alpha{\rm s}^{\square}_{ij} \hat{f}_{i,\alpha}^{\dagger}\hat{f}^{\pdagger}_{j,\alpha}+\Delta^{\pdagger}_{\square}{\nu}^{\square}_{ij} \big(\hat{f}_{i,\uparrow}^{\dagger}\hat{f}^{\dagger}_{j,\downarrow}+{\rm h.c.}\big)\right] \nonumber \\
&+\mu\sum_{i,\alpha}\hat{f}_{i,\alpha}^{\dagger}\hat{f}^{\pdagger}_{i,\alpha}\,.
\label{eq:MF-Z2}
\end{align}
Here, ${\rm s}_{ij}$ and $\nu_{ij}$ are sign factors for hoppings and pairings, respectively, which take values $+1 (-1)$ for bonds marked as solid (dashed) in \rev{Fig.}~\ref{fig:fig4}(b)\rev{; all the bond sums run over ordered pairs $(i,j)$, so that every term above is Hermitian and the pairing is a spin singlet}. The different colors in Fig.~\ref{fig:fig4}(b) label the symmetry-inequivalent bonds, as referenced in Eq.~\eqref{eq:MF-Z2}. We fix $\chi_{s}=1$ as the overall energy scale, and the resulting wavefunction is thus parametrized by six variational parameters:  $\chi^{}_d$, $\chi^{}_{d'}$, $\chi^{}_{\square}$, $\Delta^{}_s$, $\Delta^{}_{\square}$, and the onsite chemical potential, $\mu$.

 \begin{figure}[t]
 \includegraphics[width=\linewidth]{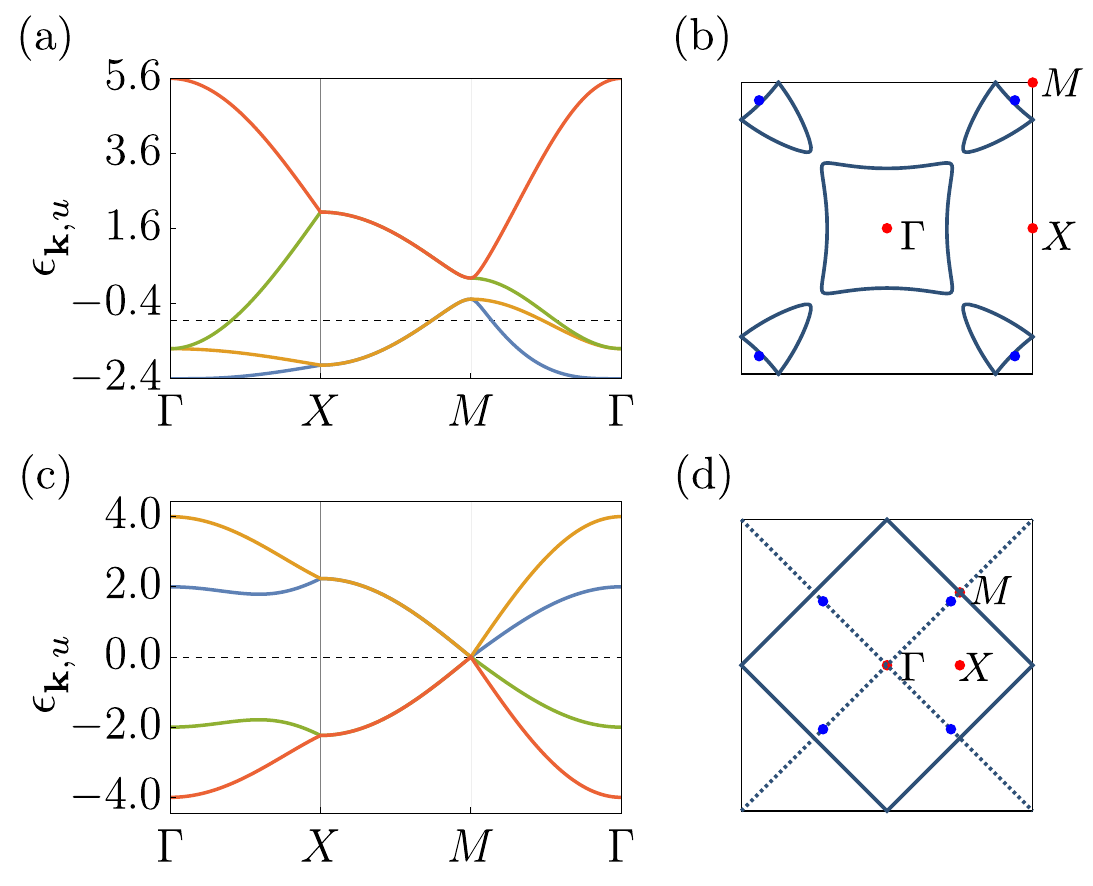}
	\caption{(a) Dispersion of the parent U100 state and (b) illustration of its Fermi surfaces.  The parameters here are chosen to be the VMC-optimized parameters at $J_s/J_d=0.80$ on a $14\times14$ cluster, given by $\chi^{}_s=1.0$, $\chi^{}_d=0.53$, $\chi^{}_{d'}=0.13$, and $\chi^{}_{\square}=0.47$. The dashed line in the band structure marks the Fermi level. The blue dots in (b) are the VMC\rev{-}optimized \rev{locations of the} Dirac nodes of the Z3000 state. (c) The dispersion of another parent state of Z3000---namely, U800---in the gauge with uniform hopping $\chi^{}_s=1.0$ and a $d^{}_{x^2-y^2}$ pairing which is set to $\Delta_s=0.5$. In (d), solid  and dashed lines indicate the zero-energy contours of $\chi^{}_s$ and the nodal lines of $\Delta_s$ ($d$-wave pairing) terms, respectively. The crossings at $M(\pm{\pi}/{2},\pm{\pi}/{2})$ give the positions of the Dirac nodes. The blue dots indicate the locations of the Dirac nodes in the Z3000 state resulting from $\Delta_{\square}$ pairing and remaining hopping amplitudes.}
	\label{fig:u100_dis_fermi}
\end{figure}

 \begin{figure*}
 \includegraphics[width=1.0\linewidth]{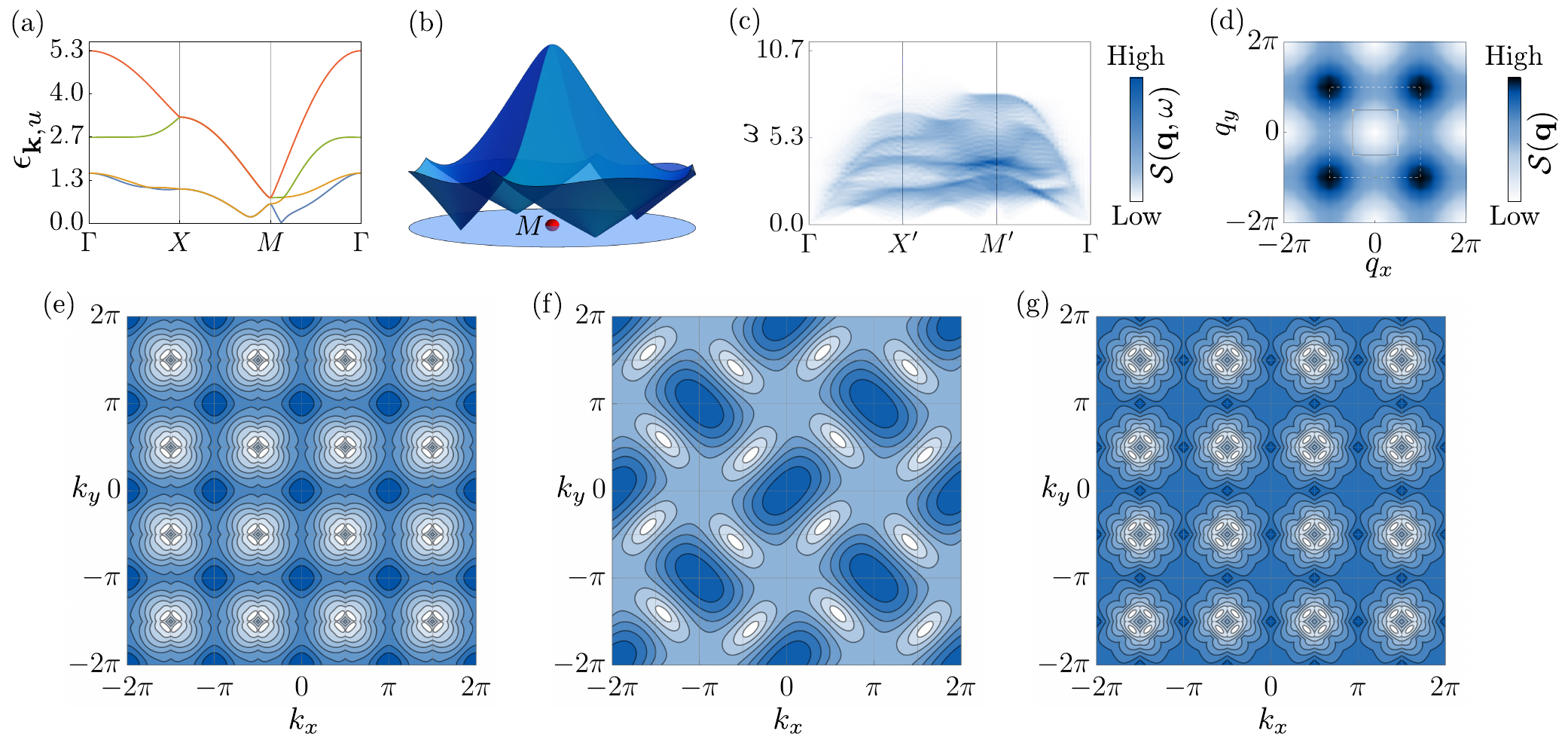}
	\caption{(a) Spinon band structure of the Z3000 state for the variationally optimized parameters at $J_{s}/J_{d}=0.80$, given by $\chi^{}_s=1.0$, $\chi^{}_d=0.46$, $\chi^{}_{d'}=-0.09$, $\chi^{}_{\square}=0.14$, $\Delta^{}_s=0.39$, $\Delta^{}_{\square}=0.38$, and $\mu=0.35$. (b) Spinon dispersion of the optimized Z3000 state, showing the four Dirac cones residing at $M(\pi/2,\pi/2)+(\pm\varepsilon,\pm\varepsilon)$. (c) Dynamical spin structure factor, and (d) equal-time spin structure factor of this state. The solid and dashed squares denote the first Brillouin zone (FBZ) and the extended Brillouin zone (EBZ), respectively. The lower panel demonstrates the equivalence of the Dirac dispersion in Z3000 and Z2A$zz$13. (e) Contour plot of the lowest eigenenergy for Z3000, yielding the positions of the Dirac points which are located at $(\pm {\pi}/{2}\pm\varepsilon,\pm {\pi}/{2}\pm\varepsilon)$. (f) The same for Z2A$zz$13, where one can identify the Dirac cones at  $\pm({\pi}/{2}+\varepsilon,{\pi}/{2}+\varepsilon)$ and $\pm(-{\pi}/{2}+\varepsilon,{\pi}/{2}-\varepsilon)$ which lead to $C_4$ symmetry around $(\pi,0)$ and $(0,\pi)$, and $C_2$ symmetry around $(0,0)$. The correspondence between the Dirac dispersions in Z3000 and Z2A$zz$13 is manifest upon adopting a gauge for Z2A$zz$13 with an enlarged $(2\times2)$ unit cell (g). Note that the back-folding of the Brillouin zone leads to the similar splitting of Dirac cones around $(\pm {\pi}/{2},\pm {\pi}/{2})$. The color scheme is \rev{the} same as that in panel (c).}
	\label{fig:z3000_dis_ssf_dsf}
\end{figure*}

Although the Z3000 \textit{Ansatz} can be obtained by lowering the U(1) IGG---via singlet pairing terms---of three distinct U(1) states, namely, U100, U300, and U800 in Table~\ref{table:z2_ansatze}, it is most straightforward to view it as a descendant of the parent U100 state (see also \rev{App.}~\ref{app:parent}). Indeed, \rev{because symmetries are realized linearly (nonprojectively) in the U100 state, this state} allows for uniform real hopping amplitudes on \emph{all} diagonal and square-lattice bonds. Thus, it can be retrieved from the Z3000 \textit{Ansatz}~\eqref{eq:MF-Z2} by setting the pairings $\Delta_{s}=\Delta_{\square}=0$. The band structure of the parent U100 state is shown in Fig.~\ref{fig:u100_dis_fermi}(a)\rev{,} and\rev{,} at half filling\rev{,} it features a Fermi surface [Fig.~\ref{fig:u100_dis_fermi}(b)]. On adding a nonzero $\Delta_s$, the nearly square-shaped Fermi surface surrounding the $\Gamma$ point is completely \rev{gapped out}, and the other Fermi surfaces in the vicinity of the $M$ \rev{$(\pm\pi/2,\pm\pi/2)$} points shrink to isolated points, forming Dirac cones at $M(\pi/2,\pi/2)+(\pm\varepsilon,\pm\varepsilon)$. The Dirac points always reside on the line $\overline{\Gamma M}$. This is illustrated in Fig.~\ref{fig:dirac_points_parameters} [\rev{App.}~\ref{app:u1_gen_prop}], where we show the contour plot of the energies of the lowest eigenmode of Z3000 in the $k^{}_x$--$k^{}_y$ plane. The dependence of the magnitude of $\varepsilon$ on all the variational parameters is illustrated in Fig.~\ref{fig:dirac_points_parameters_GM} [\rev{App.}~\ref{app:u1_gen_prop}]. In Fig.~\ref{fig:z3000_dis_ssf_dsf}(a), we plot the dispersion of the Z3000 state for the VMC-optimized parameters at $J_s/J_d=0.8$ while the four Dirac cones surrounding the $M$ point are shown in Fig.~\ref{fig:z3000_dis_ssf_dsf}(b). We also present the spin correlations of this mean-field state, as reflected in its dynamical and equal-time spin structure factors, in Figs.~\ref{fig:z3000_dis_ssf_dsf}(c) and (d), respectively.

It is worthwhile to discuss the splitting of the Dirac nodes in the Z3000 \textit{Ansatz}, particularly in comparison with those in the Z2A$zz$13 state on the square lattice~\cite{Wen-2002}. When one considers only uniform hopping $\chi_s$ and $d_{x^2-y^2}$ pairing $\Delta_s$ on the nearest-neighbor bonds\footnote{One can additionally include pairing on the \rev{third}-nearest-neighbor bonds (which are twice the nearest-neighbor \rev{distance}), i.e., connecting $(0,0)$ to $(\pm 2,0)$ and $(0,\pm 2)$.}, both the square and Shastry-Sutherland lattices exhibit Dirac nodes at the commensurate momenta $M(\pm{\pi}/{2},\pm{\pi}/{2})$ [see Figs.~\ref{fig:u100_dis_fermi}(c) and~\ref{fig:u100_dis_fermi}(d)]. However, the inclusion of any of the additional hopping amplitudes $\chi_d$, $\chi_{d'}$, $\chi_\square$, or of the onsite term $\mu$ in the Z3000 \textit{Ansatz} splits the Dirac node at each $M$ point into four nodes residing at incommensurate momenta $(\pm {\pi}/{2}\pm\varepsilon,\pm {\pi}/{2}\pm\varepsilon)$, with the nodes always located along $\overline{\Gamma M}$ [marked by blue dots in Fig.~\ref{fig:u100_dis_fermi}(d)].
The node positions can in fact be obtained in closed form: as shown in  \rev{App.}~\ref{app:epsilon}, the gapless points are the zeros of a quartic polynomial whose coefficients are explicit functions of all the mean-field amplitudes. In particular, the nodes remain exactly commensurate, $\varepsilon=0$, if and only if $\mu=\pm(\chi_{d}+\chi_{d'}-2\chi_{\square})$, independently of $\Delta_{s}$ and $\Delta_{\square}$; the empty-square pairing $\Delta_{\square}$ acting alone leaves the nodes pinned at $M$. Thus, the Z3000 state can also be viewed as descending from the U800 \textit{Ansatz}. We illustrate this behavior using a contour plot of the lowest-energy eigenmode of the Z3000 state, shown in Fig.~\ref{fig:z3000_dis_ssf_dsf}(e).

A similar phenomenon occurs in the square-lattice Z2A$zz$13 state: including additional mean-field amplitudes corresponding to $d_{xy}$ pairings on the diagonals of the squares shifts the Dirac nodes from $(\pm{\pi}/{2},\pm{\pi}/{2})$ to $\pm({\pi}/{2}+\varepsilon,{\pi}/{2}+\varepsilon)$ and $\pm(-{\pi}/{2}+\varepsilon,{\pi}/{2}-\varepsilon)$, as shown in Fig.~\ref{fig:z3000_dis_ssf_dsf}(f). In this case, the band structure exhibits $C_4$ symmetry about $(\pi,0)$ and $(0,\pi)$, and $C_2$ symmetry about $(0,0)$. Therefore, although the Z3000 and Z2A$zz$13 \textit{Ans\"atze} are adiabatically connected, the equivalence of their band structures is not immediately apparent. The origin of this difference lies in the gauge structure: the Z2A$zz$13 \textit{Ansatz} in Ref.~\cite{Wen-2002} is defined in the geometrical unit cell, whereas the Z3000 \textit{Ansatz} is realized in a $2\times2$ unit cell. To facilitate a direct comparison, we redefine the Z2A$zz$13 \textit{Ansatz} in a $2\times2$ unit cell and back-fold the Brillouin zone accordingly. This procedure leads to an identical splitting of Dirac nodes in the spinon dispersion of the lowest eigenmode [see Fig.~\ref{fig:z3000_dis_ssf_dsf}(g)], as observed for the Z3000 state. Consequently, one expects the gapless points in the spin excitations to appear at $(\pi\pm\delta, \pi\pm\delta)$ (where $\delta=2\varepsilon$), as well as at $(\pi,\pm\delta)$ and $(\pm\delta, \pi)$, consistent with the two-spinon excitation spectra of the Z2A$zz$13 state reported in Ref.~\cite{Wen-2002}. Signatures of these gapless excitations are already visible in the mean-field dynamical structure factor shown in Fig.~\ref{fig:z3000_dis_ssf_dsf}(c). They become more pronounced beyond the mean-field level upon including gauge fluctuations, as will be demonstrated in the dynamical structure factor obtained from Keldysh pf-FRG.

It is also instructive to compare these results with previous studies of the Z2A$zz$13 state on the square lattice, which has been proposed as a candidate QSL ground state of the $J_1$--$J_2$ model in Refs.~\cite{Hu-2013,Ferrari-2020}. In those works, a version of the Z2A$zz$13 \textit{Ansatz} \rev{modified relative to} the original proposal in Ref.~\cite{Wen-2002} \rev{is adopted} in order to realize the $\mathbb{Z}_2$ gauge structure. Specifically, the $d_{xy}$ pairing is defined on bonds connecting $(0,0)$ to $(\pm2,\pm2)$---which are twice the length of the diagonals inside the squares, corresponding to the fifth-nearest-neighbor bonds---instead of the bonds connecting $(0,0)$ to $(\pm1,\pm1)$, i.e., the diagonals inside the squares as originally considered in Ref.~\cite{Wen-2002}. Restricting the $d_{xy}$ singlet pairing amplitudes to the $(\pm2,\pm2)$ bonds ensures that the Dirac points remain located at the commensurate momenta $(\pm{\pi}/{2},\pm{\pi}/{2})$ in the mean-field spectrum. As a consequence, the two-spinon excitation spectrum contains gapless excitations at $(\pi,\pi)$, $(\pi,0)$, and $(0,\pi)$, rather than at the incommensurate points $(\pi\pm\delta, \pi\pm\delta)$, $(\pi,\pm\delta)$, and $(\pm\delta, \pi)$ that would arise if the $d_{xy}$ pairings were defined on the $(\pm1,\pm1)$ bonds (see Ref.~\cite{Wen-2002}).

On the Shastry-Sutherland lattice, however, we find that it is not possible to define the Z2A$zz$13 \textit{Ansatz} in an analogous gauge that preserves the Dirac points at $(\pm{\pi}/{2},\pm{\pi}/{2})$. Indeed, even when the singlet pairing $\Delta_\square$ is restricted to the $(\pm2,\pm2)$ diagonals of the empty squares---rather than the $(\pm1,\pm1)$ diagonals---each Dirac point at $(\pm{\pi}/{2},\pm{\pi}/{2})$ still splits into four nodes located at $(\pm{\pi}/{2}\pm\varepsilon,\pm{\pi}/{2}\pm\varepsilon)$. As a result, the gapless points in the spin excitation spectrum appear at the same incommensurate momenta $(\pi\pm\delta, \pi\pm\delta)$, $(\pi,\pm\delta)$, and $(\pm\delta, \pi)$ as discussed in Ref.~\cite{Wen-2002}, with $\delta=2\varepsilon$ (see  \rev{App.}~\ref{app:epsilon} for the exact values of $\varepsilon$). This behavior is manifest in the mean-field dynamical spin structure factor of Fig.~\ref{fig:z3000_dis_ssf_dsf}(\rev{c}); the Keldysh pf-FRG result of Fig.~\ref{fig:SSfactor}(a) is consistent with it, although the small values of $\delta$ obtained here lie below the momentum resolution of the lattices accessible to pf-FRG.

Therefore, to construct an \textit{Ansatz} that naturally interpolates between the square-lattice $J_1$--$J_2$ model and the Shastry-Sutherland geometry, we use the short-range representative of the Z2A$zz$13 PSG introduced in Ref.~\cite{Wen-2002}, with $d_{xy}$ pairings on the $(\pm 1,\pm 1)$ diagonals. This choice produces split Dirac nodes and low-frequency spin excitations at incommensurate momenta. It should be distinguished from the square-lattice variational representative of Ref.~\cite{Hu-2013}, in which the $d_{xy}$ pairing is placed on fifth-nearest-neighbor $(\pm 2,\pm 2)$ bonds; that representative keeps the Dirac nodes pinned at $(\pm \pi/2,\pm \pi/2)$ and gives the lower variational energy on the square-lattice $J_1$--$J_2$ axis. Away from this high-symmetry limit, however, the fifth-neighbor pairing does not give a detectable energy gain in the Shastry-Sutherland geometry, while the short-range representative is energetically preferred in our optimized
Gutzwiller-projected calculations. The two wavefunctions are therefore distinct
microscopic representatives of the same Z2A$zz$13/Z3000 PSG class, rather than
gauge-equivalent forms of a single \textit{Ansatz}.

After performing a particle-hole transformation on the  spinons with flavor $\downarrow$,
\begin{equation}
\hat{f}^{\dagger}_{i,\downarrow} \to \hat{f}^{\pdagger}_{i,\downarrow}, \quad
\hat{f}^{\dagger}_{i,\uparrow} \to \hat{f}^{\dagger}_{i,\uparrow},
\label{eq:ph_transformation}
\end{equation}
the mean-field Hamiltonian~(\ref{eq:MF-Z2}) commutes with the total number of particles. Thus, $|\phi^{\pdagger}_{\rm MF}\rangle$ is
defined by filling suitable single-particle orbitals. In this process, the boundary conditions should \rev{be chosen so that} they lead to a unique state (by filling \textit{all} orbitals in a shell with the same mean-field energy). Periodic (P) and antiperiodic (A) boundary conditions along the ${\bf a}_1$ and ${\bf a}_2$ lattice-vector directions [see Fig.~\ref{fig:schematic}(\rev{b})] can be considered, leading to four choices: [P,P], [P,A], [A,P], and [A,A]. Here, \rev{for the Z3000 state,} we impose antiperiodic and periodic boundary conditions along the ${\bf a}_1$ and ${\bf a}_2$ directions, respectively, i.e., [A,P], which leads to a closed shell (unique state) on $L\times L$ clusters with \rev{any even $L$}.

\subsection{Gutzwiller-projected \textit{Ansatz}}

A legitimate QSL wavefunction, residing in the correct sector with one fermion per site (corresponding to the physical
Hilbert space of the spin-$1/2$ model), is then obtained by applying the Gutzwiller projector to the mean-field state $|\phi^{\pdagger}_{\rm MF}\rangle$:
\begin{equation}
|\Psi^{\pdagger}_{{\rm QSL}}\rangle = {\cal P}^{\pdagger}_{\text G} |\phi^{\pdagger}_{\rm MF}\rangle\,,
\end{equation}
where ${\cal P}^{\pdagger}_{\text G}= \prod_i \big ( \hat{n}^{\pdagger}_{i,\uparrow}-\hat{n}^{\pdagger}_{i,\downarrow} \big )^2$, $\hat{n}^{\pdagger}_{i,\alpha} = \hat{f}_{i,\alpha}^{\dagger}\hat{f}^{\pdagger}_{i,\alpha}$
being the fermionic density of flavor $\alpha$ on site $i$. The variational energy and correlation functions for $|\Psi_{{\rm QSL}}\rangle$ can
be computed in a straightforward fashion using Monte Carlo sampling, by virtue of which the constraint is imposed exactly~\cite{Becca_Sorella_2017}. Furthermore, a stochastic reconfiguration (SR) optimization enables us to obtain accurate estimates of the variational parameters contained in Eq.~(\ref{eq:MF-Z2}) with small statistical uncertainties~\cite{Sorella-2005,Yunoki-2006}, even when a large number of parameters are involved~\cite{Iqbal-2015,Iqbal-2021,Hering-2022,Ferrari-2023,muller-2024_ancilla}.

A standard technique to systematically improve the accuracy of variational wavefunctions and approach the true ground state involves the application of a few L\'anczos steps on a given variational state $|\Psi_{\rm QSL}\rangle$~\cite{Sorella-2001,Becca_2015}:
\begin{equation}\label{eq:lanczos}
|\Psi_{p\text{-}\rm{LS}}\rangle =  \left ( 1+\sum_{k=1}^{p}\lambda_{k}{\cal \hat{H}}^{k} \right )|\Psi_{{\rm QSL}}\rangle\, ,
\end{equation}
where the $\lambda_{k}$s are variational parameters. For large enough $p$, $|\Psi_{p\text{-}\rm{LS}}\rangle$ is guaranteed to converge to the exact ground state $|\Psi_{\rm ex}\rangle$ provided the initial trial state $|\Psi_{\rm QSL}\rangle$ is not orthogonal to $|\Psi_{\rm ex}\rangle$, i.e., for $\langle\Psi_{\rm ex}|\Psi_{\rm QSL}\rangle \neq 0$. However, on large clusters, only a few L\'anczos steps can be efficiently executed, and here, we consider the \rev{cases} $p=1$ and $p=2$ ($p=0$ corresponds to the initial trial wavefunction $|\Psi_{\rm QSL}\rangle$). Based on this sequence, an estimate of the exact ground-state energy can be obtained by the method of zero-variance extrapolation. In fact, for sufficiently accurate states, $E-E_{\rm ex} \approx {\rm constant}\times\sigma^{2}$, where $E=\langle {\cal \hat{H}} \rangle/N$ and
$\sigma^{2}=(\langle {\cal \hat{H}}^{2}\rangle{-}\langle {\cal \hat{H}}\rangle^{2})/N$ are the energy and variance per site, respectively. Thus, the exact ground-state energy $E_{\rm ex}$ can be extracted by fitting $E$ vs $\sigma^{2}$ for $p=0$, $1$, and $2$. The energy, its variance, and the correlation functions after the application of a few L\'anczos steps are calculated using the standard variational Monte Carlo method.

\begin{table}
\caption{Energies per site $E/J_{d}$ of all symmetric $\mathbb{Z}_2$ \textit{Ans\"atze} listed in Table~\ref{table:z2_ansatze_extended} computed at $J_{s}/J_{d}=0.80$ on a $14\times 14$ cluster. The wavefunction includes all symmetry-allowed amplitudes on the $u^{}_s$, $u^{}_d$, $u^{}_{d'}$\rev{,} and $u^{}_{\square}$ bonds and onsite terms. The boundary condition (BC) \rev{that} gives the lowest energy is \rev{indicated; e.g.,} [A,P] \rev{denotes} antiperiodic BC along the $\mathbf{a}_1=(1,0)$ direction and periodic BC along the $\mathbf{a}_2=(0,1)$ direction.}
\begin{ruledtabular}
\begin{tabular}{ccc}
Label &Energy & BC \\
			\hline
Z2000& $-0.37500(1)$ & [P,A]  \\
Z2100& $-0.37500(1)$& [A,P]  \\
Z2001& $-0.37500(1)$& [A,A]  \\
Z2101& $-0.33509(5)$& [A,A]  \\
Z2011& $-0.43440(1)$& [P,A]  \\
Z2111& $-0.43452(5)$ & [A,P]  \\
\hline
Z3000& $-0.43728(1)$ & [A,P]  \\
Z3100& $-0.39336(1)$ & [A,A]  \\
Z3010& $-0.42129(1)$ & [A,A]  \\
Z3110& $-0.39340(1)$ & [A,A]  \\
Z3001& $-0.37973(1)$ & [A,P]  \\
Z3101& $-0.43529(2)$ & [A,A] \\
Z3011& $-0.43436(1)$ & [P,A]  \\
Z3111& $-0.34089(3)$ & [A,A]  \\
\hline
Z5001& $-0.43321(1)$ & [A,A] \\
Z5101& $-0.40872(1)$ & [A,A] \\
Z5011& $-0.42974(1)$ & [A,A] \\
Z5111& $-0.43493(1)$ & [A,A] \\
		\end{tabular}
	\end{ruledtabular}
	\label{table:z2_ansatze_energies}
\end{table}

\floatsetup[table]{capposition=bottom}
\begin{table}[b]
\centering
\begin{tabular}{lllllll}
 \hline \hline
       \multicolumn{1}{c}{$L$}
    & \multicolumn{1}{c}{$\Delta^{}_{s}$}
    & \multicolumn{1}{c}{$\Delta^{}_{\square}$}
    & \multicolumn{1}{c}{$\chi^{}_{d}$}
    & \multicolumn{1}{c}{$\chi^{}_{d'}$}
    & \multicolumn{1}{c}{$\chi^{}_{\square}$}
    & \multicolumn{1}{c}{$\mu$}
      \\ \hline

\multirow{1}{*}{$6$} & $0.256(1)$ & $0.085(1)$ & $0.376(1)$ & $+0.123(1)$ & $0.181(1)$ & $0.931(1)$  \\

\multirow{1}{*}{$8$} & $0.355(1)$ & $0.261(1)$ & $0.386(1)$ & $-0.034(2)$ & $0.126(1)$ & $0.605(1)$  \\

\multirow{1}{*}{$10$} & $0.371(1)$ & $0.320(1)$ & $0.423(1)$ & $-0.062(1)$ & $0.131(1)$ & $0.475(1)$  \\

\multirow{1}{*}{$12$} & $0.381(2)$ & $0.353(2)$ & $0.446(1)$ & $-0.076(2)$ & $0.136(2)$ & $0.399(2)$  \\

\multirow{1}{*}{$14$} & $0.387(3)$ & $0.381(2)$ & $0.462(2)$ & $-0.088(2)$ & $0.139(2)$ & $0.345(2)$  \\

\multirow{1}{*}{$20$} & $0.400(2)$ & $0.428(2)$ & $0.490(2)$ & $-0.110(2)$ & $0.143(2)$ & $0.254(3)$  \\

\multirow{1}{*}{$30$} & $0.414(3)$ & $0.465(2)$ & $0.512(1)$ & $-0.128(3)$ & $0.147(2)$ & $0.187(3)$  \\

\multirow{1}{*}{$40$} & $0.422(3)$ & $0.476(3)$ & $0.522(3)$ & $-0.136(3)$ & $0.148(2)$ & $0.155(4)$  \\

\multirow{1}{*}{$50$} & $0.426(4)$ & $0.480(4)$ & $0.533(4)$ & $-0.137(3)$ & $0.153(4)$ & $0.135(4)$  \\

\multirow{1}{*}{$60$} & $0.425(4)$ & $0.478(4)$ & $0.533(4)$ & $-0.138(4)$ & $0.152(4)$ & $0.126(5)$  \\

 \hline \hline

\end{tabular}
\caption{The singlet pairing and hopping amplitudes of the $\mathbb{Z}_{2}$ Dirac QSL (Z3000 state) on different cluster sizes obtained after VMC optimization for $J_s/J_d=0.80$. The hopping on the square bonds $\chi_{s}$ is set to one.}
\label{tab:pairing-optimized}
\end{table}

\begin{SCfigure*}
        \centering
        \includegraphics[width=0.325\textwidth]{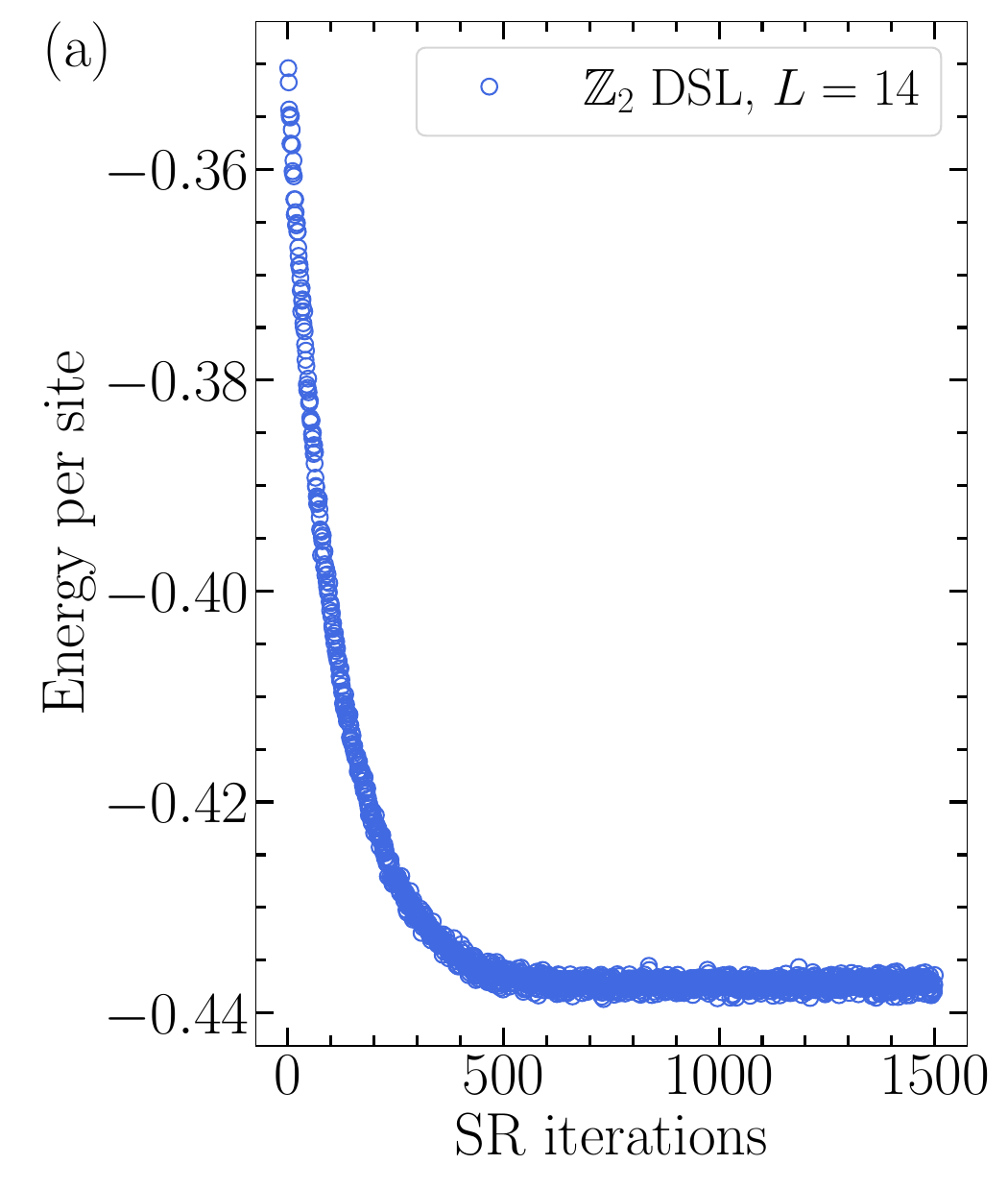}
        \includegraphics[width=0.325\textwidth]{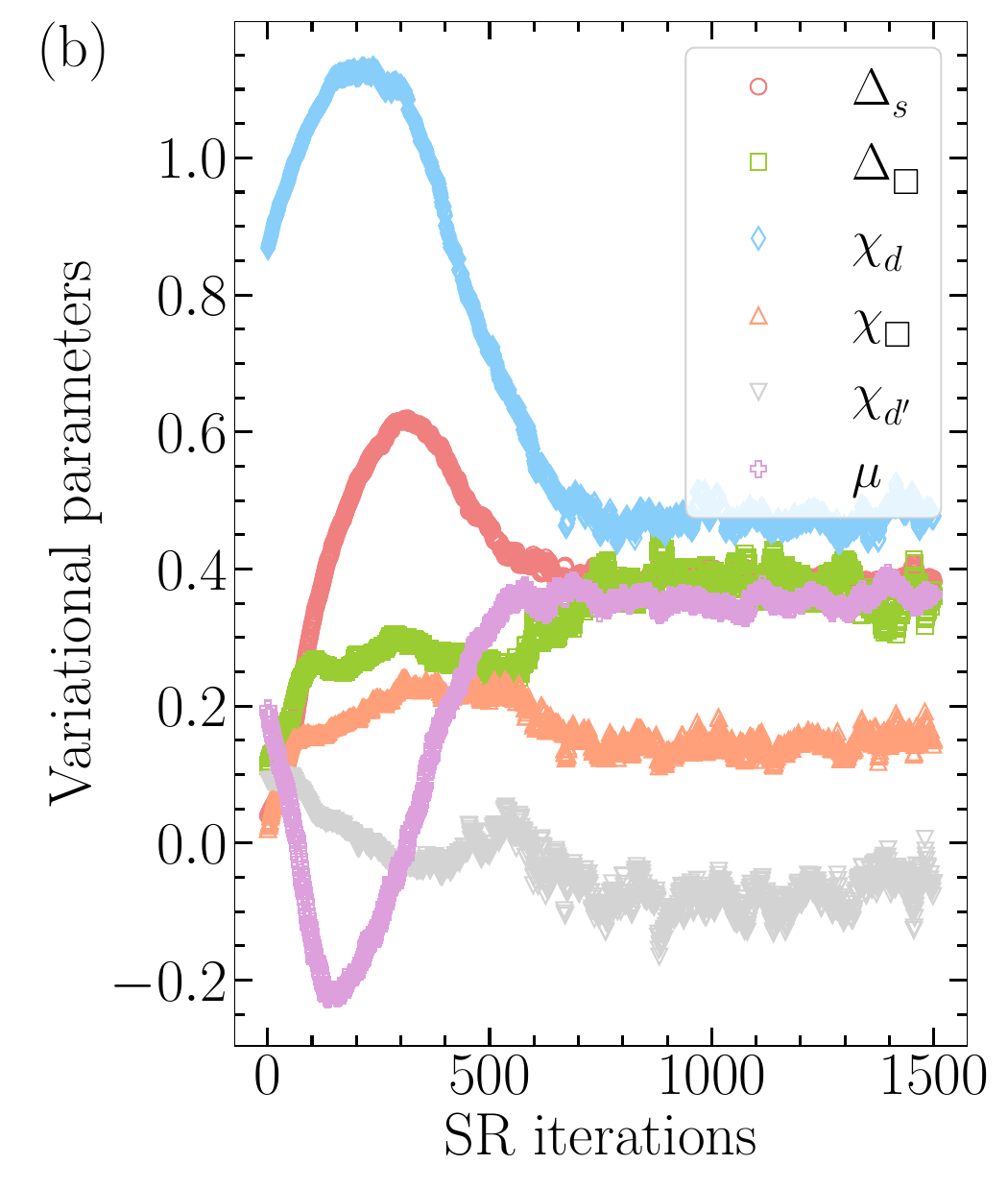}
        \caption{A typical variational Monte Carlo optimization run for the Z$3000$ state on a $14\times14$\rev{-}site cluster for $J_{s}/J_{d}=0.80$. (a) Energy per site in units of $J_{d}$, and (b) variational parameters $\Delta^{}_s$, $\Delta^{}_{\square}$, $\chi^{}_{d}$, $\chi^{}_{\square}$, $\chi^{}_{d'}$, and $\mu$, as a function of the number of stochastic reconfiguration (SR) iterations. In (b), the initial parameters are taken to be $\Delta^{}_{s}=0.04$, $\Delta^{}_{\square}=0.11$, $\chi^{}_{d}=0.87$, $\chi^{}_{\square}=0.02$, $\chi^{}_{d'}=0.09$, $\mu=0.2$, and we set $\chi^{}_{s}=1$ as the reference. The optimized parameter values presented in Table~\ref{tab:pairing-optimized} are obtained by averaging over a much larger number of converged SR iterations than shown here. In comparison, the energy per site of the optimal parent U(1) state (defined by uniform $\chi_s$ and $d_{x^2-y^2}$ $\Delta_{s}$ pairing) is $E/J_{d}=-0.43321(1)$.}
        \label{fig:vmc_optimization}
\end{SCfigure*}

\begin{figure*}
\includegraphics[width=0.325\linewidth]{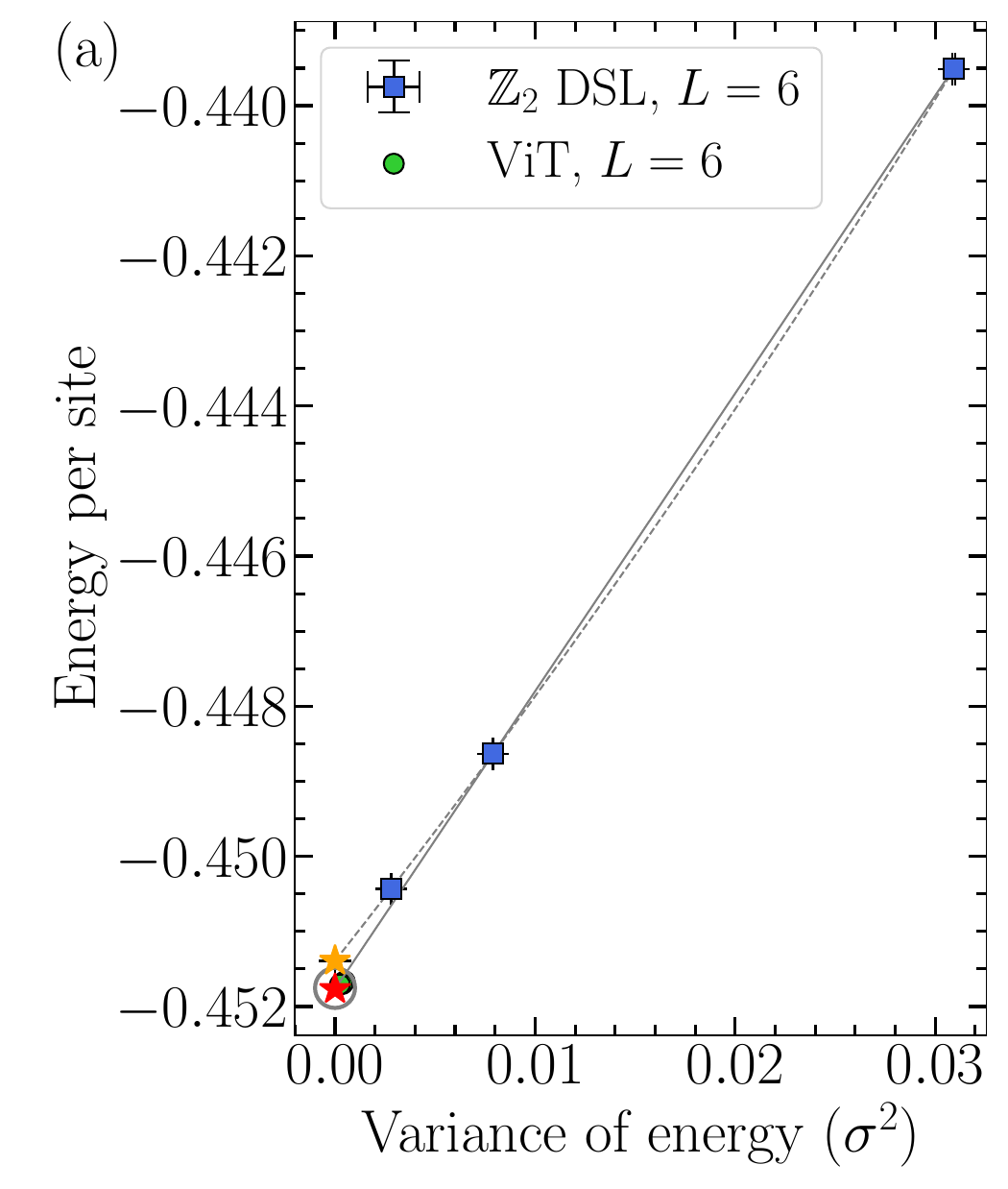}
\includegraphics[width=0.325\linewidth]{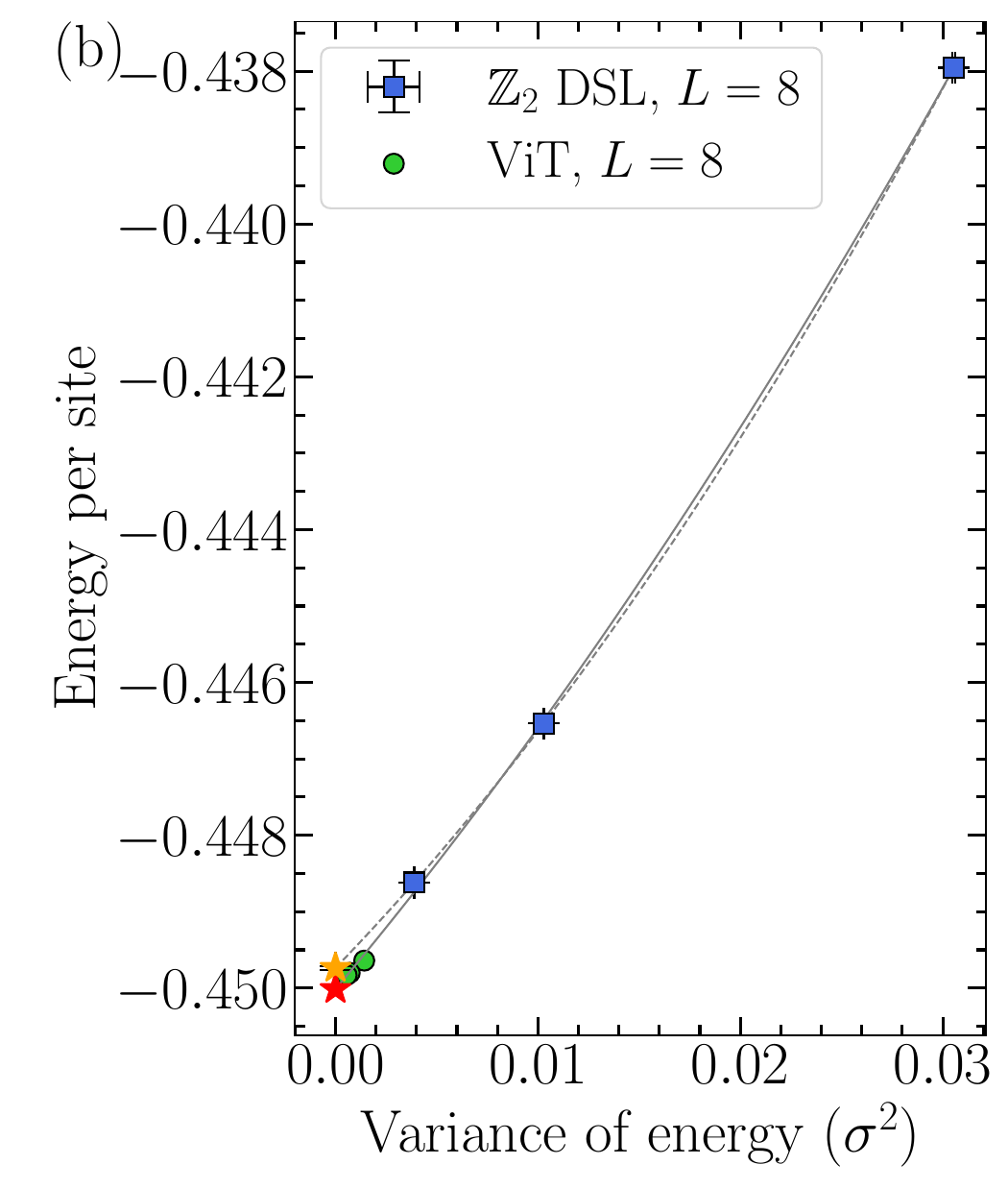}
\includegraphics[width=0.325\linewidth]{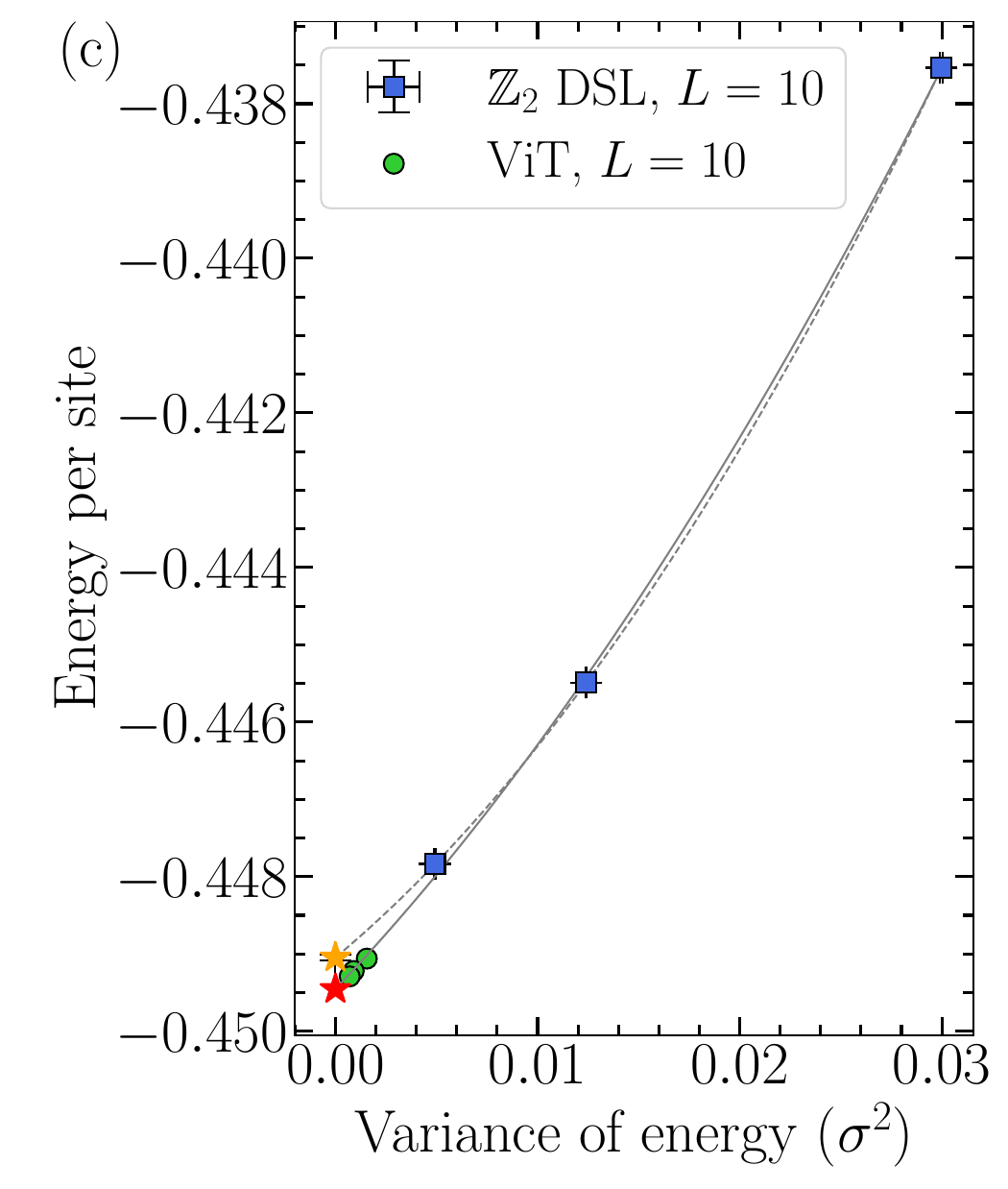}\\ \vspace{0.2cm}
\includegraphics[width=0.325\linewidth]{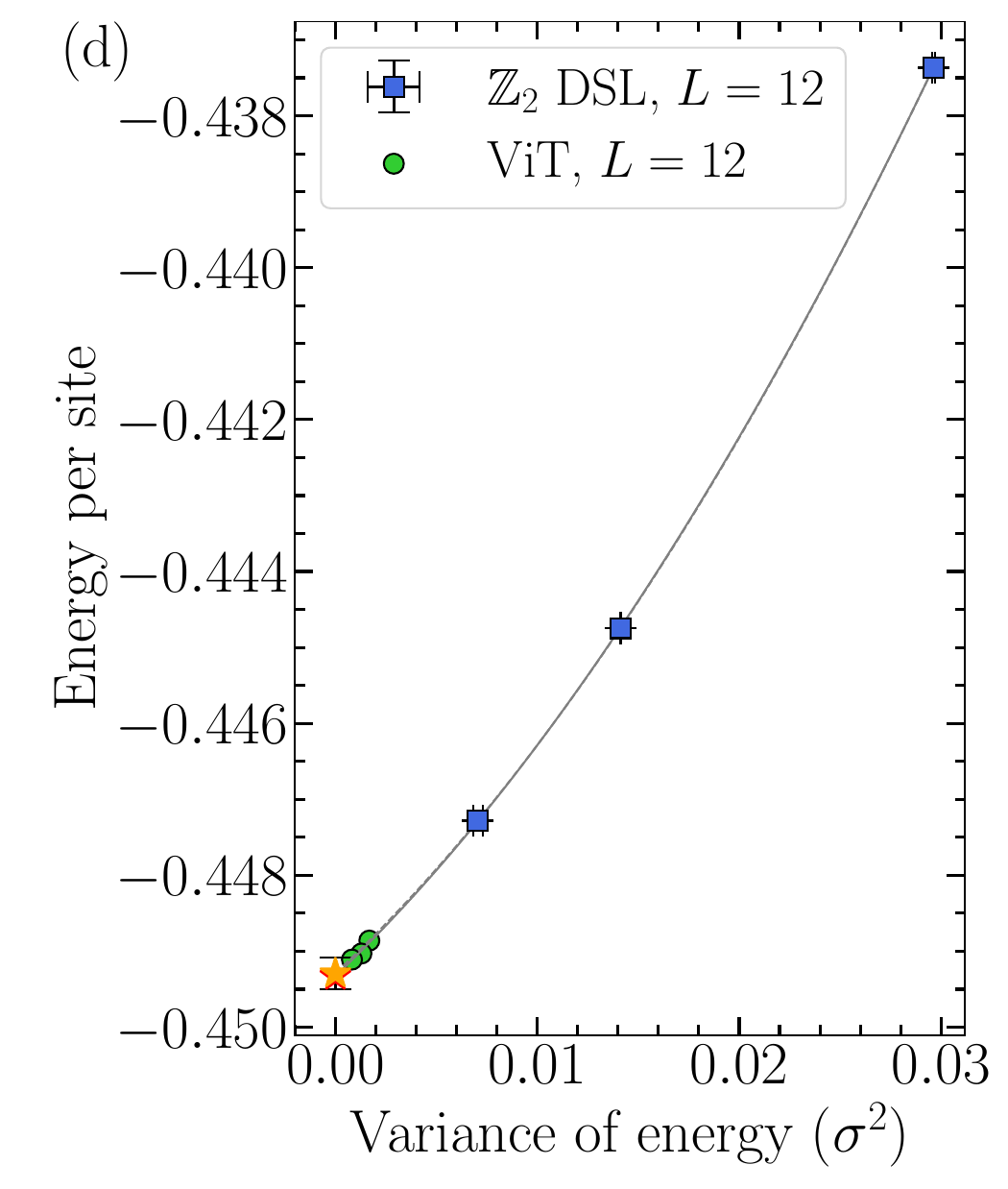}
\includegraphics[width=0.325\linewidth]{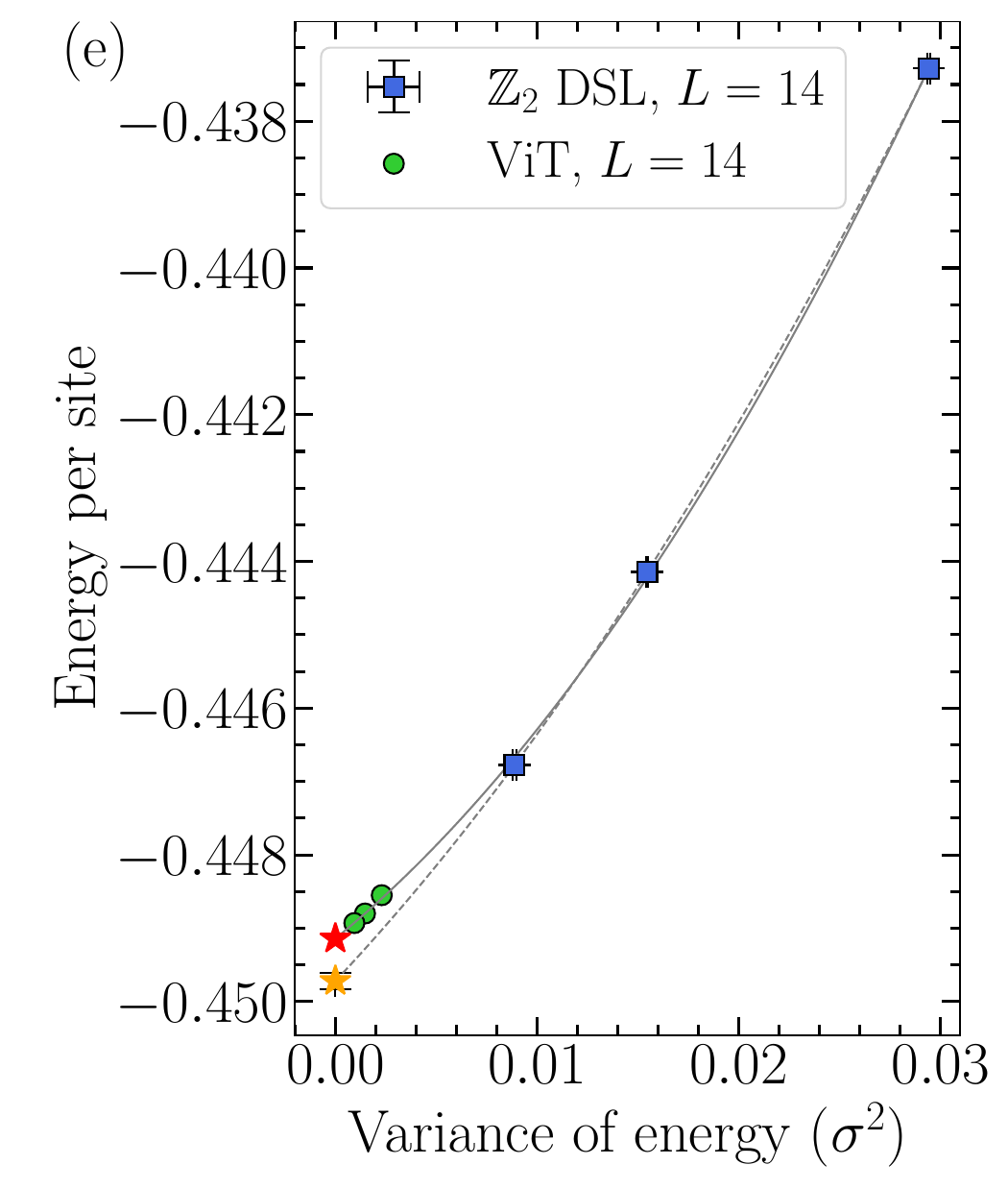}
\includegraphics[width=0.325\linewidth]{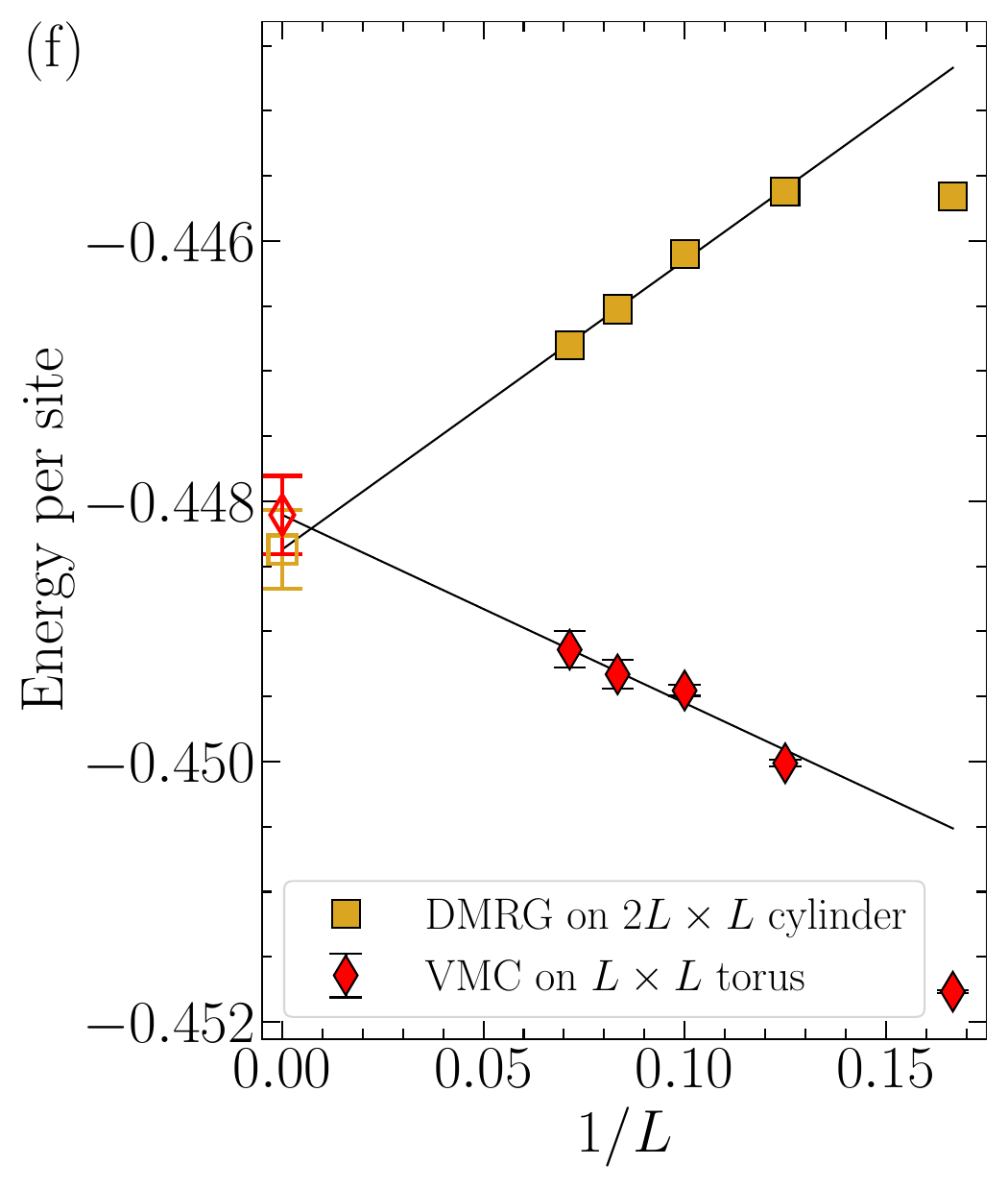}
	\caption{Variational energies for the $\mathbb{Z}_{2}$ Dirac QSL (Z3000 state) at $J_{s}/J_{d}=0.80$, as a function of the variance of energy, for zero, one, and two L\'anczos steps. (a)--(e) An accurate estimate of the ground-state energy on the $L = 6, 8, 10, 12,$ and $14$ clusters is obtained by extrapolating the three variational results from VMC (blue squares) together with the ViT data (green circles) at smaller variance~\cite{Luciano-2024} to the zero-variance limit (red star) by a quadratic fit (solid curve). The estimate obtained by employing a quadratic fit (dashed curve) to only the three VMC points (blue squares) leads to less reliable values of the ground-state energy (yellow star), especially for larger system sizes, as seen \rev{from its} nonmonotonic behavior with respect to system size (see Fig.~\ref{fig:fss_ls} and Table~\ref{tab:en-lanczos}, and the main text for a description of the extrapolation method). For $L=6$, the ED energy is marked by an \rev{open} circle. (f) The thermodynamic estimate of the ground-state energy, as obtained by a finite-size scaling of the estimated energies from VMC $+$ ViT on the $L=8, 10, 12$\rev{,} and $14$ clusters using a $1/L$ scaling form and a linear fit. The yellow squares provide a comparison with the DMRG estimates from Ref.~\cite{Yang-2022}.}
	\label{fig:lanczos}
\end{figure*}

\floatsetup[table]{capposition=bottom}
\begin{table}[b]
\begin{footnotesize}
\centering
\begin{tabular}{lllll}
 \hline \hline
       \multicolumn{1}{c}{$L$}
    & \multicolumn{1}{c}{$0$-LS}
    & \multicolumn{1}{c}{$1$-LS}
    & \multicolumn{1}{c}{$2$-LS}
    & \multicolumn{1}{c}{Extrapolation}
      \\ \hline

\multirow{1}{*}{$6$} & $-0.439512(1)$ & $-0.448633(1)$ & $-0.450432(1)$ & $-0.451391(7)$    \\

\multirow{1}{*}{$8$} & $-0.437955(1)$ & $-0.446537(1)$ & $-0.448619(1)$ & $-0.449735(23)$    \\

\multirow{1}{*}{$10$} & $-0.4375346(6)$ & $-0.4454888(6)$ & $-0.447838(3)$ & $-0.449046(33)$     \\

\multirow{1}{*}{$12$} & $-0.4373669(5)$ & $-0.444746(1)$ & $-0.447279(4)$ & $-0.44929(21)$    \\

\multirow{1}{*}{$14$} & $-0.4372820(6)$ & $-0.444144(1)$ & $-0.446776(4)$ &  $-0.44972(11)$    \\

 \hline \hline

\end{tabular}
\end{footnotesize}
\caption{The energies per site $E/J_d$ of the $\mathbb{Z}_{2}$ Dirac QSL (Z3000 state) with $p=0$, $1$, and $2$ L\'anczos steps, and the zero-variance extrapolation estimate on different cluster sizes as obtained by VMC, for $J_s/J_d=0.80$.}
\label{tab:en-lanczos}
\end{table}

\begin{figure}	\includegraphics[width=1.0\linewidth]{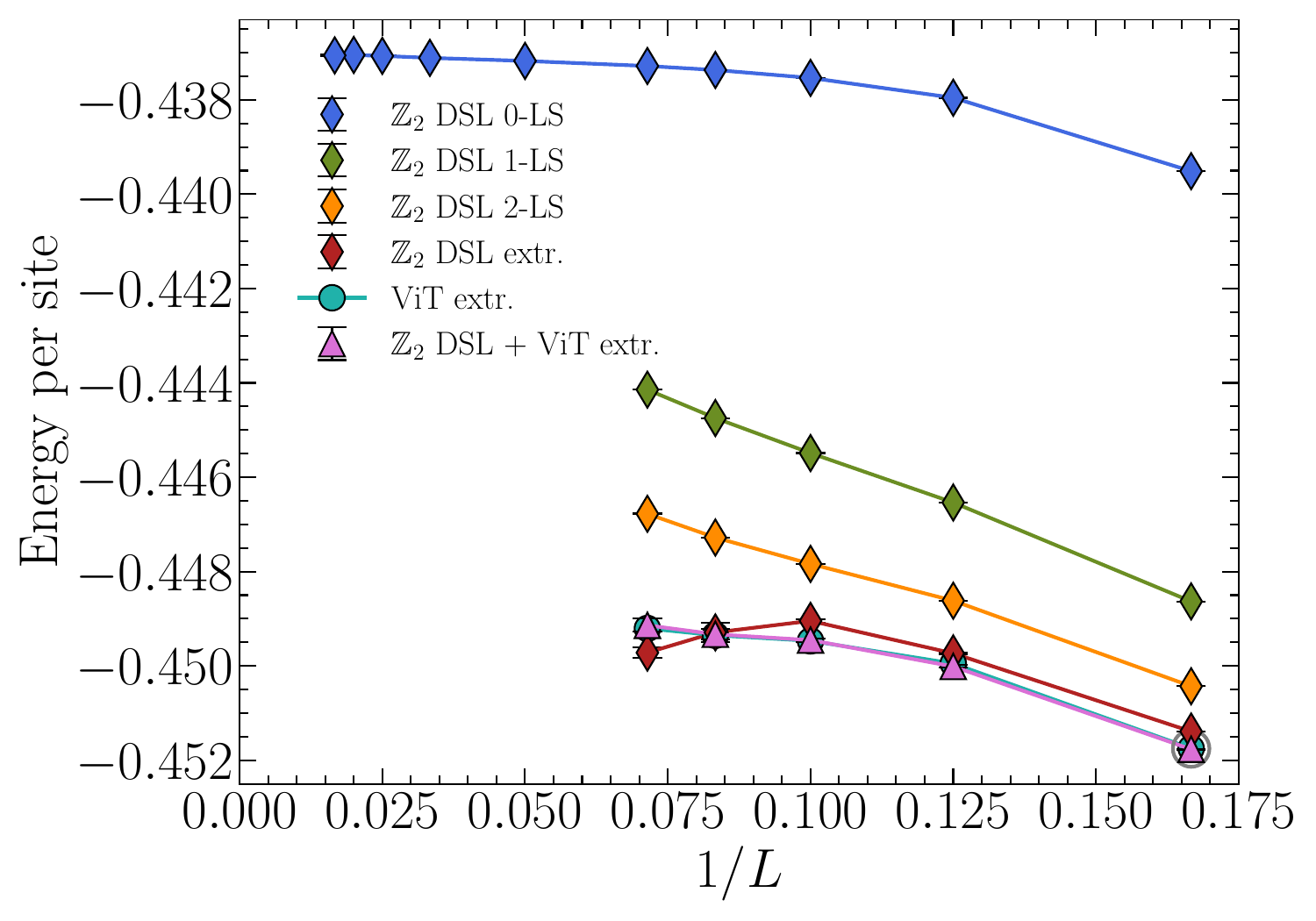}
	\caption{The dependence on system size of the energies of the $\mathbb{Z}_{2}$ Dirac spin liquid (DSL), with $J_{s}/J_{d}=0.8$, for $p=0$, $1$, \rev{and} $2$ L\'anczos steps and zero-variance extrapolated estimates using (i) only VMC data ($\mathbb{Z}_{2}$ DSL extr.), and (ii) combining VMC with lower-variance ViT data ($\mathbb{Z}_{2}$ DSL + ViT extr.)~\cite{Luciano-2024}. The latter is also consistent with ViT-only extrapolation estimates~\cite{Luciano-2024}. Also marked (gray circle) is the ED energy on a 36-site cluster.}
	\label{fig:fss_ls}
\end{figure}

\floatsetup[table]{capposition=bottom}
\begin{table}[b]
\begin{footnotesize}
\centering
\begin{tabular}{lllll}
 \hline \hline
       \multicolumn{1}{c}{$L$}
    & \multicolumn{1}{c}{VMC}
    & \multicolumn{1}{c}{ViT}
    & \multicolumn{1}{c}{VMC + ViT}
    & \multicolumn{1}{c}{DMRG}
      \\ \hline

\multirow{1}{*}{$6$} & $-0.451391(7)$ & $-0.451750$ & $\bm{-0.451768(8)}$ & $-0.44565886$  \\

\multirow{1}{*}{$8$} & $-0.449735(23)$ & $-0.44995$ & $\bm{-0.450013(27)}$ & $-0.44562053$  \\

\multirow{1}{*}{$10$} & $-0.449046(33)$ & $-0.449470$ & $\bm{-0.449454(41)}$ & $-0.44609839$  \\

\multirow{1}{*}{$12$} & $-0.44929(10)$ & $-0.449352$ & $\bm{-0.44933(11)}$ & $-0.44652433$  \\

\multirow{1}{*}{$14$} & $-0.44972(11)$ & $-0.449207$ & $\bm{-0.44914(14)}$ & $-0.44680369$  \\

\multirow{1}{*}{$\infty$} & $*$ & $-0.4486(2)$ & $\bm{-0.4481(3)}$ & $-0.44837(30)$  \\

 \hline \hline

\end{tabular}
\end{footnotesize}
\caption{The ground-state energies of the spin-$1/2$ Heisenberg model at $J_s/J_d=0.80$ for different cluster sizes, as obtained from zero-variance extrapolation. The extrapolation is performed using data on $L\times L$ tori from VMC, neural\rev{-}network quantum states (ViT)~\cite{Luciano-2024}, and combined extrapolation with both VMC and ViT energies (VMC $+$ ViT)\rev{,} or from DMRG~\cite{Yang-2022} on $2L\times L$ cylinders with periodic boundary conditions along the shorter direction. Our most accurate estimate of the ground-state energy on any given cluster is obtained from VMC $+$ ViT extrapolation and is marked in bold. \rev{The asterisk indicates that no VMC-only value is quoted in the thermodynamic limit, because of the nonmonotonic size dependence of the VMC-only extrapolation discussed in the text.} The exact-diagonalization (ED) value for $L=6$ is $-0.4517531$~\cite{Luciano-2024}.}
\label{tab:methods}
\end{table}

\begin{figure*}[tbp]
\includegraphics[width=\linewidth]{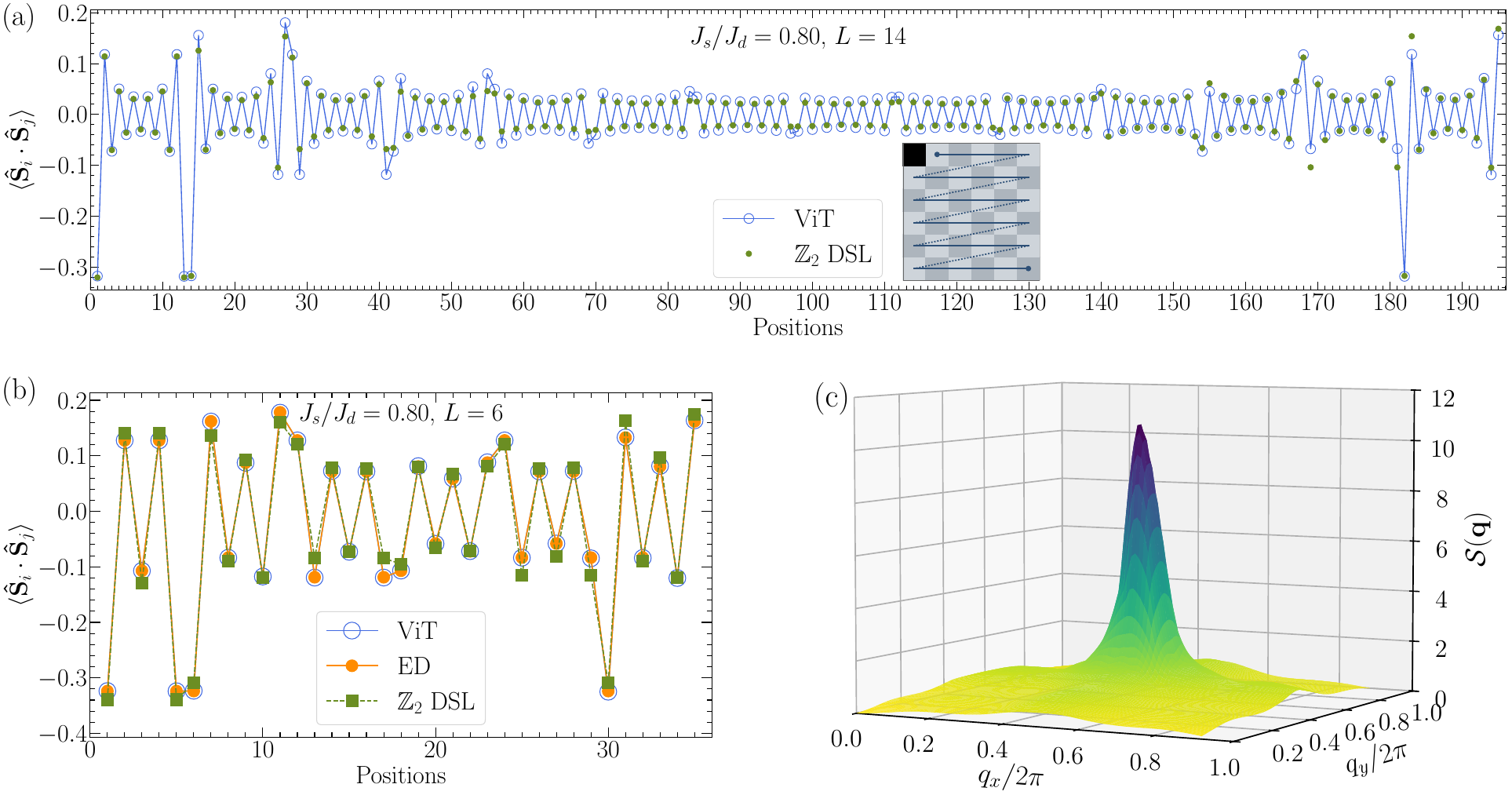}
	\caption{(a) The equal-time isotropic spin-spin correlations $\langle {\mathbf {\hat S}}_{i}\cdot \mathbf{{\hat S}}_{j}\rangle$ in real space for the projected $\mathbb{Z}_{2}$ DSL (Z3000 state) supplemented with \rev{two L\'anczos steps (LS)}, as computed using VMC (\rev{filled} green dots) and compared with the ViT wavefunction (empty blue circles) from Ref.~\cite{Luciano-2024} on a $14\times14$ lattice at $J_{s}/J_{d}=0.80$. The error bars in VMC are $\sim \mathcal{O}(10^{-4})$.  Inset: the blue line schematically illustrates the path on the $L=14$ and $L=6$ clusters along which the spin-spin correlations (with respect to the site marked in black) are ordered in the panel. Note that the ViT wavefunction explicitly breaks SU(2) spin-rotation symmetry and the quantity plotted is $\langle {\hat S}^{x}{\hat S}^{x} + {\hat S}^{y}{\hat S}^{y} + {\hat S}^{z}{\hat S}^{z}\rangle$; our wavefunction is manifestly SU(2) symmetric, and we thus plot $3\times\langle {\hat S}^{z}{\hat S}^{z}\rangle$. (b) The same on an $L=6$ cluster, now including a comparison with ED (orange circles). (c) The equal-time spin structure factor for the $\mathbb{Z}_{2}$ DSL (supplemented by 2 LS and a subsequent zero-variance extrapolation) obtained with VMC on a $14\times14$ lattice; its peak at $(\pi,\pi)$ is precisely the value $\mathcal{S}(\pi,\pi)\approx10.19$ quoted in the text.
}
	\label{fig:corr_L14}
\end{figure*}

\subsection{Variational Monte Carlo}
Our variational calculations are performed on $L\times L$ tori, which respect all symmetries of the $p4g$ wallpaper group, with periodic boundary conditions imposed on the spin Hamiltonian~\eqref{eq:mod-ham}. We focus our analysis on the coupling ratio $J_{s}/J_{d}=0.80$\rev{,} which is located in the middle of the QSL regime seen in Refs.~\cite{Yang-2022, Wang-2022, Keles-2022, Luciano-2024}, and for which the energy and correlation functions have been reported with different numerical approaches, thereby facilitating detailed comparisons.

\subsubsection{Energetics}
Among the \textit{Ans\"atze} listed in Table~\ref{table:z2_ansatze_square_shastry}, which can be continuously connected to the states on the square lattice, we find that the VMC optimization of the projected Z3000 \textit{Ansatz} consistently yields the lowest variational energy on all clusters, with sizes ranging from $L=6$ to $40$, while the other \textit{Ans\"atze} (including those which are not connected to the square lattice) have higher energies (see Table~\ref{table:z2_ansatze_energies} for energies of all $\mathbb{Z}_{2}$ \textit{Ans\"atze}). For the optimized Z3000 state, crucially, we observe that the singlet pairing amplitudes $\Delta_{s}$ and $\Delta_{\square}$ converge to appreciable values across all cluster sizes, approaching their thermodynamic values (together with the remaining amplitudes) around a cluster size of $L=50$ (refer to Table~\ref{tab:pairing-optimized}). This provides strong evidence for a robust $\mathbb{Z}_2$ gauge structure in the thermodynamic limit. In particular, the $\Delta_{\square}$ pairing (responsible for the $\mathbb{Z}_{2}$ gauge structure) leads to an energy gain $(E_{\mathbb{Z}_{2}}-E_{{\rm U}(1)})/E_{{\rm U}(1)}\sim0.01$ on top of the parent U(1) state (constructed from uniform $\chi_{s}$ and $d_{x^2-y^2}$ $\Delta_{s}$ amplitudes \emph{only}). This is similar to the $J_{1}$--$J_{2}$ square-lattice Heisenberg antiferromagnet\rev{,} where the energy gain of the Z2A$zz$13 state (due to $d_{xy}$ singlet pairing terms on fifth-neighbor bonds\footnote{On the Shastry-Sutherland lattice, a nonzero singlet pairing amplitude on fifth-\rev{neighbor} bonds---connecting $(0,0)$ to sites $(\pm2,\pm2)$---does not give any detectable energy gain.}) is $\sim1\%$ on top of the parent U(1) state (which shares the same parent U(1) state as the Z3000 \textit{Ansatz}\rev{,} being likewise composed of uniform $\chi_{s}$ and $d_{x^2-y^2}$ $\Delta_{s}$ amplitudes \emph{only})~\cite{Hu-2013}. In Fig.~\ref{fig:vmc_optimization}, we present a typical VMC optimization run showing the evolution of the energy and the parameters of the Z3000 state starting from a generic\rev{, unbiased initial} point. We verify the convergence by initiating many optimization runs with different random initial values\rev{ and find} that the energy and variational parameters converge neatly to the same value (within error bars) after averaging over a sufficient number of Monte Carlo steps. It is important to note that while the energy converges after $\sim 500$ iterations, the parameters converge to their final values only after a much larger number of SR iterations (see Fig.~\ref{fig:vmc_optimization}). This is because in optimizing the former, the steepest-descent forces are calculated using correlated sampling \rev{rather than} via energy differences~\cite{Sorella-2005,Iqbal-2011}.

Having established that the Z3000 \textit{Ansatz} represents the lowest-energy variational wavefunction (within the class of Gutzwiller-projected fermionic states) for all the system sizes investigated---and that the singlet pairing amplitudes characterizing it as well as the resulting energy gains remain size-consistent---we consider it our starting $(p=0)$ state and improve \rev{upon it} by applying L\'anczos steps. At $J_{s}/J_{d}=0.80$, the effect of two L\'anczos steps for different cluster sizes is shown in Fig.~\ref{fig:lanczos} (see also Table~\ref{tab:en-lanczos} for the actual values of the energies of the Z3000 state). To refine the accuracy of the extrapolation, we also include the smaller-variance ViT data from Ref.~\cite{Luciano-2024}. A zero-variance extrapolation is then performed via a quadratic fit of the six points (three VMC and three ViT), $E = E_{\rm ex} + \mathcal{A}\times\sigma^{2} + \mathcal{B}\times(\sigma^{2})^2$. The inclusion of the ViT data in the fit proves crucial in curing the apparent nonmonotonic behavior with system size in the zero-variance-extrapolated estimates of the ground-state energies (compare ``$\mathbb{Z}_{2}$ DSL extr.'' versus ``$\mathbb{Z}_{2}$ DSL \rev{+ }ViT extr.'' in Fig.~\ref{fig:fss_ls}, and see Table~\ref{tab:methods} for the corresponding energies), thus enabling a reliable estimation in the thermodynamic limit. A caveat of energy--variance extrapolation is that the relation between the variational energy $E$ and the variance $\sigma^2 = \langle H^2 \rangle - \langle H \rangle^2$ is not universal across unrelated variational families. In general, both the slope and possible curvature of $E(\sigma^2)$ depend on how the trial state distributes its residual weight over excited eigenstates. As a consequence, different \textit{Ans\"atze} may exhibit distinct finite-variance behavior even when they target the same ground state. For this reason, combining \rev{VMC and ViT} data in a single extrapolation is justified only if both datasets populate the same smooth low-variance trend and extrapolate to a consistent zero-variance intercept. To verify this explicitly, we performed separate variance extrapolations for the VMC and ViT data, as well as a constrained fit allowing for method-dependent slopes (and curvature) but enforcing a common intercept $E_0$. All procedures yield extrapolated ground-state energies that agree within uncertainties.

For our present purposes, the computationally demanding $p=2$ calculations
have been performed only for the $L=6,8,10,12$, and $14$ clusters. It is worth emphasizing that this zero-variance extrapolation procedure yields size-consistent estimates of the energy per site \rev{not only for} the ground state~\cite{Iqbal-2013,Iqbal-2016} but also \rev{for} excited states~\cite{Iqbal-2014}.  In fact, it is well known that the L\'anczos procedure (for a fixed $p$) becomes progressively inefficient upon increasing the system size, yet the extrapolation remains accurate~\cite{Becca_2015}. This can be observed from the gain in the energy and the variance with respect to $p=0$, which decreases with $L$ (see Fig.~\ref{fig:lanczos} and Table~\ref{tab:en-lanczos}), but the extrapolation is unaffected since the slope is essentially unchanged~\cite{Hu-2013,Iqbal-2016}. Indeed, our estimate of the ground-state energy on the $6\times6$ cluster is in excellent agreement with the exact result, and equal (within error bars) to the corresponding estimate from ViT~\cite{Luciano-2024} (see Fig.~\ref{fig:lanczos}(a) and Table~\ref{tab:methods}). Similarly, on the $8\times8$ and larger clusters, the estimate of the ground-state energy from VMC is equal (within error bars) to the ViT estimate (see Figs.~\ref{fig:lanczos}(b)--(e) and Table~\ref{tab:methods}). By using the ground-state energy estimates on different cluster sizes $(L=8, 10, 12, 14)$, we can also perform a finite-size extrapolation; see Fig.~\ref{fig:lanczos}(f). At $J_{s}/J_{d}=0.80$, our estimate for the energy per site of the infinite two-dimensional system is
\begin{equation}
E^{\mathrm{2D}}_{\infty}/J_{d}=-0.4481(3)\,.
\end{equation}
This is equal (within error bars) to the ViT~\cite{Luciano-2024} and DMRG~\rev{\cite{Yang-2022}} estimates (see Fig.~\ref{fig:lanczos}(f) and Table~\ref{tab:methods}), and we obtain this competitive value by optimizing \emph{only} eight parameters [six from Eq.~\eqref{eq:MF-Z2} and two from the L\'anczos procedure in Eq.~\eqref{eq:lanczos}], as opposed to the very large number of parameters that need to be optimized in the ViT~\cite{Luciano-2024} and DMRG~\rev{\cite{Yang-2022}} approaches. Note, moreover, that our energies are obtained for a state that has all the symmetries of the lattice by construction, while the DMRG states are computed on cylinders with open boundaries.

In summary, our energetic analysis lends support to the scenario of a $\mathbb{Z}_{2}$ Dirac state providing a good variational description of the gapless QSL reported in earlier studies~\cite{Wang-2022,Luciano-2024}. In the next section, based on an analysis of the correlation functions, we bolster the case for a $\mathbb{Z}_{2}$ QSL ground state and argue that it is the same phase as the QSL in the $J_{1}$--$J_{2}$ square-lattice Heisenberg antiferromagnet~\cite{Hu-2013}.

\subsubsection{Correlations}
We begin by comparing the real-space, isotropic, equal-time, spin-spin correlations of the $\mathbb{Z}_{2}$ Dirac state (supplemented with 2 LS) with those obtained from the unbiased neural\rev{-}network ViT wavefunction, which has recently been diagnosed to feature gapless excitations~\cite{Luciano-2024}. In Fig.~\ref{fig:corr_L14}(a), for the $L$\,$=$\,$14$ cluster, we notice that the Z3000 state correctly reproduces the sign of \emph{all} ViT correlators\rev{,} with their magnitudes also \rev{in} excellent agreement. A similar level of concurrence between VMC, ViT, and ED correlators is also observed on the $6\times6$ cluster [see Fig.~\ref{fig:corr_L14}(b)]. In Fourier space, the equal-time spin structure factor is
\begin{equation}
\label{eq:spin_sf}
\mathcal{S}(\mathbf{k})=\sum_{\mathbf r}e^{\rev{\dot\iota}\mathbf{k}\cdot\mathbf r}\,\langle\hat{\mathbf S}_0\cdot\hat{\mathbf S}_{\mathbf r}\rangle,
\end{equation}
whose peak at the antiferromagnetic wavevector $\mathbf k_m=(\pi,\pi)$ defines the squared magnetization $m^2=\mathcal{S}(\pi,\pi)/N$. For the projected $\mathbb{Z}_{2}$ Dirac state obtained from VMC, $\mathcal{S}(\mathbf k)$ displays \rev{a soft maximum} at $(\pi,\pi)$ [see Fig.~\ref{fig:corr_L14}(c)], with its value on the $L=14$ cluster being $\mathcal{S}(\pi,\pi)\approx10.19$; this is to be compared with $\mathcal{S}(\pi,\pi)\approx 9.89$ obtained from ViT on the same cluster\footnote{The VMC result is the zero-variance extrapolated estimate for the ground state\rev{,} which is obtained by supplementing the projected $\mathbb{Z}_{2}$ Dirac state with one and two L\'anczos steps followed by a zero-variance extrapolation of $\mathcal{S}(\pi,\pi)$ via a quadratic fit, while the ViT result is extrapolated with the number of layers.}. These results are in agreement and provide evidence that the gapless spin liquid found in Refs.~\cite{Luciano-2024,Yang-2022} is a $\mathbb{Z}_{2}$ Dirac QSL described by the Z3000 state.

\begin{figure}[t]
\includegraphics[width=\linewidth]{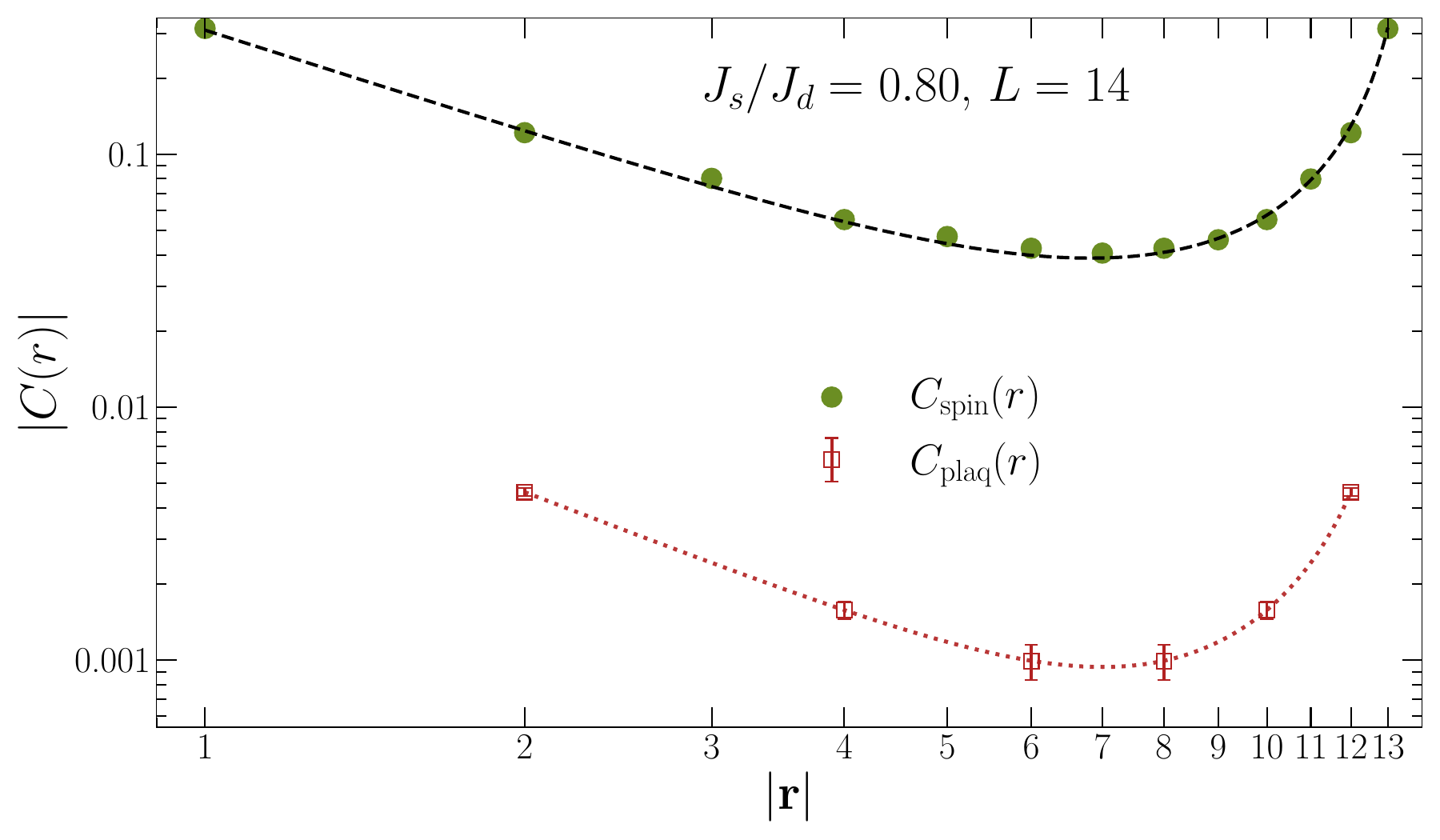}
	\caption{
Magnitude of equal-time real-space correlations for the $\mathbb{Z}_{2}$ DSL, plotted on a log-log scale along the $\mathbf a_{1}=(1,0)$ direction on a $14\times14$ Shastry-Sutherland lattice at $J_s/J_d=0.80$. We show the spin-spin correlator
$C_{\rm spin}(r)=|\langle \hat{\mathbf S}_{i}\cdot\hat{\mathbf S}_{i+r}\rangle|$ (where $r=|\mathbf{r}|$) together with the connected plaquette-plaquette correlator
$C_{\rm plaq}(r)=|\langle \hat Q_{0}\hat Q_{r}\rangle-\langle \hat Q_{0}\rangle\langle \hat Q_{r}\rangle|$,
where $\hat P_{r}$ is the four-spin cyclic permutation operator on the empty Shastry-Sutherland plaquette \rev{anchored at $r$ and $\hat Q_{r}=\frac{1}{2}(\hat P_{r}+\hat P_{r}^{-1})$ is its time-reversal-symmetric combination}. The correlations are obtained by extrapolating the zero-, one-, and two-L\'anczos-step estimates to the zero-variance limit using a quadratic fit; the corresponding error bars are of order $10^{-4}$. To account for periodic boundary conditions, which produce the upturn at large distances, the data are fitted to the periodic power-law form~\cite{Morita-2015,Nomura-2021}
$C(r)\propto |r|^{-p}+\sum_{n\neq0}\left(|r-nL|^{-p}-|nL|^{-p}\right)$
along the one-dimensional cut. The spin-spin fit gives $p_{\rm spin}=z+\eta=1.35(3)$, while the plaquette-plaquette correlations fitted \rev{with} the same form give $p_{\rm plaquette}\rev{{}={}}1.62(5)$\rev{, which is the analogous exponent for the plaquette channel}.
}
	\label{fig:corr_L14_decay}
\end{figure}

\begin{figure*}[tb]
\includegraphics[width=1.0\linewidth]{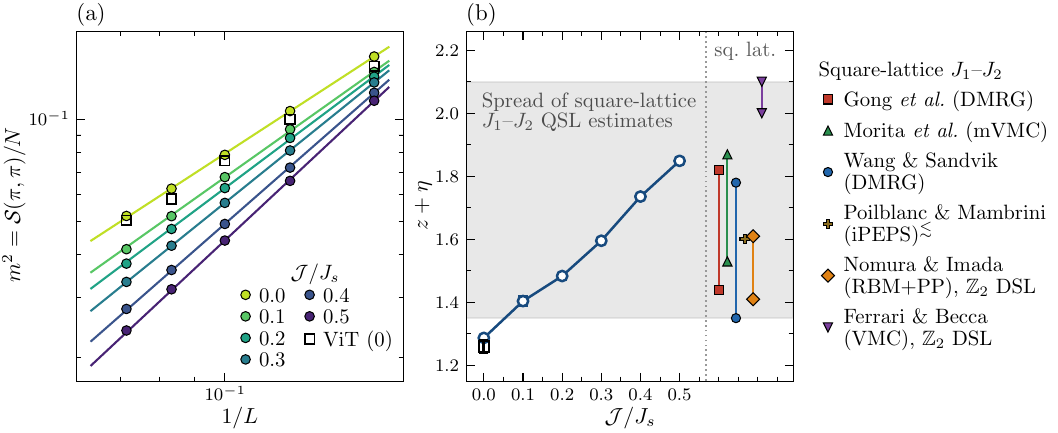}
\caption{Evolution of the spin-spin correlation decay exponent $z+\eta$ along
the river of liquidity, from the Shastry-Sutherland limit
($\mathcal{J}/J_{s}=0$) towards the square-lattice $J_{1}$--$J_{2}$ limit.
(a)~Log-log plot of the squared magnetization $m^{2}=\mathcal{S}(\pi,\pi)/N$
versus $1/L$ for $L=6$--$14$ at the values of $\mathcal{J}/J_{s}$ marked in
Figs.~\ref{fig:crossings} and~\ref{fig:overlap}; solid lines are fits to the
critical form $m^{2}(L)=A\,L^{-(z+\eta)}$. As in
Fig.~\ref{fig:m2vsL}(b), each $m^{2}(L)$ is the zero-variance extrapolated
estimate obtained from zero, one, and two L\'anczos steps. Open squares are the
independent ViT data~\cite{Luciano-2024} at the Shastry-Sutherland point.
(b)~The fitted exponent $z+\eta$ as a function of $\mathcal{J}/J_{s}$
(blue circles; the open square is the ViT cross-check at
$\mathcal{J}/J_{s}=0$). To the right of the dotted line we collect, for
comparison, the values reported for the gapless QSL window of the
square-lattice $J_{1}$--$J_{2}$ model: Gong \textit{et al.}~\cite{Gong-2014},
Morita \textit{et al.}~\cite{Morita-2015}, Wang and
Sandvik~\cite{Wang-2018}, Poilblanc and Mambrini~\cite{Poilblanc-2017}
(an upper bound, $z+\eta=1+\eta_{s}\lesssim1.6$), Nomura and
Imada~\cite{Nomura-2021}, and Ferrari and Becca~\cite{Ferrari-2020}; the last
two, like the present work, explicitly identify the state as a
$\mathbb{Z}_{2}$ Dirac spin liquid, with Ref.~\cite{Ferrari-2020} evaluating the projected state without the application of L\'anczos steps. Each vertical bar spans the range of values quoted by that study across the nonmagnetic window, illustrating both that the
exponent is coupling dependent and that the various methods do not agree on a
single value. The shaded band marks the overall spread of these estimates.}
\label{fig:exponent_river}
\end{figure*}

\begin{figure}
\includegraphics[width=0.49\linewidth]{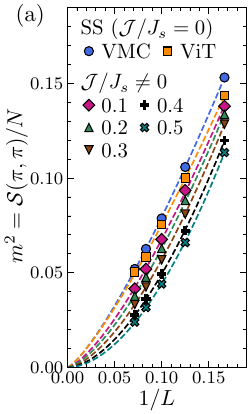}
\includegraphics[width=0.49\linewidth]{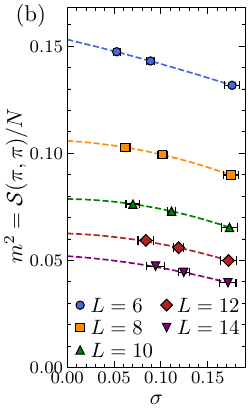}
	\caption{(a) Finite-size scaling of the squared magnetization $m^2=\mathcal{S}(\pi,\pi)/N$ (where $N=L\times L$) as a function of $1/L$ from $L=6$ up to $L=14$ obtained from VMC, and compared to ViT data from Ref.~\cite{Luciano-2024}. The values reported for each size $L$ are obtained by extrapolating the values of $m^2$ at zero, one, and two L\'anczos steps to zero variance via a quadratic fit [see panel (b)], which yields their estimates in the ground state. The error bars of the extrapolated values in the thermodynamic limit are estimated via a resampling technique with Gaussian noise. The fits in (a) associated \rev{with the} dashed curves are obtained using the critical form $m^2=aL^{-(z+\eta)}$, giving $z+\eta=1.30(3)$.}
	\label{fig:m2vsL}
\end{figure}

In Fig.~\ref{fig:corr_L14_decay}, we show the decay of real-space spin-spin correlations $|\langle\mathbf{\hat S}_{i}\cdot\mathbf{\hat S}_{j}\rangle|$ \rev{along} the $(1,0)$ direction on a $14\times14$ cluster at $J_{s}/J_{d}=0.80$. Given the gapless nature of the phase, one expects that $|\langle\mathbf{\hat S}_{i}\cdot\mathbf{\hat S}_{j}\rangle|\propto|\mathbf{r}|^{-(z+\eta)}$, manifesting the critical behavior due to gapless triplet excitations. In fact, we find such a power-law decay with an exponent of $z+\eta=1.35(3)$\rev{,} indicating the critical nature of N\'eel antiferromagnetic correlations in the QSL phase, in accordance with the findings of DMRG and ViT~\cite{Yang-2022,Luciano-2024}. Since estimates from a single system size alone could have ambiguities, we corroborate our finding by also estimating the exponent from the system size dependence of $m^2 (L)=\mathcal{S}(\pi,\pi)/N\rev{\propto} L^{-(z+\eta)}$ [see Fig.~\ref{fig:m2vsL}(a)], which yields $z+\eta=1.30(3)$; thus, both our estimates are in agreement with the DMRG and ViT estimate of $\sim1.3$~\rev{\cite{Yang-2022,Luciano-2024}}. Given the $\mathbb{Z}_{2}$ nature of the QSL, one expects that\rev{,} upon tuning $\mathcal{J}/J_{s}$ (and the other couplings that interpolate to the square lattice)\rev{,} the exponent varies continuously across the river of liquidity. We have verified this explicitly. Repeating the above analysis---i.e., the zero-variance extrapolation of $m^{2}(L)$ from zero, one, and two L\'anczos steps, followed by a fit to the critical form $m^{2}(L)\propto L^{-(z+\eta)}$---at the points along the river marked in Figs.~\ref{fig:crossings} and~\ref{fig:overlap}, we obtain $z+\eta=1.29(2)$, $1.40(2)$, $1.48(1)$, $1.60(2)$, $1.74(2)$, and $1.85(2)$ for $\mathcal{J}/J_{s}=0$, $0.1$, $0.2$, $0.3$, $0.4$, and $0.5$, respectively [see Fig.~\ref{fig:exponent_river}]. The exponent thus evolves smoothly and monotonically, without any discontinuity, from the Shastry-Sutherland value towards the range reported for the gapless QSL of the square-lattice $J_{1}$--$J_{2}$ antiferromagnet. We stress that no single value is agreed upon for the latter: published estimates span $z+\eta\simeq1.35$--$1.87$ from DMRG and many-variable variational Monte Carlo~\cite{Gong-2014,Morita-2015,Wang-2018}, $z+\eta\simeq1.41$--$1.61$ from a neural\rev{-}network many-body solver~\cite{Nomura-2021}, an upper bound $z+\eta\lesssim1.6$ from iPEPS~\cite{Poilblanc-2017}, and $z+\eta\simeq2.0$--$2.1$ from Gutzwiller-projected \rev{wavefunctions} (without the application of L\'anczos steps)~\cite{Ferrari-2020,Ferrari-2021}; moreover, all of the studies that surveyed the frustration ratio find the exponent itself to be coupling dependent~\cite{Gong-2014,Morita-2015,Wang-2018,Nomura-2021}. The values obtained here fall within this spread, \rev{the quoted uncertainties are fit-covariance estimates over $L=6$--$14$, and the exponents} should be understood as effective \rev{ones} on the accessible clusters. These findings provide evidence that the QSL region in the Shastry-Sutherland antiferromagnet is described by a $\mathbb{Z}_{2}$ Dirac QSL and could putatively be in the same phase as the QSL in the square-lattice Heisenberg antiferromagnet. Furthermore, this suggests a mechanism common to both lattices whereby N\'eel order melts due to frustration to give way to a gapless spin liquid~\cite{Hu-2013}.

Going beyond such equal-time correlations, next, we discuss the \textit{dynamical} spin structure factors obtained using the Keldysh pseudo-fermion functional renormalization group (pf-FRG).

\subsection{Keldysh pf-FRG}
To compare the proposed \textit{Ans\"atze} with numerical results, we employ a novel methodological approach: the \rev{pf-FRG} within the Keldysh formalism. This allows for direct computation of the dynamical spin structure factor for the Heisenberg model\footnote{Some early attempts have been made at obtaining  $\mathcal{S}({\mathbf q},\omega)$ within the dimer phase of the SSM based on variants of perturbative approaches~\cite{Knetter-2004}.}. The pf-FRG is already a well-established method for investigating the ground-state properties of Heisenberg systems, especially for frustrated and higher-dimensional lattices, where conventional techniques such as DMRG, quantum Monte Carlo (QMC), and VMC face difficulties, e.g., the sign problem or challenges due to dimensionality~\cite{Mueller-2023}.

The use of the Keldysh formalism makes it possible to calculate dynamical properties directly on the real-frequency axis, avoiding the need for analytical continuation associated with Matsubara frequencies (which is known to be \rev{unreliable because of} numerical inaccuracies~\cite{Reuther-2014}). The drawback of the Keldysh pf-FRG is that we have to resort to a finite temperature for the calculations, which, in this case, is given by $T/J_d=0.1$. This is because the FRG flow shows nonmonotonic features below this temperature scale---possibly due to the numerical integration schemes involved---which then induce artifacts in the structure factor. The Keldysh formalism also increases the computational cost, restricting our analysis to a $9 \times 9$ lattice.
Furthermore, the starting point for the FRG is a quantum paramagnetic state, which, together with the finite temperature, suppresses effects around $(0,0)$ that other techniques might predict more reliably. Thus, we cannot exclude a spectral branch emanating from the $\Gamma(0,0)$ point at small frequencies, as obtained at the mean-field level [see Fig.~\ref{fig:z3000_dis_ssf_dsf}(c)]. All details concerning the Keldysh pf-FRG are discussed in a companion publication~\cite{Potten-Prep}.

\begin{figure}[t]
\includegraphics[width=1.0\linewidth]{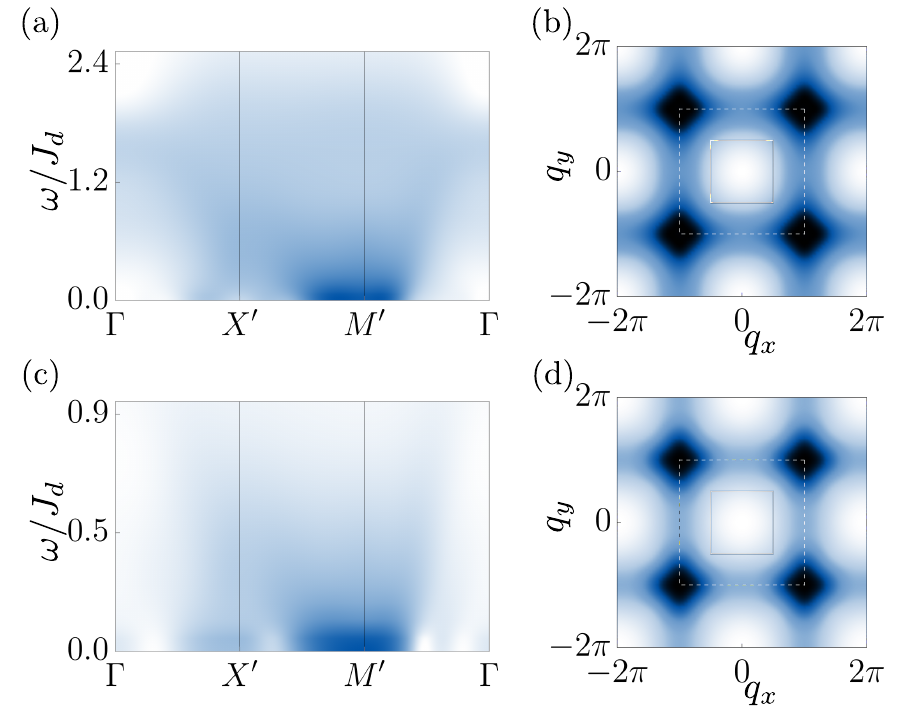}
	\caption{(a) Dynamical and (b) equal-time spin structure factors obtained from Keldysh pf-FRG for $J_{s}/J_{d}=0.80$ at $T/J_{d}=0.1$ on a $9\times9$ lattice. (c, d) The corresponding structure factors upon introduction of an XXZ anisotropy of $J_{zz}/J_{xy}=2.5$\rev{,} relevant for Pr$_2$Ga$_2$BeO$_7$. The conventions for the color maps are adopted from those used in \rev{Fig.}~\ref{fig:z3000_dis_ssf_dsf}(c) and (d), respectively.}
	\label{fig:SSfactor}
\end{figure}

In practice, we use the Keldysh pf-FRG to calculate the retarded dynamical spin susceptibility $\chi^\text{ret}(\mathbf{q},\omega)$, which can be related to the spin structure factor by
\begin{align}\label{eq:structureFactor}
\mathcal{S}^{}(\mathbf{q},\omega)=\frac{1}{\pi}(1-\mathrm{e}^{-\beta\omega})^{-1}\text{Im}\big(\chi^\text{ret}(\mathbf{q},\omega)\big)\,,
\end{align}
\rev{where $\beta=1/T$.}
Naturally, we exclude small frequencies for which numerical artifacts would be \rev{spuriously} enhanced.
Our results for the dynamical spin structure factor are presented in Fig.~\ref{fig:SSfactor}(a). Due to the finite temperature, we see \rev{strongly} broadened branches\rev{,} but still two distinct lobes of maximal intensity around $(\pi,\pi)$ [at $(\pi\pm0.4,\pi\pm0.4)$] are discernible, while the $(\pi,\pi)$ point itself also features appreciable intensity. This matches the expectation for the scenario of a $\mathbb{Z}_{2}$ Dirac QSL with nodal points at $M(\pi/2,\pi/2)+(\pm\varepsilon,\pm\varepsilon)$ [see Fig.~\ref{fig:z3000_dis_ssf_dsf}(b)]. Indeed, scattering from thus-situated Dirac cones would result in the principal intensity at low energy being observed at $(\pi,\pi)\pm(2\varepsilon,2\varepsilon)$ with \rev{lower} intensity at $(\pi,\pi)$ due to the presence of a larger number of scattering channels (wavevectors connecting different Dirac cones) for the former. Furthermore, we see a gradual decrease in intensity with increasing $\omega$ and a broad continuum consistent with deconfined fractionalized excitations, as expected for a $\mathbb{Z}_{2}$ QSL.

To facilitate a direct comparison with future neutron-scattering measurements on Pr$_2$Ga$_2$BeO$_7$~\cite{Li-2024}, we evaluate the dynamical structure factor $\mathcal{S}({\bf q},\omega)$ at $J_{s}/J_{d}=0.80$ in the presence of XXZ anisotropy $J_{zz}/J_{xy}=2.5$ on both $J_{s}$ and $J_{d}$ bonds\rev{; this model has} been proposed in Ref.~\cite{Li-2024} as a candidate minimal effective Hamiltonian. In Fig.~\ref{fig:SSfactor}(c), the calculated $\mathcal{S}({\bf q},\omega)$ exhibits two prominent lobes of spectral weight characteristic of the Dirac spin liquid. We further predict enhanced spectral intensity extending to higher $\omega$, including a relative enhancement at the $M'$\rev{$(\pi,\pi)$} point \rev{of the extended Brillouin zone (EBZ)} compared to neighboring maxima [see Fig.~\ref{fig:SSfactor}(d)], together with vertical fountain-like streaks of scattering. The presence or absence of these features provides a direct and falsifiable benchmark for future neutron-scattering measurements probing the $(hk0)$ plane.

It is important to note that the corresponding classical $(S\to\infty)$ model hosts an incommensurate spiral ground state for $J_{s}/J_{d}<1$~\cite{Shastry-1981,Low-2002,Grechnev-2013}. The corresponding equal-time spin structure factor would show Bragg peaks at the incommensurate \rev{wavevectors} $(\pi,\Theta)$ or $(\Theta,\pi)$~\cite{Chung-2001} with $\Theta=\pi\pm\arccos(J_{s}/J_{d})$. Indeed, one of the double maxima observed along the segment $\overline{X'M'}$ \rev{[$X'(\pi,0)$]} (at a shifted value of $\Theta$ due to quantum fluctuations~\cite{Chubukov-1984}) reflects the underlying spiral correlations\footnote{Such a spiral phase was also shown to have a higher energy within the coupled-cluster approach~\cite{Darradi-2005}.}, while the additional maxima along the segment $\overline{\Gamma M'}$ at points $(\pi\pm\rev{\delta},\pi\pm\rev{\delta})$ \rev{reflect} the presence of Dirac cones and \rev{are} not expected classically. Given that the Keldysh pf-FRG results have been obtained without any assumptions on the wavefunction or symmetry constraints and, in this sense, \rev{are unbiased and follow} from a correlator-based approach, they further corroborate the picture of a $\mathbb{Z}_{2}$ Dirac QSL ground state.

Unlike its dynamical counterpart, the equal-time spin structure factor provided by Keldysh pf-FRG shows broad diffuse maxima centered around $(\pi,\pi)$. This is due to the similar cumulative spectral weights obtained at and around $(\pi,\pi)$ when $\mathcal{S}(\mathbf{q}, \omega)$ is integrated over $\omega$, which washes out the underlying distribution of intensities at small $\omega$. Moreover, this is consistent with the $\mathcal{S}(\mathbf{q})$ computed using VMC, which also shows a single broadened \rev{maximum} at $(\pi,\pi)$ [see Fig.~\ref{fig:corr_L14}(c)]. It is interesting to note that the equal-time $\mathcal{S}(\mathbf{q})$ obtained from both VMC [Fig.~\ref{fig:corr_L14}(c)] and Keldysh pf-FRG [Fig.~\ref{fig:SSfactor}(b)] \rev{does} not qualitatively differ from the mean-field profile [see Fig.~\ref{fig:z3000_dis_ssf_dsf}(d)]. This is likely due to the fact that the $\mathbb{Z}_{2}$ gauge field is gapped \rev{and mediates} only short-range interactions between spinons, \rev{so} it does not significantly alter the mean-field expectation~\cite{Wen-2007}.
In contrast, for gapless states with a U(1) IGG, gauge fluctuations can lead to drastic differences between the mean-field and projected $\mathcal{S}(\mathbf{q})$  (cf. Ref.~\cite{Kiese-2023} for the kagome U(1) Dirac QSL).

\section{Phase diagram}
\label{sec:phasediagram}

The location of the phase boundaries that flank the $\mathbb{Z}_2$ Dirac spin liquid, and the continuity of the corresponding transitions, are established here in two complementary and mutually consistent ways. We first determine the boundaries in an unbiased manner from the finite-size scaling of the N\'eel and plaquette-VBC order parameters and from a correlation-ratio analysis (Sec.~\ref{sec:orderparam}), computed within a single variational family able to represent the ordered phases and the spin liquid on the same footing (Sec.~\ref{sec:pfaffian}). We then corroborate these boundaries with the energetic competition of restricted trial states (Sec.~\ref{sec:energetics}).

\subsection{Unified pairing--magnetic variational wavefunctions}
\label{sec:pfaffian}

The mean-field description of the $\mathbb{Z}_2$ Dirac spin liquid in
Eq.~\eqref{eq:MF-Z2} contains singlet hopping ($\chi$) and singlet pairing
($\Delta$) amplitudes only, and is therefore, by construction, incapable of
developing magnetic order. To locate the phase boundaries in an unbiased
manner and to characterize the nature of the transitions, we require a
\emph{single} variational family in which the N\'eel order parameter can be
switched on and tuned continuously to zero while the $d$-wave spin-liquid
pairing remains active. To this end we augment the spinon Hamiltonian by an
in-plane (coplanar) magnetic field,
\begin{equation}
\label{eq:mf_pfaffian}
\begin{aligned}
\hat H^{\pdagger}_{\rm MF}
={}&
\underbrace{
\begin{aligned}
&\sum_{\langle ij\rangle,\alpha}
 \chi_{ij}\hat f^\dagger_{i\alpha}\hat f^{\pdagger}_{j\alpha}
+\sum_{ij}\Delta_{ij}
 \hat f^\dagger_{i\uparrow}\hat f^\dagger_{j\downarrow}
+\text{H.c.}
\\[-2pt]
&\qquad
+\mu\sum_{i,\alpha}
 \hat f^\dagger_{i\alpha}\hat f^{\pdagger}_{i\alpha}
\end{aligned}
}_{\substack{\mathbb Z_2\ \text{Dirac }\rev{\textit{Ansatz}}\\
\text{Eq.~\eqref{eq:MF-Z2}}}}
\\[-2pt]
&+h\sum_i\mathbf M_i\!\cdot\hat{\mathbf S}_i,
\end{aligned}
\end{equation}
where $\mathbf{M}^{}_{i}=\big(\cos(\mathbf q\!\cdot\!\mathbf R^{}_{i}),
\sin(\mathbf q\!\cdot\!\mathbf R^{}_{i}),0\big)$ encodes N\'eel order with pitch
vector $\mathbf q=(\pi,\pi)$, and $h$ is a variational parameter whose magnitude
controls the strength of the magnetic order [Eq.~\eqref{eq:mf_pfaffian}
generalizes the magnetic \textit{Ansatz} of Ref.~\cite{Iqbal-2016} [their Eq.~(9)] by
retaining the singlet-pairing channel of Eq.~\eqref{eq:MF-Z2}]. Setting $h=0$
recovers the fully symmetric $\mathbb{Z}_2$ Dirac spin liquid; setting the
pairings to zero recovers a magnetically ordered state of the type used to
build the Jastrow wavefunction of Eq.~\eqref{eqn:var-cl} \rev{below (Sec.~\ref{sec:energetics})}.

The coexistence of singlet pairing and a \emph{transverse} magnetic field
requires a wavefunction more general than a Slater determinant. Taken
separately, each channel is tractable: the magnetic term $h\,\mathbf M_i\!\cdot\!
\hat{\mathbf S}_i$ is, in the original spin basis, a number-conserving spin-flip
hopping, while the pairing term becomes number conserving after a particle-hole
transformation on the down spins~\cite{Iqbal-2016}, so that in
either case the Gutzwiller-projected amplitude reduces to an ordinary
determinant. When both terms are present, however, no single-particle basis
renders Eq.~\eqref{eq:mf_pfaffian} number conserving---the transverse field
becomes an anomalous (pairing) term precisely in the basis that resolves the
BCS pairing---and the projected amplitude is no longer a determinant.

We therefore adopt the general Bogoliubov--de Gennes (BdG) form of
Eq.~\eqref{eq:mf_pfaffian}. Writing the $2L^2$ spin-orbitals collectively as
$\hat f^{}_{I}$ (with $I$ running over sites and spin), the uncorrelated ground
state is the antisymmetrized geminal (pair) product
\begin{equation}
\label{eq:pfaffian_state}
|\Phi^{}_{0}\rangle=\exp\!\Big(\tfrac{1}{2}\sum_{IJ}F^{}_{IJ}\,
\hat f^{\dagger}_{I}\hat f^{\dagger}_{J}\Big)|0\rangle,\qquad
F=V\,U^{-1},
\end{equation}
where $U,V$ are the particle and hole blocks of the BdG Bogoliubov rotation and
$F=-F^{\rm T}$ is the antisymmetric pairing matrix. The Gutzwiller-projected
variational state is $|\Psi^{}_{\rm var}\rangle=\mathscr J_z\,\mathcal
P_{G}|\Phi^{}_{0}\rangle$, and the amplitude of a spin configuration $x$ (one
fermion per site) is a \emph{Pfaffian},
\begin{equation}
\label{eq:pfaffian_amp}
\langle x|\Psi^{}_{\rm var}\rangle \;\propto\; {\rm Pf}\big(F^{}_{x}\big),
\end{equation}
$F_x$ being the submatrix of $F$ on the occupied spin-orbitals of
$x$~\cite{Becca_Sorella_2017}.

In the Monte Carlo sampling, each elementary spin move $x\!\to\!x'$ is accepted
with a probability set by the ratio ${\rm Pf}(F_{x'})/{\rm Pf}(F_{x})$, evaluated in $\mathcal
O(N)$ per move through a low-rank (Sherman--Morrison-type) update of the inverse
pairing matrix $F_x^{-1}$; the variational parameters
$\{\chi,\Delta,\mu,h\}$ are optimized by stochastic reconfiguration using the
logarithmic derivatives $\partial_\alpha\ln\langle x|\Psi_{\rm var}\rangle=
\tfrac12\,{\rm Tr}\!\big(F_x^{-1}\partial_\alpha F_x\big)$. We validate the
Pfaffian implementation against the determinant code in the two limits $h\!\to\!0$
(pure $\mathbb{Z}_2$ Dirac spin liquid) and $\Delta\!\to\!0$ (pure magnetic
state), recovering the determinant amplitudes in both cases. This unified
family allows the N\'eel order parameter to develop and vanish continuously as a
function of $J_s/J_d$ within one wavefunction, which makes the
order-parameter finite-size scaling of Sec.~\ref{sec:orderparam} an internally
consistent probe of the QSL--N\'eel transition. For the plaquette VBC we retain
the descendant P5A \rev{\textit{Ansatz}} [Eq.~\eqref{eq:z3000_plaquette}], which likewise
contains the fully symmetric Z3000 state as the submanifold obtained when the
symmetry-inequivalent amplitudes are set equal, so that the plaquette
distortion too can be tuned continuously to zero.

\subsection{Order-parameter and correlation-ratio finite-size scaling}
\label{sec:orderparam}

\begin{figure*}[tb]
\includegraphics[width=1.0\linewidth]{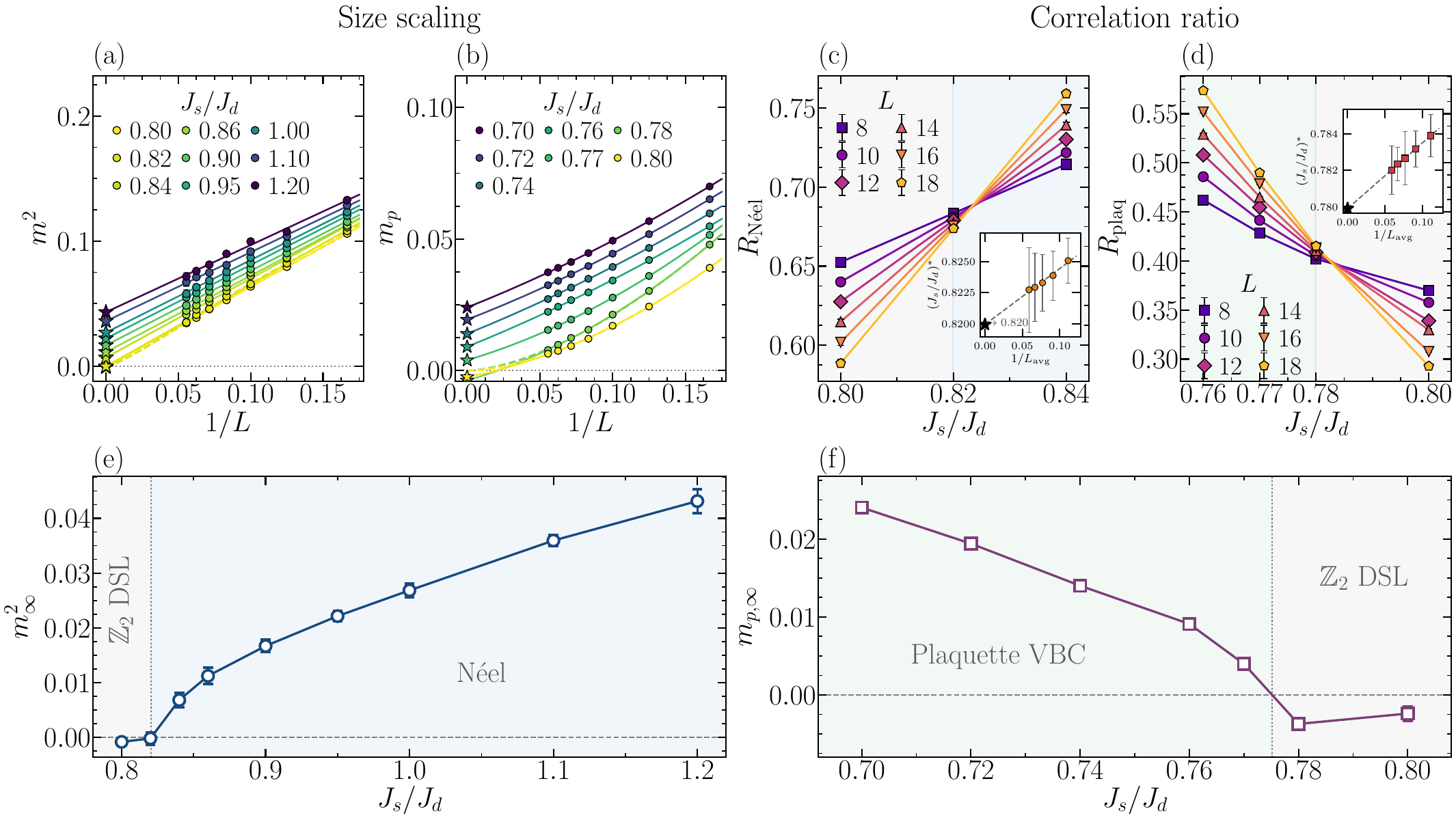}
\caption{Unbiased determination of the phase boundaries flanking the
$\mathbb{Z}_2$ Dirac spin liquid in the pure Shastry-Sutherland model
($\mathcal{J}=0$), computed within the unified variational family
of Sec.~\ref{sec:pfaffian}.
\emph{Top row, left pair\rev{---}size scaling.}
(a)~Squared N\'eel magnetization $m^2=\mathcal{S}(\pi,\pi)/N$ versus $1/L$ for
$L=6$--$18$ and coupling ratios $J_s/J_d=0.80$--$1.20$ (color scale); solid
lines are equal-weight linear fits $m^2(L)=m^2_\infty+a/L$, and the dashed
curves for the two near-critical ratios are the critical form
$m^2(L)=a\,L^{-(z+\eta)}$. (b)~Plaquette-VBC order parameter $m_p(L)$ versus
$1/L$ for $J_s/J_d=0.70$--$0.80$; solid lines are quadratic fits
$m_p(L)=m_{p,\infty}+b/L+c/L^2$, \rev{and the} dashed curves the bare critical form for the two
near-critical ratios. \emph{Top row, right pair\rev{---}correlation ratio.} (c)~N\'eel
correlation ratio $R_{\rm N\acute{e}el}=1-\mathcal S(\mathbf k_m+\delta\mathbf
k)/\mathcal S(\mathbf k_m)$, $\mathbf k_m=(\pi,\pi)$, and (d)~plaquette
correlation ratio $R_{\rm plaq}$, for $L=8$--$18$; the curves cross at the
respective transitions. Insets: crossing point $(J_s/J_d)^\ast$ of consecutive
sizes versus $1/L_{\rm avg}$, extrapolating (star) to
$(J_s/J_d)_{\rm N\acute{e}el}\simeq 0.820$ and
$(J_s/J_d)_{\rm PVBC}\simeq 0.780$.
\emph{Bottom row\rev{---}thermodynamic limit.} (e)~$m^2_\infty$ and
(f)~$m_{p,\infty}$ versus $J_s/J_d$; both order parameters vanish continuously
at $0.82$ and $0.775$\rev{,}
respectively, delimiting the intermediate $\mathbb{Z}_2$ Dirac spin-liquid
window. Statistical error bars are smaller than the symbols unless shown.}
\label{fig:orderparam}
\end{figure*}

To determine the phase boundaries of the spin-liquid regime without relying on
the energetic competition of separately optimized trial states, we compute the
N\'eel and plaquette-VBC order parameters, and their correlation ratios, as
functions of system size within the unified variational family introduced in
Sec.~\ref{sec:pfaffian}. This is the strategy employed for the
Shastry-Sutherland and $J_1$--$J_2$ models
in Refs.~\cite{Wang-2022,Yang-2022,Luciano-2024,Corboz-2025}, and it allows both
the location and the continuity of the transitions to be assessed on the same
footing.
\emph{Observables.} We probe the two competing orders through their static
structure factors and correlation ratios, following the conventions established
for the Shastry-Sutherland and $J_1$--$J_2$ models in
Refs.~\cite{Luciano-2024,Wang-2022}. N\'eel order is diagnosed by the spin structure factor $\mathcal{S}(\mathbf k)$ \rev{of Eq.~\eqref{eq:spin_sf} and the squared magnetization $m^{2}=\mathcal{S}(\pi,\pi)/N$ defined below it}. Plaquette-VBC order is
diagnosed by the four-spin cyclic permutation operator $\hat P_{\mathbf r}$, which
cyclically permutes the four spins of the empty Shastry-Sutherland plaquette anchored at the
site $\mathbf r$~\cite{Luciano-2024}. In terms of its time-reversal-symmetric
combination $\hat Q_{\mathbf r}=\tfrac12(\hat P_{\mathbf r}+\hat P_{\mathbf r}^{-1})$,
the connected plaquette--plaquette correlation and its Fourier transform read
\begin{align}
\label{eq:plaq_corr}
C_{\rm plaq}(\mathbf r)&=\langle \hat Q_{\mathbf r}\hat Q_0\rangle
-\langle \hat Q_{\mathbf r}\rangle\langle \hat Q_0\rangle ,\\
\label{eq:plaq_sf}
\mathcal{C}_{\rm plaq}(\mathbf{k})&=\sum_{\mathbf r}e^{\rev{\dot\iota}\mathbf{k}\cdot\mathbf r}\,
C_{\rm plaq}(\mathbf r) ,
\end{align}
with $\mathcal{C}_{\rm plaq}(\mathbf k)$ developing a peak at the plaquette
ordering wavevector $\mathbf k_p=(\pi,0)$ [equivalently $(0,\pi)$]. Following
Ref.~\cite{Luciano-2024}, we take as the plaquette order parameter the difference
of the diagonal correlation at its two largest distances,
\begin{equation}
\label{eq:mp_def}
m_p(L)=\big|\,C_{\rm plaq}(L/2,L/2)-C_{\rm plaq}(L/2-1,L/2-1)\,\big| ,
\end{equation}
which stays finite when plaquette order is present and vanishes otherwise.
Finally, the correlation ratios of the two orders are
\begin{equation}
\label{eq:corr_ratio}
R_{\rm N\acute{e}el}=1-\frac{\mathcal{S}(\mathbf k_m+\delta\mathbf k)}{\mathcal{S}(\mathbf k_m)},
\qquad
R_{\rm plaq}=1-\frac{\mathcal{C}_{\rm plaq}(\mathbf k_p+\delta\mathbf k)}{\mathcal{C}_{\rm plaq}(\mathbf k_p)},
\end{equation}
with $\lVert\delta\mathbf k\rVert=2\pi/L$. In an ordered phase\rev{,} the corresponding
structure factor develops a diverging peak and the ratio approaches unity in the
thermodynamic limit, whereas it flows to zero in the disordered phase; the
crossing of the $R(L)$ curves for different system sizes therefore locates the
transition~\cite{Luciano-2024,Wang-2022}.

\emph{N\'eel transition.} The squared magnetization
$m^2=\mathcal{S}(\pi,\pi)/N$ [Eq.~\eqref{eq:spin_sf}] is evaluated on $L\times L$ tori up to $L=18$ and
extrapolated to the thermodynamic limit. As shown in
Fig.~\ref{fig:orderparam}(a), a linear-in-$1/L$ extrapolation yields an
intercept $m^2_\infty$ that is zero within error throughout the spin-liquid
regime and turns on continuously as the N\'eel side is entered
[Fig.~\ref{fig:orderparam}(e)], vanishing at
$(J_s/J_d)_{\rm N\acute{e}el}\simeq 0.82$. Near the
transition the data are equally well described by the critical form
$m^2(L)\sim L^{-(z+\eta)}$ [dashed curves in Fig.~\ref{fig:orderparam}(a)], with
an anomalous exponent $z+\eta=1.30(3)$ consistent with the value obtained from
the long-distance decay of the real-space spin correlations
(Fig.~\ref{fig:corr_L14_decay}) and with DMRG and neural\rev{-}network
results~\cite{Yang-2022,Luciano-2024}. As an independent and manifestly
size-convergent probe, the N\'eel correlation ratio
$R_{\rm N\acute{e}el}$ [Fig.~\ref{fig:orderparam}(c)] displays curves for
different $L$ that cross at a single coupling. The consecutive-size crossings
drift smoothly with $1/L_{\rm avg}$ and extrapolate to
$(J_s/J_d)_{\rm N\acute{e}el}\simeq 0.820$ [inset],
in \rev{close} agreement with the order-parameter extrapolation.

\emph{Plaquette-VBC transition.} An analogous analysis for the plaquette-VBC
order parameter $m_p$ [Eq.~\eqref{eq:mp_def}], computed within the descendant P5A \rev{\textit{Ansatz}}
[Eq.~\eqref{eq:z3000_plaquette}], is shown in
\rev{Fig.}~\ref{fig:orderparam}(b) \rev{and} (f). A quadratic-in-$1/L$ extrapolation gives an
order parameter $m_{p,\infty}$ that is finite in the VBC phase and vanishes
continuously at $(J_s/J_d)_{\rm PVBC}\simeq 0.775$\rev{.}
The critical decay of the plaquette correlations themselves is not extracted from
this order-parameter scaling, but separately from the long-distance behavior of
the plaquette-plaquette correlation function $\rev{C}_{\rm plaq}(r)$ on the
$L=14$ cluster (Fig.~\ref{fig:corr_L14_decay}). The plaquette correlation ratio
$R_{\rm plaq}$ [Fig.~\ref{fig:orderparam}(d)] again exhibits a well-defined
crossing, extrapolating to
$(J_s/J_d)_{\rm PVBC}\simeq 0.780$ [inset], consistent
with the order-parameter determination.

Two conclusions follow. First, the two boundaries obtained here---$(J_s/J_d)_{\rm PVBC}\simeq 0.78$ and
$(J_s/J_d)_{\rm N\acute{e}el}\simeq0.82$---agree,
within error, with the independent estimates from DMRG, ED, neural\rev{-}network, and
tensor-network studies (Table~\ref{tab:transitions}), and with the energetic
consistency check of Sec.~\ref{sec:energetics} below. The intermediate
$\mathbb{Z}_2$ Dirac spin-liquid window is thus delimited by two mutually
consistent, unbiased diagnostics rather than by the crossing of restricted
variational energies alone. Second, on both sides the order parameter is found
to vanish \emph{continuously}, and each correlation ratio has a single crossing
point rather than the two-branch structure characteristic of a first-order
transition. Within the resolution of our data we therefore find no indication of
first-order behavior at either boundary; the results are consistent with the
continuous (or pseudocritical) scenario expected from the fermionic
deconfined-criticality description of the companion field
theory~\cite{Feuerpfeil-2026}, although a definitive determination of the
critical universality class lies beyond the present system sizes.

\subsection{Energetic competition: a consistency check}
\label{sec:energetics}
As an independent corroboration of the phase boundaries determined above, we compare the optimized variational energies of restricted trial states tailored to the competing N\'eel, plaquette-VBC, and spin-liquid phases. We emphasize that these energy crossings serve only as an energetic consistency check and are not used to determine the order or universality class of the transitions, which are addressed by the order-parameter and correlation-ratio analysis of Sec.~\ref{sec:orderparam}.

Having identified the nature of the QSL, we now investigate its energetic competition with the neighboring N\'eel antiferromagnetic and plaquette\rev{-}VBC phases that flank the spin-liquid regime. \rev{As already noted in Sec.~\ref{sec:pfaffian}, the QSL wavefunction is the ground state of a spin-rotation-invariant mean-field Hamiltonian [Eq.~\eqref{eq:MF-Z2}] and therefore cannot develop magnetic order}; in particular, the AFM order parameter vanishes identically. Consequently, to address the transition between the QSL and the N\'eel AFM phase, it is necessary to construct a variational wavefunction that explicitly supports magnetic order. By employing two restricted classes of wavefunctions---one constrained to describe the QSL and the other to capture AFM order---we independently evaluate the corresponding variational energies and obtain an energetic estimate of the phase competition, which we use to corroborate the order-parameter determination of Sec.~\ref{sec:orderparam}.
These two trial states are the limiting cases of a common Gutzwiller-projected fermionic construction---the spin liquid is recovered when the magnetic field is switched off, and a magnetically ordered state when the singlet pairing is switched off [cf.\ the unified family of Sec.~\ref{sec:pfaffian}]. An analogous energy-crossing determination of the spin-liquid--to--N\'eel boundary, in which the energy of a projected spin-liquid state crosses that of a Jastrow-projected magnetic state, was employed for the spin-$1/2$ kagome $J_1$--$J_2$ antiferromagnet in Ref.~\cite{Iqbal-2015}.

A simple yet accurate variational wavefunction for magnetically ordered phases can be constructed directly in terms of the spin degrees of freedom~\cite{Manousakis-1991}:
\begin{equation}
\label{eqn:var-cl}
|\Psi_{\rm Magnetic}\rangle = \mathscr{J}_z \, \mathcal{P}_{S^z_{\rm tot}=0} \, |{\rm SW}\rangle ,
\end{equation}
where $|{\rm SW}\rangle$ is a spin-wave state characterized by a wavevector ${\bf q}$ and a set of phase shifts $\{\rev{\vartheta}_i\}$, one for each site in the unit cell,
\begin{equation}
|{\rm SW}\rangle =
\prod_{i} \left( |\downarrow\rangle_i + e^{\rev{\dot\iota}({\bf q}\cdot{\bf R}_i + \rev{\vartheta}_i)} |\uparrow\rangle_i \right).
\end{equation}
The state $|{\rm SW}\rangle$ is equivalent to a classical configuration in which each spin points along a fixed direction in the XY plane. The operator $\mathcal{P}_{S^z_{\rm tot}=0}$ projects onto the subspace with vanishing total magnetization along the $z$-axis. Quantum fluctuations beyond the classical spin configuration are incorporated through a long-range Jastrow factor,
\begin{equation}
\mathscr{J}^{}_z = \exp\!\left( \frac{1}{2} \sum_{ij} v^{}_{ij} \hat S^z_i \hat S^z_j \right),
\end{equation}
where, in a translationally invariant system, the pseudopotential $v_{ij}$ depends only on the distance $|{\bf R}_i-{\bf R}_j|$ between sites. In the following, we focus on the N\'eel ordering vector ${\bf q}=(\pi,\pi)$, for which the four spins in the unit cell form a \rev{two-sublattice} antiparallel arrangement. All independent parameters entering the pseudopotential $v_{ij}$ are optimized using variational Monte Carlo simulations.

\begin{figure}[!t]
\includegraphics[width=1.0\linewidth]{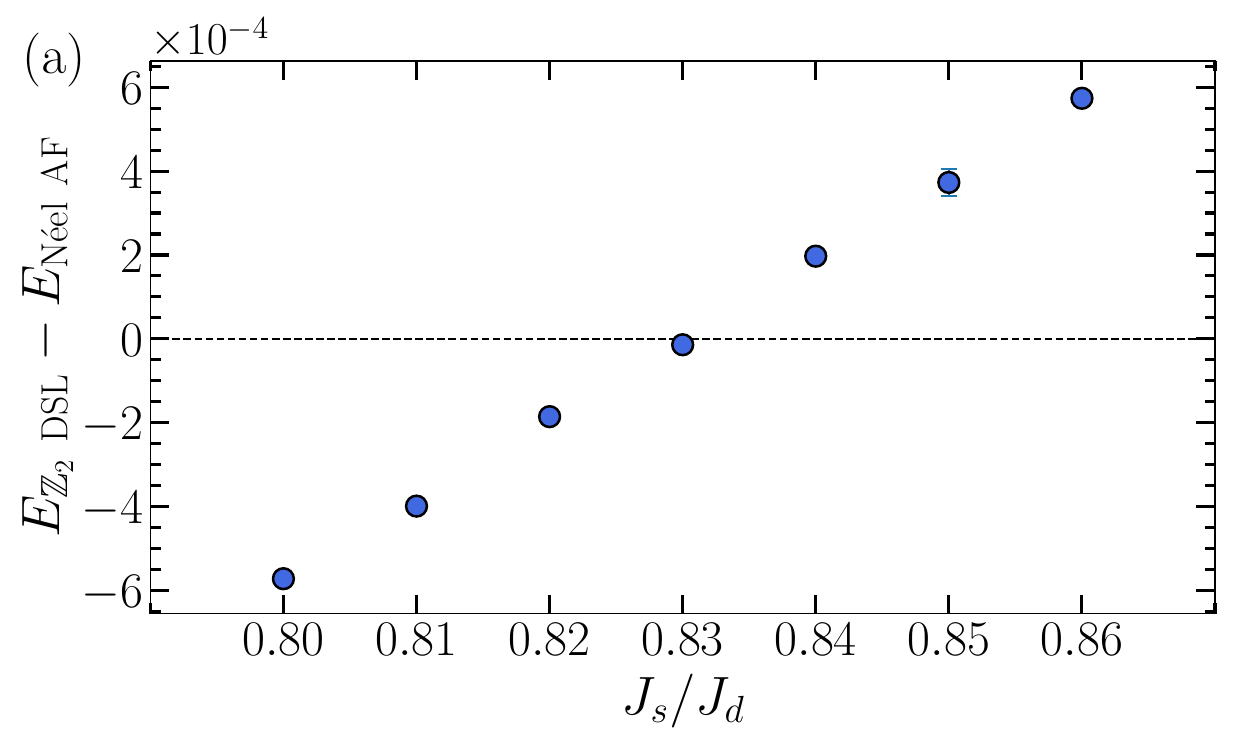}
\includegraphics[width=1.0\linewidth]{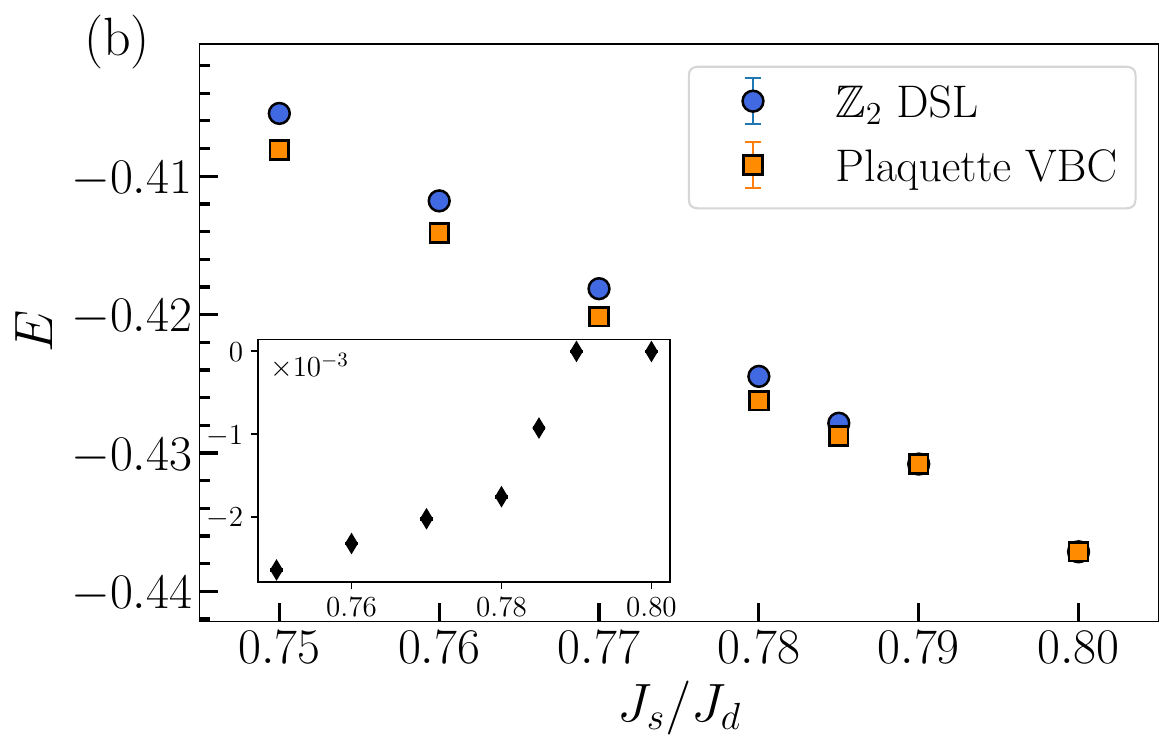}
	\caption{(a) The net energy difference per site, $(E_{\mathbb{Z}_2\,{\rm DSL}}-E_{\rm N\acute{e}el\,AFM})/J_d$, on an $L=30$ ($900$ sites) cluster of the pure Shastry-Sutherland model ($\mathcal{J}=0$), as a function of $J_s/J_d$. The transition point is identified from the intersection with the dashed line corresponding to zero difference in energies, yielding an estimate of $J_s/J_d \approx 0.83$. No systematic shift of this crossing is observed within numerical error upon increasing the system size. The resulting energy difference, of order $10^{-4}$, is consistent with previous findings~\cite{Corboz-2025}. (b) Evolution with $J_s/J_d$ of the energy per site, $E/J_d$, of the $\mathbb{Z}_2$ Dirac QSL and plaquette\rev{-}VBC states on the same $L=30$ cluster. Inset: energy difference $E_{\rm PVBC}-E_{\mathbb{Z}_2\,{\rm DSL}}$. For $J_s/J_d \lesssim 0.78$, the plaquette VBC has a lower variational energy than the $\mathbb{Z}_2$ DSL, with an energy gain that increases as $J_s/J_d$ is further reduced. In both panels, the statistical error bars are of order $10^{-6}$.}
	\label{fig:qsl_af}
\end{figure}

\floatsetup[table]{capposition=bottom}
\begin{table}
\centering
\begin{tabular}{lll}
 \hline \hline
       \multicolumn{1}{c}{Method}
    & \multicolumn{1}{c}{$(J_{s}/J_{d})_{\rm PVBC}$}
    & \multicolumn{1}{c}{$(J_{s}/J_{d})_{\rm N\acute{e}el}$}
      \\ \hline

VMC (energy crossing) & $\approx 0.78$ & $\approx 0.83$  \\
VMC (order param.\ / corr.\ ratio) & $0.775$ & $0.820$  \\

DMRG (Yang \emph{et al.}~\cite{Yang-2022}) & $0.788(2)$ & $0.820(2)$    \\

ED (Wang \emph{et al.}~\cite{Wang-2022}) & $0.789(4)$ & $0.824(8)$  \\

pf-FRG (Kele\c{s} \emph{et al.}~\cite{Keles-2022}) & $\approx 0.77$ & $\approx 0.82$  \\

ViT (Viteritti \emph{et al.}~\cite{Luciano-2024}) & $0.775(5)$ & $0.815(5)$   \\

iPEPS (Corboz \emph{et al.}~\cite{Corboz-2025}) & $0.785(5)$ & $0.82(1)$   \\

 \hline \hline
\end{tabular}
\caption{Critical coupling ratios for the transitions between the QSL and the plaquette VBC, $(J_{s}/J_{d})_{\rm PVBC}$, and between the QSL and the N\'eel antiferromagnet, $(J_{s}/J_{d})_{\rm N\acute{e}el}$, as obtained from \rev{VMC} and compared against results from other numerical methods.}
\label{tab:transitions}
\end{table}

We perform VMC calculations to determine the optimized energies of the Z3000 quantum spin liquid and the N\'eel antiferromagnetic state using the wavefunctions introduced above. The energy difference between these two competing states on an $L=30$ cluster is shown in Fig.~\ref{fig:qsl_af}(a). The crossing with the dashed line defined by $E_{\mathbb{Z}_{2}\,{\rm DSL}} - E_{\rm N\acute{e}el\,AFM} = 0$ yields an estimate of the transition point $(J_{s}/J_{d})_{\rm N\acute{e}el} \approx 0.83$, in \rev{close} agreement with results obtained from other numerical approaches (see Table~\ref{tab:transitions}). Upon increasing the system size, we observe no systematic shift of the crossing within statistical error bars, indicating that this estimate provides a reliable determination of the QSL--N\'eel AFM transition in the thermodynamic limit.

\begin{figure}
	\includegraphics[width=0.5\linewidth]{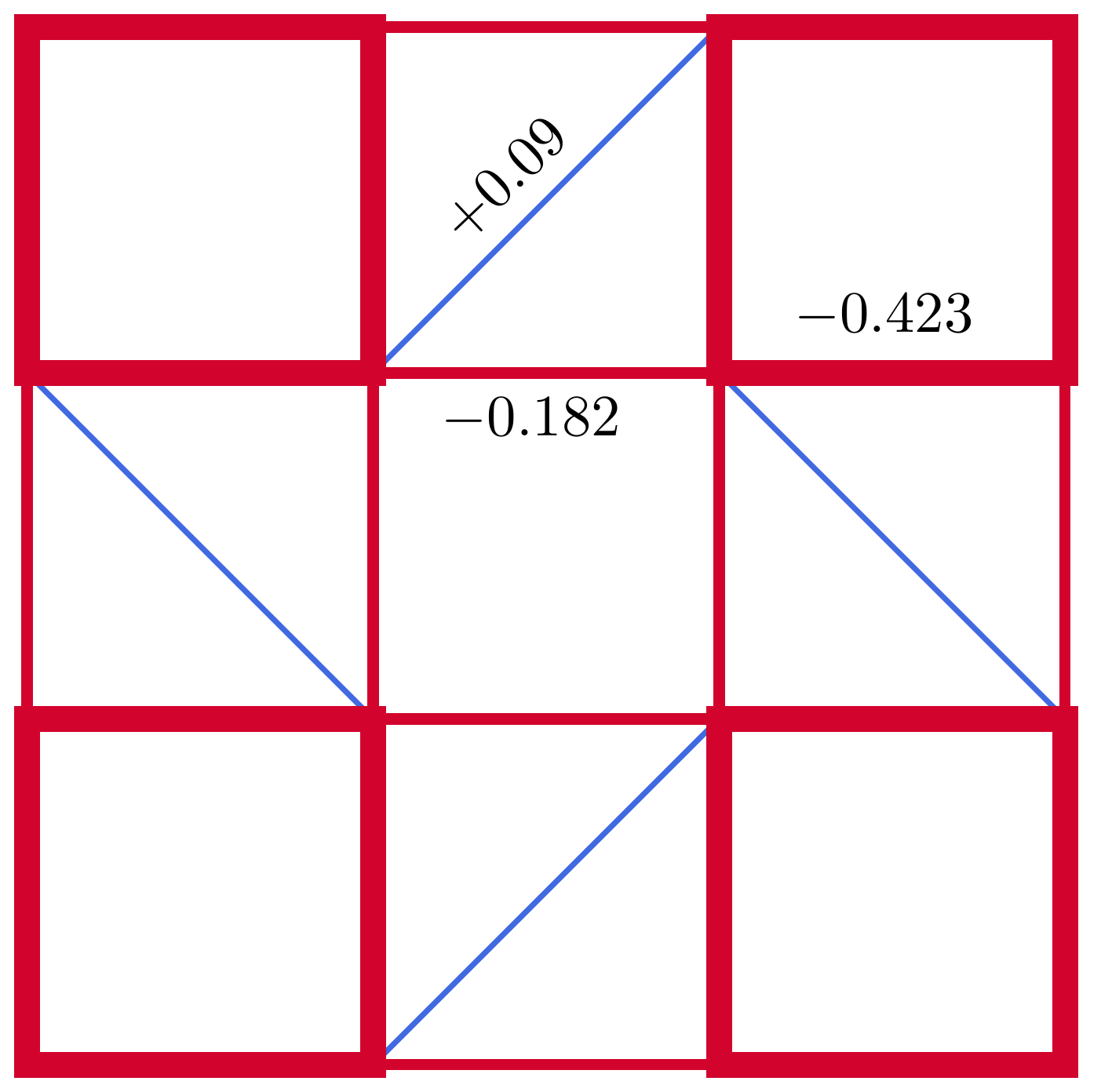}
	\caption{Real-space spin-spin correlations $\langle \hat{\mathbf S}_{i} \cdot \hat{\mathbf S}_{j} \rangle$ obtained from a representative VMC snapshot taken from the center of an $L=30$ cluster at $J_s/J_d = 0.76$. The width of each bond is proportional to the magnitude of $\langle \hat{\mathbf S}_{i} \cdot \hat{\mathbf S}_{j} \rangle$, while red (blue) bonds indicate antiferromagnetic (ferromagnetic) correlations. The numerical values of the correlations are indicated on the corresponding bonds. Clear singlet formation is observed on the empty square plaquettes, characteristic of the plaquette\rev{-}VBC phase. Statistical error bars are of order $10^{-5}$.}
	\label{fig:sisj_pvbc}
\end{figure}

To construct variational wavefunctions for the plaquette VBC---a phase that preserves translational symmetry while breaking glide $G_x$ and mirror $\sigma_{xy}$ symmetries---we classify projective symmetry groups compatible with this reduced symmetry set (see App.~\ref{app:nematic}). We identify the PSG of the plaquette VBC that continuously descends from the Z3000 QSL and construct the corresponding fermionic mean-field Hamiltonian [Eq.~\eqref{eq:z3000_plaquette} and Fig.~\ref{fig:plaquette_ansatz_p5a}]. Importantly, this Hamiltonian is smoothly connected to that of the fully symmetric Z3000 QSL [Eq.~\eqref{eq:MF-Z2}], which is recovered when symmetry-inequivalent hopping and pairing amplitudes are tuned to be equal.

Starting from this mean-field description, we generate the ground state and optimize the corresponding Gutzwiller-projected variational wavefunction within VMC. We find that for $J_{s}/J_{d} \lesssim 0.78$, the optimized plaquette\rev{-}VBC wavefunction attains a lower energy than the Z3000 QSL on the largest system sizes studied ($L=30$), with an energy gain that remains finite upon increasing system size [see Fig.~\ref{fig:qsl_af}(b)]. The associated real-space spin-spin correlations, shown in Fig.~\ref{fig:sisj_pvbc} for $J_s/J_d = 0.76$, clearly reveal singlet formation on the empty square plaquettes. By contrast, for $J_{s}/J_{d} \gtrsim 0.78$, the plaquette\rev{-}VBC wavefunction flows back to the fully symmetric QSL upon variational optimization. These results offer an estimate of the QSL--plaquette\rev{-}VBC transition at $(J_{s}/J_{d})_{\rm PVBC} \approx 0.78$, consistent with other numerical studies (Table~\ref{tab:transitions}), and establish the stability of the plaquette\rev{-}VBC phase in the thermodynamic limit.

\section{Conclusions and outlook}
\label{sec:conclusions}
A central pillar of our identification of the quantum spin liquid on the Shastry-Sutherland lattice is provided by symmetry-resolved exact diagonalization. By tracking the evolution of low-energy states across total-spin, reflection, and glide symmetry sectors, we identify a finite intermediate regime in which neither the Anderson tower of states associated with N\'eel order nor the quasidegenerate singlet manifold characteristic of plaquette or columnar symmetry breaking is present. The composite gap crossings introduced in Ref.~\cite{Wang-2022} remain well separated under finite-size scaling, demonstrating that this regime is not a fine-tuned point but a finite-width phase.

Importantly, as the Hamiltonian is continuously deformed from the Shastry-Sutherland limit to the square-lattice $J_1$--$J_2$ model, the symmetry-resolved spectrum evolves smoothly without qualitative reorganization. The resulting river of liquidity establishes adiabatic continuity between the two models. Wave\rev{-}function overlap diagnostics independently confirm that the same ground-state manifold persists throughout this interpolation. Taken together, these exact-diagonalization results provide controlled spectral evidence that the intermediate phase is a genuine quantum-disordered state extending across distinct frustrated Hamiltonians.

To identify the precise nature of the QSL, \rev{we chart} out an atlas of QSLs on the Shastry-Sutherland lattice along with their spectroscopic signatures. This is accomplished using the framework of projective symmetry groups, which, for fermionic spinons, yields 46 U(1) and 80 $\mathbb{Z}_{2}$ QSL states. Using this information, we investigate the QSL regime of the generalized spin-$1/2$ Shastry-Sutherland Heisenberg antiferromagnet and its interpolation to the square-lattice limit. We identify the subset of PSGs \rev{that} can be continuously connected to the ones on the square lattice. In particular, we pinpoint the counterpart of the $\mathbb{Z}_{2}$ Dirac QSL on the square lattice (more precisely, the $d$-wave pairing Z2A$zz$13 state in Ref.~\cite{Wen-2002}), which is a prime candidate for the QSL ground state of the $S$\,$=$\,$1/2$ $J_{1}$--$J_{2}$ square-lattice Heisenberg antiferromagnet. Our equivalent $\mathbb{Z}_{2}$ Dirac QSL (called the Z3000 state in this work) upon Gutzwiller projection and optimization over only a few parameters---using variational Monte Carlo---produces energies \rev{that agree, within error bars, with those obtained from ViT for all cluster sizes~\cite{Luciano-2024} and from DMRG~\cite{Yang-2022} in the thermodynamic limit}. Similarly, the equal-time correlators computed with VMC for this $\mathbb{Z}_{2}$ Dirac QSL show remarkable qualitative and \textit{quantitative} agreement across system sizes with those found from unconstrained ViT wavefunctions. The long-distance decay of the real-space spin-spin correlators follows a power law with a critical exponent $z+\eta=1.30(3)$, which is in agreement with that obtained from DMRG and ViT~\cite{Yang-2022,Luciano-2024}, thus providing compelling evidence for the $\mathbb{Z}_{2}$ Dirac QSL as the prime candidate for the gapless QSLs reported in these studies. We also apply the newly developed Keldysh pseudo-fermion functional renormalization group to evaluate the dynamical spin structure factor, which shows \rev{a maximum} along the segment connecting $(0,0)$ and $(\pi,\pi)$ and subdominant intensity at $(\pi,\pi)$, as would be expected from scattering by Dirac cones at $(\pi/2,\pi/2)+(\pm\varepsilon,\pm\varepsilon)$. This observation adds to the  evidence for a $\mathbb{Z}_{2}$ Dirac QSL from an independent correlator-based approach, which is blind to specific \textit{Ans\"atze} and is not dictated by symmetry constraints. Further, given the rapid  experimental and theoretical developments in exploring putative QSL behavior on the Shastry-Sutherland lattice, these predictions for the dynamical structure factors set the stage for directly identifying such exotic states in materials and models from spectroscopic signatures.

Our classification of QSLs compatible with the reduced symmetry of the proximate plaquette\rev{-}VBC order, i.e., preserving $C_{4}$ but breaking glide and mirror symmetries (see App.~\ref{app:nematic}), allows us to study the phase transitions of fully symmetric QSLs to their \rev{descendant} plaquette\rev{-}VBC orders. In particular, we demonstrate an instability of the Z3000 state to its plaquette\rev{-}VBC counterpart at $J_{s}/J_{d}\approx0.78$, in agreement with estimates from other studies. Notably, a different class of gapped lattice-nematic $\mathbb{Z}_{2}$ QSLs, which instead break $90^{\circ}$ rotational symmetry, \rev{has} previously been identified within the Schwinger-boson formalism~\cite{Chung-2001}. Such a nematic state has also been studied with Abrikosov fermions but for the square lattice~\cite{Thomson-2018}; it would be interesting to extend this classification to the Shastry-Sutherland lattice in future work.
To locate the boundaries of the spin-liquid regime\rev{,} we perform a finite-size scaling of the N\'eel and plaquette-VBC order parameters and a correlation-ratio analysis within a unified pairing--magnetic (Pfaffian) variational family. Both order parameters vanish continuously, placing the transitions at $(J_{s}/J_{d})_{\rm PVBC}\simeq0.78$ and $(J_{s}/J_{d})_{\rm N\acute{e}el}\simeq0.82$, in agreement with the energetic competition of restricted trial states and with previous studies.

While the microscopic mean-field representation of the optimal state on the Shastry-Sutherland lattice involves several symmetry-inequivalent hopping and pairing amplitudes, its long-wavelength content is much simpler~\cite{Feuerpfeil-2026}: it realizes a gapless $\mathbb{Z}_2$ Dirac spin liquid (Z3000) that can be viewed as a Higgs descendant of the SU$(2)$ $\pi$-flux parent via an intermediate U$(1)$ staggered-flux state. In this sense, the role of the PSG is not primarily to provide ``insight by nomenclature\rev{,}'' but to uniquely determine the projective action of lattice symmetries on the fractionalized spinons and Higgs fields, thereby fixing the symmetry-allowed structure of the continuum gauge theory. In this way, the PSG framework encodes the quantum order of the spin liquid and establishes its universality class. In a companion work~\cite{Feuerpfeil-2026}, we exploited this PSG analysis to construct an explicit continuum description on the Shastry-Sutherland lattice and showed that---despite the reduced microscopic symmetry relative to the square lattice---the leading Dirac-spinon Lagrangian and the Higgs potential governing the mean-field phase diagram are identical to the square-lattice theory~\cite{Shackleton-2021} up to additional irrelevant operators. This provides a unified field-theoretic framework for Dirac spin liquids on both lattices and motivates interpreting the proximate N\'eel--$\mathbb{Z}_2$--VBC phase structure through the lens of fermionic deconfined quantum criticality. The agreement between the two lattices extends to critical exponents and the mechanism of pseudocriticality (via the relevant Yukawa coupling between spinons and Higgs fields), offering a controlled theoretical setting for the numerical signatures observed in both square-lattice and Shastry-Sutherland models.

Given the rather narrow region of stability for the Shastry-Sutherland QSL along the Heisenberg axis, it would also be interesting to identify directions in Hamiltonian parameter space \rev{that could} widen this regime. To this end, there is evidence from DMRG~\cite{Li-2024} that the introduction of an XY-type anisotropy enlarges the window of stability of the QSL, and it is worth examining whether the same $\mathbb{Z}_{2}$ Dirac QSL as the one on the Heisenberg axis remains stable \rev{under} XXZ anisotropy or whether a new QSL phase occupies this region. Moreover, the inclusion of an explicit scalar spin-chirality term in the Hamiltonian has recently been shown to induce a chiral spin liquid (CSL) of the Kalmeyer-Laughlin type~\cite{Yang-2024}. Since there exist myriad distinct Kalmeyer-Laughlin CSLs---differing in their anyon content---within parton constructions, in order to uncover the microscopic nature of the CSL, it would be necessary to obtain the chiral representation classes for the algebraic PSGs~\cite{Bieri-2016} of the Shastry-Sutherland lattice. Assessing the energetics and correlation functions of the corresponding Gutzwiller-projected chiral \textit{Ans\"atze} would then help pin down the precise CSL that provides the best variational description of the ground state; as a first step in this direction, the chiral \textit{Ans\"atze} that descend from Z3000 by lowering its symmetry, together with their mean-field spectra, are given in App.~\ref{app:chiral}. Closely tied to this is the issue of identifying, within this classification scheme, the CSLs \rev{that} serve as natural candidates for continuous transitions out of our $\mathbb{Z}_{2}$ Dirac QSL. These could include the rare class of gapless CSLs~\cite{Dusel-2024,Oliviero-2022} (such as Dirac~\cite{Kim-2024} and spinon\rev{-}Fermi\rev{-}surface~\cite{Bieri-2015} chiral QSLs) or conventional gapped CSLs.

\rev{Unlike their gapped counterparts, gapless spin liquids support low-energy quasiparticle branches, with degrees of freedom that exist arbitrarily close to the ground state. This can lead to rather peculiar low-temperature behavior: such a spectrum would, for instance, imply a nonvanishing thermal conductivity $\kappa^{}_{xx}/T$ down to arbitrarily low temperatures---a residual value conventionally denoted $\kappa^{}_{0}/T$---and, if time-reversal symmetry is broken (either intrinsically or by an external perturbation), a finite thermal Hall response as well. Doping such a state also leads to a subtle intertwining of charge and spin degrees of freedom.}
Finally, from a thermodynamic stability standpoint, \rev{gapless states} could, in principle, be \rev{favored} at finite temperature because their low-energy excitations contribute additional entropy.  Consider a \textit{Gedankenexperiment} with \rev{two spin liquids, one gapped and one gapless,} that have the same internal energy at $T=0$.  At finite temperature, the \rev{latter} acquires a larger entropy, so its free energy is \rev{reduced relative to that of the former, which renders it} thermodynamically preferred.  Such entropy-driven stabilization has indeed been reported in recent experimental studies of candidate quantum spin-liquid materials.

With regard to experimental realizations of the Shastry-Sutherland lattice, the field of quantum materials is currently witnessing the arrival of several new candidates whose magnetic lattices are based on rare-earth ions of Ce$^{3+}$\rev{~}\cite{Ma-2024}, Pr$^{3+}$\rev{~}\cite{Li-2024}, Yb$^{3+}$\rev{~}\cite{Pula-2024}, and Er$^{3+}$\rev{~}\cite{Brassington-2024}. Some of these compounds offer tantalizing evidence of nonmagnetic \rev{behavior} at low temperatures and could thus host exotic QSL phases, including in the presence of magnetic fields. Recent works~\cite{Duan-2024,Liu-2024_the} have shown that these spin-orbit-coupled systems are effectively described by generalized XYZ models. In particular, Pr$_2$Ga$_2$BeO$_7$ has been argued to be described by an XXZ model to a first approximation~\cite{Li-2024}. In light of thermodynamic measurements \rev{revealing} a $T^{2}$ dependence of the specific heat, a finite $\kappa_{0}/T$ (and an associated $T$-linear term in the specific heat) from a residual density of states ascribed to disorder, and gapless inelastic neutron spectra, the likely scenario is that of a gapless QSL being realized. Indeed, it has been speculated~\cite{Li-2024} that a $d$-wave pairing $\mathbb{Z}_{2}$ QSL would be consistent with these experimental findings. Our $\mathbb{Z}_{2}$ QSL Z3000 is, in fact, a $d$-wave pairing state, and evidence from our Keldysh pf-FRG calculations indicates that it persists upon the introduction of XXZ anisotropy (similar to the kagome U(1) QSL, which is stable to XXZ anisotropy in the Hamiltonian~\cite{Hu-2015}). Furthermore, our $\mathcal{S}({\bf q},\omega)$ obtained by Keldysh pf-FRG for the XXZ model description of Pr$_2$Ga$_2$BeO$_7$ \rev{reproduces} all the principal features of the neutron\rev{-}scattering profile. Therefore, our work sheds light on the precise microscopic character of the suspected QSL phase in Pr$_2$Ga$_2$BeO$_7$ as well. The methodological development we present\rev{,} together with the dynamical spin structure factors of candidate quantum spin liquids on the Shastry-Sutherland lattice\rev{,} will serve as a valuable guide to identifying the nature of spin liquids in a variety of candidate QSL materials \rev{that} are currently being investigated experimentally.

From symmetry considerations alone, a Dzyaloshinskii-Moriya interaction is also allowed~\cite{Romhanyi-2011,Chen-2020} and likely plays an important role in modeling the behavior of Shastry-Sutherland materials. Hence, to expand the applicability of our approach to these spin-orbit-coupled rare-earth systems, one has to extend the \textit{Ans\"atze} described herein by incorporating triplet amplitudes and evaluating their energetic competitiveness for the relevant anisotropic models~\cite{Yang-2019_ssm}. Finally, QSL behavior has also been observed in a magnetic field, e.g., in Yb$_2$Be$_2$GeO$_7$~\cite{Pula-2024}, so it would be important to determine the stability of the proposed $\mathbb{Z}_{2}$ Dirac QSL in a magnetic field, along the lines of the analysis performed for the kagome U(1) Dirac QSL in Ref.~\cite{Ran-2009}.

\acknowledgments

We thank Federico Becca, Bryan Clark, Chunxiao Liu, Yukitoshi Motome, Yusuke Nomura, Rico Pohle, Mingpu Qin, Subir Sachdev, Anders Sandvik, Leyna Shackleton, Yanting Teng, Chandra Varma, and Luciano Viteritti for helpful discussions. Y.I. thanks Luciano  Viteritti and Federico Becca for providing us with the data from Ref.~\cite{Luciano-2024} and discussing the detailed comparison between ViT and VMC results. R.S. was supported by the Princeton Quantum Initiative Fellowship. The work of Y.I., J.Y.L., K.P., and R.S. was performed, in part, at the Aspen Center for Physics, which is supported by National Science Foundation Grant No.~PHY-2210452 and a grant from the Simons Foundation (1161654, Troyer). This research was supported in part by grant NSF PHY-2309135 to the Kavli Institute for Theoretical Physics (KITP). Y.I. acknowledges support from the ICTP through the Associates Programme, from the Simons Foundation through Grant No.~284558FY19, IIT Madras through the Institute of Eminence (IoE) program for establishing QuCenDiEM (Project No.~SP22231244CPETWOQCDHOC), and the International Centre for Theoretical Sciences (ICTS), Bengaluru  during a visit for participating in the program: Kagome off-scale (ICTS/KAGOFF2024/08). Y.I. also acknowledges the use of the computing resources at HPCE, IIT Madras. S.C. thanks CALMIP (grant 2025-P0677) and GENCI (project A0190500225) for computing resources.
The work in W\"urzburg is supported by the Deutsche Forschungsgemeinschaft (DFG, German Research Foundation) through Project-ID 258499086\rev{--}SFB 1170 and through the W\"urzburg-Dresden Cluster of Excellence on Complexity, Topology and Dynamics in Quantum Matter - ctd.qmat Project-ID 390858490\rev{--}EXC 2147. R.T., F.F., and K.P. thank IIT Madras for an IoE Visiting Scientist position \rev{that} enabled the completion of this work.

\section*{Data Availability Statement} The data generated during the current study are available from the corresponding author upon reasonable request.

\appendix

\section{Exact diagonalization for 20-, 24-, 28-, and 40-site clusters}
\label{sec:ed_others}

Here, we provide additional details on our exact diagonalization simulations performed on finite-size periodic clusters of up to $N$\,$=$\,$40$ sites. We refer to standard reviews, e.g., Ref.~\cite{Sandvik_review2010}, for additional technical details about the implementation of symmetries and the diagonalization of low-energy states using the L\'anczos algorithm. Exploiting lattice symmetries is useful not only to reduce the size of the Hilbert space but also to provide crucial information about quantum numbers of each eigenstate, which \rev{allows} us to discuss specific level crossings in the main text.

\begin{figure}[!t]
\centering
\includegraphics[width=0.95\linewidth]{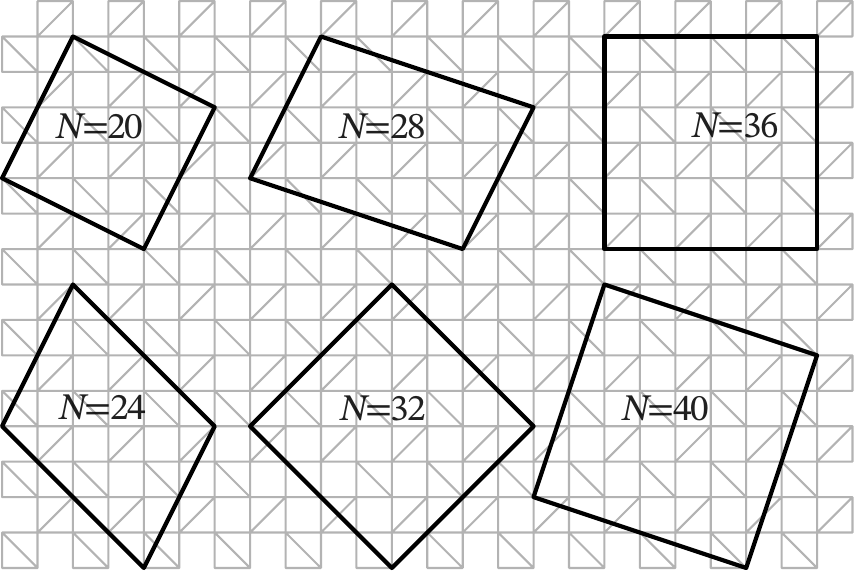}
\caption{Geometry of the periodic clusters used in our exact diagonalization
calculations, with $N=20$, $24$, $28$, $32$, $36$, and $40$ sites. In each case\rev{,}
the thick black lines delimit the simulation cell, drawn on top of the
Shastry-Sutherland lattice (gray), with periodic boundary conditions identifying
opposite edges. Because of the four-site unit cell of the lattice, only $N/4$
translations are available on a periodic $N$-site cluster. The $N=32$ and $N=36$
clusters possess the full $D_4$ point group, the $N=20$ and $N=40$ clusters have
$C_4$ symmetry, and the $N=24$ and $N=28$ clusters retain only $C_2$. These
clusters coincide with those employed in Ref.~\cite{Wang-2022}.}
\label{fig:ed_clusters}
\end{figure}

Due to the four-site unit cell of the Shastry-Sutherland lattice, as well as the interpolation, one can only use $N/4$ translations for a periodic $N$-site cluster. As noted in the analysis by \citet{Wang-2022}, the relevant low-energy levels are all located at the $\Gamma$ point of the Shastry-Sutherland lattice.
Concerning point-group symmetries, among the accessible finite-size clusters, only $N=16$, $32$, and $36$ possess the full $D_4$ symmetry group, which includes a fourfold rotation axis about the center of a fully crossed plaquette with $\mathcal{J}$ Heisenberg couplings, together with an additional mirror symmetry; we exclude the $N=16$ cluster because of its small size. Some other clusters have smaller point groups: $C_4$ for $N$\,$=$\,$20$ and $N$\,$=$\,$40$, and only $C_2$ for $N=24$ and $N=28$. The geometry of each of these clusters is displayed in Fig.~\ref{fig:ed_clusters}. Additional details can be found in Fig.~2 of Ref.~\cite{Wang-2022} and Table~\ref{tab:symmetry}.

\begin{figure}[t]
\centering
\includegraphics[width=0.95\linewidth]{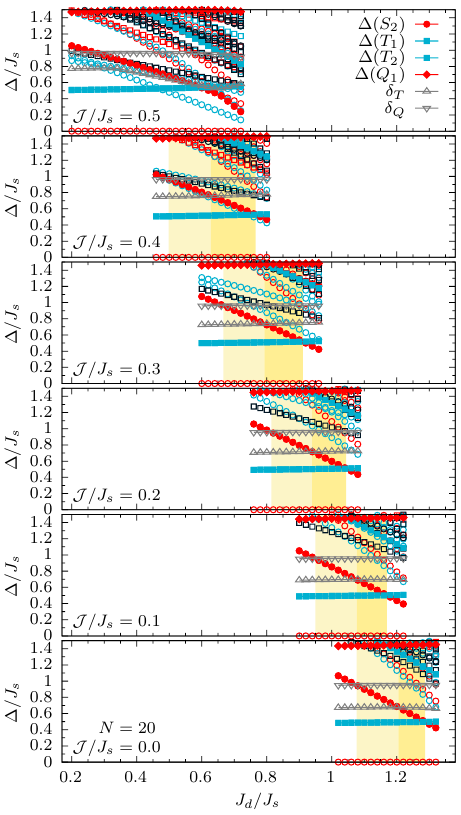}
\caption{Symmetry-resolved low-energy excitation spectrum obtained from L\'anczos exact diagonalization for the $N=20$ cluster as the Hamiltonian is interpolated from the Shastry-Sutherland limit toward the $J_1$--$J_2$ model. The notation and symmetry labeling conventions are the same as in Fig.~\ref{fig:spectra32}.}
\label{fig:spectra20}
\end{figure}

\begin{figure}[t]
\centering
\includegraphics[width=0.95\linewidth]{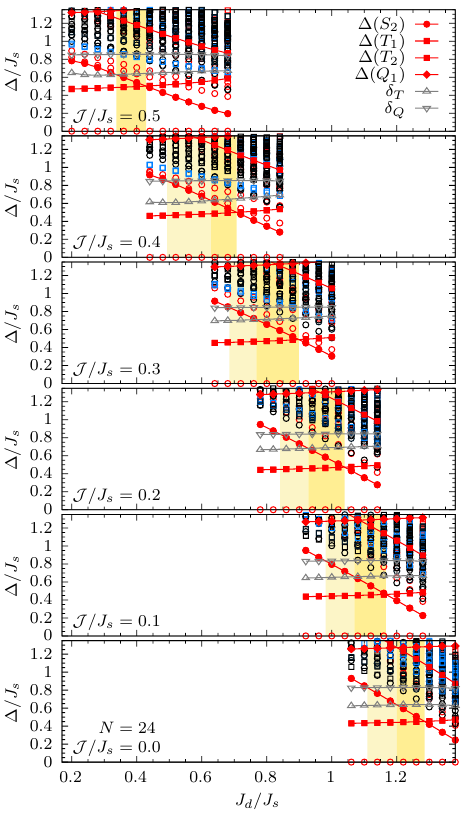}
\caption{Symmetry-resolved low-energy excitation spectrum obtained from L\'anczos exact diagonalization for the $N=24$ cluster as the Hamiltonian is interpolated from the Shastry-Sutherland limit toward the $J_1$--$J_2$ model. The notation and symmetry labeling conventions are the same as in Fig.~\ref{fig:spectra32}.}
\label{fig:spectra24}
\end{figure}

\begin{figure}[t]
\centering
\includegraphics[width=0.95\linewidth]{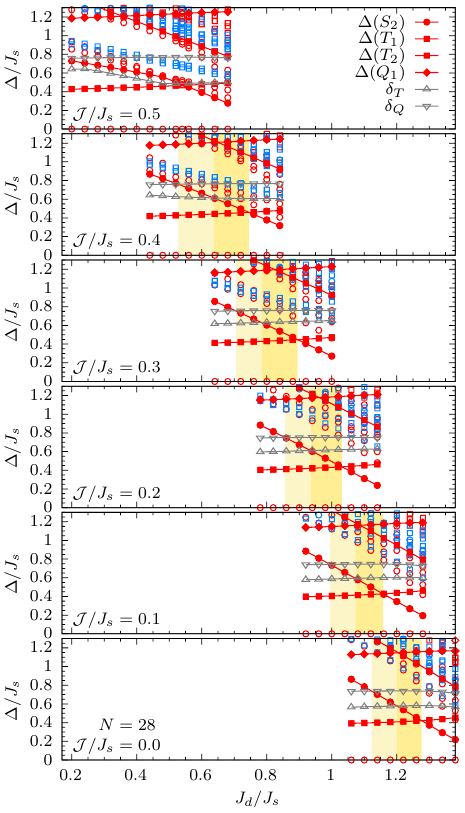}
\caption{Symmetry-resolved low-energy excitation spectrum obtained from L\'anczos exact diagonalization for the $N=28$ cluster as the Hamiltonian is interpolated from the Shastry-Sutherland limit toward the $J_1$--$J_2$ model. The notation and symmetry labeling conventions are the same as in Fig.~\ref{fig:spectra32}.}
\label{fig:spectra28}
\end{figure}

\begin{figure}[!t]
\centering
\includegraphics[width=0.95\linewidth]{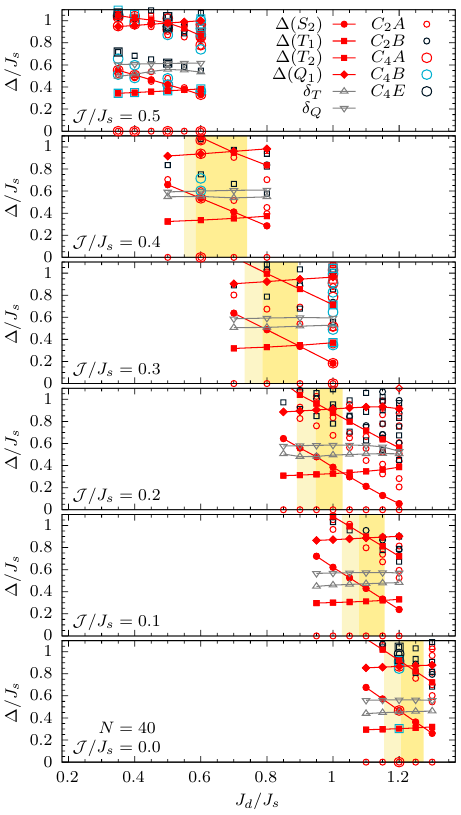}
\caption{Symmetry-resolved low-energy excitation spectra obtained from L\'anczos exact diagonalization for the $N=40$ cluster as the Hamiltonian is interpolated from the Shastry-Sutherland limit toward the $J_1$--$J_2$ model. The notation and symmetry labeling conventions are the same as in Fig.~\ref{fig:spectra32}. Most calculations exploit only the $C_2$ symmetry. Due to their higher computational cost, calculations implementing the $C_4$ symmetry of the cluster were performed \rev{only} at selected points (larger symbols) to identify the corresponding irreducible representations.}
\label{fig:spectra40}
\end{figure}

The symmetry-resolved spectra obtained on these clusters are shown in Figs.~\ref{fig:spectra20}, \ref{fig:spectra24}, \ref{fig:spectra28}, and~\ref{fig:spectra40}.

For the largest cluster considered  ($N$\,$=$\,$40$), the Hilbert-space dimension exceeds $3\times10^{9}$ states in the zero-magnetization sector when spin-reversal symmetry is used, and $5\times10^{9}$ states in the $S_z^{\mathrm{tot}}=2$ sector used to target quintuplet excitations. For such large matrices, we have obtained quicker results using only the $C_2$ subgroup of \rev{the cluster's} $C_4$ point group, thanks to a sublattice coding algorithm~\cite{Wietek2018}. Since this does not allow one to fully reconstruct all quantum numbers, we have also performed more expensive simulations for \rev{a} few parameters using the full $C_4$ point group. As shown in Fig.~\ref{fig:spectra40}, both sets of data agree very well, when available, and the slightly larger discrepancies observed for some excited states originate from their slower convergence, which is a well-known feature of the L\'anczos method. We also note that spurious eigenstates, commonly referred to as \emph{ghosts}, \rev{can} appear during the simulations.

\clearpage

\section{Gauge-enriched symmetry relations}
\label{sec:genric_gauge_con}
The relations between the different symmetry elements of the Shastry-Sutherland lattice are specified by the identities listed in Eq.~\eqref{eq:id_relation}. The projective extensions of these symmetries, $W_{\mathcal{O}}$, necessarily follow the same algebraic relations, which read as:
\begin{widetext}
\begin{subequations}
\label{eq:id_gauge_relation}
\begin{alignat}{2}
 W^{-1}_{T_x}(T^{\pdagger}_x(\mathbf{r}))W^{-1}_{T_{y}}(T^{\pdagger}_xT^{\pdagger}_y(\mathbf{r}))W^{\pdagger}_{T_x}(T^{\pdagger}_xT^{\pdagger}_y(\mathbf{r}))W^{\pdagger}_{T_{y}}(T^{\pdagger}_y(\mathbf{r}))&=F^{\pdagger}_{3}(\xi^{}_{y})&&\mbox{ or }\,\,\eta^{\pdagger}_{y}\tau^0, \\
 W^{-1}_{G_x}(G^{\pdagger}_x(\mathbf{r}))W^{-1}_{T_{x}}(G^{\pdagger}_xT^{\pdagger}_x(\mathbf{r}))W^{\pdagger}_{G_x}(G^{\pdagger}_xT^{\pdagger}_x(\mathbf{r}))W^{\pdagger}_{T_{x}}(T^{\pdagger}_x(\mathbf{r}))&=F^{\pdagger}_{3}(\xi^{}_{G_{xx}})&&\mbox{ or }\,\,\eta^{\pdagger}_{G_{xx}}\tau^0, \\
W^{-1}_{G_x}(G^{\pdagger}_x(\mathbf{r}))W^{\rev{\pdagger}}_{T_{y}}(G^{\pdagger}_xT^{\pdagger}_y(\mathbf{r}))W^{\pdagger}_{G_x}(G^{\pdagger}_xT^{\pdagger}_y(\mathbf{r}))W^{\pdagger}_{T_{y}}(T^{\pdagger}_y(\mathbf{r}))&=F^{\pdagger}_{3}(\xi^{}_{G_{xy}})&&\mbox{ or }\,\,\eta^{\pdagger}_{G_{xy}}\tau^0, \\
W^{-1}_{T_{x}}(T^{\pdagger}_x(\mathbf{r}))W^{\pdagger}_{G_x}(T^{\pdagger}_x(\mathbf{r}))W^{\pdagger}_{G_x}(G^{\pdagger}_x(\mathbf{r}))&=F^{\pdagger}_{3}(\xi^{}_{G_{x}T_x})&&\mbox{ or }\,\,\eta^{\pdagger}_{G_{x}T_x}\tau^0, \\
W^{-1}_{T_y}(T^{\pdagger}_y(\mathbf{r}))W^{-1}_{\sigma_{xy}}(T^{\pdagger}_x\sigma^{\pdagger}_{xy}(\mathbf{r}))W^{\pdagger}_{T_x}(T^{\pdagger}_x\sigma^{\pdagger}_{xy}(\mathbf{r}))W^{\pdagger}_{\sigma_{xy}}(\sigma^{\pdagger}_{xy}(\mathbf{r}))&=F^{\pdagger}_{3}(\xi^{}_{\sigma_{xy}T_x})&&\mbox{ or }\,\,\eta^{\pdagger}_{\sigma_{xy}T_x}\tau^0, \\
W^{-1}_{T_x}(T^{\pdagger}_x(\mathbf{r}))W^{-1}_{\sigma_{xy}}(T^{\pdagger}_y\sigma^{\pdagger}_{xy}(\mathbf{r}))W^{\pdagger}_{T_y}(T^{\pdagger}_y\sigma^{\pdagger}_{xy}(\mathbf{r}))W^{\pdagger}_{\sigma_{xy}}(\sigma^{\pdagger}_{xy}(\mathbf{r}))&=F^{\pdagger}_{3}(\xi^{}_{\sigma_{xy}T_y})&&\mbox{ or }\,\,\eta^{\pdagger}_{\sigma_{xy}T_y}\tau^0, \\
W^{\pdagger}_{\sigma_{xy}}(\mathbf{r})W^{\pdagger}_{G_x}(\sigma^{-1}_{xy}(\mathbf{r}))W^{\pdagger}_{\sigma_{xy}}((\sigma^{\pdagger}_{xy}G^{\pdagger}_x)^3(\mathbf{r}))W^{\pdagger}_{G_x}(G^{\pdagger}_x(\sigma^{\pdagger}_{xy}G^{\pdagger}_x)^2(\mathbf{r}))W^{\pdagger}_{\sigma_{xy}}((\sigma^{\pdagger}_{xy}G^{\pdagger}_x)^2(\mathbf{r}))&\ \; \notag \\
\times W^{\pdagger}_{G_x}(G^{\pdagger}_x\sigma^{\pdagger}_{xy}G^{\pdagger}_x(\mathbf{r}))W^{\pdagger}_{\sigma_{xy}}(\sigma^{\pdagger}_{xy}G^{\pdagger}_x(\mathbf{r}))W^{\pdagger}_{G_x}(G^{\pdagger}_x(\mathbf{r}))&=F^{\pdagger}_{3}(\xi^{}_{\sigma_{xy}G_x})&&\mbox{ or }\,\,\eta^{\pdagger}_{\sigma_{xy}G_x}\tau^0, \\
W^{\pdagger}_{\sigma_{xy}}(\sigma^{\pdagger}_{xy}(\mathbf{r}))W^{\pdagger}_{\sigma_{xy}}(\mathbf{r})&=F^{\pdagger}_{3}(\xi^{}_{\sigma_{xy}})&&\mbox{ or }\,\,\eta^{\pdagger}_{\sigma_{xy}}\tau^0 , \\
W^{\pdagger}_\mathcal{T}(\mathbf{r})W^{\pdagger}_\mathcal{T}(\mathbf{r})&=F^{\pdagger}_{3}(\xi^{}_{\mathcal{T}})&&\mbox{ or }\,\,\eta^{\pdagger}_{\mathcal{T}}\tau^0 , \\
W^{\pdagger}_\mathcal{T}(\mathbf{r}) W^{}_\mathcal{O}(\mathbf{r}) W^{-1}_\mathcal{T}(\mathcal{O}^{-1}(\mathbf{r}))  W^{-1}_\mathcal{O}(\mathbf{r})&=F^{\pdagger}_{3}(\xi^{}_{\mathcal{TO}})&&\mbox{ or }\,\,\eta^{\pdagger}_{\mathcal{TO}}\tau^0\
.
\end{alignat}
\end{subequations}
\end{widetext}
On the right-hand side of the equations above, the first expression defines the relations for the case of a U(1) IGG, where $0 \le \xi_{\ldots} < 2\pi$ is a global U(1) phase and,  as defined in the main text, $F_3(\xi)\equiv\exp({\rev{\dot\iota}\xi\tau^z})$. The second set of expressions, following the ``or\rev{,}'' corresponds to a $\mathbb{Z}_2$ IGG, with $\eta_{\ldots} = \pm1$ being a global $\mathbb{Z}_2$ sign. \rev{For now}, the different $\xi$ and $\eta$ are free parameters, but in the next two sections, we will see how different symmetry-related constraints fix their values, leading to  distinct PSG solutions.

\section{$\mathbf{U(1)}$ PSG classification}
\label{app:u1_psg_derivation}

The canonical form of a U(1) \textit{Ansatz} is given by
\begin{equation}
    \label{eq:canonical_u1}
    u^{\pdagger}_{ij}=\dot{\iota}\chi^0_{ij} \tau^0 + \chi^3_{ij} \tau^z.
    \end{equation}
The structure of the gauge transformations that keep this canonical form intact is given by
\begin{equation}
    \label{eq:canonical_u1_gauge_structure}
    W^{}_\mathcal{O}(\rev{m,n},u)=F^{\pdagger}_3(\phi^{\pdagger}_\mathcal{O}(m,n,u))(\rev{\dot\iota}\tau^x)^{w^{\pdagger}_\mathcal{O}},
\end{equation}
where $\mathcal{O}\in\{T_x,T_y,G_x,\sigma_{xy},\mathcal{T}\}$, and $w^{}_\mathcal{O}$ can take values 0 or 1. Now, we will systematically work out the possibilities for $w^{}_{\mathcal{O}}$ and $\phi_{\mathcal{O}}$ one by one for the different symmetries.

\subsection{Space-group symmetries}
First, let us analyze the solutions for the space-group symmetries of the Shastry-Sutherland lattice.
Starting with translations, \rev{the relation in Eq.}~\eqref{eq:canonical_u1_gauge_structure} for $\mathcal{O}\in\{T_x,T_y\}$ can be explicitly written as:
\begin{equation}
\left.\begin{aligned}
&W^{\pdagger}_{T_x}(m,n,u)=F^{\pdagger}_3(\phi^{\pdagger}_{T_x}(m,n,u))(\rev{\dot\iota}\tau^x)^{w^{\pdagger}_{T_x}},\\
&W^{\pdagger}_{T_y}(m,n,u)=F^{\pdagger}_3(\phi^{\pdagger}_{T_y}(m,n,u))(\rev{\dot\iota}\tau^x)^{w^{\pdagger}_{T_y}}.\\
\end{aligned}\right.
\end{equation}
\rev{Depending on} the values of $w^{}_{T_{x/y}}$, three cases may arise:
\begin{enumerate}
    \item $w^{\pdagger}_{T_x}=0$, $w^{\pdagger}_{T_y}=0$\,,
    \item $w^{\pdagger}_{T_{x}}=1$, $w^{\pdagger}_{T_{y}}=0$\,,
    \item $w^{\pdagger}_{T_{x}}=1$, $w^{\pdagger}_{T_{y}}=1$\,.
\end{enumerate}
It is readily seen that cases 2 and 3 can be excluded as these do not satisfy the conditions~\eqref{eq:id_gauge_relation}(d--f). Hence, we need to consider \rev{only} the first, and we call this choice the ``uniform gauge\rev{.}'' Now, we can always choose a gauge transformation with the form $W(m,n,u)=F_3(\theta(m,n,u))$ such that we end up with:
\begin{equation}
\label{eq:B4}
W^{\pdagger}_{T_x}(m,0,u)=W^{\pdagger}_{T_y}(m,n,u)=\tau^0.\\
\end{equation}
With these settings, inserting Eq.~\eqref{eq:B4} into the symmetry condition~\eqref{eq:id_gauge_relation}(a) leads to the solutions for \textit{all} $\mathbf{r}$:
\begin{equation}
\label{eq:tran_sol_1}
\left.\begin{aligned}
&W^{\pdagger}_{T_x}(m,n,u)=F^{\pdagger}_3(n\,\xi^{}_y),\\
&W^{\pdagger}_{T_y}(m,n,u)=\tau^0.\\
\end{aligned}\right.
\end{equation}
These two equations define the allowed projective actions of translational symmetries.

Next, we consider the glide symmetry.
With the notation $\Delta^{}_i\phi^{}_\mathcal{O}(m,n,u)\equiv\phi^{}_\mathcal{O}(m,n,u)-\phi^{}_\mathcal{O}[T^{-1}_i(m,n,u)]$ for $i=x,y$, Eqs.~\eqref{eq:id_gauge_relation}(b, c) tell us that
\begin{equation}
\label{eq:gx_1}
\left.\begin{aligned}
\Delta^{\pdagger}_i\phi^{\pdagger}_{G_x}(m,n,u)&=\phi^{\pdagger}_{T_i}(m,n,u)\\
&+(-1)^{w^{\pdagger}_{G_x}}(\xi^{\pdagger}_{G_{x\rev{i}}}-\phi^{\pdagger}_{T_i}[G^{-1}_x(m,n,u)]),
\end{aligned}\right.
\end{equation}
which, upon substitution of $\phi^{}_{T_i}$ from Eq.~\eqref{eq:tran_sol_1}, yields
\begin{equation}
\label{eq:gx_2}
\begin{aligned}
\Delta^{\pdagger}_x\phi^{\pdagger}_{G_x}(m,n,u)&= (-1)^{w^{\pdagger}_{G_x}}\xi^{\pdagger}_{G_{xx}} +[n(1+(-1)^{w^{\pdagger}_{G_x}})] \xi^{\pdagger}_y\\
&+\big[(-1)^{w^{\pdagger}_{G_x}}(\delta^{\pdagger}_{u,3}+\delta^{\pdagger}_{u,4})\big]\xi^{\pdagger}_y,\\
\Delta_y\phi^{\pdagger}_{G_x}(m,n,u)&=(-1)^{w^{\pdagger}_{G_x}}\xi^{\pdagger}_{G_{xy}}.
\end{aligned}
\end{equation}
We also have to satisfy the additional consistency condition:
\begin{equation}
\label{eq:consistent}
\left.\begin{aligned}
\Delta^{\pdagger}_x&\phi^{\pdagger}_{G_x}(m,n,u)+\Delta^{\pdagger}_y\phi^{\pdagger}_{G_x}[T^{-1}_x(m,n,u)]\\
&=\Delta^{\pdagger}_y\phi^{\pdagger}_{G_x}(m,n,u)+\Delta^{\pdagger}_x\phi^{\pdagger}_{G_x}[T^{-1}_y(m,n,u)],\\
\end{aligned}\right.
\end{equation}
which enforces that
\begin{equation}
    \label{eq:gx_consistency}
    \big[1+(-1)^{w^{\pdagger}_{G_x}}\big]\xi^{\pdagger}_y=0.
\end{equation}
Thus, $w^{}_{G_x}=0$ implies that $2\xi^{\pdagger}_y=0$, while there is no restriction if $w^{}_{G_x}=1$. Consequently, we obtain the general solution
\begin{alignat}{1}
    \label{eq:gx_sol_1}
    \phi^{\pdagger}_{G_x}(m,n,u)&=(-1)^{w^{\pdagger}_{G_x}}(m\,\xi^{\pdagger}_{G_{xx}}+n\,\xi^{\pdagger}_{G_{xy}})\\
    \nonumber
    &+(-1)^{w^{\pdagger}_{G_x}}m\,\xi^{\pdagger}_y(\delta^{\pdagger}_{u,3}+\delta^{\pdagger}_{u,4})+\bar{\phi}^{\pdagger}_{G_x,u},
    \end{alignat}
using the shorthand $\bar{\phi}_{G_x,u}\equiv\phi_{G_x}(0,0,u)$. Once $\phi_{G_x}$ is determined, we can plug it back into Eq.~\eqref{eq:canonical_u1_gauge_structure}, together with the associated value of $w^{}_{G_x}$, to obtain the set of PSG elements.

Having dealt with translation and glide symmetries, we now turn to the mirror symmetry $\sigma_{xy}$. The gauge transformations accompanying reflections once again have the general form
\begin{equation}
\label{eq:Wxy}
W^{\pdagger}_{\sigma_{xy}}=F^{\pdagger}_3(\phi^{\pdagger}_{\sigma_{xy}}(m,n,u))(\rev{\dot\iota}\tau^x)^{w^{\pdagger}_{\sigma_{xy}}}.
\end{equation}
Using Eqs.~\eqref{eq:id_gauge_relation}(\rev{e}) and~\eqref{eq:id_gauge_relation}(\rev{f}), the relations analogous to Eq.~\eqref{eq:gx_2} above read as
\begin{equation}
\label{eq:sigma_1}
\left.\begin{aligned}
\Delta^{\pdagger}_x\phi^{\pdagger}_{\sigma_{xy}}(m,n,u)=&-(-1)^{w^{\pdagger}_{\sigma_{xy}}}\xi^{\pdagger}_{\sigma_{xy}T_x}+n\,\xi^{\pdagger}_y\\
\Delta^{\pdagger}_y\phi^{\pdagger}_{\sigma_{xy}}(m,n,u)=&-(-1)^{w^{\pdagger}_{\sigma_{xy}}}(\xi^{\pdagger}_{\sigma_{xy}T_y}+m\,\xi^{\pdagger}_y).\\
\end{aligned}\right.
\end{equation}
\rev{The consistency condition analogous to Eq.~\eqref{eq:consistent} yields $\big[1+(-1)^{w^{\pdagger}_{\sigma_{xy}}}\big]\xi^{\pdagger}_y=0$, i.e., precisely the same requirement as Eq.~\eqref{eq:gx_consistency}; hence, it imposes no \emph{further} restrictions on $\xi^{\pdagger}_y$ beyond those already found above.} The final solution for $\rev{\phi}_{\sigma_{xy}}$ is given by
\begin{equation}
    \label{eq:sigma_sol_1}
    \left.\begin{aligned}
    \phi^{\pdagger}_{\sigma_{xy}}(m,n,u)=&-(-1)^{w^{\pdagger}_{\sigma_{xy}}}(m\,\xi^{\pdagger}_{\sigma_{xy}T_x}+n\,\xi^{\pdagger}_{\sigma_{xy}T_y})\\
&+mn\,\xi^{\pdagger}_{y}+\bar{\phi}^{\pdagger}_{\sigma_{xy},u},\\
    \end{aligned}\right.
\end{equation}
which we can put back into Eq.~\eqref{eq:Wxy} to determine $W_{\sigma_{xy}}$.

While Eqs.~\eqref{eq:tran_sol_1}, \eqref{eq:gx_sol_1}, and \eqref{eq:sigma_sol_1} provide the general solutions for $\phi_{\mathcal{O}}$, these may not be gauge\rev{-}independent. In fact, the U(1) phases derived above can be greatly simplified or fixed by the use of a local gauge transformation $W(m,n,u)$ under which \[W^{\pdagger}_\mathcal{O}(m,n,u)\rightarrow W^\dagger(m,n,u)W^{\pdagger}_\mathcal{O}(m,n,u)W[\mathcal{O}^{-1}(m,n,u)].\] To demonstrate this, let us choose a gauge transformation (which is independent of the sublattice index),
\begin{equation}\label{eq:local_u1}    W(m,n,u)=F^{\pdagger}_3(m\theta_x+n\theta_y).
\end{equation}
With the proper choices of $\theta_x$ and $\theta_y$, one can then set
\begin{equation}\label{eq:local_fixing_u1}
    \left.\begin{aligned}
    \xi^{\pdagger}_{\sigma_{xy}T_x}&=0,\\
    \text{for }w^{\pdagger}_{G_x}=0:\;\xi^{\pdagger}_{G_{xy}}&=0,\\
    \text{for }w^{\pdagger}_{G_x}=1:\;\xi^{\pdagger}_{G_{xx}}&=0.\\
    \end{aligned}\right.
\end{equation}
Thereafter, using a sublattice-dependent gauge transformation $W(m,n,u)$\,$=$\,$F_3(\theta_u)$ with $\theta_2$\,$=$\,$(-1)^{w^{}_{G_x}}(\theta_1-\bar{\phi}_{G_x,1})$, $\theta_3=\bar{\phi}_{G_x,3}+(-1)^{w^{}_{G_x}}\theta_4$\rev{,} and $\theta_4=(-1)^{w^{}_{\sigma^{}_{xy}}}(\theta_2-\bar{\phi}_{\sigma^{}_{xy},2})$, we can fix the following:
\begin{equation}\label{eq:sublat_fixing_u1}
\bar{\phi}^{\pdagger}_{G_x,1}=\bar{\phi}^{\pdagger}_{G_x,3}=\bar{\phi}^{\pdagger}_{\sigma_{xy},2}=0=\bar{\phi}^{\pdagger}_{\sigma_{xy},1}\delta_{w^{}_{\sigma_{xy}},1}\,,
\end{equation}
which drastically simplifies Eqs.~\eqref{eq:gx_sol_1} and \eqref{eq:sigma_sol_1}.

Continuing in this spirit, we shall exploit the remaining conditions from Eq.~\eqref{eq:id_gauge_relation} to further fix the free parameters. For instance, using Eqs.~\eqref{eq:local_fixing_u1} and \eqref{eq:sublat_fixing_u1}, Eq.~\eqref{eq:id_gauge_relation}(d) yields
\begin{equation}\label{eq:gx_tx_2}
    \left.\begin{aligned}
    \text{for }w^{\pdagger}_{G_x}=0:\;&\xi^{\pdagger}_y=\xi^{\pdagger}_{G_{xx}}=\xi^{\pdagger}_{G_{xy}}=0,\\
    &\bar{\phi}^{\pdagger}_{G_x,2}=\bar{\phi}^{\pdagger}_{G_x,4}=\xi^{\pdagger}_{G_xT_x};\\
    \text{for }w^{\pdagger}_{G_x}=1:\;&\xi^{\pdagger}_{G_{xx}}=\xi^{\pdagger}_y+2\xi^{\pdagger}_{G_{xy}}=0,\\
    &\bar{\phi}^{\pdagger}_{G_x,2}=n^{\pdagger}_{G_xT_x}\pi=\bar{\phi}^{\pdagger}_{G_x,4}+\xi^{\pdagger}_{G_{xy}}.\\
    \end{aligned}\right.
\end{equation}
Likewise, Eq.~\eqref{eq:id_gauge_relation}(h) asserts that
\begin{align}
\xi^{\pdagger}_{\sigma_{xy}T_y}=0,\;\big[1+(-1)^{w^{\pdagger}_{\sigma_{xy}}}\big]\xi^{\pdagger}_y&=0,\label{eq:sigma_cyclic_1}\\
\bar{\phi}^{\pdagger}_{\sigma_{xy},u}+(-1)^{w^{\pdagger}_{\sigma_{xy}}}\bar{\phi}^{\pdagger}_{\sigma_{xy},\sigma_{xy}(u)}&=\xi^{\pdagger}_{\sigma_{xy}}\label{eq:sigma_cyclic_2}.
\end{align}
Then, putting the content of Eq.~\eqref{eq:sublat_fixing_u1} into Eq.~\eqref{eq:sigma_cyclic_2}, one arrives at the following result for $\bar{\phi}^{\pdagger}_{\sigma_{xy},u}$:
\begin{align}
w^{\pdagger}_{\sigma_{xy}}=0:\;&\bar{\phi}^{\pdagger}_{\sigma_{xy},u}=\{\bar{\phi}^{\pdagger}_{\sigma_{xy}},0,\bar{\phi}^{\pdagger}_{\sigma_{xy}}+n^{\pdagger}_{\sigma_{xy}}\pi,2\bar{\phi}^{\pdagger}_{\sigma_{xy}}\},\notag\\
w^{\pdagger}_{\sigma_{xy}}=1:\;&\bar{\phi}^{\pdagger}_{\sigma_{xy},u}=\{0,0,\bar{\phi}^{\pdagger}_{\sigma_{xy}},0\}\label{eq:sigma_fix_u1}.
\end{align}
We are still left with one more lattice symmetry condition given by Eq.~\eqref{eq:id_gauge_relation}(g), which imposes further constraints in two cases:
\begin{align}
&(w^{\pdagger}_{G_x},w^{\pdagger}_{\sigma_{xy}})=(1,0):\;2\xi^{\pdagger}_{G_{xy}}=0\implies\xi^{\pdagger}_y=0,\\
&(w^{\pdagger}_{G_x},w^{\pdagger}_{\sigma_{xy}})=(0,1):\;\bar{\phi}^{\pdagger}_{\sigma_{xy}}=2\xi^{\pdagger}_{G_xT_x}+n^{\pdagger}_{\sigma_{xy}}\pi.
\end{align}
A simple gauge choice is possible for the case with $w_{G_x}=0$ as follows\footnote{No such uniform choice of $\bar{\phi}_{G_{x},u}$ is possible for $w_{G_x}=1$, so we will continue to use the existing gauge choice in that case.}. Employing a gauge transformation of the form $W(m,n,u)=F_3(\theta_u)$ with $\theta_2=\theta_1+\xi_{G_xT_x}/2$ and $\theta_4=\theta_3+\xi_{G_xT_x}/2$, one can set
\begin{equation}
    \bar{\phi}^{\pdagger}_{G_x,u}=\xi^{\pdagger}_{G_xT_x}/2 \implies \bar{\phi}^{\pdagger}_{G_x,u}=0.
\end{equation}
On the right-hand side above, we have used the gauge freedom which allows us to set a global U(1) phase to zero. Due to this gauge transformation, $\bar{\phi}_{\sigma_{xy},u}$ will also necessarily be modified, but we can still exploit the freedom to choose $\theta_1$ and $\theta_3$. For $w^{}_{\sigma_{xy}}=0$, choosing $\theta_3=\theta_1+\bar{\phi}^{\pdagger}_{\sigma_{xy}}$, we obtain
\begin{align}
\bar{\phi}^{\pdagger}_{\sigma_{xy},u}&=\{\bar{\phi}^{\pdagger}_{\sigma_{xy}},\bar{\phi}^{\pdagger}_{\sigma_{xy}},\bar{\phi}^{\pdagger}_{\sigma_{xy}}+n^{\pdagger}_{\sigma_{xy}}\pi,\bar{\phi}^{\pdagger}_{\sigma_{xy}}\},\notag\\
\implies\bar{\phi}^{\pdagger}_{\sigma_{xy},u}&=\{0,0,n^{\pdagger}_{\sigma_{xy}}\pi,0\}.
\end{align}
On the other hand, for $w_{\sigma_{xy}}=1$, we choose $\theta_1=0,\;\theta_3=-\xi_{G_xT_x}$, which leads to
\begin{align}
&\bar{\phi}^{\pdagger}_{\sigma_{xy},u}=\{0,0,\bar{\phi}^{\pdagger}_{\sigma_{xy}}+2\xi^{\pdagger}_{G_xT_x},0\}\notag,
\end{align}
and, for simplicity, we will redefine $\bar{\phi}_{\sigma_{xy}}+2\xi_{G_xT_x}\rightarrow\bar{\phi}_{\sigma_{xy}}$.

The upshot of this sequence of operations is that we have now specified all the values of $\bar{\phi}_{\mathcal{O}}$ and $\xi_{\ldots}$ that determine $\phi_{\mathcal{O}}$ for all space-group symmetries $\mathcal{O}$. In turn, $\phi_{\mathcal{O}}$---together with different combinations of $w^{}_{\mathcal{O}}$---then feeds into Eq.~\eqref{eq:canonical_u1_gauge_structure} and defines the PSGs.

\subsection{Time-reversal symmetry}

Per Eq.~\eqref{eq:canonical_u1_gauge_structure}, the gauge transformation associated with time reversal takes the form
\begin{equation}
      W^{\pdagger}_\mathcal{T}(\rev{m,n},u)=F^{\pdagger}_3(\phi^{\pdagger}_\mathcal{T}(m,n,u))(\rev{\dot\iota}\tau^x)^{w^{\pdagger}_\mathcal{T}}.
\end{equation}
Following Eq.~\eqref{eq:id_gauge_relation}(j) for $\mathcal{O}\in\{T_x,T_y\}$, we find (for both $w^{}_\mathcal{T}=0$ and $w^{}_\mathcal{T}=1$) that
\begin{equation}\label{eq:time_sol_1_u1}
\phi^{\pdagger}_\mathcal{T}(m,n,u)=m\,\xi^{\pdagger}_{\mathcal{T}x}+n\,\xi^{\pdagger}_{\mathcal{T}y}
+\bar{\phi}^{\pdagger}_{\mathcal{T},u}.
\end{equation}
As for Eq.~\eqref{eq:consistent} earlier, the consistency condition yields
\begin{equation}
    \big[1-(-1)^{w^{\pdagger}_{\mathcal{T}}}\big]\xi^{\pdagger}_y=0,
\end{equation}
which implies that for $w^{}_{\mathcal{T}}=1$, $2\xi_y=0$.

To fix the rest of the parameters, it is convenient to consider the two possible values of $w^{}_\mathcal{T}$ separately.

\subsubsection{$w^{\pdagger}_\mathcal{T}=0$}
In this case, the symmetry relation~\eqref{eq:id_gauge_relation}(i) gives
\begin{equation}
    \label{eq:tr_wt0_1_u1}
    2\xi^{\pdagger}_{\mathcal{T}x}=2\xi^{\pdagger}_{\mathcal{T}y}=0,\;\bar{\phi}^{\pdagger}_{\mathcal{T},u}=\bar{\phi}^{\pdagger}_\mathcal{T}+n^{\pdagger}_u\pi=n^{\pdagger}_u\pi,
\end{equation}
where the gauge freedom can be utilized to set $\bar{\phi}_\mathcal{T}=0$. Now, using the equation above and Eq.~\eqref{eq:id_gauge_relation}(j) for $\mathcal{O}=G_x,\sigma_{xy}$, one can fix:
\begin{equation}
    \label{eq:tr_wt0_3_u1}    \xi^{\pdagger}_{\mathcal{T}x}=\xi^{\pdagger}_{\mathcal{T}y}=0.
\end{equation}
Note that having an \textit{Ansatz} with nonvanishing amplitudes on NN bonds requires that
\begin{equation}\label{eq:tr_wt0_2_u1}
   \bar{\phi}^{\pdagger}_{\mathcal{T},u}=u\,\pi.
\end{equation}
However, since diagonal bonds connect sites \rev{whose sublattice indices $u$ have the same parity}, this also implies that choosing such a gauge for time reversal results in  vanishing \textit{Ans\"atze} on the NNN bonds.

\subsubsection{$w^{\pdagger}_\mathcal{T}=1$}
When $w^{}_\mathcal{T}=1$, it is always possible to choose a gauge such that $\phi_\mathcal{T}(m,n,u)=0$. The advantage of such a choice is that the mean-field parameters on all the bonds then comprise only real hopping terms.

Here, we demonstrate this process for the case of $w^{}_{G_x}$\,$=$\,$w^{}_{\sigma_{xy}}$\,$=$\,$0$. First, we perform a gauge transformation given by
\begin{equation}\label{eq:time_u1_wt1_gauge}
    W^{\pdagger}_\mathcal{T}(m,n,u)=F^{\pdagger}_3\left(m\frac{\xi^{\pdagger}_{\mathcal{T}x}}{2}+n\frac{\xi^{\pdagger}_{\mathcal{T}y}}{2}+\frac{\bar{\phi}^{\pdagger}_{\mathcal{T},u}}{2}\right)\, .
\end{equation}
This immediately sets
\begin{equation}
\phi^{\pdagger}_{\mathcal{T}}(m,n,u)\rightarrow\Tilde{\phi}^{\pdagger}_{\mathcal{T}}(m,n,u)=0.
\end{equation}
The price we pay for this is that $\phi^{\pdagger}_{G_x}$ and $\phi^{\pdagger}_{\rev{\sigma_{xy}}}$ will be modified simultaneously to
\begin{alignat}{2}
\Tilde{\phi}^{\pdagger}_{G_x}&=&&-n\,\xi^{\pdagger}_{\mathcal{T}y}+\frac{\bar{\phi}^{\pdagger}_{\mathcal{T},2}-\bar{\phi}^{\pdagger}_{\mathcal{T},1}}{2}(\delta^{\pdagger}_{u,1}-\delta^{\pdagger}_{u,2})\notag\\
& &&+\frac{\bar{\phi}^{\pdagger}_{\mathcal{T},4}-\bar{\phi}^{\pdagger}_{\mathcal{T},3}}{2}(\delta^{\pdagger}_{u,3}-\delta^{\pdagger}_{u,4})\notag\\
& &&-\frac{\xi^{\pdagger}_{\mathcal{T}x}}{2}(\delta^{\pdagger}_{u,1} + \delta^{\pdagger}_{u,4})
-\frac{\xi^{\pdagger}_{\mathcal{T}y}}{2} (\delta^{\pdagger}_{u,3} + \delta^{\pdagger}_{u,4}),\\
\Tilde{\phi}^{\pdagger}_{\sigma_{xy}}&= &&(n-m)\frac{\xi^{\pdagger}_{\mathcal{T}x}-\xi^{\pdagger}_{\mathcal{T}y}}{2}+p^{}_{\sigma_{xy}}\pi\delta^{\pdagger}_{u,3}\notag\\
& &&+\frac{\bar{\phi}^{\pdagger}_{\mathcal{T},4}-\bar{\phi}^{\pdagger}_{\mathcal{T},2}}{2}(\delta^{\pdagger}_{u,2}-\delta^{\pdagger}_{u,4}),
\end{alignat}
with $p_{\sigma_{xy}}$\,$=$\,$0,1$. For notational convenience, we drop the tilde on $\phi$ hereafter.

Lastly, making use of Eq.~\eqref{eq:id_gauge_relation}(j) produces two other constraints:
\begin{align}
&\xi^{}_{\mathcal{T}G_x}=-2{\phi}^{}_{G_x}(m,n,u), \\
&\xi^{}_{\mathcal{T}\sigma_{xy}}=-2{\phi}^{}_{\sigma_{xy}}(m,n,u),
\end{align}
leading to the relations:
\begin{align}
2\xi^{\pdagger}_{\mathcal{T}y}=0&\Rightarrow \xi^{\pdagger}_{\mathcal{T}y}=p^{}_{G_{xy}}\pi,\\
\xi^{\pdagger}_{\mathcal{T}x}=\xi^{\pdagger}_{\mathcal{T}y}&=p^{\pdagger}_{G_{xy}}\pi,\\
\bar{\phi}^{\pdagger}_{\mathcal{T},2}=\bar{\phi}^{\pdagger}_{\mathcal{T},4},\;\bar{\phi}^{\pdagger}_{\mathcal{T},1}&=\bar{\phi}^{\pdagger}_{\mathcal{T},3},\\
2(\bar{\phi}^{\pdagger}_{\mathcal{T},2}-\bar{\phi}^{\pdagger}_{\mathcal{T},1})&=p^{\pdagger}_{G_{xy}}\pi, \\
&\Rightarrow\bar{\phi}^{\pdagger}_{\mathcal{T},2}=\bar{\phi}^{\pdagger}_{\mathcal{T},1}+\frac{p^{\pdagger}_{G_{xy}}\pi}{2}+p^{\pdagger}_{G_xT_{x}}\pi, \notag
\end{align}
where both $p^{\pdagger}_{G_{xy}}$ and $p^{\pdagger}_{G_xT_{x}}$ take values 0 or 1.
Combining these equations and performing an appropriate global phase shift, the end result is
\begin{align}
&{\phi}^{}_{G_x}(m,n,u)=p^{}_{G_{xy}}\pi(n+\delta^{}_{u,4})+p^{}_{G_{x}T_x}\pi(\delta^{}_{u,2}+ \delta^{}_{u,4}),\\
&{\phi}^{}_{\sigma_{xy}}(m,n,u)=p^{}_{\sigma_{xy}}\pi\delta^{}_{u,3},
\end{align}
for the class $w^{}_{G_x}=w^{}_{\sigma_{xy}}=0$, $w^{}_\mathcal{T}=\rev{1}$.

The PSGs for the other classes can be straightforwardly worked out in the same manner. We summarize all the U(1) PSGs thus obtained in Eqs.~\eqref{eq:g_translation_u}--\eqref{eq:g_time_u} and Table~\ref{table:u1_psg}.

\section{$\mathbf{\mathbb{Z}_2}$ PSG classification}
\label{app:z2_psg_derivation}

The PSGs for a $\mathbb{Z}_2$ IGG can be derived in \rev{a} similar fashion to the U(1) case. We proceed, as before, by considering the spatial and time-reversal symmetries separately.

\subsection{Space-group symmetries}
Exploiting the local gauge redundancy, we first fix:
\begin{equation}
W^{\pdagger}_{T_x}(m,0,u)=W^{\pdagger}_{T_y}(m,n,u)=\tau^0.\\
\end{equation}
With these settings, the symmetry condition~\eqref{eq:id_gauge_relation}(a) directly leads to the following solutions for the translational PSGs:
\begin{equation}
\label{eq:tran_sol_z1}
\left.\begin{aligned}
&W^{\pdagger}_{T_x}(m,n,u)=\eta^{\rev{n}}_y\tau^0,\\
&W^{\pdagger}_{T_y}(m,n,u)=\tau^0,\\
\end{aligned}\right.
\end{equation}
where $\eta_y$ (and all other parameters $\eta_{\ldots}$ introduced hereafter) can be $\pm 1$.

Let us introduce the notation $\Omega^{}_i[W^{}_\mathcal{O}(m,n,u)] \equiv W^{}_\mathcal{O}(m,n,u)W^{}_\mathcal{O}[T^{-1}_i(m,n,u)]$ for $i=x,y$. In this language, after substituting Eq.~\eqref{eq:tran_sol_z1} into Eqs.~\eqref{eq:id_gauge_relation}(b, c), we find
\begin{equation}
\left.\begin{aligned}
\Omega_x[W^{\pdagger}_{G_x}(m,n,u)]=&\eta^{\pdagger}_{G_{xx}}\eta^{-(\delta_{u,3}+\delta_{u,4})}_y\tau^0,\\
\Omega_y[W^{\pdagger}_{G_x}(m,n,u)]=&\eta^{\pdagger}_{G_{xy}}\tau^0.\\
\end{aligned}\right.
\end{equation}
Akin to the U(1) case, we can identify a consistency equation  as
\begin{equation}
	\label{eq:consistent_z2}
	\left.\begin{aligned}
		\Omega^{\pdagger}_x&[W^{\pdagger}_{G_x}(m,n,u)]\Omega^{\pdagger}_{y}[W^{\pdagger}_{G_x}[T^{-1}_x(m,n,u)]]=\\
		&\Omega^{\pdagger}_{y}[W^{\pdagger}_{G_x}(m,n,u)]\Omega^{\pdagger}_{x}[W^{\pdagger}_{G_x}[T^{-1}_{y}(m,n,u)]];\\
	\end{aligned}\right.
\end{equation}
however, this does not impose any new constraints. \rev{Consequently}, the solution for $W^{\pdagger}_{G_x}$ takes the generic form
\begin{equation}
    \label{eq:gx_sol_z1}
    \left.\begin{aligned}
    W^{\pdagger}_{G_x}(m,n,u)=&\eta^m_{G_{xx}}\eta^n_{G_{xy}}
\eta^{-m(\delta_{u,3}+\delta_{u,4})}_y\mathcal{W}^{\pdagger}_{G_x,u},\\
    \end{aligned}\right.
\end{equation}
where $\mathcal{W}_{G_x,u}$ is a sublattice-dependent \rev{SU(2)} matrix in pseudospin space.

Likewise\rev{,} the solution for $W^{\pdagger}_{\sigma_{xy}}$ can be directly obtained from Eqs.~\eqref{eq:id_gauge_relation}(e, f) as
\begin{equation}
    \label{eq:sig_sol_z1}
    \left.\begin{aligned}
    W^{\pdagger}_{\sigma_{xy}}(m,n,u)=&\eta^m_{\sigma_{xy}T_x}\eta^n_{\sigma_{xy}T_y}
\eta^{mn}_y\mathcal{W}^{\pdagger}_{\sigma_{xy},u}.\\
    \end{aligned}\right.
\end{equation}

Once again, we can try to simplify the PSG solutions by a clever choice of gauge transformations.
Note that a local transformation of the form $W(m,n,u)=\eta^m_{\rev{1}}\eta^n_{\rev{2}}\tau^0$ does not impact the projective operation of translations. However, \rev{it does} modify the spatial dependencies of $W_{\sigma_{xy}}$ and $W_{G_x}$. On choosing $\eta^{\pdagger}_{\rev{1}}$ and $\eta^{\pdagger}_{\rev{2}}$ appropriately, one can set
\begin{equation}\label{eq:local_fixing_z1}
 \eta^{\pdagger}_{\sigma_{xy}T_x}  = +1.
\end{equation}
Combining this with a sublattice-dependent gauge transformation of the form $W(m,n,u)=W_u$, we fix
\begin{equation}\label{eq:sublat_fixing_z1}
\mathcal{W}^{\pdagger}_{G_x,1}=\mathcal{W}^{\pdagger}_{G_x,3}=\mathcal{W}^{\pdagger}_{\sigma_{xy},2}=\tau^0.
\end{equation}
Then, using Eq.~\eqref{eq:sublat_fixing_z1} in Eq.~\eqref{eq:id_gauge_relation}(\rev{d}), we find
\begin{equation}\label{eq:gx_tx_z2}
    \left.\begin{aligned}
    &\eta^{\pdagger}_y=\eta^{\pdagger}_{G_{xx}}=\rev{+1},\\    &\mathcal{W}^{\pdagger}_{G_x,2}=\eta^{\pdagger}_{G_{xy}}\mathcal{W}^{\pdagger}_{G_x,4}=\eta^{\pdagger}_{G_xT_x}\rev{\tau^0}\,.\\
    \end{aligned}\right.
\end{equation}
At the same time, Eq.~\eqref{eq:id_gauge_relation}(h) gives
\begin{equation}\label{eq:sig_fix_z0}
\eta^{\pdagger}_{\sigma_{xy}T_y}  = +1  ,\;
\mathcal{W}^{\pdagger}_{\sigma_{xy},u} \mathcal{W}^{\pdagger}_{\sigma_{xy},\sigma_{xy}(u)}=\eta^{\pdagger}_{\sigma_{xy}}\tau^0.
\end{equation}
In addition, we are left with one more lattice\rev{-}symmetry condition given by Eq.~\eqref{eq:id_gauge_relation}\rev{(g)}. Using this together with the information from Eqs.~\eqref{eq:sublat_fixing_z1} and \eqref{eq:sig_fix_z0}, we obtain
\begin{equation}\label{eq:sig_fix_z}
\mathcal{W}^{\pdagger}_{\sigma_{xy},u}=
\begin{cases}
\big\{\eta^{\pdagger}_{\sigma_{xy},1}\tau^0,\tau^0,\eta^{\pdagger}_{\sigma_{xy},3}\tau^0,\tau^0\big\}; \quad &\text{for }\eta^{\pdagger}_{\sigma_{xy}}=+1\\
\big\{\dot\iota\tau^z,\tau^0,\eta^{\pdagger}_{\sigma_{xy},3}\dot\iota\tau^z,-\tau^0\big\}; \quad &\text{for }\eta^{\pdagger}_{\sigma_{xy}}=-1
\end{cases}.
\end{equation}
At this point, yet another set of gauge transformations, defined by
$W(m,n,u)=(\eta_{\sigma_{xy},1})^{(\delta_{u,1}+\delta_{u,2})}$ for $\eta^{\pdagger}_{\sigma_{xy}}=+1$ and $W(m,n,u)=(\dot\iota\tau^z)^{(\delta_{u,3}+\delta_{u,4})}$ for $\eta^{\pdagger}_{\sigma_{xy}}=-1$,
\rev{leads} to
\begin{equation}\label{eq:sig_fix_z1}
\mathcal{W}^{\pdagger}_{\sigma_{xy},u}=
\begin{cases}
\big\{1,1,\eta^{\pdagger}_{\sigma_{xy},3}\eta^{\pdagger}_{\sigma_{xy},1},1\big\}\eta^{\pdagger}_{\sigma_{xy},1}\tau^0; \quad &\text{for }\eta^{\pdagger}_{\sigma_{xy}}=+1\\
\big\{1,1,\eta^{\pdagger}_{\sigma_{xy},3},1\big\}\dot\iota\tau^z; \quad &\text{for }\eta^{\pdagger}_{\sigma_{xy}}=-1
\end{cases}.
\end{equation}
Now, by virtue of the IGG freedom, we can drop the global sign parameter $\eta_{\sigma_{xy},1}$ in the first case above. Furthermore, we define \rev{the} new parameter $\eta_{\sigma}$ \rev{as} $\eta_{\sigma}\equiv \eta_{\sigma_{xy},3}\eta_{\sigma_{xy},1}$ and $\eta_{\sigma}\equiv \eta_{\sigma_{xy},3}$ for $\eta_{\sigma_{xy}}$\,$=$\,$+1$ and $\eta_{\sigma_{xy}}$\,$=$\,$-1$, respectively. This results in the compact expression
\begin{equation}\label{eq:sig_fix_z_final}
\mathcal{W}^{\pdagger}_{\sigma_{xy},u}=(\eta^{\pdagger}_{\sigma})^{\delta^{}_{u,3}}\mathcal{W}^{\pdagger}_{\sigma},
\end{equation}
where $\mathcal{W}_{\sigma}=\tau^0$ ($\dot\iota\tau^z$) for positive (negative) $\eta_{\sigma_{xy}}$.

\subsection{Time-reversal symmetry}
To compute the PSG solutions for time reversal, we can again use Eq.~\eqref{eq:id_gauge_relation}(j) for $\mathcal{O}\in\{T_x,T_y\}$, which directly yields
\begin{equation}
    W^{\pdagger}_\mathcal{T}(m,n,u)=\eta^m_{\mathcal{T}T_x}\eta^n_{\mathcal{T}T_y}\mathcal{W}^{\pdagger}_{\mathcal{T},u}.
\end{equation}
It is easy to see that Eqs.~\eqref{eq:id_gauge_relation}(i,\,j) with $\mathcal{O}\in \{G_x,\sigma_{xy}\}$ will now introduce four extra conditions, namely,
\begin{subequations}
\label{eq:z2_time_conditions}
\begin{align}
&\eta^{\pdagger}_{\mathcal{T}T_x}=\eta^{\pdagger}_{\mathcal{T}T_y}\ , \; \\
&\mathcal{W}^2_{\mathcal{T},u}=\eta^{\pdagger}_{\mathcal{T}}\ , \; \\
&\mathcal{W}^{\pdagger}_{\mathcal{T},u}\mathcal{W}^{\pdagger}_{\sigma_{xy},u}=\eta^{\pdagger}_{\sigma_{xy}\mathcal{T}}\mathcal{W}^{\pdagger}_{\sigma_{xy},u}\mathcal{W}^{\pdagger}_{\mathcal{T},\sigma^{-1}_{xy}(u)}\ , \\
&\mathcal{W}^{\pdagger}_{\mathcal{T},u}\mathcal{W}^{\pdagger}_{G_{x},u}=\eta^{\pdagger}_{G_{x}\mathcal{T}}\eta^{(u \,\mathrm{mod} \,2)}_{\mathcal{T}T_{x}}\mathcal{W}^{\pdagger}_{G_{x},u}\mathcal{W}^{\pdagger}_{\mathcal{T},G^{-1}_{x}(u)}\ .
\end{align}
\end{subequations}

A summary of all the $\mathbb{Z}_2$ PSGs constructed from these relations is provided by  Eqs.~\eqref{eq:solution_translationj_z2}--\eqref{eq:solution_time_z2} and Table~\ref{table:z2_psg}.

\section{Mapping to square-lattice PSGs}
\label{app:mapping_square_lattice}
As discussed in Sec.~\ref{sec:mapping_square_lattice}, the square-lattice wallpaper group, $p4m$, can be generated from that of the Shastry-Sutherland lattice, $p4g$, by incorporating the symmetry $\sigma_x$ of reflections about the $x$-axis. This imposes certain additional symmetry relations, which are as follows:
\begin{subequations}
\label{eq:id_relation_sq}
\begin{align}
\sigma^{-1}_xT^{-1}_{x}\sigma^{\pdagger}_xT^{\pdagger}_{x}&=\mathds{1}\ , \; \label{eq:id_tx_sx}\\
\sigma^{-1}_xT^{-1}_{y}\sigma^{\pdagger}_xT^{\pdagger}_{y}&=\mathds{1}\ , \; \label{eq:id_ty_sx}\\
G^{-1}_x\sigma^{\pdagger}_{x}G^{\pdagger}_x\sigma^{\pdagger}_{x}&=\mathds{1}\ , \; \label{eq:id_gx_sx}\\
(\sigma^{\pdagger}_x \sigma^{\pdagger}_{xy})^4&=\mathds{1}\ , \; \label{eq:id_sx_sxy}\\
\sigma^2_{x} & = \mathds{1} \ , \; \label{eq:id_sx}\\
(\sigma^{\pdagger}_{xy}\sigma^{-1}_xG^{-1}_x)^2(\sigma^{\pdagger}_{xy}G^{\pdagger}_x\sigma^{\pdagger}_{x})^2 & = \mathds{1} \ , \; \label{eq:id_sx_sxy_gx}\\
(\sigma^{\pdagger}_{x}G^{\pdagger}_x\sigma^{\pdagger}_{xy}\sigma^{\pdagger}_{x}\sigma^{\pdagger}_{xy})^2 & = \mathds{1} \ , \; \label{eq:id_sx_sxy_gx_2}\\
\mathcal{T} \sigma^{\pdagger}_{x} \mathcal{T}^{-1} \sigma^{-1}_{x} &= \mathds{1} \
. \label{eq:id_sx_t}
\end{align}
\end{subequations}

Using this information, we can now deduce the PSGs for the enlarged space group \rev{that} includes $\sigma_x$.
For example, from Eqs.~\eqref{eq:id_tx_sx} and~\eqref{eq:id_ty_sx}, we find the projective implementation
\begin{equation}
 W^{\pdagger}_{\sigma_x}(m,n,u)=\eta^m_{\sigma_xT_x} \eta^n_{\sigma_xT_y} \mathcal{W}^{\pdagger}_{\sigma_x,u}.
\end{equation}
Similarly, the rest of the relations in Eq.~\eqref{eq:id_relation_sq} can be harnessed to impose further constraints such as $\eta_{\sigma_xT_x}=\eta_{\sigma_xT_y}=+1$ in the solution above, as well as $\eta_{G_{xy}}=+1$ in Eq.~\eqref{eq:solution_glide_z2}. Systematically considering all the gauge-inequivalent choices of $\mathcal{W}_{\sigma_x,u}$ that satisfy these symmetry constraints, we are led from the PSGs listed in Table~\ref{table:z2_psg} to the set in Table~\ref{table:z2_psg_square_1}. Note that there are  four $\eta$ parameters and a total of 17 rows in the table. Thus, the total number of PSGs for the $p4m$ symmetry group is $2^4\times17=272$; this is indeed the number of PSGs enumerated  in the square-lattice classification by \citet{Wen-2002}.

\begin{table*}
\caption{All gauge-inequivalent PSGs corresponding to the full $p4m$ space-group symmetry. The choices of $\mathcal{W}_{G_x,u}$ are the same for each row and are given by $\mathcal{W}_{G_x,u}=\{\tau^0,\eta_{G_xT_x}\tau^0,\tau^0,\eta_{G_xT_x}\tau^0\}$. \rev{For $a,b\in\{x,y,z\}$, we define $\tau^{ab}\equiv(\tau^a+\tau^b)/{\sqrt{2}}$ and $\tau^{a\bar{b}}\equiv(\tau^a-\tau^b)/{\sqrt{2}}$.}}
\begin{ruledtabular}
\begin{tabular}{cccc}
PSG No.&$\mathcal{W}^{\pdagger}_{\sigma_{xy},u}$&$\mathcal{W}^{\pdagger}_{\mathcal{T},u}$ & $\mathcal{W}^{\pdagger}_{\sigma_x,u}$ \\
			\hline
1(a)&$\{\tau^0,\tau^0,\eta^{\pdagger}_{\sigma}\tau^0,\tau^0\}$&$\{\tau^0,\eta^{\pdagger}_{G_x\mathcal{T}}\tau^0,\tau^0,\eta^{\pdagger}_{G_x\mathcal{T}}\tau^0\}$ &$\{\tau^0,\eta^{\pdagger}_{\sigma_xG_x}\tau^0,\tau^0,\eta^{\pdagger}_{\sigma_xG_x}\tau^0\}$ \\
1(b)&$\{\tau^0,\tau^0,\eta^{\pdagger}_{\sigma}\tau^0,\tau^0\}$&$\{\tau^0,\eta^{\pdagger}_{G_x\mathcal{T}}\tau^0,\tau^0,\eta^{\pdagger}_{G_x\mathcal{T}}\tau^0\}$ &$\{\dot\iota\tau^z,\eta^{\pdagger}_{\sigma_xG_x}\dot\iota\tau^z,\dot\iota\tau^z,\eta^{\pdagger}_{\sigma_xG_x}\dot\iota\tau^z\rev{\}}$ \\
\hline
2(a)&$\{\tau^0,\tau^0,\eta^{\pdagger}_{\sigma}\tau^0,\tau^0\}$&$\{\dot\iota\tau^y,\eta^{\pdagger}_{G_x\mathcal{T}}\dot\iota\tau^y,\dot\iota\tau^y,\eta^{\pdagger}_{G_x\mathcal{T}}\dot\iota\tau^y\}$ &$\{\tau^0,\eta^{\pdagger}_{\sigma_xG_x}\tau^0,\tau^0,\eta^{\pdagger}_{\sigma_xG_x}\tau^0\}$\\
2(b)&$\{\tau^0,\tau^0,\eta^{\pdagger}_{\sigma}\tau^0,\tau^0\}$&$\{\dot\iota\tau^y,\eta^{\pdagger}_{G_x\mathcal{T}}\dot\iota\tau^y,\dot\iota\tau^y,\eta^{\pdagger}_{G_x\mathcal{T}}\dot\iota\tau^y\}$ &$\{\dot\iota\tau^z,\eta^{\pdagger}_{\sigma_xG_x}\dot\iota\tau^z,\dot\iota\tau^z,\eta^{\pdagger}_{\sigma_xG_x}\dot\iota\tau^z\}$\\
2(c)&$\{\tau^0,\tau^0,\eta^{\pdagger}_{\sigma}\tau^0,\tau^0\}$&$\{\dot\iota\tau^y,\eta^{\pdagger}_{G_x\mathcal{T}}\dot\iota\tau^y,\dot\iota\tau^y,\eta^{\pdagger}_{G_x\mathcal{T}}\dot\iota\tau^y\}$ &$\{\dot\iota\tau^y,\eta^{\pdagger}_{\sigma_xG_x}\dot\iota\tau^y,\dot\iota\tau^y,\eta^{\pdagger}_{\sigma_xG_x}\dot\iota\tau^y\}$\\
\hline
3(a)&$\{\dot\iota\tau^z,\dot\iota\tau^z,\eta^{\pdagger}_{\sigma}\dot\iota\tau^z,\dot\iota\tau^z\}$&$\{\dot\iota\tau^y,\eta^{\pdagger}_{G_x\mathcal{T}}\dot\iota\tau^y,\dot\iota\tau^y,\eta^{\pdagger}_{G_x\mathcal{T}}\dot\iota\tau^y\}$ &$\{\tau^0,\eta^{\pdagger}_{\sigma_xG_x}\tau^0,\tau^0,\eta^{\pdagger}_{\sigma_xG_x}\tau^0\}$ \\
3(b)&$\{\dot\iota\tau^z,\dot\iota\tau^z,\eta^{\pdagger}_{\sigma}\dot\iota\tau^z,\dot\iota\tau^z\}$&$\{\dot\iota\tau^y,\eta^{\pdagger}_{G_x\mathcal{T}}\dot\iota\tau^y,\dot\iota\tau^y,\eta^{\pdagger}_{G_x\mathcal{T}}\dot\iota\tau^y\}$ &$\{\dot\iota\tau^z,\eta^{\pdagger}_{\sigma_xG_x}\dot\iota\tau^z,\dot\iota\tau^z,\eta^{\pdagger}_{\sigma_xG_x}\dot\iota\tau^z\}$ \\
3(c)&$\{\dot\iota\tau^z,\dot\iota\tau^z,\eta^{\pdagger}_{\sigma}\dot\iota\tau^z,\dot\iota\tau^z\}$&$\{\dot\iota\tau^y,\eta^{\pdagger}_{G_x\mathcal{T}}\dot\iota\tau^y,\dot\iota\tau^y,\eta^{\pdagger}_{G_x\mathcal{T}}\dot\iota\tau^y\}$ &$\{\dot\iota\tau^y,\eta^{\pdagger}_{\sigma_xG_x}\dot\iota\tau^y,\dot\iota\tau^y,\eta^{\pdagger}_{\sigma_xG_x}\dot\iota\tau^y\}$ \\
3(d)&$\{\dot\iota\tau^z,\dot\iota\tau^z,\eta^{\pdagger}_{\sigma}\dot\iota\tau^z,\dot\iota\tau^z\}$&$\{\dot\iota\tau^y,\eta^{\pdagger}_{G_x\mathcal{T}}\dot\iota\tau^y,\dot\iota\tau^y,\eta^{\pdagger}_{G_x\mathcal{T}}\dot\iota\tau^y\}$ &$\{\dot\iota\tau^x,\eta^{\pdagger}_{\sigma_xG_x}\dot\iota\tau^x,\dot\iota\tau^x,\eta^{\pdagger}_{\sigma_xG_x}\dot\iota\tau^x\}$ \\
3(e)&$\{\dot\iota\tau^z,\dot\iota\tau^z,\eta^{\pdagger}_{\sigma}\dot\iota\tau^z,\dot\iota\tau^z\}$&$\{\dot\iota\tau^y,\eta^{\pdagger}_{G_x\mathcal{T}}\dot\iota\tau^y,\dot\iota\tau^y,\eta^{\pdagger}_{G_x\mathcal{T}}\dot\iota\tau^y\}$ &$\{\dot\iota\tau^{zx},\eta^{\pdagger}_{\sigma_xG_x}\dot\iota\tau^{zx},\dot\iota\tau^{zx},\eta^{\pdagger}_{\sigma_xG_x}\dot\iota\tau^{zx}\}$ \\
\hline
4(a)&$\{\dot\iota\tau^z,\dot\iota\tau^z,\eta^{\pdagger}_{\sigma}\dot\iota\tau^z,\dot\iota\tau^z\}$&$\{\dot\iota\tau^z,\eta^{\pdagger}_{G_x\mathcal{T}}\dot\iota\tau^z,\dot\iota\tau^z,\eta^{\pdagger}_{G_x\mathcal{T}}\dot\iota\tau^z\}$& $\{\tau^0,\eta^{\pdagger}_{\sigma_xG_x}\tau^0,\tau^0,\eta^{\pdagger}_{\sigma_xG_x}\tau^0\}$ \\
4(b)&$\{\dot\iota\tau^z,\dot\iota\tau^z,\eta^{\pdagger}_{\sigma}\dot\iota\tau^z,\dot\iota\tau^z\}$&$\{\dot\iota\tau^z,\eta^{\pdagger}_{G_x\mathcal{T}}\dot\iota\tau^z,\dot\iota\tau^z,\eta^{\pdagger}_{G_x\mathcal{T}}\dot\iota\tau^z\}$& $\{\dot\iota\tau^z,\eta^{\pdagger}_{\sigma_xG_x}\dot\iota\tau^z,\dot\iota\tau^z,\eta^{\pdagger}_{\sigma_xG_x}\dot\iota\tau^z\}$ \\
4(c)&$\{\dot\iota\tau^z,\dot\iota\tau^z,\eta^{\pdagger}_{\sigma}\dot\iota\tau^z,\dot\iota\tau^z\}$&$\{\dot\iota\tau^z,\eta^{\pdagger}_{G_x\mathcal{T}}\dot\iota\tau^z,\dot\iota\tau^z,\eta^{\pdagger}_{G_x\mathcal{T}}\dot\iota\tau^z\}$& $\{\dot\iota\tau^y,\eta^{\pdagger}_{\sigma_xG_x}\dot\iota\tau^y,\dot\iota\tau^y,\eta^{\pdagger}_{\sigma_xG_x}\dot\iota\tau^y\}$ \\
\hline
5(a)&$\{\dot\iota\tau^z,\dot\iota\tau^z,\eta^{\pdagger}_{\sigma}\dot\iota\tau^z,\dot\iota\tau^z\}$&$\{\tau^0,\eta^{\pdagger}_{G_x\mathcal{T}}\tau^0,\tau^0,\eta^{\pdagger}_{G_x\mathcal{T}}\tau^0\}$ &$\{\tau^0,\eta^{\pdagger}_{\sigma_xG_x}\tau^0,\tau^0,\eta^{\pdagger}_{\sigma_xG_x}\tau^0\}$ \\
5(b)&$\{\dot\iota\tau^z,\dot\iota\tau^z,\eta^{\pdagger}_{\sigma}\dot\iota\tau^z,\dot\iota\tau^z\}$&$\{\tau^0,\eta^{\pdagger}_{G_x\mathcal{T}}\tau^0,\tau^0,\eta^{\pdagger}_{G_x\mathcal{T}}\tau^0\}$ &$\{\dot\iota\tau^z,\eta^{\pdagger}_{\sigma_xG_x}\dot\iota\tau^z,\dot\iota\tau^z,\eta^{\pdagger}_{\sigma_xG_x}\dot\iota\tau^z\}$ \\
5(c)&$\{\dot\iota\tau^z,\dot\iota\tau^z,\eta^{\pdagger}_{\sigma}\dot\iota\tau^z,\dot\iota\tau^z\}$&$\{\tau^0,\eta^{\pdagger}_{G_x\mathcal{T}}\tau^0,\tau^0,\eta^{\pdagger}_{G_x\mathcal{T}}\tau^0\}$ &$\{\dot\iota\tau^y,\eta^{\pdagger}_{\sigma_xG_x}\dot\iota\tau^y,\dot\iota\tau^y,\eta^{\pdagger}_{\sigma_xG_x}\dot\iota\tau^y\}$ \\
5(d)&$\{\dot\iota\tau^z,\dot\iota\tau^z,\eta^{\pdagger}_{\sigma}\dot\iota\tau^z,\dot\iota\tau^z\}$&$\{\tau^0,\eta^{\pdagger}_{G_x\mathcal{T}}\tau^0,\tau^0,\eta^{\pdagger}_{G_x\mathcal{T}}\tau^0\}$ &$\{\dot\iota\tau^{zx},\eta^{\pdagger}_{\sigma_xG_x}\dot\iota\tau^{zx},\dot\iota\tau^{zx},\eta^{\pdagger}_{\sigma_xG_x}\dot\iota\tau^{zx}\}$ \\
		\end{tabular}
	\end{ruledtabular}
	\label{table:z2_psg_square_1}
\end{table*}

\begin{table}
\caption{The same PSGs as \rev{listed} in Table~\ref{table:z2_psg_square_1}, but recast in terms of the symmetry group $\{T_{1},T_{2},\sigma_{x},\sigma_{y},\sigma_{xy},\mathcal{T}\}$ [see Eqs.~\eqref{eq:solution_square_z2_2}\rev{--}\eqref{eq:solution_square_z2_22}]. \rev{The composite symbols $\tau^{ab}$ and $\tau^{a\bar{b}}$ are defined as in Table~\ref{table:z2_psg_square_1}.}}
\begin{ruledtabular}
\begin{tabular}{ccccccc}
PSG No.&$\mathcal{W}^{\pdagger}_{T_1}$&$\mathcal{W}^{\pdagger}_{T_2}$ & $\mathcal{W}^{\pdagger}_{\sigma_y}$ & $\mathcal{W}^{\pdagger}_{\sigma_x}$ & $\mathcal{W}^{\pdagger}_{\sigma_{xy}}$&$\mathcal{W}^{\pdagger}_\mathcal{T}$ \\
			\hline
1(a)&$\tau^0$&$\tau^0$ &$\tau^0$&$\tau^0$ &$\tau^0$ &$\tau^0$ \\
1(b)&$\dot\iota\tau^z$&$\dot\iota\tau^z$ &$\dot\iota\tau^z$&$\dot\iota\tau^z$ &$\tau^0$ &$\tau^0$ \\
\hline
2(a)&$\tau^0$&$\tau^0$ &$\tau^0$&$\tau^0$ &$\tau^0$ &$\dot\iota\tau^y$ \\
2(b)&$\dot\iota\tau^z$&$\dot\iota\tau^z$ &$\dot\iota\tau^z$&$\dot\iota\tau^z$  &$\tau^0$&$\dot\iota\tau^y$ \\
2(c)&$\dot\iota\tau^y$&$\dot\iota\tau^y$ &$\dot\iota\tau^y$&$\dot\iota\tau^y$  &$\tau^0$&$\dot\iota\tau^y$ \\
\hline
3(a)&$\tau^0$&$\tau^0$ &$\tau^0$&$\tau^0$ &$\dot\iota\tau^z$ &$\dot\iota\tau^y$ \\
3(b)&$\dot\iota\tau^z$&$\dot\iota\tau^z$ &$\dot\iota\tau^z$&$\dot\iota\tau^z$ &$\dot\iota\tau^z$ &$\dot\iota\tau^y$ \\
3(c)&$\dot\iota\tau^y$&$\dot\iota\tau^y$ &$\dot\iota\tau^y$&$\dot\iota\tau^y$ &$\dot\iota\tau^z$ &$\dot\iota\tau^y$ \\
3(d)&$\dot\iota\tau^x$&$\dot\iota\tau^x$ &$\dot\iota\tau^x$&$\dot\iota\tau^x$ &$\dot\iota\tau^z$ &$\dot\iota\tau^y$ \\
3(e)&$\dot\iota\tau^{zx}$&$\dot\iota\tau^{z\bar{x}}$ &$\dot\iota\tau^{z\bar{x}}$&$\dot\iota\tau^{zx}$ &$\dot\iota\tau^z$ &$\dot\iota\tau^y$ \\
\hline
4(a)&$\tau^0$&$\tau^0$ &$\tau^0$&$\tau^0$ &$\dot\iota\tau^z$ &$\dot\iota\tau^z$ \\
4(b)&$\dot\iota\tau^z$&$\dot\iota\tau^z$ &$\dot\iota\tau^z$&$\dot\iota\tau^z$ &$\dot\iota\tau^z$ &$\dot\iota\tau^z$ \\
4(c)&$\dot\iota\tau^y$&$\dot\iota\tau^y$ &$\dot\iota\tau^y$&$\dot\iota\tau^y$ &$\dot\iota\tau^z$ &$\dot\iota\tau^z$ \\
\hline
5(a)&$\tau^0$&$\tau^0$ &$\tau^0$&$\tau^0$ &$\dot\iota\tau^z$ &$\tau^0$ \\
5(b)&$\dot\iota\tau^z$&$\dot\iota\tau^z$ &$\dot\iota\tau^z$&$\dot\iota\tau^z$ &$\dot\iota\tau^z$ &$\tau^0$ \\
5(c)&$\dot\iota\tau^y$&$\dot\iota\tau^y$ &$\dot\iota\tau^y$&$\dot\iota\tau^y$ &$\dot\iota\tau^z$ &$\tau^0$ \\
5(d)&$\dot\iota\tau^{zx}$&$\dot\iota\tau^{z\bar{x}}$ &$\dot\iota\tau^{z\bar{x}}$&$\dot\iota\tau^{zx}$ &$\dot\iota\tau^z$ &$\tau^0$ \\
		\end{tabular}
	\end{ruledtabular}
	\label{table:z2_psg_square_2}
\end{table}

\begin{table}
\caption{List of PSGs written in the square-lattice notation  of Eq.~\eqref{eq:solution_square_z2_4}.  \rev{For $a,b\in\{x,y,z\}$, we define $\tau^{ab}\equiv(\tau^a+\tau^b)/{\sqrt{2}}$ and $\tau^{a\bar{b}}\equiv(\tau^a-\tau^b)/{\sqrt{2}}$.} On the right side, we present the solutions in a different gauge, which is the same as the one used in Ref.~\cite{Wen-2002}. }
\begin{ruledtabular}
\begin{tabular}{ccccc|cccc}
PSG No\rev{.} & $\mathcal{W}^{\pdagger}_{\sigma_y}$ & $\mathcal{W}^{\pdagger}_{\sigma_x}$ & $\mathcal{W}^{\pdagger}_{\sigma_{xy}}$&$\mathcal{W}^{\pdagger}_\mathcal{T}$ & $\mathcal{W}^{\pdagger}_{\sigma_y}$ & $\mathcal{W}^{\pdagger}_{\sigma_x}$ & $\mathcal{W}^{\pdagger}_{\sigma_{xy}}$&$\mathcal{W}^{\pdagger}_\mathcal{T}$ \\
			\hline
1(a) &$\tau^0$&$\tau^0$ &$\tau^0$ &$\tau^0$&$\tau^0$&$\tau^0$ &$\tau^0$ &$\tau^0$ \\
1(b) &$\dot\iota\tau^z$&$\dot\iota\tau^z$ &$\tau^0$ &$\tau^0$ &$\dot\iota\tau^z$&$\dot\iota\tau^z$ &$\tau^0$ &$\tau^0$ \\
\hline
2(a) &$\tau^0$&$\tau^0$ &$\tau^0$ &$\dot\iota\tau^y$   &$\tau^0$&$\tau^0$ &$\tau^0$ &$\dot\iota\tau^z$ \\
2(b) &$\dot\iota\tau^z$&$\dot\iota\tau^z$  &$\tau^0$&$\dot\iota\tau^y$ &$\dot\iota\tau^x$&$\dot\iota\tau^x$  &$\tau^0$&$\dot\iota\tau^z$ \\
2(c)&$\dot\iota\tau^y$&$\dot\iota\tau^y$  &$\tau^0$&$\dot\iota\tau^y$ &$\dot\iota\tau^z$&$\dot\iota\tau^z$  &$\tau^0$&$\dot\iota\tau^z$ \\
\hline
3(a) &$\tau^0$&$\tau^0$ &$\dot\iota\tau^z$ &$\dot\iota\tau^y$  &$\tau^0$&$\tau^0$ &$\dot\iota\tau^x$ &$\dot\iota\tau^z$ \\
3(b)&$\dot\iota\tau^z$&$\dot\iota\tau^z$ &$\dot\iota\tau^z$ &$\dot\iota\tau^y$   &$\dot\iota\tau^x$&$\dot\iota\tau^x$ &$\dot\iota\tau^x$ &$\dot\iota\tau^z$ \\
3(c)&$\dot\iota\tau^y$&$\dot\iota\tau^y$ &$\dot\iota\tau^z$ &$\dot\iota\tau^y$   &$\dot\iota\tau^z$&$\dot\iota\tau^z$ &$\dot\iota\tau^x$ &$\dot\iota\tau^z$ \\
3(d)&$\dot\iota\tau^x$&$\dot\iota\tau^x$ &$\dot\iota\tau^z$ &$\dot\iota\tau^y$   &$\dot\iota\tau^y$&$\dot\iota\tau^y$ &$\dot\iota\tau^x$ &$\dot\iota\tau^z$ \\
3(e)&$\dot\iota\tau^{z\bar{x}}$&$\dot\iota\tau^{zx}$ &$\dot\iota\tau^z$ &$\dot\iota\tau^y$ &$\dot\iota\tau^{\rev{x}}$&$\dot\iota\tau^{y}$ &$\dot\iota\tau^{xy}$ &$\dot\iota\tau^z$\\
\hline
4(a)&$\tau^0$&$\tau^0$ &$\dot\iota\tau^z$ &$\dot\iota\tau^z$ &$\tau^0$&$\tau^0$ &$\dot\iota\tau^z$ &$\dot\iota\tau^z$ \\
4(b)&$\dot\iota\tau^z$&$\dot\iota\tau^z$ &$\dot\iota\tau^z$ &$\dot\iota\tau^z$ &$\dot\iota\tau^z$&$\dot\iota\tau^z$ &$\dot\iota\tau^z$ &$\dot\iota\tau^z$\\
4(c)&$\dot\iota\tau^y$&$\dot\iota\tau^y$ &$\dot\iota\tau^z$ &$\dot\iota\tau^z$ &$\dot\iota\tau^x$&$\dot\iota\tau^x$ &$\dot\iota\tau^z$ &$\dot\iota\tau^z$\\
\hline
5(a)&$\tau^0$&$\tau^0$ &$\dot\iota\tau^z$ &$\tau^0$ &$\tau^0$&$\tau^0$ &$\dot\iota\tau^z$ &$\tau^0$ \\
5(b)&$\dot\iota\tau^z$&$\dot\iota\tau^z$ &$\dot\iota\tau^z$ &$\tau^0$ &$\dot\iota\tau^z$&$\dot\iota\tau^z$ &$\dot\iota\tau^z$ &$\tau^0$ \\
5(c)&$\dot\iota\tau^y$&$\dot\iota\tau^y$ &$\dot\iota\tau^z$ &$\tau^0$ &$\dot\iota\tau^x$&$\dot\iota\tau^x$ &$\dot\iota\tau^z$ &$\tau^0$ \\
5(d)&$\dot\iota\tau^{z\bar{x}}$&$\dot\iota\tau^{zx}$ &$\dot\iota\tau^z$ &$\tau^0$ &$\dot\iota\tau^{\rev{x}}$&$\dot\iota\tau^{y}$ &$\dot\iota\tau^{xy}$ &$\tau^0$\\
		\end{tabular}
	\end{ruledtabular}
	\label{table:z2_psg_square_3}
\end{table}

To illustrate the one-to-one mapping between each row of  Table~\ref{table:z2_psg_square_1}
and the PSG structure obtained in Ref.~\cite{Wen-2002}, we need to rewrite the PSGs in terms of the projective realization of \{$T_1$, $T_2$, $\sigma_x$, $\sigma_y$, $\sigma_{xy}$\}\footnote{In Ref.~\cite{Wen-2002}, $\sigma_x$, $\sigma_y$\rev{,} and $\sigma_{xy}$ are denoted as $P_y$, $P_x$\rev{,} and $P_{xy}$, respectively.} and $\mathcal{T}$. As $T_1$, $T_2$, and $\sigma_y$ can be generated using only $G_x$, $\sigma_x$, and $\sigma_{xy}$, the projective extensions $W_{T_1}(m,n,u)$, $W_{T_2}(m,n,u)$\rev{,} and $W_{\sigma_y}(m,n,u)$ can be derived using $W_{G_x}(m,n,u)$, $W_{\sigma_x}(m,n,u)$, and $W_{\sigma_{xy}}(m,n,u)$. Recalling the relations compiled in Eq.~\eqref{eq:square_gen}, we write the corresponding projective realizations as
\begin{align}
W^{\pdagger}_{T_{1}}(m,n,u)=\eta^{\pdagger}_{T_1}&W^{\pdagger}_{G_{x}}(m,n,u)W^{\pdagger}_{\sigma_{x}}[G^{-1}_{x}(m,n,u)],\\
W^{\pdagger}_{T_{2}}(m,n,u)=\eta^{\pdagger}_{T_2}&W^{\pdagger}_{\sigma_{xy}}(m,n,u)W^{\pdagger}_{T_1}[\sigma^{\pdagger}_{xy}(m,n,u)]\notag\\
&W^{\pdagger}_{\sigma_{xy}}[T^{-1}_1\sigma^{\pdagger}_{xy}(m,n,u)],\\
W^{\pdagger}_{\sigma_{y}}(m,n,u)=\eta^{\pdagger}_{\sigma_y}&W^{\pdagger}_{\sigma_{xy}}(m,n,u)W^{\pdagger}_{\sigma_x}[\sigma^{\pdagger}_{xy}(m,n,u)]\notag\\
&W^{\pdagger}_{\sigma_{xy}}[\sigma^{-1}_x\sigma^{\pdagger}_{xy}(m,n,u)].
\end{align}
Note that the global sign parameters $\eta_{T_1}$, $\eta_{T_2}$, and $\eta_{\sigma_y}$ can be set to $+1$ using the gauge freedom. This results in the solution:
\begin{align}
W^{\pdagger}_{T_{1}}(m,n,u)&=\big\{\eta^{\pdagger}_{\sigma_xG_x},\eta^{\pdagger}_{G_xT_x},\eta^{\pdagger}_{\sigma_xG_x},\eta^{\pdagger}_{G_xT_x}\big\}\mathcal{W}^{\pdagger}_{T_{1}},\label{eq:solution_square_z2_2}\\
W^{\pdagger}_{T_{2}}(m,n,u)&=\big\{\eta^{\pdagger}_{\sigma_xG_x},\eta^{\pdagger}_\sigma\eta^{\pdagger}_{G_xT_x},\eta^{\pdagger}_\sigma\eta^{\pdagger}_{\sigma_xG_x},\eta^{\pdagger}_{G_xT_x}\big\}\mathcal{W}^{\pdagger}_{T_{2}},\\
W^{\pdagger}_{\sigma_{y}}(m,n,u)&=\big\{1,\eta^{\pdagger}_{\sigma_xG_x},1,\eta^{\pdagger}_{\sigma_xG_x}\big\}\mathcal{W}^{\pdagger}_{\sigma_{y}},\\
W^{\pdagger}_{\sigma_{x}}(m,n,u)&=\big\{1,\eta^{\pdagger}_{\sigma_xG_x},1,\eta^{\pdagger}_{\sigma_xG_x}\big\}\mathcal{W}^{\pdagger}_{\sigma_{x}},\\
W^{\pdagger}_{\sigma_{xy}}(m,n,u)&=\big\{1,1,\eta^{\pdagger}_\sigma,1\big\}\mathcal{W}^{\pdagger}_{\sigma_{xy}},\\
W^{\pdagger}_{\mathcal{T}}(m,n,u)&=\big\{1,\eta^{\pdagger}_{G_x\mathcal{T}},1,\eta^{\pdagger}_{G_x\mathcal{T}}\big\}\mathcal{W}^{\pdagger}_{\mathcal{T}}.\label{eq:solution_square_z2_22}
\end{align}
The associated matrices $\mathcal{W}$ are listed in Table~\ref{table:z2_psg_square_2}.

To simplify these expressions even further, we can now implement a gauge transformation with $W(m,n,u)=(\eta_{\sigma_xG_x}\eta_{G_xT_x})^{m+n}\big\{\eta_{\sigma_xG_x}\eta_{G_xT_x},\eta_{\sigma_xG_x},\eta_\sigma,\eta_{\sigma_xG_x}\rev{\big\}}$, which brings us to
\begin{align}\label{eq:solution_square_z2_3}
W^{\pdagger}_{T_{1}}(m,n,u)&=\big\{1,1,\eta^{\pdagger}_\sigma,\eta^{\pdagger}_\sigma\big\}\mathcal{W}^{\pdagger}_{T_{1}},\\
W^{\pdagger}_{T_{2}}(m,n,u)&=\mathcal{W}^{\pdagger}_{T_{2}},\\
W^{\pdagger}_{\sigma_{y}}(m,n,u)&=\big\{1,\eta^{\pdagger}_{G_xT_x},\eta^{\pdagger}_{G_xT_x}\eta^{\pdagger}_{\sigma_xG_x},\eta^{\pdagger}_{\sigma_xG_x}\big\}\mathcal{W}^{\pdagger}_{\sigma_{y}},\\
W^{\pdagger}_{\sigma_{x}}(m,n,u)&=\big\{1,\eta^{\pdagger}_{\sigma_xG_x},\eta^{\pdagger}_{G_xT_x}\eta^{\pdagger}_{\sigma_xG_x},\eta^{\pdagger}_{G_xT_x}\big\}\mathcal{W}^{\pdagger}_{\sigma_{x}},\\
W^{\pdagger}_{\sigma_{xy}}(m,n,u)&=\big\{1,1,\eta^{\pdagger}_\sigma,1\big\}\mathcal{W}^{\pdagger}_{\sigma_{xy}},\\
W^{\pdagger}_{\mathcal{T}}(m,n,u)&=\big\{1,\eta^{\pdagger}_{G_x\mathcal{T}},1,\eta^{\pdagger}_{G_x\mathcal{T}}\big\}\mathcal{W}^{\pdagger}_{\mathcal{T}}.
\end{align}

The square lattice, of course, has only one site per unit cell, so the solutions above  can be reformulated in the simple $(x,y)$ coordinate system as
\begin{align}
W^{\pdagger}_{T_{1}}(x,y)&=\eta^{y}_\sigma\mathcal{W}^{\pdagger}_{T_{1}},\label{eq:solution_square_z2_3_1}\\
W^{\pdagger}_{T_2}(x,y)&=\mathcal{W}^{\pdagger}_{T_{2}},\\
W^{\pdagger}_{\sigma_{y}}(x,y)&=\eta^x_{G_xT_x}\eta^y_{\sigma_xG_x}\mathcal{W}^{\pdagger}_{\sigma_{y}},\\
W^{\pdagger}_{\sigma_{x}}(x,y)&=\eta^x_{\sigma_xG_x}\eta^y_{G_xT_x}\mathcal{W}^{\pdagger}_{\sigma_{x}},\\
W^{\pdagger}_{\sigma_{xy}}(x,y)&=\eta^{xy}_\sigma\mathcal{W}^{\pdagger}_{\sigma_{xy}},\label{eq:solution_square_z2_3_5}\\
W^{\pdagger}_{\mathcal{T}}(x,y)&=\eta^{x+y}_{G_x\mathcal{T}}\mathcal{W}^{\pdagger}_{\mathcal{T}}.
\end{align}

\begin{figure*}[tb]
\includegraphics[width=1.0\linewidth]{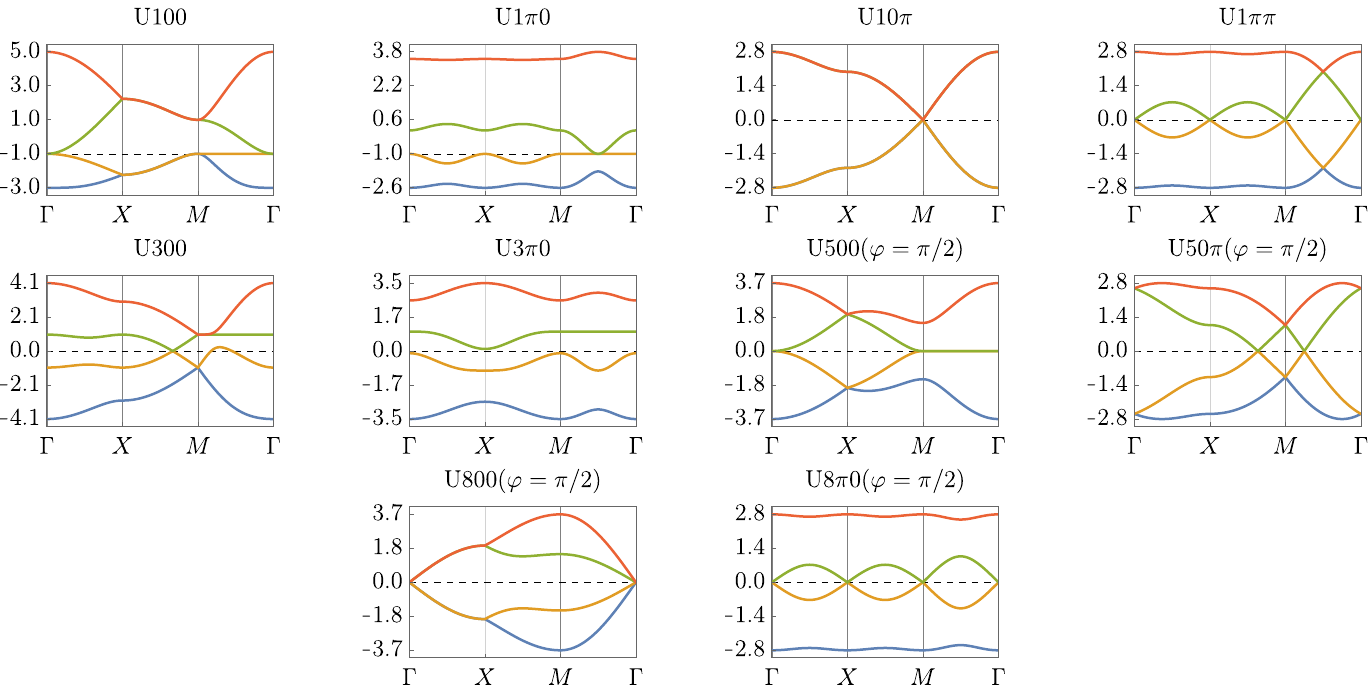}
\includegraphics[width=1.0\linewidth]{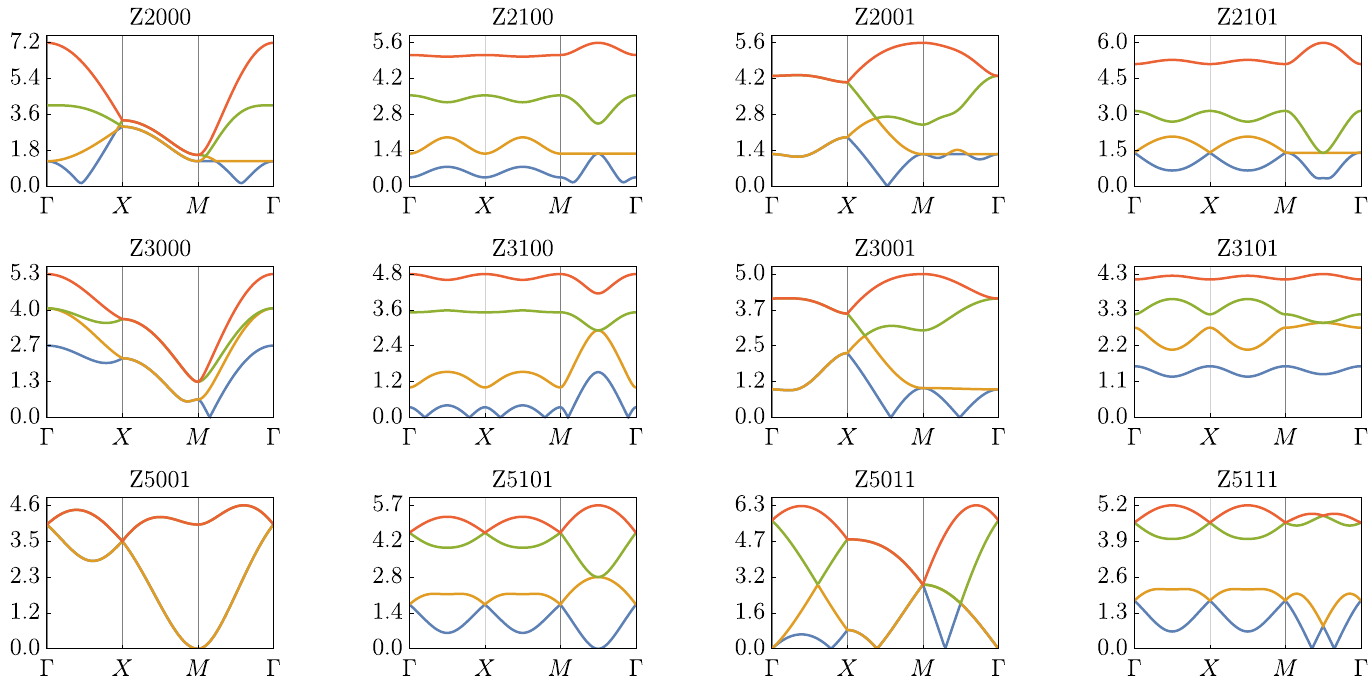}
	\caption{Spinon band structures for all the U(1) \textit{Ans\"atze} illustrated in Fig.~\ref{fig:fig2} (top), and for the $\mathbb{Z}_2$ \textit{Ans\"atze} listed in Table~\ref{table:z2_ansatze} (bottom). In both cases, we pick the mean-field parameters such that the magnitude of the symmetry-allowed hoppings is set to one. The dashed black line marks the Fermi level. For the U(1) \textit{Ans\"atze} with both imaginary and real hoppings, we define a parameter $\varphi=\text{Arg}[u^4_s]$, which is related to the flux threading the square plaquettes (see Fig.~\ref{fig:fig2}).}
	\label{fig:dispersions}
\end{figure*}
 \begin{figure}[t]
 \includegraphics[width=1.0\linewidth]{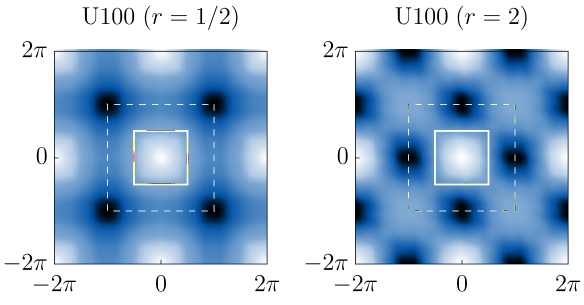}
	\caption{Equal-time structure factor for the U100 state. $r$ denotes the ratio of the hopping amplitude on the diagonals to that on the square bonds. Note that the structure factors for $r=1/2$ and $r=2$ show the same signatures as for N\'eel and stripe magnetic ordering on the square lattice, respectively. This property holds for all the U(1) and $\mathbb{Z}_2$ \textit{Ans\"atze} with nonvanishing parameters on both the square and diagonal bonds. The solid (dashed) white square marks the first (extended) Brillouin zone.}
	\label{fig:ssf_neel_stripe}
\end{figure}

\begin{figure*}[tb]
\includegraphics[width=0.85\linewidth]{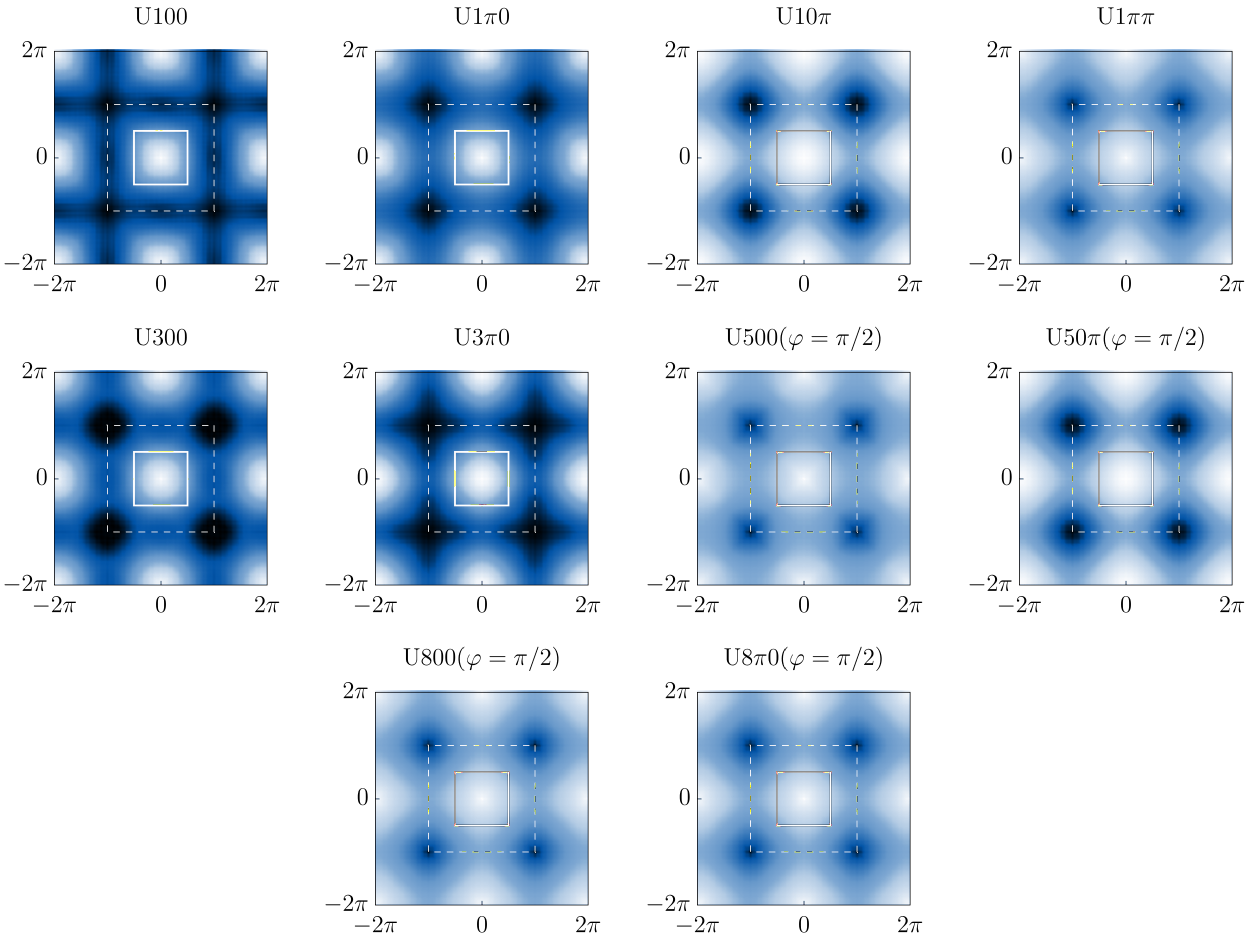}
\includegraphics[width=0.85\linewidth]{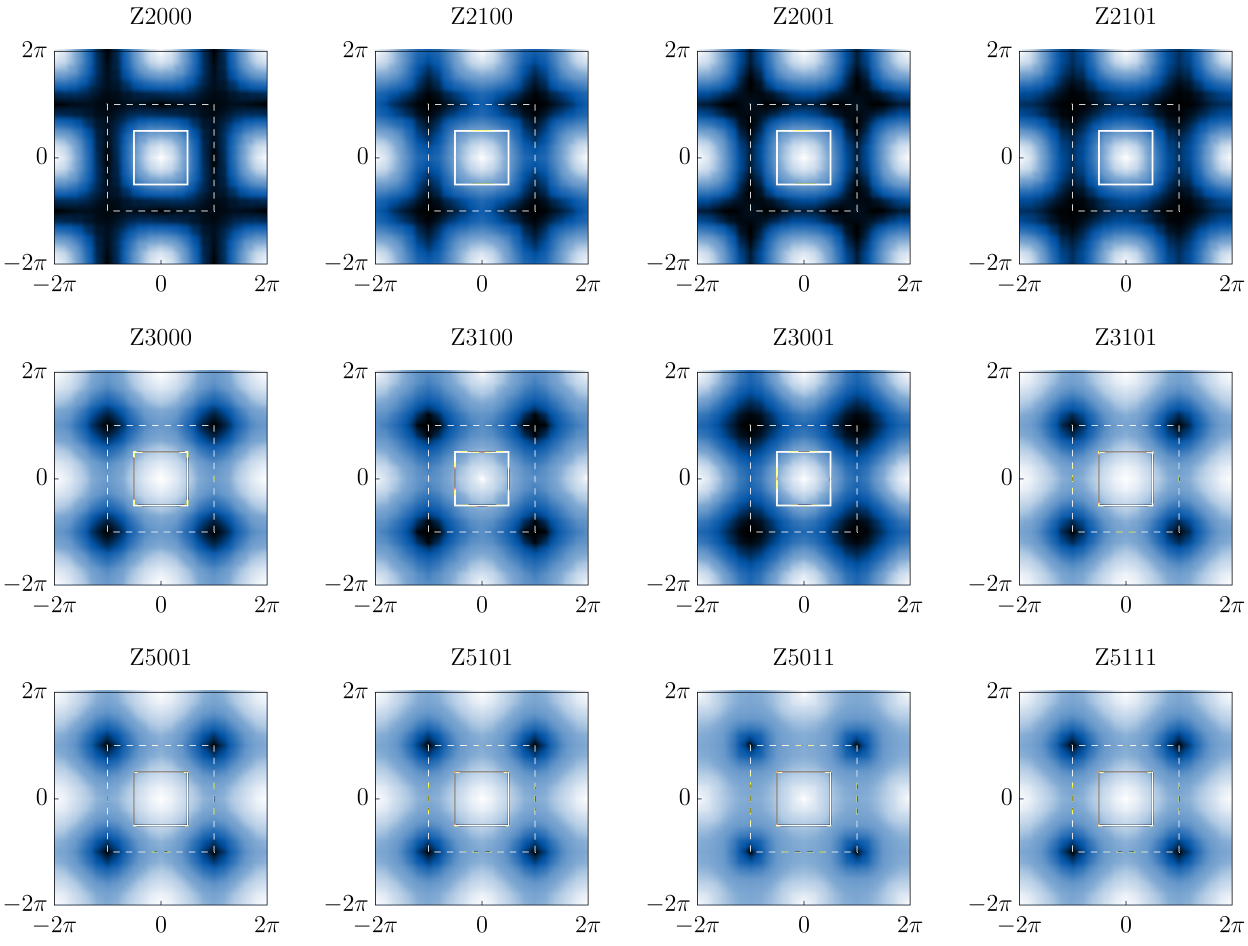}
	\caption{Equal-time structure factors for all the U(1) \textit{Ans\"atze} illustrated in Fig.~\ref{fig:fig2} (top), and for the $\mathbb{Z}_2$ \textit{Ans\"atze} listed in Table~\ref{table:z2_ansatze} (bottom), following the same conventions for the color map as in Fig.~\ref{fig:ssf_neel_stripe}. The mean-field parameters are the same as those used for Fig.~\rev{\ref{fig:dispersions}}.}
	\label{fig:U1_ssf}
\end{figure*}

\begin{figure*}[tb]
\includegraphics[width=0.85\linewidth]{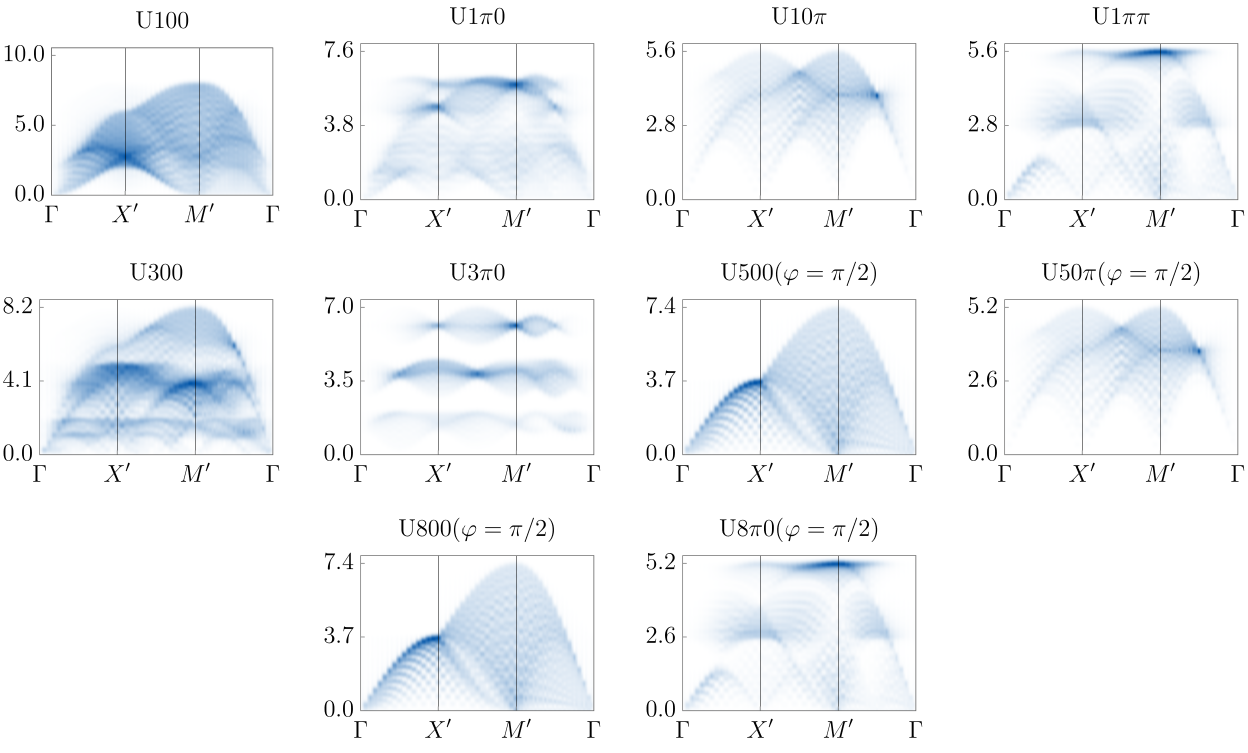}
\includegraphics[width=0.85\linewidth]{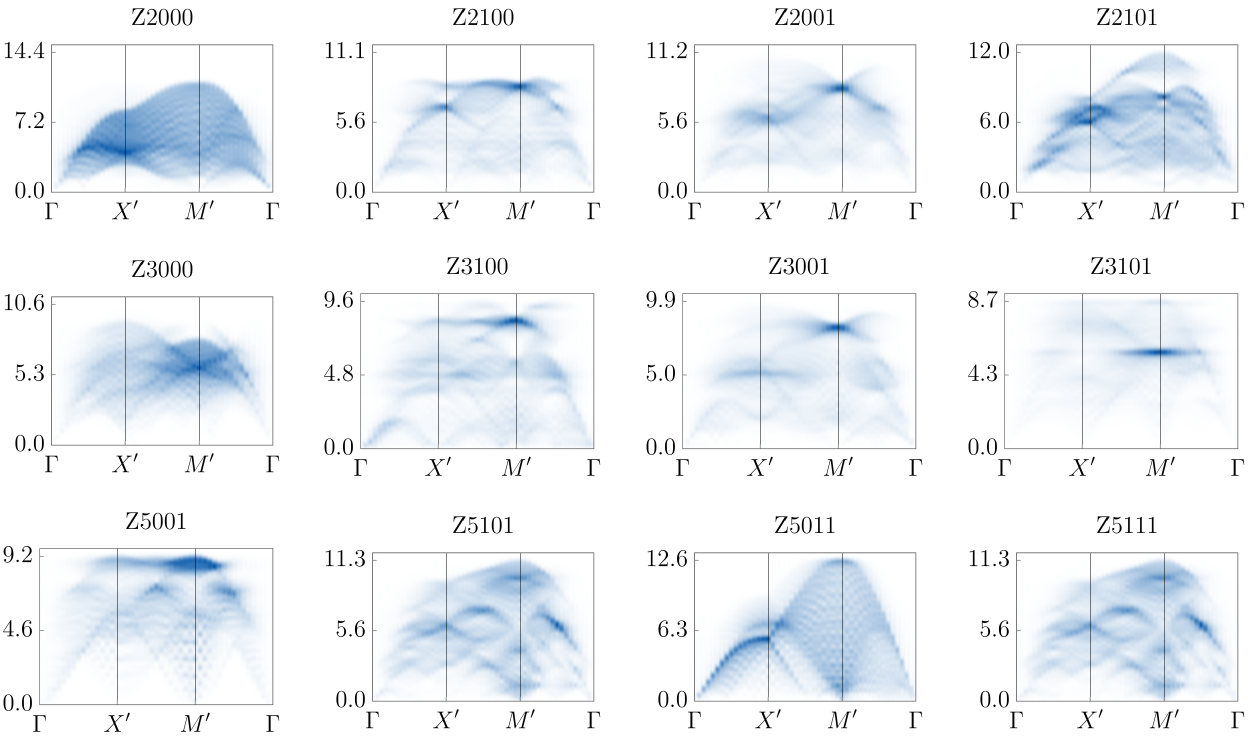}
	\caption{Dynamical structure factors for all the U(1) \textit{Ans\"atze} illustrated in Fig.~\ref{fig:fig2} (top), and for the $\mathbb{Z}_2$ \textit{Ans\"atze} listed in Table~\ref{table:z2_ansatze} (bottom). The variation in intensity from low to high is conveyed by the  gradient of the color, from white to blue (similar to Fig.~\ref{fig:SSfactor}). The mean-field parameters are the same as those used for Fig.~\rev{\ref{fig:dispersions}}.}
	\label{fig:U1_dsf}
\end{figure*}
 \begin{figure*}[!tp]
\includegraphics[width=0.85\linewidth]{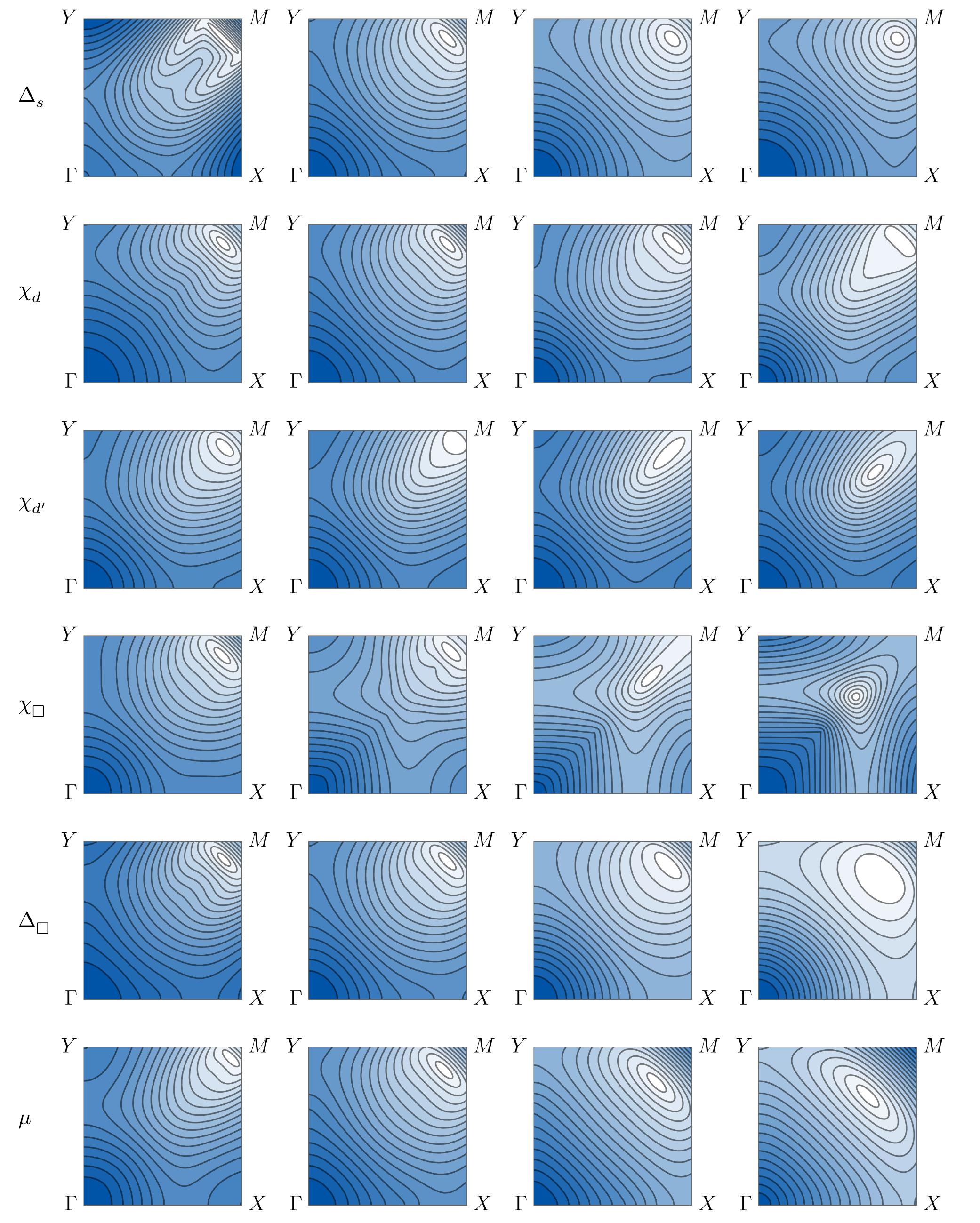}
	\caption{Contour plots of the lowest-energy mode of the Z3000 \textit{Ansatz}, illustrating the positions of the Dirac points as the variational parameters are varied. In each row, the parameter indicated on the left ($\Delta_{s}$, $\chi_{d}$, $\chi_{d'}$, $\chi_{\square}$, $\Delta_{\square}$, and $\mu$) is varied, while the remaining parameters are fixed at their optimized VMC values. The four columns correspond to parameter values $0.2$, $0.4$, $0.6$, and $0.8$, respectively. The hopping amplitude $\chi_{s}$ is fixed to $\chi_{s}=1.0$ throughout. The color scheme is the same as that used in Fig.~\ref{fig:z3000_dis_ssf_dsf}. The corresponding dispersion curves along the $\Gamma \rightarrow M$ direction are shown in Fig.~\ref{fig:dirac_points_parameters_GM}.}	\label{fig:dirac_points_parameters}
\end{figure*}

Additionally, one can conveniently set $\mathcal{W}_{T_{1}}=\mathcal{W}_{T_{2}}=\tau^0$ by applying a gauge transformation of the form $W(x,y)=\mathcal{W}^y_{T_{2}}\mathcal{W}^x_{T_{1}}$. This will not change the overall structure of the solutions in Table~\ref{table:z2_psg_square_2} up to a sign factor, which can be absorbed either using the gauge freedom (if it is global) or by redefining the  parameters $\eta$. As an example, for $\mathcal{W}_{T_{1}}=\dot\iota\tau^{zx}=\dot\iota(\tau^{z}+\tau^{\rev{x}})/\sqrt{2}$, $W_{T_{1}}(x,y)$ and $W_{\sigma_{xy}}(x,y)$ in
Eqs.~\eqref{eq:solution_square_z2_3_1} and \eqref{eq:solution_square_z2_3_5}, respectively, become $\rev{W_{T_{1}}(x,y)=(-\eta_\sigma)^{y}\mathcal{W}_{T_{1}}}$ and $W_{\sigma_{xy}}(x,y)=(-\eta_\sigma)^{xy}\mathcal{W}_{\sigma_{xy}}$. The original $x, y$ dependence can be recovered by the redefinition $\eta_\sigma\rightarrow-\eta_\sigma$.

After all these steps, we arrive at the full solution for the square-lattice PSGs, which \rev{reads}
\begin{subequations}
    \label{eq:solution_square_z2_4}
    \begin{align}
W^{\pdagger}_{T_{1}}(x,y)&=\eta^{y}_\sigma\rev{\tau^0},\label{eq:solution_square_z2_4_1}\\
W^{\pdagger}_{T_2}(x,y)&=\tau^0,\\
W^{\pdagger}_{\sigma_{y}}(x,y)&=\eta^x_{G_xT_x}\eta^y_{\sigma_xG_x}\mathcal{W}^{\pdagger}_{\sigma_{y}},\\
W^{\pdagger}_{\sigma_{x}}(x,y)&=\eta^x_{\sigma_xG_x}\eta^y_{G_xT_x}\mathcal{W}^{\pdagger}_{\sigma_{x}},\\
W^{\pdagger}_{\sigma_{xy}}(x,y)&=\eta^{xy}_\sigma\mathcal{W}^{\pdagger}_{\sigma_{xy}}\label{eq:solution_square_z2_4_5},\\
W^{\pdagger}_{\mathcal{T}}(x,y)&=\eta^{x+y}_{G_x\mathcal{T}}\mathcal{W}^{\pdagger}_{\mathcal{T}},
\end{align}
\end{subequations}
with the $\mathcal{W}$ matrices noted on the left side in Table~\ref{table:z2_psg_square_3}. On the right side of the same table, we write these matrices in a different gauge, used in Ref.~\cite{Wen-2002}. This completes the mapping of projective realizations of the symmetries from the SSM to those of the square lattice.

\section{Properties of U(1) and $\mathbb{Z}_2$ \textit{Ans\"atze}}
\label{app:u1_gen_prop}

The properties of an \textit{Ansatz}---such as its spinon excitation spectrum, the dynamical structure factor, and the equal-time structure factor---depend on the choices of the mean-field parameters, which, in principle, should be evaluated self-consistently based on specific models. Here, we take a different approach and concentrate on more generic properties rather than focusing on a particular model.

In the canonical gauge, any U(1) \textit{Ansatz} consists \rev{solely of} hopping terms, so the pseudospin $\uparrow$ and $\downarrow$ sectors are decoupled. Working in the basis defined by $\hat{f}_{\mathbf{k},\alpha} = (\hat{f}_{\mathbf{k},1,\alpha}, \hat{f}_{\mathbf{k},2,\alpha}, \hat{f}_{\mathbf{k},3,\alpha}, \hat{f}_{\mathbf{k},4,\alpha})^{\rev{\mathrm{T}}}$, the quadratic Hamiltonian in $\mathbf{k}$-space reads
\begin{equation}
    \hat{H}(\mathbf{k}) = \sum_{\alpha = \uparrow, \downarrow} \hat{f}^\dagger_{\mathbf{k},\alpha} \hat{H}^\text{U(1)}_{\mathbf{k}} \hat{f}^{\pdagger}_{\mathbf{k},\alpha}.
    \label{eq:u1_ham}
\end{equation}
Owing to spinon number conservation, the chemical potential does not need to be explicitly considered and can be ignored here. The one-particle-per-site constraint (on average) is enforced by setting the Fermi energy such that we are at half filling.

In the following, we summarize the generic spinon excitation characteristics of different U(1) \textit{Ans\"atze}, which are illustrated in the top panel of Fig.~\rev{\ref{fig:dispersions}}. The band structures are plotted along the path $\Gamma \rightarrow X \rightarrow M \rightarrow \Gamma$, where the high-symmetry points are $\Gamma(0,0)$, $X(\pi/2,0)$, and $M(\pi/2,\pi/2)$ in the first Brillouin zone\rev{; below, we also refer to $Y(0,\pi/2)$}.
\begin{itemize}
    \item The excitation spectrum for U1$00$ is quadratic and gapless at the $\Gamma$ point, while U1$\pi0$ exhibits quadratic gapless excitations at isolated $\mathbf{k}$-points. The spectrum of U1$0\pi$ consists of Dirac cones located at $(\pm\pi/2, \pm\pi/2)$, as also seen for the SU(2) Dirac spin liquid labeled as SU2B$n0$ on the square lattice~\cite{Wen-2002}. U1$\pi\pi$ features a Fermi surface with linear gapless excitations along $(k_x, -k_x)$ and $(k_x, k_x \pm \pi/2)$, forming Dirac nodal lines.

    \item The spectrum of U3$00$ is gapless with a linear dispersion, and Dirac cones  are observed along the $\overline{XM}$ and $\overline{YM}$ lines. A Fermi surface is seen for $\chi_d < 1.5\chi_s$. U3$\pi0$ has a gapped spinon spectrum. The link fields of U3$0\pi$ and U3$\pi\pi$ are the same as those of U1$0\pi$ and U1$\pi\pi$, respectively. However, the former states include a staggered chemical potential that opens up a gap. Notably, the staggered chemical potential also opens a gap for U3$00$.

    \item The excitation spectrum of U5$00$ is quadratic and gapless along  $(k_x, \pm k_x)$, forming nodal lines. For U5$0\pi$, the spectrum is also gapless but linearly dispersing, forming a Dirac nodal loop surrounding the $\Gamma$ or $M$ point.

    \item The U8$\xi0$ family encompasses a series of \textit{Ans\"atze} corresponding to different choices of $\xi = \mathcal{P}\pi/\mathcal{Q}$, for $\rev{\mathcal{P} \in \mathbb{Z}^{\geq 0}}$ \rev{and} $\rev{\mathcal{Q} \in \mathbb{Z}^{+}}$. Here, we limit our analysis to states realizable within a single unit cell, \rev{namely} $\xi = 0$ and $\xi = \pi$. Both cases exhibit a Dirac dispersion. For $\xi = 0$, the Dirac cones are anchored at the $\Gamma$ point, while for $\xi = \pi$, they reside at all high-symmetry points.
\end{itemize}

Next, we discuss the excitation spectra of the $\mathbb{Z}_2$ \textit{Ans\"atze}. In this case, the interpretation of the quasiparticles is quite different from that for a U(1) IGG because spinon number conservation is lost due to the presence of pairing terms. We begin with the generic Bogoliubov\rev{--}de Gennes (BdG) basis
\begin{equation}
    \hat{\Psi}^{\dagger}_\mathbf{k} = \big(\hat{f}^{\dagger}_{\mathbf{k},\uparrow}, (\hat{f}^{\pdagger}_{-\mathbf{k},\downarrow})^{\rev{\mathrm{T}}}\big),
    \label{eq:bdg_basis}
\end{equation}
using which, any general mean-field Hamiltonian in $\mathbf{k}$-space can be expressed as
\begin{equation}
    \hat{H}(\mathbf{k}) = \hat{\Psi}^\dagger_\mathbf{k} \hat{H}^\text{BdG}_{\mathbf{k}} \hat{\Psi}^{\pdagger}_\mathbf{k};\quad
    \hat{H}^\text{BdG}_\mathbf{k} =
\begin{bmatrix}
\hat{H}^\text{U(1)}_\mathbf{k} & \hat{H}^{\mathbb{Z}_2}_\mathbf{k} \\
(\hat{H}^{\mathbb{Z}_2}_\mathbf{k})^\dagger & \rev{-\big(\hat{H}^\text{U(1)}_{-\mathbf{k}}\big)^{\mathrm{T}}}
\end{bmatrix},
    \label{eq:bdg_ham}
\end{equation}
where $\hat{H}^\text{U(1)}_\mathbf{k}$ and $\hat{H}^{\mathbb{Z}_2}_\mathbf{k}$ denote the hopping and pairing terms in the \textit{Ansatz}, respectively. The superscript $\mathbb{Z}_2$ indicates that these terms are the ones responsible for breaking the IGG down to $\mathbb{Z}_2$.

Upon diagonalization, $\hat{H}^\text{BdG}_\mathbf{k}$ yields eigenvalues in positive and negative pairs, i.e., $\pm\epsilon^{}_{\mathbf{k},u}$. In the diagonalized basis, $\hat{f}_{\mathbf{k},u,\alpha} \rightarrow \hat{\zeta}_{\mathbf{k},u,\alpha}$, the Hamiltonian can be rewritten as
\begin{equation}
\begin{aligned}
    \hat{H}(\mathbf{k}) &= \sum_{u=1}^4 \epsilon^{\pdagger}_{\mathbf{k},u} (\hat{\zeta}^\dagger_{\mathbf{k},u,\uparrow} \hat{\zeta}^{\pdagger}_{\mathbf{k},u,\uparrow} - \hat{\zeta}^{\pdagger}_{\mathbf{k},u,\downarrow} \hat{\zeta}^\dagger_{\mathbf{k},u,\downarrow}) \\
    &= \sum_{u=1}^4 \sum_{\alpha=\uparrow,\downarrow} \epsilon^{\pdagger}_{\mathbf{k},u} (\hat{\zeta}^\dagger_{\mathbf{k},u,\alpha} \hat{\zeta}^{\pdagger}_{\mathbf{k},u,\alpha} - \rev{\tfrac{1}{2}}).
\end{aligned}
    \label{eq:z2_ham}
\end{equation}
This formulation allows us to interpret the spinon excitations of a generic symmetric $\mathbb{Z}_2$ mean-field \textit{Ansatz} as Bogoliubov quasiparticles that always have a positive energy. In other words, the Fermi level lies at zero energy, leaving the quasiparticle-free ground state as the vacuum. In this case, the explicit inclusion of Lagrange multipliers is required to enforce the one-particle-per-site constraint on average.

In the \rev{bottom panel of} Fig.~\rev{\ref{fig:dispersions}}, we showcase the spinon dispersion spectra for the $\mathbb{Z}_2$ \textit{Ans\"atze} listed in Table~\ref{table:z2_ansatze}. For purposes of illustration, all mean-field parameters are set to unity. The general properties of these states are as follows:
\begin{itemize}
    \item For Z2000, the chemical potential $\mu \neq 0$ (required to satisfy the constraint) opens up a small gap in general. Both Z2100 and Z2101 are also generically gapped. The spectrum of Z2001 always exhibits a Dirac dispersion with a Dirac cone positioned between the $X$ and $M$ points.
    \item Similarly to Z2001, Z3000 hosts a Dirac dispersion, with a Dirac cone located between $\Gamma$ and $M$. Z3100 shows a gapless spectrum with either a Fermi surface or Dirac cones, depending on the mean-field parameters. The spectrum of Z3001 generally features a Fermi surface surrounding the $M$ points, or a Dirac dispersion for specific parameter choices. Z3101 can exhibit gapped, Dirac, or Fermi-surface states.
    \item Within the Z5 class, Z5101, Z5011, and Z5111 exhibit gapless dispersions with a Fermi surface, while Z5001 has a quadratic dispersion that becomes gapless at the $M$ point.
    \item For the remaining six \textit{Ans\"atze} (Z2011, Z2111, Z3010, Z3110, Z3011, Z3111), the term breaking the IGG from U(1) to $\mathbb{Z}_2$ is the onsite piece $\mu \tau^z$. The link fields retain the original U(1) flux structures, leading these \textit{Ans\"atze} to qualitatively resemble their parent U(1) states.
\end{itemize}

To distinguish and further characterize these states, we can calculate their respective structure factors, which are also of utility for comparisons to neutron\rev{-}scattering experiments. The longitudinal component (which suffices due to spin-rotation symmetry) of the equal-time spin-spin correlation defines the equal-time spin structure factor as
\begin{equation}
\mathcal{S}^{}(\mathbf{q}) = \frac{\rev{3}}{\rev{N}} \sum_{i,j} e^{\dot{\iota}\mathbf{q} \cdot \mathbf{r}_{ij}} \langle \hat{S}^z_i \hat{S}^z_j \rangle,
    \label{eq:ssf}
\end{equation}
where $\rev{N}$ is the number of lattice sites\rev{; the factor of three restores the isotropic normalization of Eq.~\eqref{eq:spin_sf}}.
In Fig.~\ref{fig:ssf_neel_stripe}, we show the equal-time structure factor of the U(1) state labeled U100 for two choices of parameters: $r = 1/2$ and $r = 2$, where $r$ is the ratio of mean-field amplitudes on the diagonal and square bonds. With stronger square bonds, maxima appear at the $M'(\pi,\pi)$ points in the extended Brillouin zone (EBZ), while for dominant diagonal bonds, the maxima occur at $X'(\pi,0)$. These maxima are not sharp Bragg peaks due to spin-rotation symmetry but instead form a continuum. This qualitative behavior persists across all U(1) and $\mathbb{Z}_2$ \textit{Ans\"atze} with nonzero mean-field parameters on both square and diagonal bonds. For \textit{Ans\"atze} with vanishing diagonal parameters, the maxima always appear at $M'$.

For representative parameter choices (consistent with the dispersion plots in Fig.~\rev{\ref{fig:dispersions}}), the equal-time structure factors of all the U(1) and $\mathbb{Z}_2$ \textit{Ans\"atze} are shown in Fig.~\ref{fig:U1_ssf}. While most states exhibit minimal qualitative differences in their equal-time structure factor, the distinctions are clearer in the dynamical structure factor (DSF), defined as
\begin{equation}
\mathcal{S}^{}(\mathbf{q},\omega) = \rev{\frac{3}{2\pi N}} \int^{+\infty}_{-\infty} \rev{d\tau\,} e^{\dot{\iota}\omega\tau} \sum_{i,j} e^{\dot{\iota}\mathbf{q} \cdot \mathbf{r}_{ij}} \big \langle \hat{S}^z_i(\tau) \hat{S}^z_j(0) \big\rangle.
    \label{eq:dsf}
\end{equation}
The DSFs are plotted for the various \textit{Ans\"atze} along the high-symmetry path $\Gamma(0,0) \rightarrow X'(\pi,0) \rightarrow M'(\pi,\pi) \rightarrow \Gamma$ in the EBZ in Fig.~\ref{fig:U1_dsf}.

 \begin{figure*}[tb]
\includegraphics[width=0.8\linewidth]{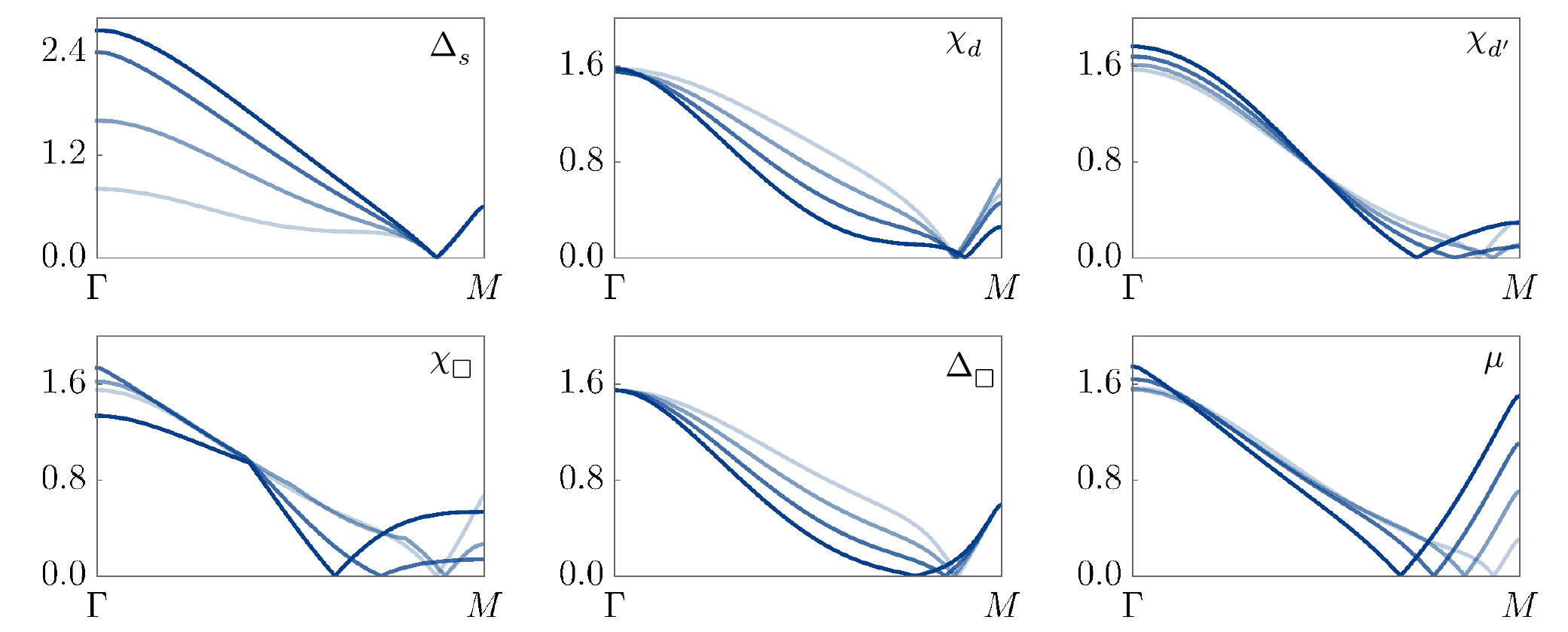}
	\caption{Dispersion of the lowest-energy mode of the Z3000 \textit{Ansatz}, illustrating the positions of the Dirac points as the variational parameters are varied. The initial parameters are set by the optimized values obtained from VMC, and small variations are introduced around these values for $\Delta_{s}$, $\chi_{d}$, $\chi_{d'}$, $\chi_{\square}$, $\Delta_{\square}$, and $\mu$. The hopping amplitude $\chi_{s}$ is fixed to $\chi_{s}=1.0$ throughout. In each panel, the opacity of the curves encodes the magnitude of the corresponding parameter, with lower (higher) opacity representing smaller (larger) parameter values.}	\label{fig:dirac_points_parameters_GM}
\end{figure*}

In addition, it is instructive to comment on the dispersion of the Z3000 state when all diagonal terms are included, namely $\chi_{d}$, $\chi_{d'}$, $\chi_{\square}$, and $\Delta_{\square}$, since this is the key \textit{Ansatz} considered in the present study. As discussed in the main text, the inclusion of any of the hopping terms $\chi_{d}$, $\chi_{d'}$, $\chi_{\square}$ (or of the onsite term $\mu$) leads to a splitting of the Dirac nodes, whereas $\Delta_{\square}$ acting alone does not; see  \rev{App.}~\ref{app:epsilon} for the exact closed-form analysis.
In Fig.~\ref{fig:dirac_points_parameters}, we present contour plots of the lowest-energy eigenmode of the Z3000 state as a function of the variational parameters. Notably, for all parameter choices explored, the Dirac nodes remain located along the $\overline{\Gamma M}$ line in momentum space. Moreover, the dependence of the Dirac-node positions on the variational parameters is illustrated explicitly in Fig.~\ref{fig:dirac_points_parameters_GM}.

\section{Exact location of the Dirac nodes of the Z3000 \textit{Ansatz}}
\label{app:epsilon}

In this appendix\rev{,} we determine the positions of the Dirac nodes of the Z3000
\textit{Ansatz}~\eqref{eq:MF-Z2} in closed form, i.e., we obtain the
displacement $\varepsilon$ of the nodes away from the commensurate momenta
$M(\pm\pi/2,\pm\pi/2)$ as an explicit function of all the mean-field
amplitudes. All the analytic statements below have been verified to machine
precision against a direct diagonalization of the Bogoliubov--de Gennes (BdG)
Hamiltonian. \rev{The auxiliary quantities $a$, $b$, $c$, $t$, $\mathcal{U}$,
$\mathcal{V}$, $y$, $\rho$, $\sigma$, $R$, $S$, $G$, $H^{}_{1,2}$, and $d^{}_{A}$
introduced below are shorthands used only within this appendix.}

\subsection{Reduction of the node condition to a $4\times4$ determinant}

After the particle-hole transformation~\eqref{eq:ph_transformation} on the
$\downarrow$ spinons, the BdG Hamiltonian associated with Eq.~\eqref{eq:MF-Z2}
reads, at momentum $\mathbf{k}$,
\begin{equation}
\rev{\mathcal{H}}(\mathbf{k})=
\begin{pmatrix}
h(\mathbf{k}) & D(\mathbf{k})\\[2pt]
D(\mathbf{k}) & -h(\mathbf{k})
\end{pmatrix},
\label{eq:bdg_block}
\end{equation}
where $h(\mathbf{k})$ (which includes the onsite term $\mu$) and
$D(\mathbf{k})$ are the $4\times4$ Hermitian Bloch matrices of the hopping and
of the singlet-pairing amplitudes in the four-site geometric unit cell. Since
$\rev{\mathcal{H}}$ anticommutes with $\tau^{y}\otimes\rev{\mathds{1}}$, its determinant factorizes,
\begin{equation}
\det \rev{\mathcal{H}}(\mathbf{k})=\bigl|\det C(\mathbf{k})\bigr|^{2},
\qquad
C(\mathbf{k})\equiv h(\mathbf{k})-\dot\iota\,D(\mathbf{k}),
\label{eq:chiral_reduction}
\end{equation}
so that the gapless points of the \textit{Ansatz} are exactly the zeros of the
$4\times4$ complex determinant $\det C(\mathbf{k})$.

\subsection{Restriction to the $\overline{\Gamma M}$ line}

On the diagonal $k^{}_{x}=k^{}_{y}=k$, which contains the nodes, the matrix $C$
takes the explicit form
\begin{equation}
C\big|^{}_{k^{}_{x}=k^{}_{y}=k}=
\begin{pmatrix}
\mu & a & \rho+\dot\iota\sigma & b\\
a & \mu & b & R+S\\
\rho-\dot\iota\sigma & b & \mu & a\\
b & R-S & a & \mu
\end{pmatrix},
\label{eq:Cline}
\end{equation}
in the sublattice basis $(u^{}_{1},u^{}_{2},u^{}_{3},u^{}_{4})$, where, with
$c\equiv\cos k$,
\begin{equation}
\begin{aligned}
a &= 2c\,(\chi^{}_{s}-\dot\iota\Delta^{}_{s}), &\qquad
b &= 2c\,(\chi^{}_{s}+\dot\iota\Delta^{}_{s}),\\
\rho &= 2\chi^{}_{\square}+(\chi^{}_{d}+\chi^{}_{d'})\cos 2k, &\qquad
\sigma &= (\chi^{}_{d}-\chi^{}_{d'})\sin 2k,\\
R &= (\chi^{}_{d}+\chi^{}_{d'})+2\chi^{}_{\square}\cos 2k, &\qquad
S &= 2\Delta^{}_{\square}\sin 2k.
\end{aligned}
\label{eq:Cline_defs}
\end{equation}
The $(u^{}_{1},u^{}_{3})$ channel collects the bonds lying along the mirror
axis $\sigma^{}_{xy}$ and carries no pairing on this line, whereas the entire
$\Delta^{}_{\square}$ dependence resides in the $(u^{}_{2},u^{}_{4})$ channel
through $S$. The antiunitary operation obtained by combining complex
conjugation with the sublattice exchange $u^{}_{1}\!\leftrightarrow\!u^{}_{3}$
leaves $C$ invariant, which forces $\det C$ to be \emph{real} along
$\overline{\Gamma M}$. The node condition is therefore a single real equation
in the single unknown $k$---this is \rev{why} the nodes are pinned to the
$\overline{\Gamma M}$ line, as observed numerically in
Fig.~\ref{fig:dirac_points_parameters}.

\subsection{Closed-form node equation}

Performing a Schur decomposition of Eq.~\eqref{eq:Cline} with respect to the
$(u^{}_{1},u^{}_{3})\oplus(u^{}_{2},u^{}_{4})$ splitting, we obtain the compact
closed-form result
\begin{equation}
\det C=\frac{\rev{\mathcal{Q}}}{d^{}_{A}},
\label{eq:detC_closed}
\end{equation}
where
\begin{equation}
\begin{split}
\rev{\mathcal{Q}}&=(\mu\,d^{}_{A}-G)^{2}-(R^{2}-S^{2})\,d_{A}^{2}-H^{}_{1}H^{}_{2}\\
&+d^{}_{A}\bigl[R\,(H^{}_{1}+H^{}_{2})+S\,(H^{}_{2}-H^{}_{1})\bigr],
\end{split}
\label{eq:detC_numerator}
\end{equation}
and
\begin{alignat}{1}
\nonumber
d^{}_{A}&=\mu^{2}-\rho^{2}-\sigma^{2},\quad
t=2\chi^{}_{s}\Delta^{}_{s},\\
\nonumber
\rev{\mathcal{U}}&=\chi_{s}^{2}+\Delta_{s}^{2},\qquad\,\,\,\,
\rev{\mathcal{V}}=\chi_{s}^{2}-\Delta_{s}^{2},\\
\nonumber
G&=8c^{2}\,(\mu\, \rev{\mathcal{V}}-\rho\, \rev{\mathcal{U}}),\\
H^{}_{1,2}&=8c^{2}\bigl(\mu\, \rev{\mathcal{U}}-\rho\, \rev{\mathcal{V}}\mp t\,\sigma\bigr).
\label{eq:aux}
\end{alignat}
Equation~\eqref{eq:detC_closed} is a quartic polynomial in $y=\cos^{2}k$ whose
coefficients are explicit polynomials in all the mean-field amplitudes. Its
roots, and hence the node displacement $\varepsilon$---defined by the node
position $k^{*}=\pi/2+\varepsilon$ through $y^{*}=\sin^{2}\varepsilon$---are
thus available in closed form.

\subsection{Exact analytic consequences}

\textit{(i) Pinning criterion.} At the commensurate point $y=0$, i.e.,
$\mathbf{k}=M(\pi/2,\pi/2)$, the quartic factorizes exactly as
\begin{equation}
\det C\big|^{}_{M}=
\bigl(\chi^{}_{d}+\chi^{}_{d'}-2\chi^{}_{\square}-\mu\bigr)^{2}
\bigl(\chi^{}_{d}+\chi^{}_{d'}-2\chi^{}_{\square}+\mu\bigr)^{2} .
\label{eq:pinning}
\end{equation}
The Dirac nodes therefore remain \emph{exactly} at $(\pm\pi/2,\pm\pi/2)$, i.e.,
$\varepsilon=0$, if and only if
\begin{equation}
\mu=\pm\bigl(\chi^{}_{d}+\chi^{}_{d'}-2\chi^{}_{\square}\bigr),
\label{eq:pinning_criterion}
\end{equation}
\textit{independently} of the pairing amplitudes $\Delta^{}_{s}$ and
$\Delta^{}_{\square}$. The BdG gap at $M$ is
$\min^{}_{\pm}\bigl|\mu\pm(\chi^{}_{d}+\chi^{}_{d'}-2\chi^{}_{\square})\bigr|$.
For the optimized parameters on the $L=14$ cluster\rev{,} this evaluates to $0.249$,
in exact agreement with the direct diagonalization; correspondingly, tuning
$\mu$ to $|\chi^{}_{d}+\chi^{}_{d'}-2\chi^{}_{\square}|=0.096$ while keeping
all the remaining amplitudes fixed at their optimal values (including
$\Delta^{}_{\square}=0.381$) closes the gap at $M$ and yields $\varepsilon=0$.

\textit{(ii) Displacement due to the chemical potential alone.} Setting
$\chi^{}_{d}=\chi^{}_{d'}=\chi^{}_{\square}=\Delta^{}_{\square}=0$, the quartic
factorizes as
$\det C=(\mu^{2}-16\chi_{s}^{2}c^{2})\,(\mu^{2}+16\Delta_{s}^{2}c^{2})$,
whence the exact and $\Delta^{}_{s}$-independent result
\begin{equation}
\varepsilon^{}_{\mu}=\arcsin\!\left(\frac{\mu}{4\chi^{}_{s}}\right).
\label{eq:eps_mu}
\end{equation}

\textit{(iii) $\Delta^{}_{\square}$ alone does not displace the nodes.} If
$\mu=\chi^{}_{d}=\chi^{}_{d'}=\chi^{}_{\square}=0$, the $(u^{}_{1},u^{}_{3})$
block of Eq.~\eqref{eq:Cline} vanishes identically and the determinant reduces
to $\det C=(a^{2}-b^{2})^{2}=-256\,\chi_{s}^{2}\Delta_{s}^{2}\,c^{4}$,
which is \emph{independent} of $\Delta^{}_{\square}$: the nodes stay pinned at
$M$. The empty-square pairing thus contributes to $\varepsilon$ only in
combination with the hopping amplitudes and $\mu$, through the $S$-dependent
terms of Eq.~\eqref{eq:detC_closed}.

\textit{(iv) Individual contributions.} Switching on a single amplitude at a
time on top of the parent state ($\chi^{}_{s}=1$, $\Delta^{}_{s}=0.387$), with
magnitudes taken from the $L=14$ optimum, yields
$\varepsilon(\mu\!=\!0.345)=0.086$,
$\varepsilon(\chi^{}_{d}\!=\!0.462)=0.285$,
$\varepsilon(\chi^{}_{d'}\!=\!-0.088)=0.057$, and
$\varepsilon(\chi^{}_{\square}\!=\!0.139)=0.175$ (in radians), whereas
$\varepsilon(\Delta^{}_{\square}\!=\!0.381)=0$, in accordance with point (iii).
In the fully optimized state\rev{,} these individual contributions largely
compensate---as dictated by Eq.~\eqref{eq:pinning}---leaving the much smaller
value $\varepsilon=0.091$.

\subsection{System-size dependence at the variational optimum}

Evaluating Eq.~\eqref{eq:detC_closed} with the optimized amplitudes of
Table~\ref{tab:pairing-optimized} gives the exact node displacement of the
variational state at $J^{}_{s}/J^{}_{d}=0.80$ for each cluster size:

\begin{table}[h]
\centering
\begin{tabular}{cccc}
\hline\hline
$L$ & $\varepsilon$ (rad) & $\varepsilon/\pi$ & $\delta/\pi=2\varepsilon/\pi$ \\
\hline
 6 & 0.246 & 0.078 & 0.156 \\
 8 & 0.158 & 0.050 & 0.101 \\
10 & 0.125 & 0.040 & 0.080 \\
12 & 0.105 & 0.033 & 0.067 \\
14 & 0.091 & 0.029 & 0.058 \\
20 & 0.066 & 0.021 & 0.042 \\
30 & 0.047 & 0.015 & 0.030 \\
40 & 0.036 & 0.011 & 0.023 \\
60 & 0.025 & 0.008 & 0.016 \\
\hline\hline
\end{tabular}
\caption{Exact displacement $\varepsilon$ of the Dirac nodes of the optimized
Z3000 \textit{Ansatz} at $J^{}_{s}/J^{}_{d}=0.80$, obtained from
Eq.~\eqref{eq:detC_closed} evaluated at the variational parameters of
Table~\ref{tab:pairing-optimized}. The last \rev{column} gives the corresponding
incommensurability $\delta=2\varepsilon$ of the gapless two-spinon momenta.}
\label{tab:epsilon_L}
\end{table}

The displacement decreases monotonically with the system size and is well
described by $\varepsilon(L)\simeq1.2/L+\varepsilon^{}_{\infty}$ with
$\varepsilon^{}_{\infty}\simeq0.005$~rad ($\approx0.002\pi$), i.e., it is
consistent with a vanishing---or at most very small---displacement in the
thermodynamic limit, within the uncertainties of the optimized parameters.
Equivalently, the variational optimization drives the amplitudes toward the
pinning manifold of Eq.~\eqref{eq:pinning_criterion}: on the largest cluster
($L=60$) one finds $\mu=0.126$ versus
$\chi^{}_{d}+\chi^{}_{d'}-2\chi^{}_{\square}=0.091$, with the mismatch still
decreasing with $L$. This behavior is natural on a finite torus, where the
closed-shell condition favors node positions that are incommensurate with the
discrete momentum mesh, whose spacing scales as $1/L$; it also rationalizes the
numerical difficulty, reported for the square lattice~\cite{Hu-2013}, of
stabilizing representatives with displaced nodes. At the sizes accessible to
our simulations\rev{,} the incommensurability of the gapless two-spinon momenta is
$\delta=2\varepsilon(L)$, e.g., $\delta\simeq0.06\pi$ on the $L=14$ cluster
used in Fig.~\ref{fig:z3000_dis_ssf_dsf}.

\section{PSGs for plaquette VBCs}
\label{app:nematic}

So far, we have focused exclusively on symmetric QSLs in our discussion. However, in order to describe the \rev{instability of} our fully symmetric QSLs \rev{toward} the neighboring plaquette\rev{-}VBC phase reported for $J_{s}/J_{d}\lesssim0.78$~\cite{Yang-2022,Wang-2022,Keles-2022,Luciano-2024,Corboz-2025,Chen-2025,Yuan-2026}, we now also perform a classification of QSLs with a lower symmetry, namely, that of the plaquette VBC. This enables us to construct fermionic mean-field states for plaquette VBCs that descend from their parent fully symmetric QSLs\rev{,} thereby allowing us to describe transitions between QSLs and plaquette VBCs. In particular, our goal is to identify the plaquette VBC \textit{Ansatz} that is continuously connected to the Z3000 state.

Realizing a plaquette VBC on the empty squares of the Shastry-Sutherland lattice (as observed in numerical studies) imprints a different symmetry group on the system: this order requires breaking both $G_x$ and $\sigma_{xy}$ but preserving the fourfold rotational, $C_4$, symmetry. Hence, the modified symmetry group is given by $\{T_x,T_y,C_4,\mathcal{T}\}$. The symmetry relations analogous to Eq.~\eqref{eq:id_relation} are
\begin{align}
T^{-1}_xT^{-1}_yT^{}_xT^{}_y&=\mathds{1},\\
T^{}_yC^{-1}_4T^{}_xC^{}_4&=\mathds{1},\\
T^{-1}_xC^{-1}_4T^{}_yC^{}_4&=\mathds{1},\\
C^4_4=\mathcal{T}^2&=\mathds{1},\\
\mathcal{T}\mathcal{O}\mathcal{T}^{-1}\mathcal{O}^{-1}&=\mathds{1},\quad \rev{\mathcal{O} \in \{T^{}_{x}, T^{}_{y}, C^{}_{4}\}}.
\end{align}

\begin{table*}
\caption{Mean-field parameters on the reference bonds, $u^{}_s$, $u^\prime_s$, $u^{}_d$, $u^{}_{d'}$, $u^{}_{\square}$\rev{,} and $u^{}_{\square'}$ for all possible \textit{Ans\"atze} taking only translations, $C_4$, and time-reversal symmetries into account. Each row stands for two \textit{Ans\"atze} labeled by classes A ($\eta_y=+1$) and B ($\eta_y=-1$). All the \textit{Ans\"atze} besides P22A, P22B, P23A, P23B, P26A, P26B, P27A, and P27B correspond to a $\mathbb{Z}_2$ IGG.}
\begin{ruledtabular}
\begin{tabular}{cccccccccc}
Label & $\{\eta^{}_y,\eta^{}_{C_4},\eta^{}_{C_4y},\eta^{}_{\mathcal{T}},\eta^{}_{\mathcal{T}C_4}\}$ & $\mathcal{W}^{}_\mathcal{T}$ & $u^{}_s$ & $u^\prime_s$ & $u^{}_d$& $u^{}_{d'}$ & $u^{}_{\square}$ & $u^{}_{\square'}$ & Onsite\\
			\hline
P1A/B & $\{\eta^{\pdagger}_y,+,+,+,+\}$& $\dot\iota\tau^y$ &$\tau^{z,x}$&$\tau^{z,x}$& $\tau^{z,x}$& $\tau^{z,x}$& $\tau^{z,x}$& $\tau^{z,x}$ & $\tau^{z,x}$\\
P2A/B & $\{\eta^{\pdagger}_y,+,-,+,+\}$& $\dot\iota\tau^y$ &$\tau^{z,x}$&$\tau^{z,x}$& $0$ & $0$& $0$& $0$ & $\tau^{z,x}$\\
P3A/B & $\{\eta^{\pdagger}_y,-,+,+,+\}$& $\dot\iota\tau^y$ &$\tau^{z,x}$&$\tau^{z,x}$& $0$ & $0$& $\tau^{z,x}$& $0$ & $\tau^{z,x}$\\
P4A/B & $\{\eta^{\pdagger}_y,-,-,+,+\}$& $\dot\iota\tau^y$ &$\tau^{z,x}$&$\tau^{z,x}$& $\tau^{z,x}$ & $\tau^{z,x}$& $0$& $\tau^{z,x}$ & $\tau^{z,x}$\\
\hline
P5A/B & $\{\eta^{\pdagger}_y,+,+,+,-\}$& $\dot\iota\tau^y$ &$\tau^{0,y}$&$\tau^{0,y}$& $\tau^{z,x}$ & $\tau^{z,x}$& $\tau^{z,x}$& $\tau^{z,x}$ & $\tau^{z,x}$\\
P6A/B & $\{\eta^{\pdagger}_y,+,-,+,-\}$& $\dot\iota\tau^y$ &$\tau^{0,y}$&$\tau^{0,y}$& $0$ & $0$& $0$& $0$ & $\tau^{z,x}$\\
P7A/B & $\{\eta^{\pdagger}_y,-,+,+,-\}$& $\dot\iota\tau^y$ &$\tau^{0,y}$&$\tau^{0,y}$& $0$ & $0$& $\tau^{z,x}$& $0$ & $\tau^{z,x}$\\
P8A/B & $\{\eta^{\pdagger}_y,-,-,+,-\}$& $\dot\iota\tau^y$ &$\tau^{0,y}$&$\tau^{0,y}$& $\tau^{z,x}$ & $\tau^{z,x}$& $0$& $\tau^{z,x}$ & $\tau^{z,x}$\\
\hline
P9A/B & $\{\eta^{\pdagger}_y,+,+,-,+\}$& $\dot\iota\tau^y$ &$\tau^{z,x}$&$\tau^{0,y}$& $\tau^{y}$ & $\tau^{y}$& $\tau^{z,x}$& $\tau^{z,x}$ & $\tau^{z,x}$\\
P10A/B & $\{\eta^{\pdagger}_y,+,-,-,+\}$& $\dot\iota\tau^y$ &$\tau^{z,x}$&$\tau^{0,y}$& $\tau^{0}$ & $\tau^{0}$& $0$& $0$ & $\tau^{z,x}$\\
P11A/B & $\{\eta^{\pdagger}_y,-,+,-,+\}$& $\dot\iota\tau^y$ &$\tau^{z,x}$&$\tau^{0,y}$& $\tau^{0}$ & $\tau^{0}$& $\tau^{z,x}$& $0$ & $\tau^{z,x}$\\
P12A/B & $\{\eta^{\pdagger}_y,-,-,-,+\}$& $\dot\iota\tau^y$ &$\tau^{z,x}$&$\tau^{0,y}$& $\tau^{y}$ & $\tau^{y}$& $0$& $\tau^{z,x}$ & $\tau^{z,x}$\\
\hline
P13A/B & $\{\eta^{\pdagger}_y,+,+,-,-\}$& $\dot\iota\tau^y$ &$\tau^{0,y}$&$\tau^{z,x}$& $\tau^{y}$ & $\tau^{y}$& $\tau^{z,x}$& $\tau^{z,x}$ & $\tau^{z,x}$\\
P14A/B & $\{\eta^{\pdagger}_y,+,-,-,-\}$& $\dot\iota\tau^y$ &$\tau^{0,y}$&$\tau^{z,x}$& $\tau^{0}$ & $\tau^{0}$& $0$& $0$ & $\tau^{z,x}$\\
P15A/B & $\{\eta^{\pdagger}_y,-,+,-,-\}$& $\dot\iota\tau^y$ &$\tau^{0,y}$&$\tau^{z,x}$& $\tau^{0}$ & $\tau^{0}$& $\tau^{z,x}$& $0$ & $\tau^{z,x}$\\
P16A/B & $\{\eta^{\pdagger}_y,-,-,-,-\}$& $\dot\iota\tau^y$ &$\tau^{0,y}$&$\tau^{z,x}$& $\tau^{y}$ & $\tau^{y}$& $0$& $\tau^{z,x}$ & $\tau^{z,x}$\\
\hline
\hline
P17A/B & $\{\eta^{\pdagger}_y,+,+,+,-\}$& $\tau^0$ &$\tau^{0,z}$&$\tau^{0,x,z}$& $0$ & $0$& $0$& $0$ & $0$\\
P18A/B & $\{\eta^{\pdagger}_y,+,-,+,-\}$& $\tau^0$ &$\tau^{0,z}$&$\tau^{0,x,z}$& $0$ & $0$& $0$& $0$ & $0$\\
P19A/B & $\{\eta^{\pdagger}_y,-,+,+,-\}$& $\tau^0$ &$\tau^{0,z}$&$\tau^{0,x,z}$& $0$ & $0$& $0$& $0$ & $0$\\
P20A/B & $\{\eta^{\pdagger}_y,-,-,+,-\}$& $\tau^0$ &$\tau^{0,z}$&$\tau^{0,x,z}$& $0$ & $0$& $0$& $0$ & $0$\\
\hline
P21A/B & $\{\eta^{\pdagger}_y,+,+,-,+\}$& $\tau^0$ &$0$&$\tau^{0,z}$& $\tau^{\rev{z,x}}$ & $\tau^{x,y,z}$ & $0$& $0$ & $0$\\
P22A/B & $\{\eta^{\pdagger}_y,+,-,-,+\}$& $\tau^0$ &$0$&$\tau^{0,z}$& $\tau^{0}$ & $\tau^{0}$ & $0$& $0$ & $0$\\
P23A/B & $\{\eta^{\pdagger}_y,-,+,-,+\}$& $\tau^0$ &$0$&$\tau^{0,z}$& $\tau^{0}$ & $\tau^{0}$ & $0$& $0$ & $0$\\
P24A/B & $\{\eta^{\pdagger}_y,-,-,-,+\}$& $\tau^0$ &$0$&$\tau^{0,z}$& $\tau^{\rev{z,x}}$ & $\tau^{x,y,z}$ & $0$& $0$ & $0$\\
\hline
P25A/B & $\{\eta^{\pdagger}_y,+,+,-,-\}$& $\tau^0$ &$\tau^{0,z}$&$0$& $\tau^{\rev{z,x}}$ & $\tau^{x,y,z}$ & $0$& $0$ & $0$\\
P26A/B & $\{\eta^{\pdagger}_y,+,-,-,-\}$& $\tau^0$ &$\tau^{0,z}$&$0$& $\tau^{0}$ & $\tau^{0}$ & $0$& $0$ & $0$\\
P27A/B & $\{\eta^{\pdagger}_y,-,+,-,-\}$& $\tau^0$ &$\tau^{0,z}$&$0$& $\tau^{0}$ & $\tau^{0}$ & $0$& $0$ & $0$\\
P28A/B & $\{\eta^{\pdagger}_y,-,-,-,-\}$& $\tau^0$ &$\tau^{0,z}$&$0$& $\tau^{\rev{z,x}}$ & $\tau^{x,y,z}$ & $0$& $0$ & $0$\\
		\end{tabular}
	\end{ruledtabular}
	\label{table:z2_ansatze_plaquette}
\end{table*}

 \begin{figure}[!t]
 \includegraphics[width=1.0\linewidth]{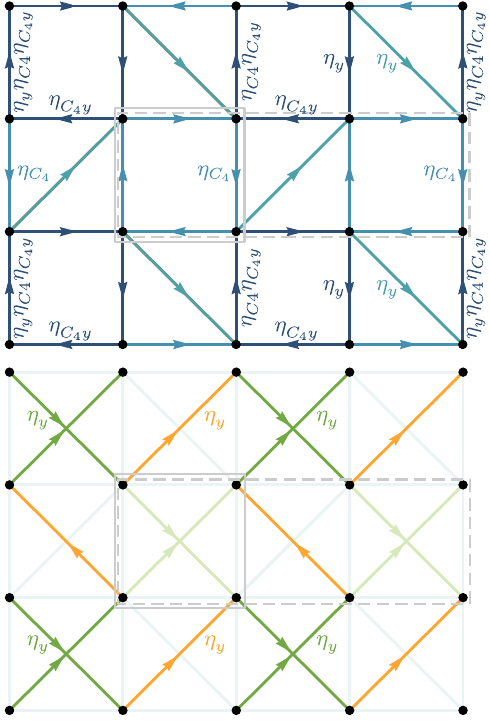}
	\caption{Representation of $\mathbb{Z}_2$ \textit{Ans\"atze} compatible with the symmetry group of the plaquette VBC. The link fields on the light blue (square), dark blue (square), teal (diagonal), orange (diagonal), light green (diagonal), and dark green (diagonal) bonds are given by $u^{}_s=\dot\iota\chi^0_{s}\tau^0+\Delta^1_{s}\tau^x+\Delta^2_{s}\tau^y+\chi^3_{s}\tau^z$, $u^\prime_{s}=\dot\iota\chi^0_{s'}\tau^0+\Delta^1_{s'}\tau^x+\Delta^2_{s'}\tau^y+\chi^3_{s'}\tau^z$, $u^{}_d=\dot\iota\chi^0_{d}\tau^0+\Delta^1_{d}\tau^x+\Delta^2_{d}\tau^y+\chi^3_{d}\tau^z$, $u^{}_{d'}=\dot\iota\chi^0_{d'}\tau^0+\Delta^1_{d'}\tau^x+\Delta^2_{d'}\tau^y+\chi^3_{d'}\tau^z$, $u^{}_{\square}=\dot\iota\chi^0_{\square}\tau^0+\Delta^1_{\square}\tau^x+\Delta^2_{\square}\tau^y+\chi^3_{\square}\tau^z$, and $u^{}_{\square'}=\dot\iota\chi^0_{\square'}\tau^0+\Delta^1_{\square'}\tau^x+\Delta^2_{\square'}\tau^y+\chi^3_{\square'}\tau^z$, respectively. The allowed mean fields for the 48 different $\mathbb{Z}_2$ \textit{Ans\"atze} can be read off from Table~\ref{table:z2_ansatze_plaquette}. The solid (dashed) gray box marks the unit cell required to realize \textit{Ans\"atze} in class A, with $\eta_y=+1$ (class B, with $\eta_y=-1$). The onsite terms are uniform. }
	\label{fig:plaquette_ansatz}
\end{figure}

\begin{figure}[!tb]
 \includegraphics[width=1.0\linewidth]{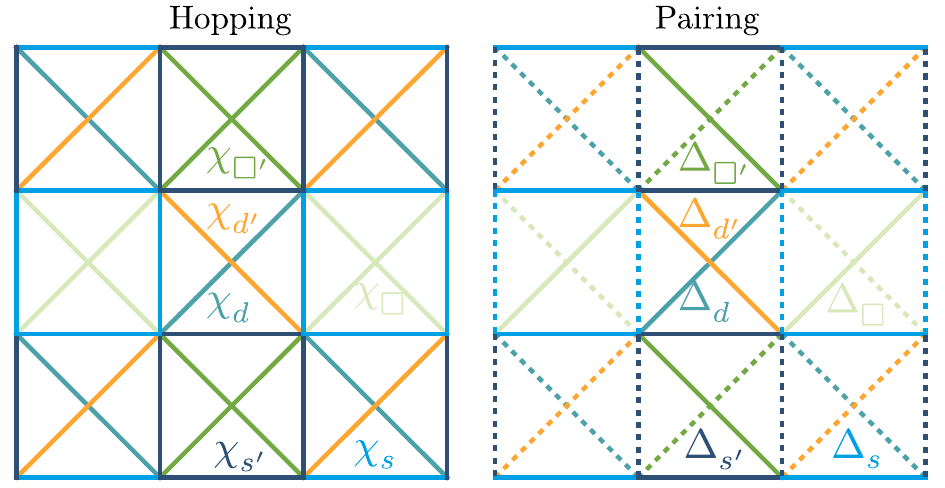}
	\caption{Graphical illustration of the sign structure of hopping and pairing fields in the \textit{Ansatz} P5A. In contrast to the mean-field parameters listed in Table~\ref{table:z2_ansatze_plaquette}, here, we choose a real gauge in order to write the \textit{Ansatz} purely in terms of real hopping and pairing parameters, thereby making the connection to Z3000 explicit. The dashed lines denote a negative sign. The onsite hopping is uniform while the onsite pairing is negative on even\rev{-}numbered sites.}
	\label{fig:plaquette_ansatz_p5a}
\end{figure}

For concreteness, we consider the center of an empty square (which the $C_4$ axis passes through) as the origin of the coordinate system, and the square itself is taken to be the new reference unit cell; see Fig.~\ref{fig:plaquette_ansatz}. The coordinates of the four sites (indexed by $u$) within the unit cell  are given by $1\rev{\,}(-a/2,a/2)$, $2\rev{\,}(-a/2,-a/2)$, $3\rev{\,}(a/2,-a/2)$, and $4\rev{\,}(a/2,a/2)$. Now, we introduce the sublattice-dependent coordinate system $(X,Y,u)$, akin to Eq.~\eqref{eq:coordinate} earlier. Under the present symmetry group elements, $(X,Y,u)$ transforms as
\begin{align}
T^{}_x:\, &(X,Y,u)\rightarrow(X+1,Y,u),\\
T^{}_y:\, &(X,Y,u)\rightarrow(X,Y+1,u),\\
C^{}_4:\, &(X,Y,u)\rightarrow(-Y,X,C^{}_4(u)),
\end{align}
where $C^{}_4(\{1, 2, 3, 4\}) = \{2,3,4,1\}$.

Following the same prescription as in App.~\ref{app:z2_psg_derivation}, we can obtain the solutions of the algebraic PSGs:
\begin{align}
 & W^{\pdagger}_{T^{}_x}(X,Y,u)=\eta^Y_y\tau^0,\;W^{\pdagger}_{T_y}(X,Y,u)=\tau^0\ , \; \\
 & W^{\pdagger}_{C_4}(X,Y,u)=\eta^Y_{C_{4y}}\eta^{XY}_y\mathcal{W}^{\pdagger}_{C_4,u}\ , \; \label{eq:solution_c4_z2}\\
& W^{\pdagger}_{\mathcal{T}}(X,Y,u)=\eta^{X+Y}_\mathcal{T}\eta^u_{\mathcal{T}C_4}\mathcal{W}^{\pdagger}_{\mathcal{T}}\
.
\end{align}
Here, $\mathcal{W}_{C_4,u}=\{\tau^0,\tau^0,\tau^0,\eta_{C_4}\tau^0\}$, and the matrix $\mathcal{W}_{\mathcal{T}}$ can be either $\tau^0$ or $\dot\iota\tau^y$. Each of the five parameters $\{\eta_y,\eta_{C_4},\eta_{C_4y},\eta_{\mathcal{T}},\eta_{\mathcal{T}C_4}\}$ \rev{takes} values $\pm1$, so together with the two options for $\mathcal{W}_{\mathcal{T}}$,  we find a total of 64 algebraic PSGs. However, for $\mathcal{W}_{\mathcal{T}}=\tau^0$, the solution with $\eta_\mathcal{T}=\eta_{\mathcal{T}C^{}_4}=+1$ has to be excluded as the mean-field parameters trivially vanish on all the bonds. Therefore, we are left with 56 PSGs. The corresponding symmetry-allowed mean-field parameters are documented in Table~\ref{table:z2_ansatze_plaquette}. Note that since we restrict ourselves to short-range \textit{Ans\"atze} with only NN and NNN bonds, only 48 of the 56 PSGs lead to QSL states with IGG $\mathbb{Z}_2$. The remaining eight (P22A, P22B, P23A, P23B, P26A, P26B, P27A, and P27B) carry a U(1) IGG since the pairing terms \rev{vanish identically} up to the neighbors we \rev{consider}. In Table~\ref{table:z2_ansatze_plaquette}, each row corresponds to two \textit{Ans\"atze}, one with $\eta_y=+1$ (class A) and the other with $\eta_y=-1$ (class B). Realization of the class-B \textit{Ans\"atze} requires doubling of the unit cell along the $x$-axis. The detailed sign structures of these \textit{Ans\"atze} are illustrated in Fig.~\ref{fig:plaquette_ansatz}.

Among the \textit{Ans\"atze} compiled  in Table~\ref{table:z2_ansatze_plaquette}, the \textit{Ansatz} \rev{labeled} P5A is a descendant of Z3000 when breaking $\sigma^{}_{xy}$ and $G^{}_{x}$ while preserving $C^{}_4$. Although this connection is not immediately apparent in the listed form, one can recast this \textit{Ansatz} in terms of real hopping and pairing amplitudes using a gauge transformation $W(X,Y,u)=(-1)^{(\delta_{u,1}+\delta_{u,4})}(\dot\iota\tau^z)^{(\delta_{u,2}+\delta_{u,4})}$ as follows:
\begin{alignat}{1}
&{\cal \hat{H}}^{\text{P5A}}_{{\rm MF}} =\sum_{i}\left[ \mu\sum^{}_{\alpha} \hat{f}_{i,\alpha}^{\dagger}\hat{f}^{\pdagger}_{i,\alpha}+\Delta\rev{\,\nu^{}_{i,1}} \big(\hat{f}_{i,\uparrow}^{\dagger}\hat{f}^{\dagger}_{i,\downarrow}+{\rm h.c.}\big)\right]\notag\\
&+\sum_{\langle ij\rangle_\text{light blue}}\left[ \chi^{}_{s}\sum^{}_{\alpha}{\rm s}^{s}_{ij} \hat{f}_{i,\alpha}^{\dagger}\hat{f}^{\pdagger}_{j,\alpha}+\Delta^{}_{s}{\nu}^{s}_{ij} \big(\hat{f}_{i,\uparrow}^{\dagger}\hat{f}^{\dagger}_{j,\downarrow}+{\rm h.c.}\big)\right]\notag\\&+\sum_{\langle ij\rangle_\text{dark blue}}\left[ \chi^{}_{s'}\sum^{}_{\alpha}{\rm s}^{s'}_{ij} \hat{f}_{i,\alpha}^{\dagger}\hat{f}^{\pdagger}_{j,\alpha}+\Delta^{}_{s'}{\nu}^{s'}_{ij} \big(\hat{f}_{i,\uparrow}^{\dagger}\hat{f}^{\dagger}_{j,\downarrow}+{\rm h.c.}\big)\right]\notag\\
&+
\sum_{\langle ij\rangle_\text{teal}}\left[ \chi^{}_{d}\sum^{}_{\alpha}{\rm s}^{d}_{ij} \hat{f}_{i,\alpha}^{\dagger}\hat{f}^{\pdagger}_{j,\alpha}+\Delta^{}_{d}{\nu}^{d}_{ij} \big(\hat{f}_{i,\uparrow}^{\dagger}\hat{f}^{\dagger}_{j,\downarrow}+{\rm h.c.}\big)\right]\notag\\
&+\sum_{\langle ij\rangle_\text{orange}}\left[ \chi^{}_{d'}\sum^{}_{\alpha}{\rm s}^{d'}_{ij} \hat{f}_{i,\alpha}^{\dagger}\hat{f}^{\pdagger}_{j,\alpha}+\Delta^{}_{d'}{\nu}^{d'}_{ij} \big(\hat{f}_{i,\uparrow}^{\dagger}\hat{f}^{\dagger}_{j,\downarrow}+{\rm h.c.}\big)\right]\notag\\
&+\sum_{\langle ij\rangle_\text{light green}}\left[ \chi^{}_{\square}\sum^{}_{\alpha}{\rm s}^{\square}_{ij} \hat{f}_{i,\alpha}^{\dagger}\hat{f}^{\pdagger}_{j,\alpha}+\Delta^{}_{\square}{\nu}^{\square}_{ij} \big(\hat{f}_{i,\uparrow}^{\dagger}\hat{f}^{\dagger}_{j,\downarrow}+{\rm h.c.}\big)\right]\notag\\
&+\sum_{\langle ij\rangle_\text{dark green}}\left[ \chi^{}_{\square'}\sum^{}_{\alpha}{\rm s}^{\square'}_{ij} \hat{f}_{i,\alpha}^{\dagger}\hat{f}^{\pdagger}_{j,\alpha}+\Delta^{}_{\square'}{\nu}^{\square'}_{ij} \big(\hat{f}_{i,\uparrow}^{\dagger}\hat{f}^{\dagger}_{j,\downarrow}+{\rm h.c.}\big)\right]\,.
\label{eq:z3000_plaquette}
\end{alignat}
The associated sign parameters can be read off \rev{from} Fig.~\ref{fig:plaquette_ansatz_p5a}\rev{, in which $\nu^{}_{i,1}=\pm1$ denotes the sign of the onsite pairing}. The \textit{Ansatz} Z3000 can be obtained from P5A by setting $\chi^{}_{s}=\chi^{}_{s'}$, $\Delta^{}_{s}=\Delta^{}_{s'}$, $\chi^{}_{\square}=\chi^{}_{\square'}$, $\Delta^{}_{\square}=\Delta^{}_{\square'}$, \rev{$\Delta=0$,} and $\Delta^{}_{d}=\Delta^{}_{d'}=0$\rev{---the last two conditions removing the onsite and diagonal pairings, respectively, which are absent in Z3000}.

\section{Parent U(1) states for Z3000}
\label{app:parent}

As mentioned in Sec.~\ref{sec:mfn}, the Z3000 \textit{Ansatz} can be obtained by starting from three \textit{distinct} U(1) states, namely, U100, U300, and U800, and breaking down the U(1) IGG to $\mathbb{Z}_2$ via singlet pairing terms. While we earlier focused on the viewpoint that the Z3000 state is a descendant of U100, here, for completeness, we illustrate how both U300 and U800 can equivalently be regarded as \rev{parent states}.

 This is easier to see first for U800. Concretely, the connection of U800 to Z3000 can be found by setting the mean-field parameters on all the diagonal bonds in Z3000 to zero, followed by an appropriate gauge transform. In the hopping-only gauge, the U800 state exhibits a Dirac cone anchored at the $\Gamma$ point (see Fig.~\rev{\ref{fig:dispersions}}).

\begin{figure}[t]
 \includegraphics[width=\linewidth]{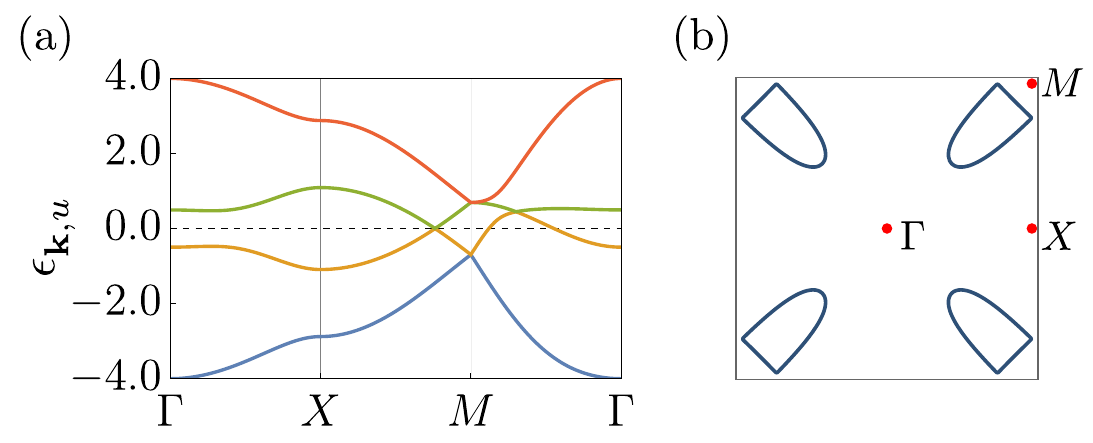}
	\caption{(a) Dispersion of the U300 state, and (b) illustration of its Fermi surfaces. The mean-field parameters here are set to $\chi^{}_s=1.0$, $\chi^{}_d=0.5$, $\chi^{}_{d'}=0.4$, and $\chi^{}_{\square}=0.3$. The dashed line indicates the Fermi level.}
	\label{fig:u300_dis_fermi}
\end{figure}

 The U300 \textit{Ansatz} has a nontrivial sign structure. \rev{Taking all the diagonals into account, it} is given by
\begin{align}
&u^{\pdagger}_{12}=u^{\pdagger}_{43}=u^{\pdagger}_{21}=u^{\pdagger}_{34}=u^{\pdagger}_{14}=u^{\pdagger}_{23}=u^{\pdagger}_{41}=u^{\pdagger}_{32}=\chi^{}_s\tau^z,\notag\\
&u^{\pdagger}_{13}=-u^{\pdagger}_{42}=\chi^{}_{d}\tau^z,\;u^{\pdagger}_{d_1}=-u^{\pdagger}_{d_2}=\chi^{}_{d'}\tau^z,\notag\\
&u^{\pdagger}_{d_3}=-u^{\pdagger}_{d_4}=u^{\pdagger}_{d_5}=-u^{\pdagger}_{d_6}=\chi^{}_{\square}\tau^z,\notag\\
&\sum_\gamma a^{}_\gamma(m,n,u)\tau^\gamma=(\rev{-1})^u\mu\tau^z.
\end{align}
At first glance, the connection to Z3000 is far from apparent. However, this can be established by setting $\chi_{s}=0$ and $\Delta_{\square}=0$ in the Z3000 \textit{Ansatz}, followed by a gauge transformation of the form
$$W(m,n,u)=(-1)^{(\delta_{u,1}+\delta_{u,2})}(\dot\iota\tau^y)^{(\delta_{u,2}+\delta_{u,4})}.$$
The dispersion of the U\rev{300} state is sketched in Fig.~\ref{fig:u300_dis_fermi}(a), and features clear Fermi surfaces as evidenced by Fig.~\ref{fig:u300_dis_fermi}(b).

\section{Chiral spin\rev{-}liquid descendants of Z3000}
\label{app:chiral}

An \textit{Ansatz} belongs to the class of chiral spin liquids (CSLs) when time-reversal symmetry $\mathcal{T}$ is broken while spin-rotation symmetry remains intact. In contrast to the classification discussed previously---which assumes full symmetry of the lattice space group together with time reversal---a CSL respects lattice symmetries only up to a possible combination with $\mathcal{T}$. Consequently, the relevant symmetry group in the present case is
\begin{equation}
\mathcal{O}' \in \{T_x, T_y, G_x \mathcal{T}^{\upsilon_{G_x}}, \sigma_{xy}\mathcal{T}^{\upsilon_{\sigma_{xy}}}\},
\end{equation}
where $\upsilon_{\mathcal{O}} \in \{0,1\}$. When $\upsilon_{\mathcal{O}}=0$, the lattice symmetry $\mathcal{O}$ is preserved while $\mathcal{T}$ is broken. In contrast, when $\upsilon_{\mathcal{O}}=1$, both $\mathcal{O}$ and $\mathcal{T}$ are individually broken, but their product $\mathcal{O}\mathcal{T}$ remains a symmetry of the system.

The next step is to determine the projective realization of this symmetry group. Since time reversal does not change the position of lattice sites, the projective symmetry group (PSG) of $\mathcal{O}\mathcal{T}$ coincides with that of $\mathcal{O}$. Therefore, the PSG elements associated with $\mathcal{O}'$ are identical to those obtained previously for the symmetric case and are given by Eqs.~\eqref{eq:solution_translationj_z2}, \eqref{eq:solution_glide_z2}, and \eqref{eq:solution_sigma_z2}.

The corresponding \textit{Ans\"atze} must then satisfy the symmetry condition
\begin{align}
\label{eq:sym_con_gauge_csl}
W_{\mathcal{O}}^{\dagger}(\mathcal{O}(i))\,
u^{\pdagger}_{\mathcal{O}(i)\mathcal{O}(j)}\,
\rev{W_{\mathcal{O}}^{\pdagger}(\mathcal{O}(j))}
=
(-1)^{\upsilon_{\mathcal{O}}}\, u^{\pdagger}_{ij},
\end{align}
which ensures invariance under the combined operation $\mathcal{O}\mathcal{T}^{\upsilon_{\mathcal{O}}}$.

In the following, we focus on CSL states obtained by breaking time-reversal symmetry in the $\mathbb{Z}_2$ spin-liquid \textit{Ansatz} labeled Z3000. The corresponding PSG is given by
\begin{align}
W^{\pdagger}_{T_x}(m,n,u) &= W^{\pdagger}_{T_y}(m,n,u) = \tau^0, \\
W^{\pdagger}_{G_x}(m,n,u) &= \tau^0, \qquad
W^{\pdagger}_{\sigma_{xy}}(m,n,u) = \rev{\dot\iota}\tau^z.
\end{align}

With this PSG, four distinct CSL realizations are possible, corresponding to the four choices of $(\upsilon_{G_x}, \upsilon_{\sigma_{xy}})$: $(0,0)$, $(0,1)$, $(1,0)$, and $(1,1)$. We label these states using the notation Z3000$_{\upsilon_{G_x},\upsilon_{\sigma_{xy}}}$. The explicit forms of the four CSL \textit{Ans\"atze} are presented below\rev{, where we have abbreviated $\eta^{}_{\upsilon^{}_{\mathcal{O}}} \equiv (-1)^{\upsilon^{}_{\mathcal{O}}}$ and, as before, $p$\,$=$\,$0(1)$ for $\mathcal{W}^{}_{\sigma_{xy},u}\propto\tau^{0(z)}$ (so $p$\,$=$\,$1$ for Z3000)}.

\begin{figure}[tb]
\includegraphics[width=1.0\linewidth]{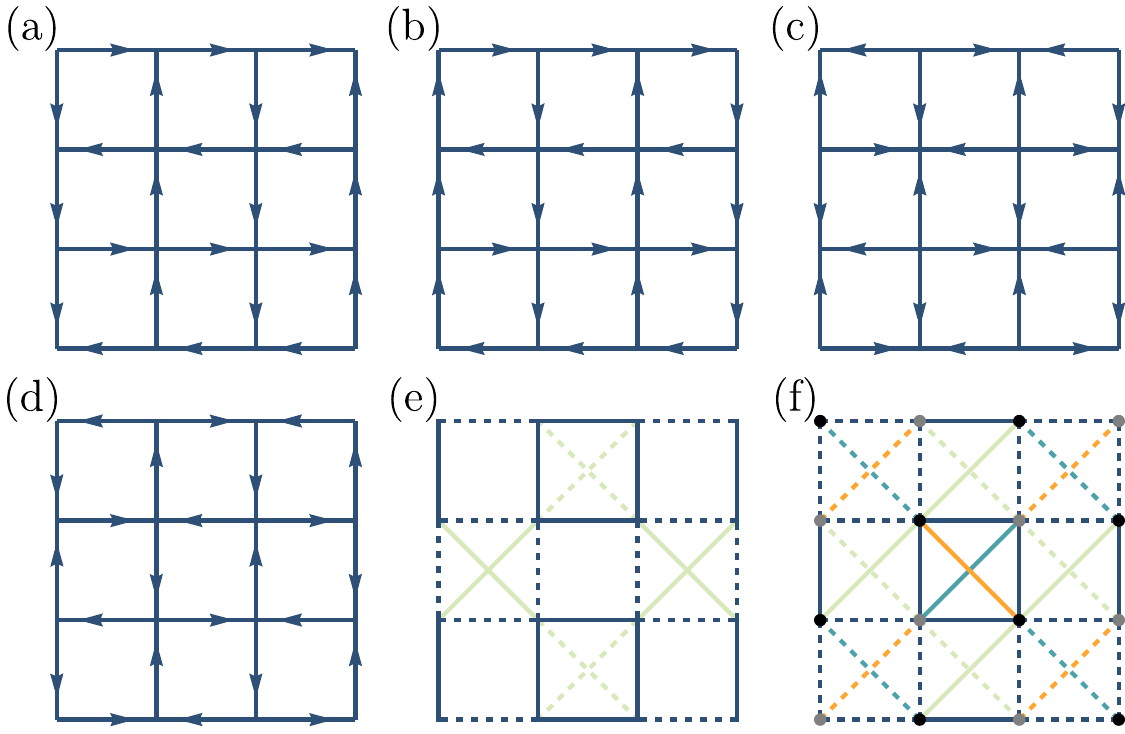}
	\caption{Graphical illustration of the sign structures ($s_{ij}$ and $\nu_{ij}$) for the CSL states obtained by lowering the symmetry of the Z3000 \textit{Ansatz}. Panels (a)--(d) show the configurations of $s^{s}_{ij,0}$ for the \textit{Ans\"atze} labeled Z3000$_{00}$, Z3000$_{01}$, Z3000$_{10}$, and Z3000$_{11}$, respectively. Along the \rev{arrow} direction\rev{,} $s^{s}_{ij,0}=+1$, while $s^{s}_{ij,0}=-1$ elsewhere. Panel (e) illustrates the configurations of $\nu^{s}_{ij,2}$ and $\nu^{\square}_{ij,2}$ for Z3000$_{10}$, shown in blue and green, respectively; solid (dashed) bonds denote positive (negative) signs. Panel (f) shows the configurations of $\nu^{s}_{ij,2}$, $\nu^{d}_{ij,1}$, $\nu^{d'}_{ij,1}$, and $\nu^{\square}_{ij,2}$ for Z3000$_{11}$, represented in blue, teal, orange, and green, respectively. In addition, this state exhibits a staggered onsite pairing, where the positive ($\nu_{i,1}=+1$) and negative ($\nu_{i,1}=-1$) signs are indicated by gray and black sites.}
	\label{fig:chiral_z2_sign}
\end{figure}

\begin{widetext}
\begin{align}
&u^{\pdagger}_{12} = u^\dagger_{43} = \eta^{\pdagger}_{\upsilon^{}_{G^{}_{x}}}u^{\pdagger}_{21} = \eta^{\pdagger}_{\upsilon^{}_{G^{}_{x}}}u^\dagger_{34} = u^{\pdagger}_s= \dot\iota\chi^0_{s}\tau^0 + \Delta^1_{s}\tau^x +\Delta^2_{s}\tau^y + \chi^3_{s}\tau^z, \\
&\eta^{\pdagger}_{\upsilon^{}_{\sigma^{}_{xy}}}u^{\pdagger}_{14} = \eta^{\pdagger}_{\upsilon^{}_{\sigma^{}_{xy}}} u^\dagger_{23}= \eta^{\pdagger}_{\upsilon^{}_{G^{}_{x}}}\eta^{\pdagger}_{\upsilon^{}_{\sigma^{}_{xy}}}u^{\pdagger}_{41} = \eta^{\pdagger}_{\upsilon^{}_{G^{}_{x}}}\eta^{\pdagger}_{\upsilon^{}_{\sigma^{}_{xy}}}u^\dagger_{32} = \tau^z u^{\pdagger}_s \tau^z, \\
&u^{\pdagger}_{13} =\eta^{}_{\upsilon^{}_{G^{}_{x}}} u^{\pdagger}_{42} = u^{\pdagger}_{d}=( \Delta^1_{d}\tau^x + \Delta^2_{d}\tau^y)\delta^{}_{\upsilon^{}_{\sigma^{}_{xy}},1} + \chi^3_{d}\tau^z\delta^{}_{\upsilon^{}_{\sigma^{}_{xy}},0},\\
&u^{\pdagger}_{d^{}_1} =\eta^{}_{\upsilon^{}_{G^{}_{x}}} u^{\pdagger}_{d^{}_2} =u^{\pdagger}_{d'}= ( \Delta^1_{d'}\tau^x + \Delta^2_{d'}\tau^y)\delta^{}_{\upsilon^{}_{\sigma^{}_{xy}},1} + \chi^3_{d'}\tau^z\delta^{}_{\upsilon^{}_{\sigma^{}_{xy}},0},\\
& u^{\pdagger}_{d_3} =\eta^{}_{\upsilon^{}_{G^{}_{x}}}\eta^{\pdagger}_{\upsilon^{}_{\sigma^{}_{xy}}} (\tau^z)^p u^{\pdagger}_{d_4} (\tau^z)^p =\eta^{\pdagger}_{\upsilon^{}_{\sigma^{}_{xy}}} (\tau^z)^p u^{\pdagger}_{d_5} (\tau^z)^p =\eta^{}_{\upsilon^{}_{G^{}_{x}}} u^{\pdagger}_{d_6}=u^{\pdagger}_{\square} \equiv \Delta^1_{\square} \tau^x + \Delta^2_{\square} \tau^y + \chi^3_{\square} \tau^z,\\
&\rev{\sum_\gamma} a^{\pdagger}_\gamma(m,n,u+1)\tau^\gamma=\eta^{u+1}_{\upsilon^{}_{G^{}_{x}}}\left(( \Delta^{1}\tau^x + \Delta^2\tau^y)\delta^{}_{\upsilon^{}_{\sigma^{}_{xy}},1} + \mu\tau^z\delta^{}_{\upsilon^{}_{\sigma^{}_{xy}},0}\right)\;.
\end{align}
\end{widetext}
Among these four possibilities, the connection between the Z3000 state and the \textit{Ansatz} corresponding to $(\upsilon_{G_x},\upsilon_{\sigma_{xy}})=(0,0)$, denoted Z3000$_{00}$, is explicit. The correspondence for the remaining three cases can be revealed through appropriate gauge transformations. The resulting gauge representations are given as follows:
\begin{widetext}
\begin{alignat}{2}
\label{eq:MF-Z2_chiral_00}
{\cal \hat{H}}^{\text{Z3000}_{\text{00}}}_{{\rm MF}} &={\cal \hat{H}}^{\text{Z3000}}_{{\rm MF}}&&+
\dot\iota\chi^{0}_{s}\sum_{\langle ij\rangle_\text{blue},\alpha}{\rm s}^{s}_{ij,0} \hat{f}_{i,\alpha}^{\dagger}\hat{f}^{\pdagger}_{j,\alpha}+
\sum_{\langle ij\rangle_\text{green}} \dot\iota\rev{\Delta^{2}_{\square}}{\nu}^{\square}_{ij,2} \big(\hat{f}_{i,\uparrow}^{\dagger}\hat{f}^{\dagger}_{j,\downarrow}+{\rm h.c.}\big)\,,\\
{\cal \hat{H}}^{\text{Z3000}_{\text{01}}}_{{\rm MF}} &={\cal \hat{H}}^{\text{Z3000}}_{{\rm MF}}&&+
\sum_{\langle ij\rangle_\text{blue}}\left[ \dot\iota\chi^{0}_{s}\sum^{}_{\alpha}{\rm s}^{s}_{ij,0} \hat{f}_{i,\alpha}^{\dagger}\hat{f}^{\pdagger}_{j,\alpha}+\dot\iota\Delta^{2}_{s}{\nu}^{s}_{ij,2} \big(\hat{f}_{i,\uparrow}^{\dagger}\hat{f}^{\dagger}_{j,\downarrow}+{\rm h.c.}\big)\right]+\sum_{\langle ij\rangle_\text{teal}} \dot\iota\Delta^{2}_{d}{\nu}^{d}_{ij,2} \big(\hat{f}_{i,\uparrow}^{\dagger}\hat{f}^{\dagger}_{j,\downarrow}+{\rm h.c.}\big)\notag\\
& &&+\sum_{\langle ij\rangle_\text{orange}} \dot\iota\Delta^{2}_{d'}{\nu}^{d'}_{ij,2} \big(\hat{f}_{i,\uparrow}^{\dagger}\hat{f}^{\dagger}_{j,\downarrow}+{\rm h.c.}\big)+
\sum_{\langle ij\rangle_\text{green}} \dot\iota\Delta^{2}_{\square}{\nu}^{\square}_{ij,2} \big(\hat{f}_{i,\uparrow}^{\dagger}\hat{f}^{\dagger}_{j,\downarrow}+{\rm h.c.}\big)\,,
\label{eq:MF-Z2_chiral_01}
\end{alignat}
\begin{alignat}{2}
\label{eq:MF-Z2_chiral_10}
{\cal \hat{H}}^{\text{Z3000}_{\text{10}}}_{{\rm MF}} =&{\cal \hat{H}}^{\text{Z3000}}_{{\rm MF}}&&+
\sum_{\langle ij\rangle_\text{blue}}\left[ \dot\iota\chi^{0}_{s}\sum^{}_{\alpha}{\rm s}^{s}_{ij,0} \hat{f}_{i,\alpha}^{\dagger}\hat{f}^{\pdagger}_{j,\alpha}+\dot\iota\Delta^{2}_{s}{\nu}^{s}_{ij,2} \big(\hat{f}_{i,\uparrow}^{\dagger}\hat{f}^{\dagger}_{j,\downarrow}+{\rm h.c.}\big)\right]+
\sum_{\langle ij\rangle_\text{green}} \dot\iota\Delta^{2}_{\square}{\nu}^{\square}_{ij,2} \big(\hat{f}_{i,\uparrow}^{\dagger}\hat{f}^{\dagger}_{j,\downarrow}+{\rm h.c.}\big)\,,
\\
{\cal \hat{H}}^{\text{Z3000}_{\text{11}}}_{{\rm MF}} =&{\cal \hat{H}}^{\text{Z3000}}_{{\rm MF}}&&+
\sum_{\langle ij\rangle_\text{blue}}\left[ \dot\iota\chi^{0}_{s}\sum^{}_{\alpha}{\rm s}^{s}_{ij,0} \hat{f}_{i,\alpha}^{\dagger}\hat{f}^{\pdagger}_{j,\alpha}+\dot\iota\Delta^{2}_{s}{\nu}^{s}_{ij,2} \big(\hat{f}_{i,\uparrow}^{\dagger}\hat{f}^{\dagger}_{j,\downarrow}+{\rm h.c.}\big)\right]+\sum_{\langle ij\rangle_\text{teal}} \Delta^{1}_{d}{\nu}^{d}_{ij,1} \big(\hat{f}_{i,\uparrow}^{\dagger}\hat{f}^{\dagger}_{j,\downarrow}+{\rm h.c.}\big)\notag\\
& &&+\sum_{\langle ij\rangle_\text{orange}} \Delta^{1}_{d'}{\nu}^{d'}_{ij,1} \big(\hat{f}_{i,\uparrow}^{\dagger}\hat{f}^{\dagger}_{j,\downarrow}+{\rm h.c.}\big)+
\sum_{\langle ij\rangle_\text{green}} \Delta^{2}_{\square}{\nu}^{\square}_{ij,2} \big(\hat{f}_{i,\uparrow}^{\dagger}\hat{f}^{\dagger}_{j,\downarrow}+{\rm h.c.}\big)+\Delta^{1}\sum_{i}\nu^{}_{i,1}\big(\hat{f}_{i,\uparrow}^{\dagger}\hat{f}^{\dagger}_{i,\downarrow}+{\rm h.c.}\big)\,.
\label{eq:MF-Z2_chiral_11}
\end{alignat}
\end{widetext}
Here, the parameters $s$ and $\nu$ specify the sign configurations of the corresponding amplitudes\rev{, and their second subscript labels the Pauli component to which the amplitude belongs ($0$ for $\tau^0$, $1$ for $\tau^x$, and $2$ for $\tau^y$)}. The symmetry properties of the resulting four states are summarized in Table~\ref{table:z2_ansatze_chiral}. For Z3000$_{00}$, the pairing term $\nu^{\square}_{ij,2}$ follows the same sign pattern as $\nu^{\square}_{ij}$ in Eq.~\eqref{eq:MF-Z2} (see also Fig.~\ref{fig:fig4}(\rev{b})). In the case of Z3000$_{01}$, all additional pairing terms are uniform, namely $\nu^{s}_{ij,2}=\nu^{d}_{ij,2}=\nu^{d'}_{ij,2}=\nu^{\square}_{ij,2}=+1$. The remaining sign configurations for the other chiral \textit{Ans\"atze} are illustrated in Fig.~\ref{fig:chiral_z2_sign}.\\

 \begin{figure}[tb]
 \includegraphics[width=1.0\linewidth]{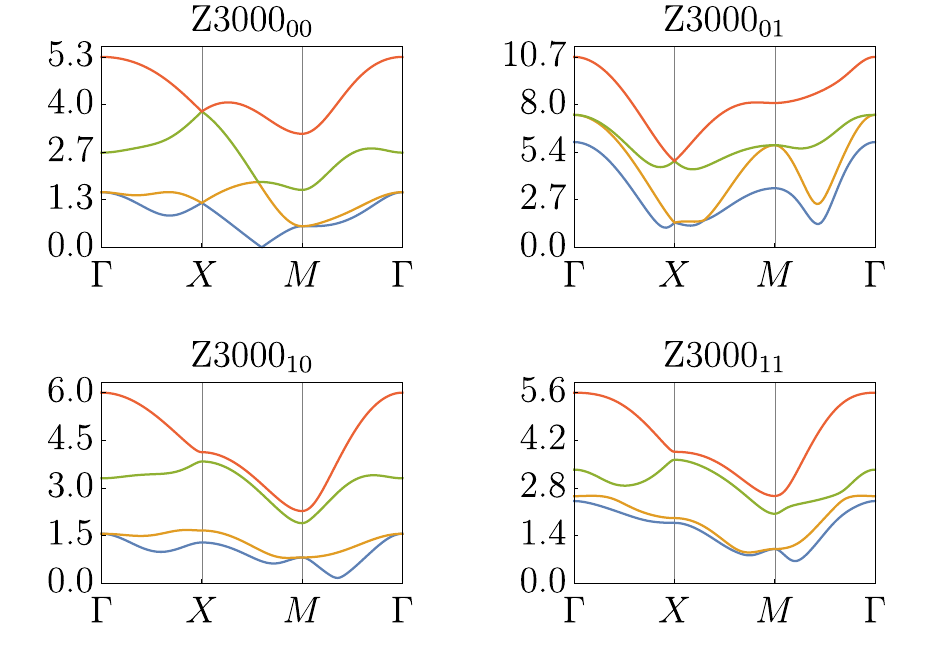}
	\caption{Mean-field dispersion of the four chiral spin-liquid states derived from the Z3000 \textit{Ansatz}.}
	\label{fig:chiral_z2_dis}
\end{figure}

\begin{table}[H]
\caption{Symmetry properties of the four CSLs \rev{that} are symmetry-lowered descendants of Z3000.}
\begin{ruledtabular}
\begin{tabular}{cccccc}
Label & $G^{}_{x}$ & $\sigma^{}_{xy}$ & $\mathcal{T}$ & $\mathcal{T}G^{}_{x}$ & $\mathcal{T}\sigma^{}_{xy}$\\
			\hline
Z3000$_\text{00}$& \cmark & \cmark & \xmark & \xmark & \xmark\\
Z3000$_\text{01}$& \cmark & \xmark & \xmark & \xmark & \cmark\\
Z3000$_\text{10}$& \xmark & \cmark & \xmark & \cmark & \xmark\\
Z3000$_\text{11}$& \xmark & \xmark & \xmark & \cmark & \cmark\\
		\end{tabular}
	\end{ruledtabular}
	\label{table:z2_ansatze_chiral}
\end{table}

The CSL \textit{Ansatz} labeled Z3000$_{00}$ exhibits gapless excitations with Dirac points located along either the $\overline{\rev{\Gamma M}}$ or $\overline{\rev{XM}}$ directions, depending on the values of the parameters $\chi^{0}_{s}$ and $\Delta^{2}_{\square}$. In contrast, the Z3000$_{01}$ \textit{Ansatz} \rev{develops a gap once} the pairing amplitudes $\Delta^{2}_{d}$ and $\Delta^{2}_{d'}$ \rev{are} sufficiently large. The remaining two CSL \textit{Ans\"atze}, Z3000$_{10}$ and Z3000$_{11}$, typically exhibit fully gapped excitation spectra. The corresponding dispersions for these four CSL states are shown in Fig.~\ref{fig:chiral_z2_dis}.

\bibliography{references}

\end{document}